\documentclass[prd,aps,showpacs,floats,letterpaper,floatfix,groupedaddress,eqsecnum]{revtex4}

\usepackage{amssymb,amsmath}
\usepackage[dvips]{graphicx}

\usepackage{dcolumn,epsfig}

\def\nn{\nonumber}

\def\lisa{{\em LISA}}

\def\be{\begin{equation}}
\def\ee{\end{equation}}
\def\beq{\begin{eqnarray}}
\def\eeq{\end{eqnarray}}

\def\ii{{\rm i}}

\def\IL{\relax{\rm I\kern-.18em L}}

\def\nn{\nonumber}
\def\f{\frac}
\def\om{\omega_{lmn}}
\def\flm{f_{lmn}}
\def\ph{\phi_{lmn}}
\def\ta{\tau_{lmn}}
\def\Qlm{Q_{lmn}}
\def\Slm{S_{lmn}}

\newlength{\sizeonefig}
\newlength{\sizetwofig}
\begin{document}

\title{On gravitational-wave spectroscopy
of massive black holes
with the space interferometer \lisa}

\author{Emanuele Berti} \email[Email: ]{berti@wugrav.wustl.edu}

\author{Vitor Cardoso} \email[Email: ]{vcardoso@phy.olemiss.edu}
\altaffiliation{Present address: Department of Physics
and Astronomy, The University of Mississippi, University, MS
38677-1848, USA}

\author{Clifford M. Will} \email[Email: ]{cmw@wuphys.wustl.edu}

\affiliation{McDonnell Center for the Space Sciences, Department of
Physics, Washington University, St.  Louis, Missouri 63130, USA}

\date{\today}

\begin{abstract}
Newly formed black holes are expected to emit characteristic radiation
in the form of quasi-normal modes, called ringdown waves, with
discrete frequencies.  
\lisa~should be able to detect the ringdown waves emitted by
oscillating supermassive black holes throughout the observable
Universe.   We 
develop a multi-mode formalism, applicable to any interferometric detectors,
for detecting ringdown signals, for
estimating
black hole parameters from those signals, and for testing the no-hair
theorem of general relativity.  Focusing on \lisa, we use current
models of its sensitivity to compute the expected
signal-to-noise ratio for ringdown events, the relative parameter
estimation accuracy, and the resolvability of different modes.  
We also discuss the  extent to which uncertainties on
physical parameters, such as the black hole spin and the energy
emitted in each mode, will affect our ability to do black hole
spectroscopy.
\end{abstract}


\maketitle

\section{Introduction}
\label{intro}

The Laser Interferometer Space
Antenna (\lisa) is being designed to
observe gravitational waves in the low-frequency regime, between
$10^{-5}$ and $10^{-1}$ Hz.  A leading candidate source of detectable
waves is the inspiral and merger of pairs of supermassive black holes
(SMBHs). 
The signal should comprise three pieces: an inspiral waveform,
a merger waveform and a ringdown
waveform.  The inspiral waveform, originating from that part of the
decaying orbit leading up to the innermost stable orbit, has been
analyzed using post-Newtonian theory and black-hole perturbation
theory, and extensive studies of the detectability of this phase of
the signal have been carried out (see eg. \cite{BBW,BC} and references
therein).  
The nature of the merger waveform is largely unknown at present, and is the
subject of work in numerical relativity.

The ringdown waveform originates from the distorted final black hole,
and consists of a superposition of quasi-normal modes (QNMs).  Each
mode has a complex frequency, whose real part is the oscillation
frequency and whose imaginary part is the inverse of the damping
time, that is uniquely determined by the mass $M$ and angular momentum
$J$ of the black hole.  The amplitudes and phases 
of the various modes are
determined by the specific process that formed the final hole.

The uniqueness of the modes' frequencies and damping times is directly
related to the ``no hair'' theorem of general relativistic black
holes, and thus a reliable detection and accurate
identification of QNMs could provide the
``smoking gun'' for black holes and an important test of general
relativity in the strong-field regime~\cite{dreyer}.

In a pioneering analysis,
Flanagan and Hughes (\cite{fh}, henceforth
FH) showed that, independently of
uncertainties in the black hole spin and in the relative efficiency of
radiation into various modes,
the signal-to-noise ratio (SNR) for black hole ringdown could be 
comparable to that for
binary inspiral.  
Consequently, both ringdown and inspiral
radiation from SMBHs should be  
sufficiently strong relative to
the proposed sensitivity of \lisa~that they may both be detectable
with high SNR
throughout the observable universe.  With high SNR comes high
accuracy, and hence the potential to {\em measure} ringdown QNMs and
to test general relativity. 

The FH analysis provided some insight into the issue of {\it
detectability} of ringdown waves. However, to our knowledge, the
problem of {\it parameter estimation} from black hole ringdown with
\lisa~has not been discussed in depth in the literature to date.  Most
existing studies have referred specifically to high-frequency ringdown
sources and Earth-based 
interferometers~\cite{echeverria,finn,kaa,creighton,nakano,tsunesada}. 

The main purpose
of this paper is 
to discuss the {\em measurability} of ringdown QNM frequencies using \lisa~by
carefully developing a framework for analyzing QNM radiation, and then
applying the standard
``Fisher matrix'' formalism for parameter estimation \cite{finn}.
We will treat both single-mode and multi-mode cases.

From a mathematical point of view the 
excitation amplitude of QNMs is an ill-defined concept, because QNMs are not
complete~\cite{Beyer,Beyer2,NollertPrice,Nollert,Szpak}.   
Following~\cite{leaver2,A94,Nils97,GA} 
we will associate with
each QNM an ``excitation coefficient'' that quantifies in a pragmatic
(as opposed to rigorous)
way the response
of a black hole to perturbations with some given angular dependence.
We will also define useful energy coefficients to characterize the
energy deposited into various QNMs.
Unfortunately, there is only sketchy information at present from
numerical and perturbative simulations of distorted black holes,
gravitational collapse or head-on collisions of two black
holes as to what might be expected for the amplitudes, phases or energies of
QNMs.  It is clear that the QNM content of
the waveform will depend strongly on the initial conditions and on the
details of the distortion.  In the absence of such information we will
consider appropriate ranges of energy coefficients, and ranges of
relative QNM amplitudes and phases as a way to assess the measurability of
ringdown modes in some generality.  

Although we will consider a range of SMBH masses from 
$10^5$ to $10^8 \,M_\odot$, we note that, for masses smaller than 
about $10^6~M_\odot$, the damping time of the
waves may be shorter than the light-travel time along
the \lisa~arms, and as a consequence the number of observable oscillations
will be so small that a Fisher matrix approach may not be fully reliable.
We will also consider the full
range of SMBH angular momenta, from zero to near extremal.  

\begin{figure*}[t]
\begin{center}
\epsfig{file=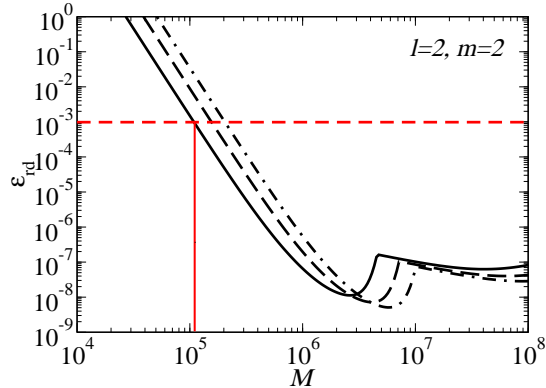,width=6cm,angle=-90}
\caption{Value of $\epsilon_{\rm rd}$ required to detect the
fundamental mode with $l=m=2$ (detection being defined by a SNR of
$10$) at $D_L=3~$Gpc. For illustrative purposes here we pick the
fundamental mode with $l=m=2$, but the dependence on $(n,l,m)$ is very
weak.  The three curves correspond to $j=0$ (solid), $j=0.8$ (dashed)
and $j=0.98$ (dot-dashed), where $j=J^2/M=a/M$ is the dimensionless angular
momentum parameter of the hole.  
The ``pessimistic'' prediction from numerical simulations of
head-on collisions is
$\epsilon_{\rm rd}=10^{-3}$ (as marked by the dashed horizontal line),
so we should be able to see all equal-mass mergers with a final black
hole mass larger than about $\sim 10^5~M_\odot$ (the vertical line is
just a guide to the eye).  The dip in the curves is a consequence of
white-dwarf confusion noise in the \lisa~noise curve.
\label{epsrd}}
\end{center}
\end{figure*}

One of our conclusions is
that the prospects for detection of ringdown radiation by
\lisa~are quite encouraging.  Figure \ref{epsrd} shows the value of
the 
fraction of ``ringdown energy'' $\epsilon_{\rm rd}$ 
(defined as the fraction of the black hole mass radiated in
ringdown gravitational waves) 
deposited in the fundamental ``bar mode'' with
$l=m=2$ (assuming that mode dominates) that is required for the mode
to be detectable by \lisa~with a SNR of 10 from a distance of 3 Gpc.  
Three values of the dimensionless
angular momentum parameter $j \equiv J/M^2 = a/M$ are
shown: zero, 0.8 (an astrophysically interesting value), and 0.98.  
Recall
that $0 \le j \le 1$, spanning the range from Schwarzschild to extremal Kerr
black holes.
For SMBH masses between
$10^6$ and $10^7 \,M_\odot$, deposition energies as small as $10^{-7}$
of the mass
should be detectable.  We show that this conclusion is not strongly
dependent on $(l,m)$, or on the overtone index $n$.

\begin{figure*}[t]
\begin{center}
\epsfig{file=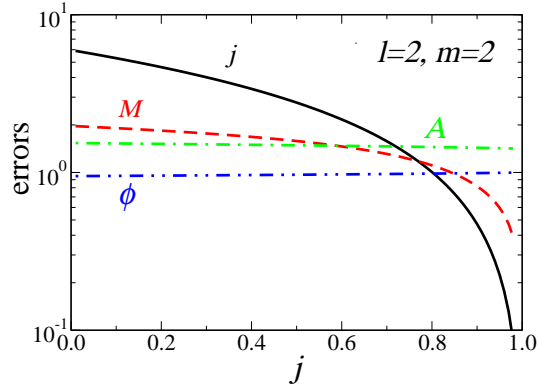,width=6cm,angle=-90} 
\caption{Errors (multiplied by the signal-to-noise ratio $\rho$)
in measurements of different parameters for the
fundamental $l=m=2$ mode as functions of the
angular momentum parameter $j$. Solid (black) lines give $\rho
\sigma_j$, dashed (red) lines $\rho \sigma_M/M$, dot-dashed (green)
lines $\rho \sigma_A/A$, dot-dot-dashed (blue) lines $\rho
\sigma_{\phi}$, where $\sigma_k$ denotes the estimated rms error for variable
$k$, $M$ denotes the mass of the
black hole, and $A$ and $\phi$ denote the amplitude and phase of the wave.
\label{errs-intro}}
\end{center}
\end{figure*}

We also find that accurate measurements of SMBH mass and angular
momentum may be possible.  For detection of the fundamental $l=m=2$
bar mode, for example, Figure \ref{errs-intro} shows the estimated
error
(multiplied by the SNR, $\rho$) in measuring
the SMBH mass $M$, angular momentum parameter $j$,
QNM amplitude $A$, and phase $\phi$ (see Sec. \ref{onemode} for
definitions).  
For example for an energy deposition of $10^{-4} M$ into the fundamental
mode of a $10^6 \,M_\odot$ SMBH with $j=0.8$ at 3 Gpc ($\rho \sim 200$), $M$
and $j$ could be measured to levels of a percent; if
the energy deposition is only $10^{-6}$, they could still be measured to 10
percent. 
Generalizing to multi-mode detection (specifically to detection of two
modes with a range of relative amplitudes), we find similar results.

\begin{figure*}[t]
\begin{center}
\epsfig{file=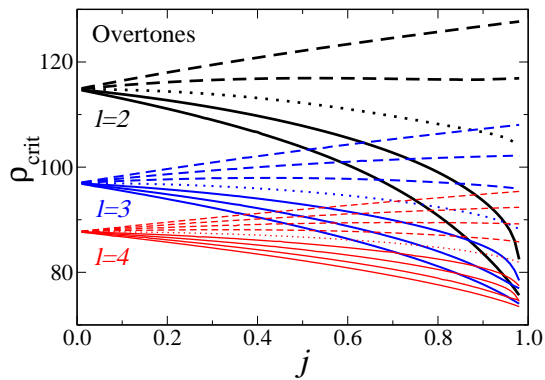,width=6cm,angle=-90}
\caption{``Critical'' SNR $\rho_{\rm crit}$ required to resolve the
fundamental mode ($n=0$) from the first overtone ($n'=1$) with the
same angular dependence ($l=l'$, $m=m'$). We assume the amplitude of
the overtone is one tenth that of the fundamental mode. 
Solid lines refer to $m=l,..,1$ (bottom to top),
the dotted line to $m=0$, and dashed lines to $m=-1,..,-l$ (bottom to
top).
\label{milanfar-fig2}}
\end{center}
\end{figure*}

\begin{figure*}[t]
\begin{center}
\begin{tabular}{ccc}
\epsfig{file=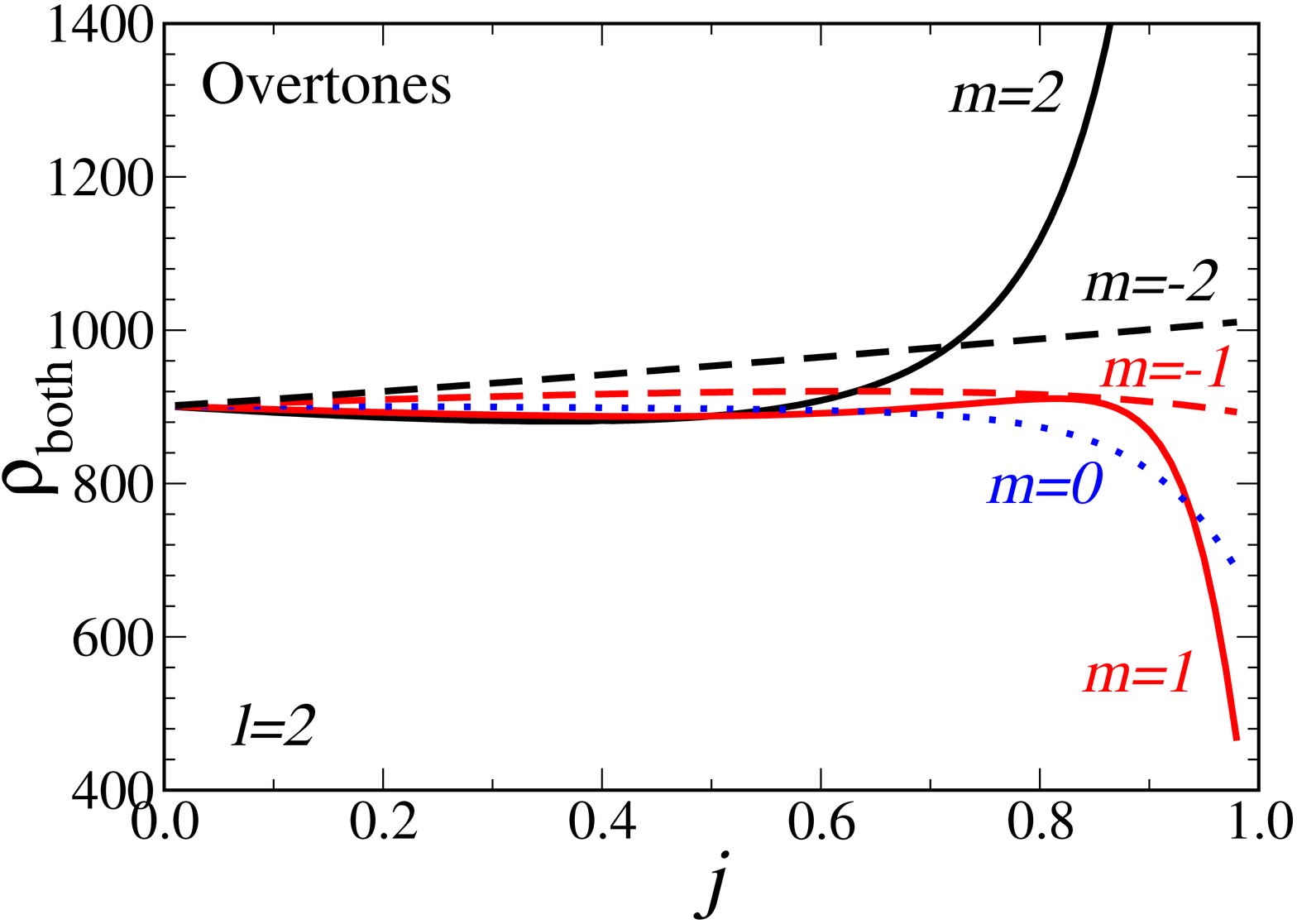,width=4.5cm,angle=-90} &
\epsfig{file=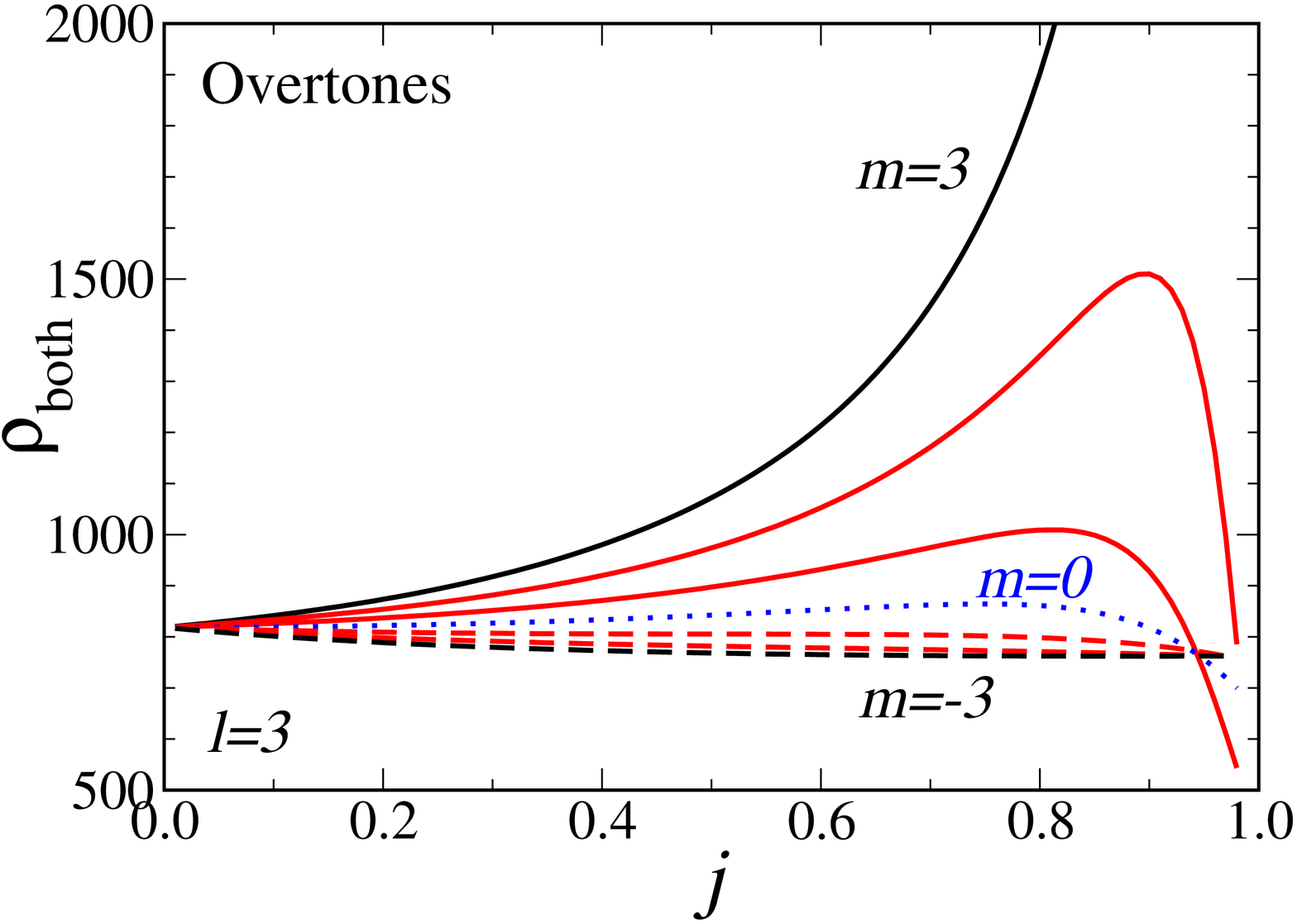,width=4.5cm,angle=-90} &
\epsfig{file=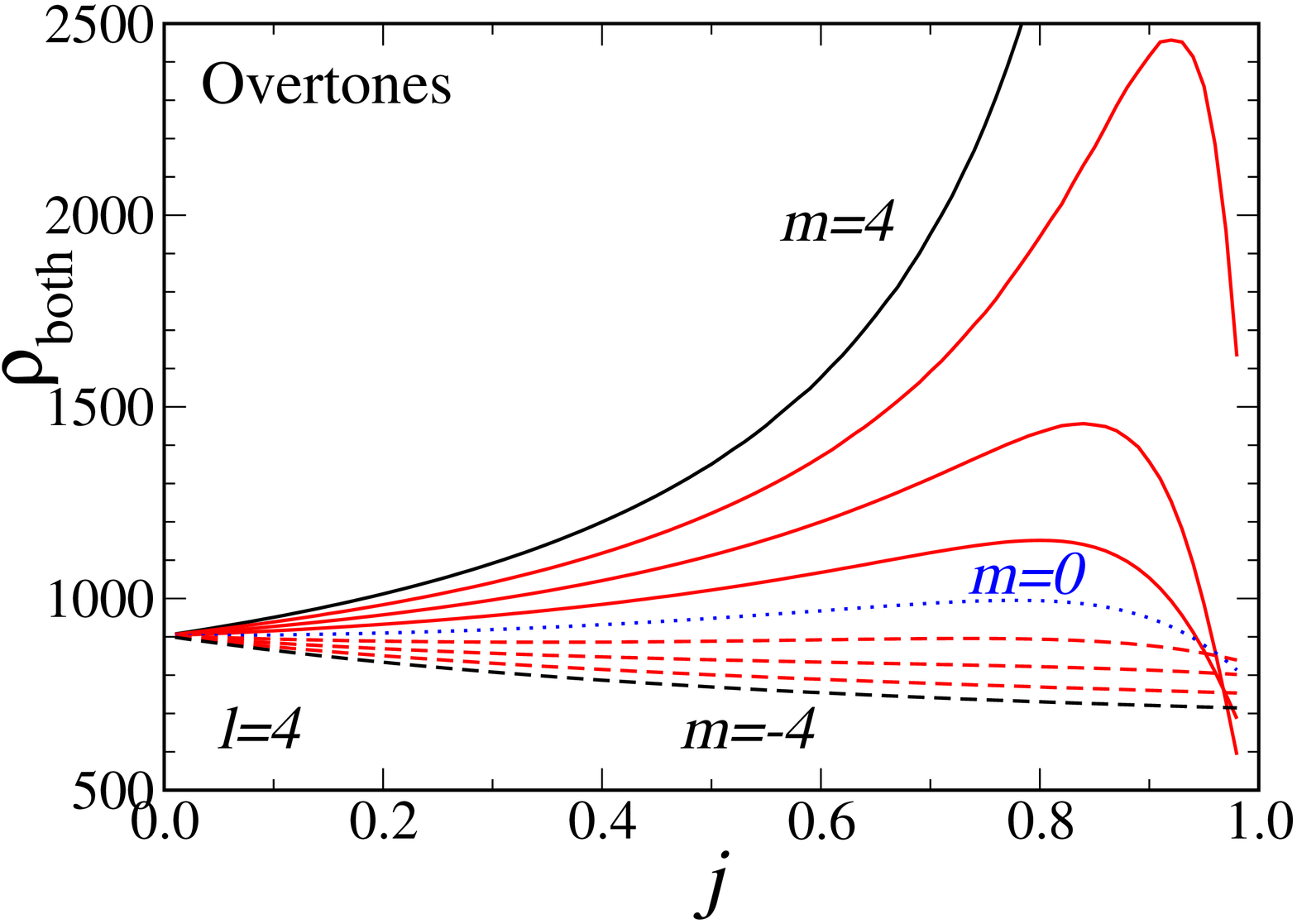,width=4.5cm,angle=-90} \\
\end{tabular}
\caption{``Critical'' SNR $\rho_{\rm both}$ required to resolve {\it
both} the frequency and the damping time of the fundamental mode
($n=0$) from the first overtone ($n'=1$) with the same angular
dependence ($l=l'$, $m=m'$). We assume the amplitude of the overtone
is one tenth that of the fundamental mode.  Solid lines refer to
$m=l,..,1$ (top to bottom), the dotted line to $m=0$, and dashed lines
to $m=-1,..,-l$ (top to bottom, unless indicated).  In the color
versions, we used black for the modes with $l=|m|$, red for those with
$0<|m|<l$ and blue for $m=0$. For $l=m$ the critical SNR grows
monotonically as $j \to 1$.
\label{milanfar-fig3}}
\end{center}
\end{figure*}

In order to test the no-hair theorem, it is necessary (though not
sufficient) to resolve two
QNMs~\cite{dreyer}; roughly speaking, 
one mode is used to measure $M$ and $j$, the other to test
consistency with the GR prediction.  Using a simple extension of the
Rayleigh criterion for resolving sinusoids, we estimate the SNR
required to resolve the frequencies and/or the damping times of
various pairs of modes, as a function of the angular momentum
parameter $j$.  For example, to resolve the fundamental $(n=0)$ mode for
a given $(l,m)$ from the first overtone $(n=1)$ for the same $(l,m)$,
the critical SNR required to resolve either frequency or damping time
is shown in Fig. \ref{milanfar-fig2}, 
while the SNR required to resolve both is
shown in Fig. \ref{milanfar-fig3}.  
These values assume that the amplitude of the
overtone is $1/10$ that of the fundamental mode.  Comparing Figs.
\ref{milanfar-fig2} and \ref{milanfar-fig3} with Fig. \ref{insp-rd} in Sec.
\ref{singlemodenum}, 
we infer that tests of the no-hair
theorem should be feasible, even under rather pessimistic assumptions
about the ringdown efficiency $\epsilon_{rd}$, as long as the first
overtone radiates a fraction $\sim 10^{-2}$ of the total ringdown
energy.  However, resolving {\em both} frequencies and damping times
typically requires a SNR greater than about $10^3$.
This is only
possible under rather optimistic assumptions about the radiative
efficiency, and it can be impossible if the dominant mode has $l=m=2$
and the black hole is rapidly spinning (solid black line in the left
panel of Fig.~\ref{milanfar-fig3}).
Requiring SNRs at least as large as $10^2$ implies that resolving QNMs
will be impossible for redshifts larger than about 10.

The remainder of this paper provides details.
In Sec.~\ref{preliminaries} we introduce our notation and formalism,
and clarify some conceptual issues related to the QNM decomposition of
gravitational waveforms.
In Sec.~\ref{onemode} we compute the single-mode SNR assuming that
only one mode dominates the ringdown.  
As a first step we update the
FH analysis of \lisa's SNR for detection of single-mode
waveforms.  With respect to FH we use a better semianalytic
approximation of the \lisa~noise curve, we consider different QNM
frequencies, and we compare with the expected SNR for inspiral as
computed in \cite{BBW}. We explore the angular momentum dependence
by considering black holes with $j\equiv a/M=0,~0.8,~0.98$, and confirm
the FH expectation that angular momentum does not have a big effect on
the SNR. Uncertainties in the ringdown efficiency $\epsilon_{\rm rd}$
have a larger impact, since $\rho\sim
\sqrt{\epsilon_{\rm rd}}$. 
In Sec.~\ref{onemode-pe} we assess the accuracy of parameter
estimation in single-mode situations, revisiting the analyses in
\cite{echeverria,finn,kaa}, and show that it is in general very good.
A more detailed analysis shows that, for
counterrotating ($m<0$) and axisymmetric ($m=0$) modes, 
rotation doesn't necessarily help parameter
estimation, and for some counterrotating modes the error can even blow
up at ``critical'' values of the black hole angular momentum.
Following preliminary remarks in Sec.~\ref{excitation}
describing mathematical issues in and model
predictions for multi-mode ringdowns, 
in Sec.~\ref{multimode} we generalize to a two-mode
analysis, computing the SNR and errors on parameter estimation. We find
that it is computationally convenient to treat separately the
case of waveforms in which the modes have different angular dependence
($l\neq l'$ or $m\neq m'$) and the case where the overtones have the
same angular dependence as the fundamental mode.  
In Sec.~\ref{nohair}
we determine the
minimum signal-to-noise ratio required to discriminate between
different quasi-normal mode pairs. 
Conclusions and perspectives for future work are presented in
Sec.~\ref{conclusions}.

In the Appendices we collect various numerical results and technical
calculations. Appendix \ref{deltafun} discusses the accuracy of a
useful semianalytic approach to the calculation of the SNR (the
$\delta$-function approximation, valid for ringdown signals with
large quality factor). In Appendix \ref{app:FisherOne} we present an
explicit calculation of the Fisher matrix using different conventions
and different numbers of free parameters in the ringdown waveform.
Appendix \ref{app:noise} describes our semianalytical model of the
\lisa~noise.  Appendix \ref{recover-rayleigh} discusses a particular
aspect of our criterion for resolving normal modes.  Finally, Appendix
\ref{app:QNM} lists for reference the values of the complex
frequencies and the angular separation constants of the spin-weighted
spheroidal harmonics for selected normal modes, and also displays
analytical fits of the QNM frequencies accurate within a few percent.

\section{Black hole ringing: preliminaries}
\label{preliminaries}

\subsection{Optimal mass range for ringdown detection by \lisa}

During the ringdown phase, perturbations of a Kerr black
hole die away as exponentially damped sinusoids, whose frequencies and
damping times are given by (complex) QNM frequencies
\cite{kokkotas}. We decompose the perturbations in spheroidal
harmonics $S_{lm}(\iota,\beta)$ of ``spin weight'' 2 
\cite{teukolsky,berticardoso}, where $l$
and $m$ are indices analogous to those for standard spherical
harmonics, and 
$\iota$ and $\beta$ are angular variables 
such that the azimuthal, or $\beta$ dependence goes like
$e^{{\ii\beta}}$.  For each ($l,m$) there is an infinity of resonant
quasi-normal frequencies
\cite{kokkotas}, which control the intermediate time behavior of the
signal.  We label each of these frequencies by an overtone index $n$
such that 
the mode
with $n=0$ has the longest damping time, followed by $n=1$ and so
on.  Thus, in the end QNM frequencies are parameterized by three
numbers: $l\,,m$ and $n$. Now, the time dependence of the signal
during ringdown is of the form $e^{\ii\omega t}$, but since
$\omega=\omega_{lmn}+ \ii/\tau_{lmn}$ is in general a complex number,
we will follow the usual conventions and write this as
$e^{-t/\tau_{lmn}} \sin\left(\omega_{lmn} t+ \varphi_{lmn} \right) $,
or
$e^{-t/\tau_{lmn}} \cos\left(\omega_{lmn} t+ \varphi_{lmn} \right) $,
where $\omega_{lmn}=2\pi f_{lmn}$ is the mode's real part and
$\tau_{lmn}$ is the damping time of the oscillation. We will also
define the quality factor of a QNM as
\be
Q_{lmn}\equiv \pi f_{lmn} \tau_{lmn}=\omega_{lmn} \tau_{lmn}/2\,.
\ee
In our analysis of the detectability of ringdown radiation, we will assume
that the signal lasts for at least 
one light-propagation time
corresponding to the \lisa~arm length $L\simeq 5\cdot 10^9$~m,  or $T_{\rm
light}=L/c\simeq 16.68~{\rm s}$ (shorter-lived signals may require
specialized detection techniques).  This places a rough lower limit on
the black hole masses that are relevant.  To see this,
we note that  
the fundamental mode of a Schwarzschild
black hole corresponds to an axially symmetric ($m=0$), quadrupolar
($l=2$) perturbation
with frequency
\be
f_{200}= \pm 1.207\cdot10^{-2} (10^6 M_\odot/M)~{\rm Hz}\,,
\label{freq0}
\ee
and damping time
\be
\tau_{200}=55.37 (M/10^6 M_\odot)~{\rm s}\,.
\ee

\begin{figure*}[t]
\begin{center}
\begin{tabular}{cc}
\epsfig{file=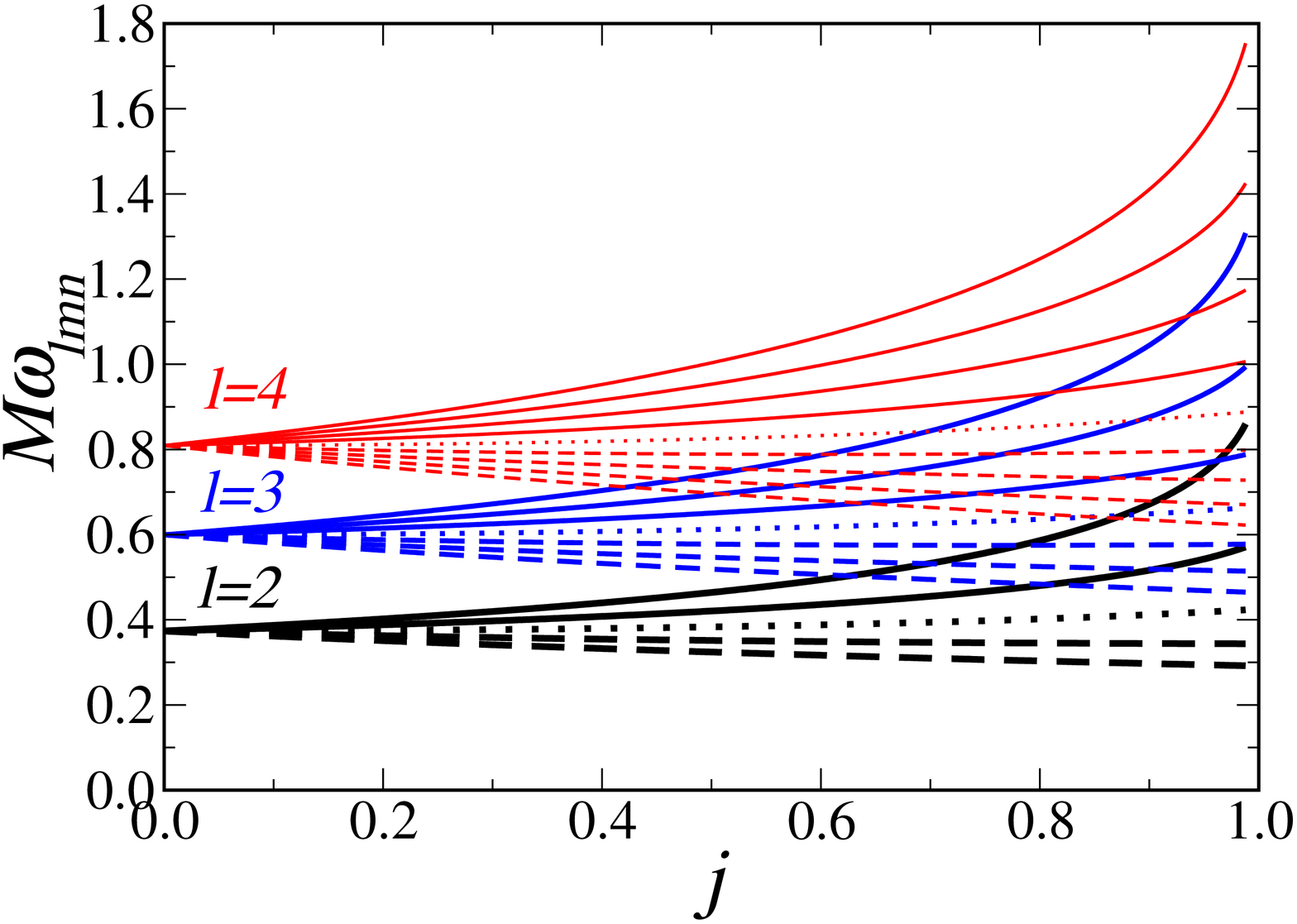,width=6cm,angle=-90} &
\epsfig{file=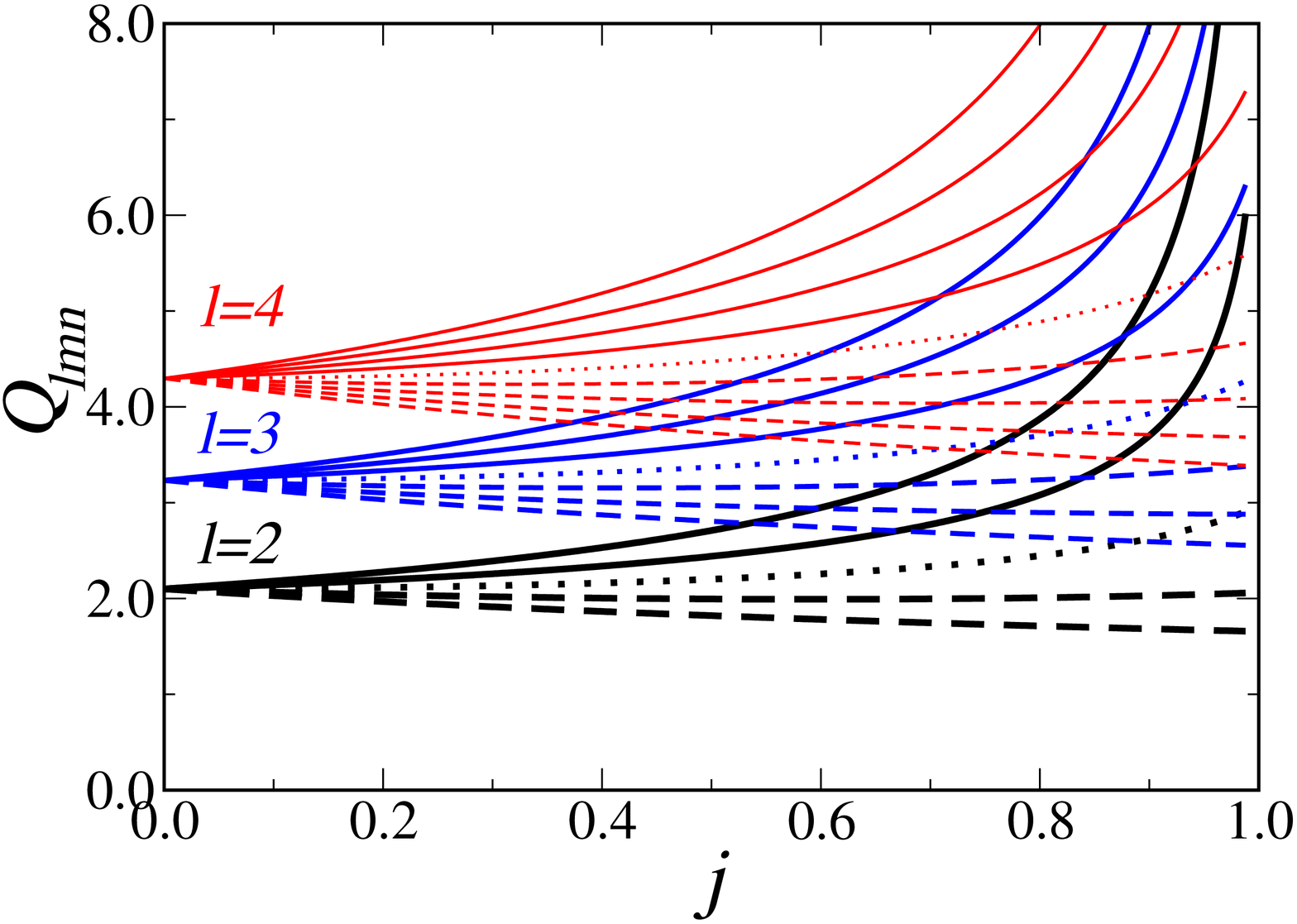,width=6cm,angle=-90} \\
\end{tabular}
\caption{Frequency $\flm$ (left) and quality factor $\Qlm$ (right) for
the fundamental modes with $l=2,~3,~4$ and different values of
$m$. Solid lines refer to $m=l,..,1$ (from top to bottom), the dotted
line to $m=0$, and dashed lines refer to $m=-1,..,-l$ (from top to
bottom). Quality factors for the higher overtones are lower than the
ones we display here. 
\label{Qmodes}}
\end{center}
\end{figure*}

For rotating holes, the dimensionless frequencies ($M\omega_{lmn}$)
and quality factors for the
fundamental modes for 
$l=2, \, 3,\,4$ are shown as a function of $j$ in Fig. \ref{Qmodes}.
Although the quality factors and damping times for corotating ($m>0$)
modes may be larger for rapidly rotating holes, the effects are not
dramatic:   
for $j=0.80$ the damping time 
$\tau_{220}=65.18\left(M/10^6 M_\odot\right)$~s, and for $j=0.98$ the
damping time $\tau_{220}=127.7\left(M/10^6 M_\odot \right)$~s.
Accordingly we will restrict our attention to masses larger than $10^6$
or a few times $10^5 M_\odot$.  

We can also estimate an upper limit for masses to be considered by
noting that {\lisa}'s low frequency noise may provide a lower cutoff
at $10^{-4}$ Hz.  Equation (\ref{freq0}) then gives a mass upper limit
of around $10^8 M_\odot$; if the \lisa~performance can be extended
down to $10^{-5}$ Hz, then masses as large as $10^9 M_\odot$
may be detectable.  Again, these rough bounds are not terribly
dependent on the black hole spin or the mode.

\subsection{Quasinormal mode decomposition and polarization of the waveform}
\label{polariz}

The plus and cross components of the gravitational waveform emitted
by a perturbed Kerr black hole can be written in terms of the radial
Teukolsky function $R_{lm\omega}$ as \cite{teukolsky}
\be \label{Fourier} h_++\ii h_{\times}
=-\f{2}{r^4}\int_{-\infty}^{+\infty}
\frac{d\omega}{\omega ^2} e^{\ii\omega t}
\sum _{lm}S_{lm} \left( \iota,\beta \right) R_{lm\omega}(r)\,.
\ee
The radial Teukolsky function $R_{lm\omega} \sim r^3 Z_{lm\omega}^{\rm
out}e^{-\ii\omega r}$ as $r\to \infty$, where $Z_{lm\omega}^{\rm out}$
is a complex amplitude \cite{NRezzolla}.

We assume that the gravitational wave signal during the ringdown phase
can be expressed as a linear superposition of exponentially decaying
sinusoids. QNMs are known not to be a complete set, and thus
such an expansion is not well defined
mathematically. However numerical simulations of a variety of
(perturbative and non-perturbative) dynamical processes involving
black holes show that, at intermediate times, the response of a black
hole {\it is} indeed well described by a linear superposition of
damped exponentials. We defer further discussion of the
meaning of the QNM expansion to Sec.~\ref{excitation}. For the time
being, we just {\it assume} that we can write the gravitational
waveform as a formal QNM expansion (rather than as a standard Fourier
expansion in the real frequency $\omega$) of the Teukolsky function,
so that we can replace Eq.~(\ref{Fourier}) by
\be h_++\ii h_{\times}=\frac{1}{r}
\sum _{lmn} e^{\ii\omega_{lmn} t}e^{-t/\ta} \Slm  \left( \iota,\beta
\right)
Z_{lmn}^{\rm out} \,,
\ee
where $n$ denotes the overtone index and from here on the coordinate
$t$ stands for the retarded time $t-r$. We write the complex wave
amplitude $Z_{lmn}^{\rm out}$ in terms of a real amplitude ${\cal A}_{lmn}$ and
a real phase $\phi_{lmn}$, and to follow the FH convention we factor
out the black hole mass $M$: $Z_{lmn}^{\rm out}=M{\cal
A}_{lmn}e^{\ii\phi_{lmn}}$. In this way we get
\be \label{QNMexp}
h_++\ii h_{\times}=\frac{M}{r}\sum _{lmn}
{\cal A}_{lmn}
e^{\ii(\omega_{lmn} t+\phi_{lmn})}e^{-t/\ta} \Slm\,.
\ee
In this expansion the spheroidal functions $\Slm=S_{lm}(a\om)$ are
evaluated at the (complex) QNM frequencies, so they are complex
numbers (henceforth we drop the $(\iota,\beta)$ 
angular dependence on the $\Slm$).

\begin{figure*}[t]
\begin{center}
\epsfig{file=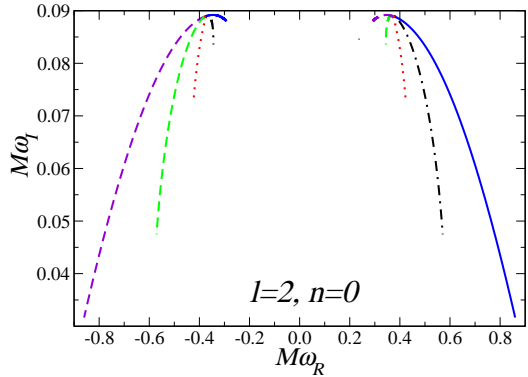,width=6cm,angle=270}
\caption{Fundamental $l=2$ QNM frequencies of the Kerr black hole in
the range $j\in[0,0.99]$.
Solid (blue) lines correspond to $m=2$,
dot-dashed (black) lines to $m=1$,
dotted (red) lines to $m=0$,
dashed (green) lines to $m=-1$,
and short dashed (violet) lines to $m=-2$.
\label{KerrRefl}}
\end{center}
\end{figure*}

One frequently finds in the literature the astrophysically reasonable
assumption that only the
$l=m=2$ mode is present in the waveform.
This viewpoint has two conceptual
flaws.  

First, QNMs of Schwarzschild and Kerr black holes always come ``in pairs''.
In the Kerr case, for a
given $(l,m)$ and a given value of $a=jM$ the eigenvalue problem
admits {\it two} solutions: one with positive (real part of) the
frequency, the other with negative real part of the frequency and
different damping time. To illustrate this property, in
Fig.~\ref{KerrRefl} we show the fundamental Kerr QNM with $l=2$ and
different values of $m$. Positive-$m$ frequencies are related to
negative-$m$ frequencies by a simple symmetry property: one can easily
see from the perturbation equations that to any QNM frequency
characterized by $(l,m,n)$ there corresponds a QNM frequency
characterized by $(l,-m,n)$ such that
\be\label{minusm}
-\om=\omega_{l-mn}\,,
\qquad
1/\ta=1/\tau_{l-mn}\,,
\qquad
A_{lmn}^*=A_{l-mn}\,,
\ee
(here $A_{lmn}$ is the angular separation constant, not to be confused
with the mode amplitude ${\cal A}_{lmn}$). In this sense, any solution
with positive $m$ is nothing but the ``mirror image'' of a solution
with negative real part and negative $m$.  For $m=0$ (and, of course,
in the Schwarzschild case) the two ``mirror solutions'' are degenerate
in modulus of the frequency and damping time. However, in general,
{\it a ``mode with a given $(l,m)$'' will always contain a
superposition of two different damped exponentials}. One of these
exponentials could be invisible in the actual waveform because its
damping time is shorter, or perhaps because it is less excited in the
given physical situation, but formally one can never have anything
like an isolated ``$l=m=2$ frequency'' with a positive real part.
This excitation of both modes is
actually observed in time-evolutions of perturbative fields in Kerr
backgrounds~\cite{krivan,dorband}.  

Second, a single-mode
expansion automatically restricts attention to circularly polarized
gravitational waves; more generally
one cannot specify the polarization state of the waveform by assuming that
it is described by a single QNM frequency.  This problem has been
overlooked in all previous treatments of the gravitational radiation
emitted during ringdown. This omission has no serious consequences for
nonrotating holes, but it is conceptually inconsistent when $j\neq
0$. Consider, for example, the starting point of the analysis in
FH. They assume that the waveform can be written as [Eq.~(3.15) in FH]
\be\label{FHwaveform}
h_++\ii h_\times=
\frac{M{\cal A}_{lmn}}{r}
e^{\ii(\omega_{lmn} t+\phi_{lmn})}e^{-t/\ta} \Slm\,,
\ee
with $l=m=2$. This is not a general assumption: it implies that the waves 
are circularly polarized.

On the contrary, the polarization of the ringdown
waveform depends on the physical process generating the distortion of
the black hole.  In fact, most studies dealing with point particles in
the vicinities of black holes show that the $h_{\times}$ component is
extremely hard to excite \cite{nakamura}. The only exception is
provided by particles in circular motion, which do not resonantly
excite QNMs anyway. To excite QNMs one needs an object passing through
the potential barrier peak, for instance during the merger phase. In
this case the motion is almost radial, so one can presumably
lean on the point
particle results and assume $h_{\times} \sim 0$ (see however
\cite{lazarus,AG} for discussion of possible circular polarizations).

Accordingly, a consistent approach to ringdown waveforms begins with a
general superposition of modes, including the
``twin'' modes with frequency $\omega'_{lmn}=-\omega_{l-mn}$ 
and a different damping
$\ta'=\tau_{l-mn}$.  Then, using the symmetry property
(\ref{minusm}) we can easily see that:
\beq
\nn
h_++\ii h_\times
&=&
\frac{M}{r}\sum_{lmn}
\biggl\{{\cal A}_{lmn} e^{\ii(\omega_{lmn} t+\phi_{lmn})}e^{-t/\ta} \Slm+
{\cal A}'_{lmn} e^{\ii(\omega'_{lmn} t+\phi'_{lmn})}e^{-t/\ta'} 
\Slm' \biggr\}
\\
&=&
\frac{M}{r}\sum_{lmn}
\biggl\{
{\cal A}_{lmn}e^{\ii(\omega_{lmn} t+\phi_{lmn})}e^{-t/\ta} \Slm+
{\cal A}'_{lmn}e^{\ii(-\omega_{l-mn} t+\phi'_{lmn})}e^{-t/\tau_{l-mn}}
S_{l-mn}^*
\biggr\}
\nn\\
&=&
\frac{M}{r}\sum_{lmn}
\biggl\{
{\cal A}_{lmn}e^{\ii(\omega_{lmn} t+\phi_{lmn})}e^{-t/\ta} \Slm+
{\cal A}'_{l-mn}e^{\ii(-\omega_{lmn} t+\phi'_{l-mn})}e^{-t/\tau_{lmn}}
S_{lmn}^*
\biggr\}
\nn\\
&=& \frac{M}{r}\sum_{lmn} \biggl\{ {\cal A}_{lmn}e^{\ii(\omega_{lmn} t+\phi_{lmn})}e^{-t/\ta} \Slm+ {\cal
A}'_{lmn}e^{\ii(-\omega_{lmn} t+\phi'_{lmn})}e^{-t/\tau_{lmn}}
S_{lmn}^* \biggr\} \,.
\label{genwave}
\eeq
In going from the second to the third line we relabeled $m\to -m$
in the second term, and in going from the third to the
fourth line we changed the labeling of the (arbitrary) constants,
replacing ${\cal A}'_{l-mn}$ by ${\cal A}'_{lmn}$ and $\phi'_{l-mn}$
by $\phi'_{lmn}$. So the general waveform depends on four arbitrary,
real constants: ${\cal A}_{lmn}$, ${\cal A}'_{lmn}$, $\phi_{lmn}$ and
$\phi'_{lmn}$ for each ($l,~m,~n$).

Thus it is clear that only by combining positive and negative values
of $m$ can we require the waveform to have any given polarization
state. In particular, if
${\cal A}_{lmn} e^{\ii\phi_{lmn}} \Slm=
{\cal A}'_{lmn} e^{\ii\phi'_{lmn}} \Slm^*$
the waveform becomes pure real (we have a ``plus'' state); if instead
${\cal A}_{lmn} e^{\ii\phi_{lmn}} \Slm=
-{\cal A}'_{lmn} e^{\ii\phi'_{lmn}} \Slm^*$
it becomes pure imaginary (we have a ``cross'' state).
In our single-mode analysis we will usually write the (real) plus and
cross components measured at the detector as damped sinusoids,
specifying arbitrarily their amplitude and relative phase. More
rigorously, when we write the waveform as a damped sinusoid we really
mean that we have performed a sum of the appropriate 
QNM components, as described
above.  

\subsection{Including cosmological redshift}

The general waveform (\ref{genwave}) is written in the rest frame of
the black hole, and thus all the quantities appearing there
($M\,,\,\omega_{lmn}$ and $\tau_{lmn}$) are measured in that
frame. However, because of cosmological effects, in the detector's
frame all dimensionful quantities should be interpreted as
redshifted. The prescription to include cosmological effects is very
simple \cite{fh,markovic}: $r$ should be replaced by the luminosity
distance $D_L(z)$, and all quantities with dimensions $[{\rm mass}]^p$
should enter the waveforms at the detector multiplied by the factor
$(1+z)^p$. So, whenever the source is at cosmological distance, our
$r$ should be replaced by $D_L(z)$, $M$ by the redshifted mass
$(1+z)M^0$, $\flm$ by the redshifted frequency $\flm ^0/(1+z)$, and
$\tau_{lmn}$ by $(1+z)\ta ^0$ (where all quantities marked by a
superscript $0$ are measured in the source frame).
In our numerical work, we use the values of cosmological parameters reported
in \cite{cosmology}.

\section{Signal-to-noise ratio for a single-mode waveform}
\label{onemode}

\subsection{Analytic Results}
\label{snranalytic}

We begin by studying the SNR for detection of a single QNM.
From Eq. (\ref{genwave}), we can express the (real) waveform
measured at the detector as a linear superposition of $h_{+}$ and
$h_\times$, where, for the given mode $(l,~m,~n)$,
\begin{subequations}
\beq
h_{+}&=&
\f{M}{r}
\Re\left[
{\cal A}^{+}_{lmn}
e^{\ii(\om t+\ph^{+})}e^{-t/\ta} \Slm (\iota,\beta)
\right] \,,\\
h_{\times}&=&
\f{M}{r}
\Im\left[
{\cal A}^{\times}_{lmn}
e^{\ii(\om t+\ph^{\times})} e^{-t/\ta} \Slm (\iota,\beta)
\right]\,,
\eeq
\label{waveform0}
\end{subequations}
where ${\cal A}^{+,\times}_{lmn}$ and $\ph^{+,\times}$ are real, and
are related to the quantities 
${\cal A}_{lmn}$,
${\cal A}^{\prime}_{lmn}$,
$\ph$, and 
$\ph^{\prime}$ of Eq. (\ref{genwave}) by  
${\cal A}^{+,\times}_{lmn} e^{\ii\ph^{+,\times}} = 
{\cal A}_{lmn} e^{\ii\ph} \pm {\cal A}^{\prime}_{lmn}
e^{-\ii\ph^{\prime}}$, where the $+(-)$ signs correspond to the $+(\times)$
polarizations, respectively.
The waveform measured at a detector is given by 
\beq
h = h_{+} F_+(\theta_S,\phi_S,\psi_S) + h_{\times}
F_\times(\theta_S,\phi_S,\psi_S) \,,
\label{detectwave}
\eeq
where $F_{+,\times}$ are pattern functions that depend on the orientation
of the detector and the direction of the source, given by
\begin{subequations}
\beq
F_+(\theta_S,\phi_S,\psi_S) &=& \frac{1}{2}(1 + \cos^2 \theta_S) \cos 2\phi_S \cos
2\psi_S
-\cos \theta_S \sin 2\phi_S \sin 2\psi_S\;, \\
F_{\times}(\theta_S,\phi_S, \psi_S) &=& \frac{1}{2}(1 + \cos^2 \theta_S) \cos
2\phi_S \sin 2\psi_S
+ \cos \theta_S \sin 2\phi_S \cos 2\psi_S\;.
\label{pattern}
\eeq
\end{subequations}

To compute the SNR we will usually follow the prescription described
in Appendix A of FH (henceforth the {\it FH convention} or {\it FH
doubling prescription}) as follows:
(1) Assume that the waveform for $t<0$ is identical to the waveform for
$t>0$ except for the sign of $t/\ta$, i.e. that we replace the decay
factor $e^{-t/\ta}$ with $e^{-|t|/\ta}$.
(2) Compute the SNR using the standard expression,
\be\label{SNRdef}
\rho^2 = 4\int_0^\infty \f{\tilde h^*(f) \tilde h(f)}{S_h(f)}df\,,
\ee
where $\tilde h(f)$ is the Fourier transform of the waveform, and 
$S_h(f)$ is the noise spectral density of the detector.
(3) Divide by a correction factor of $\sqrt{2}$ in amplitude to
compensate for the doubling-up in step (1).

In calculating the SNR, we will 
average over source directions and over detector and black-hole 
orientations, making use of the angle averages: 
$\langle F_+^2\rangle=\langle F_\times^2\rangle=1/5$, 
$\langle F_+F_\times\rangle=0$, and 
$\langle |\Slm|^2 \rangle=1/4\pi$.   This simple averaging is feasible
because the mode damping time is short compared to the orbital period
of~\lisa.

Sometimes, for comparison, we will not follow the three steps we just
described, but will calculate the Fourier transform of the waveform by
integrating only over the range $t>0$. Since
this was the method used by Echeverria \cite{echeverria} and Finn
\cite{finn}, we will refer to this procedure as {\it the Echeverria-Finn
(EF) convention}.   

In the rest of this Section we will follow the FH prescription. The
Fourier transform of the waveform can be computed 
using the elementary relation
\be\label{bom}
\int_{-\infty}^{\infty} e^{\ii \omega t}
\left(e^{\pm \ii \om t-|t|/\ta}\right) dt=
\f{2/\ta}{(1/\ta)^2+(\omega\pm \om)^2}
\equiv 2b_\pm\,.
\ee
Then the Fourier transforms of the plus and cross components become:
\begin{subequations}
\beq
\tilde h_{+}&=&
\f{M}{r}
{\cal A}^{+}_{lmn}
\left[e^{\ii\ph^{+}} \Slm b_++e^{-\ii\ph^{+}} \Slm^* b_-\right] \,,\\
\tilde h_{\times}&=&
-\ii
\f{M}{r}
{\cal A}^{\times}_{lmn}
\left[e^{\ii\ph^{\times}} \Slm b_+-e^{-\ii\ph^{\times}} \Slm^* b_-\right] \,.
\eeq
\label{hf}
\end{subequations}
We can directly plug these Fourier transforms into the definition
(\ref{SNRdef}) of the SNR to get
\beq
\rho^2(\theta_S,\phi_S,\psi_S,\iota,\beta)&=&
2\left(\f{M}{r}\right)^2
\int_0^\infty \f{df}{S_h(f)}\times \nn\\
&&\times
\biggl\{
\left(b_+^2+b_-^2\right)
\left[
{\cal A}^{+\,2}_{lmn}F_+^2
+{\cal A}^{\times\,2}_{lmn} F_\times^2
-2{\cal A}^{+}_{lmn}{\cal A}^{\times}_{lmn} F_+ F_\times
\sin(\ph^+-\ph^\times)
\right]
|\Slm|^2\nn\\
&& +2b_+ b_-
\biggl[
\Re \left[ \left(
{\cal A}^{+\,2}_{lmn}F_+^2 e^{2\ii\ph^{+}}-
{\cal A}^{\times\,2}_{lmn} F_\times^2
e^{2\ii\ph^{\times}} \right)\Slm^2 \right]
\nn\\
&&
+2{\cal A}^{+}_{lmn}{\cal A}^{\times}_{lmn} F_+ F_\times
\Im \left(e^{\ii(\ph^++\ph^{\times})} \Slm^2 \right)
\biggr]
\biggr\}\,.
\label{snrangles}
\eeq

The terms proportional to $\Slm^2$ cannot be angle-averaged analytically
in the usual way, so
to deal with this general expression one must
perform a Monte Carlo simulation.
Given randomly generated values of the angles we can compute
numerically the spin-weighted spheroidal harmonics at the QNM
frequencies, plug
the harmonics into the integrals, and finally average the resulting
SNRs.  We leave this for future work. 

However,
especially for slowly-damped modes with $m \geq 0$, the imaginary part
of $S_{lmn}$ is typically smaller than the real part:
\be\label{Slmre}
\Slm\simeq \Re (\Slm)\,,\qquad\,
\Re (\Slm) \gg \Im (\Slm)\,.
\ee
We give a quantitative discussion of the validity of this
approximation elsewhere \cite{berticardoso}.
We will henceforth assume that the $\Slm$ are real, so that
$\Slm^2 = |\Slm^2|$, and we can complete the angle averaging
analytically to obtain
\beq
\rho^2 &=&
\f{1}{10\pi} \left(\f{M}{r}\right)^2
\int_0^\infty \f{df}{S_h(f)}\times \nn\\
&&\times
\left\{
\left(b_+^2+b_-^2\right)
\left[
{\cal A}^{+\,2}_{lmn} + {\cal A}^{\times \,2}_{lmn} \right]
+2b_+ b_-
\left[
{\cal A}^{+\,2}_{lmn} \cos(2\ph^{+})-
{\cal A}^{\times\,2}_{lmn}  \cos(2\ph^{\times})
\right] \right\} \,.
\label{rhosqanalytic}
\eeq
We expect that the resulting SNR
should be reasonably close to the true angle-averaged result 
as long as the imaginary part of
the harmonics is not too large.

We make the further approximation that the damping time is
sufficiently long that the frequency-dependent functions $b_+^2+b_-^2$
and $b_+ b_-$ may be replaced by suitable $\delta$-functions, namely
in the large $Q_{lmn}$ or large $\tau_{lmn}$ limit,
\beq
b_+^2 + b_-^2 &\to& \f{\tau_{lmn}}{4} \left[ \delta(f-f_{lmn}) +
\delta(f+f_{lmn}) \right ] \,,\nn \\
b_+ b_- &\to& \f{\tau_{lmn}}{8} \f{1}{1+4Q_{lmn}^2} \left[
\delta(f-f_{lmn}) +
\delta(f+f_{lmn}) \right ] \,,
\eeq
where the normalizations are obtained by integrating over positive
frequencies only.
This approximation is mathematically, though not physically 
equivalent to assuming that the noise density
$S_h(f)$ is strictly constant.  
We then obtain the angle-averaged SNR,
\beq\label{rhoFH}
\rho^2  =
\f{\Qlm}{40\pi^2\flm (1+4\Qlm^2)S_h(f_{lmn})}
&\times&
\left\{
\left(\f{M{\cal A}^{+}_{lmn}}{r}\right)^2
\left[1+\cos(2\ph^+)+4\Qlm^2\right]
\right. \nn \\
&+&\left.
\left(\f{M{\cal A}^{\times}_{lmn}}{r}\right)^2
\left[1-\cos(2\ph^\times)+4\Qlm^2\right]
\right\} \,.
\eeq

For simplicity, FH also make an assumption about the relative amplitudes and
phases of the waves, taking
a pure cosine for the $+$-polarization ($\ph^{+} = 0$), 
a pure sine for the $\times$-polarization ($\ph^{\times} = 0$), 
and assuming 
${\cal A}^{+}_{lmn}={\cal A}^{\times}_{lmn}={\cal A}_{lmn}$. 
With these assumptions, the SNR takes the form, 
\begin{subequations}
\beq
\rho_{\rm FH}^2 &=&
\left(\f{M}{r}\right)^2
\f{{\cal A}^{2}_{lmn} }
{80 \pi^5 \tau_{lmn}^2 }
\int_0^\infty \f{df}{S_h(f)} 
\biggl\{
\f{1}{\left[(f+\flm)^2+(2\pi \ta)^{-2}\right]^2}+
\f{1}{\left[(f-\flm)^2+(2\pi \ta)^{-2}\right]^2}
\biggr\} \label{breitwigner}\\
&\simeq&
\left(\f{M}{r}\right)^2
\f{\Qlm {\cal A}^{2}_{lmn}} 
{20 \pi^2 \flm S_h(f_{lmn})}\,,
\label{rhodeltafn}
\eeq
\label{rhoFHmain}
\end{subequations}
where the second expression corresponds to the $\delta$-function limit.

It is now useful, 
following FH, to define an energy spectrum through the relation
\be
\label{rhodedf}
\rho^2 =
\f{2}{5\pi^2 r^2}
\int_0^\infty \f{1}{f^2 S_h(f)} \f{dE}{df} df\,.
\ee
From Eq. (\ref{rhosqanalytic}), we obtain
\beq\label{dEdfours}
\f{dE}{df}&=&
\f{\pi M^2 f^2}{4}
\left\{
\left(b_+^2+b_-^2\right)
\left[
{\cal A}^{+\,2}_{lmn} + {\cal A}^{\times \,2}_{lmn} \right]
+2b_+ b_-
\left[
{\cal A}^{+\,2}_{lmn} \cos(2\ph^{+})-
{\cal A}^{\times\,2}_{lmn}  \cos(2\ph^{\times})
\right] \right\} \,.
\eeq
We then define the ``radiation efficiency'' $\epsilon_{\rm rd}$, by 
\be
\epsilon_{\rm rd} \equiv \f{E_{\rm GW}}{M}
= \f{1}{M}\int_0^\infty \f{dE}{df}df\,.
\label{radeff}
\ee
Substituting Eq. (\ref{dEdfours}) into (\ref{radeff}) and integrating, and
comparing the result with Eq. (\ref{rhoFH}), we find a relation between SNR
and radiation efficiency for a given mode, in the $\delta$-function or 
constant-noise limit, 
\beq
\label{rhoanalytic}
\rho_{\rm FH} = \left (\frac{2}{5} \right )^{1/2} 
	\left (\frac{1}{\pi f_{lmn} r} \right )
	\left (\frac{\epsilon_{\rm rd} M}{S_h(f_{lmn})} \right )^{1/2}
	\frac{2Q_{lmn}}{\sqrt{1+4Q_{lmn}^2}} \,.
\eeq
independently of any condition on the relative amplitudes or phases.

With the FH choice of phases and amplitudes, 
the resulting energy spectrum
is their formula (3.18):
\beq\label{FHspec}
\left(\f{dE}{df}\right)_{\rm FH}&=&
\f{{\cal A}_{lmn}^2 M^2 f^2}{32 \pi^3 \ta^2}
\biggl\{
\f{1}{\left[(f+\flm)^2+(2\pi \ta)^{-2}\right]^2}+
\f{1}{\left[(f-\flm)^2+(2\pi \ta)^{-2}\right]^2}
\biggr\}\nn\\
&\simeq& 
\f{{\cal A}_{lmn}^2 \Qlm M^2 \flm}{8}
\delta(f-\flm)\label{deltaapp}\,. \eeq
where the second expression corresponds to the $\delta$-function limit.
Integrating the FH energy spectrum (\ref{FHspec}) explicitly, we find that 
the amplitude is
related to $\epsilon_{\rm rd}$ by
\be\label{Aepsilon}
{\cal A}_{lmn}=
\sqrt{\f{32\Qlm \epsilon_{\rm rd}}{M\flm(1+4\Qlm^2)}}
\simeq
\sqrt{\f{8 \epsilon_{\rm rd}}{M \Qlm \flm}}
\,,
\ee
where the second expression corresponds to the $\delta$-function limit.

Using our general spectrum (\ref{dEdfours}) we can relate
the polarization-phase dependent
amplitude to an efficiency per polarization $\epsilon^{+,\times}_{\rm rd}$ by
\be
{\cal A}^{+,\times}_{lmn}=
\sqrt{\f{64\Qlm \epsilon_{\rm rd}^{+,\times}}
{M\flm [1+4\Qlm^2 \pm \cos (2\ph^{+,\times})]}}
\simeq
\sqrt{\f{16 \epsilon_{\rm rd}^{+,\times}}{M \Qlm \flm}}\,,
\ee
where the upper and lower signs refer to the $+$ and $\times$ polarizations,
respectively, and
where the last step again corresponds to the $\delta$-function limit.

The  expressions used in this Section are valid for any interferometric
detector. In all of our {\em LISA} calculations we take into account
the fact that the \lisa~arms form an angle of 60 degrees; as a result, when
integrating our results with the \lisa~noise curve, we must
multiply
all amplitudes 
by a geometrical correction factor $\sqrt{3}/2$, so that  ${\cal
A}^{LISA}_{+,\times}=\sqrt{3}/2\times {\cal A_{+,\times}}$.

We now combine the expression (\ref{rhoanalytic}) with an analytic
approximation for the \lisa~noise curve (here, for simplicity, we
exclude white-dwarf confusion noise; see Appendix \ref{app:noise}) of
the form
\beq
S_h^{\rm NSA}(f)=
\left[9.18\times 10^{-52}\left(\f{f}{1~{\rm
Hz}}\right)^{-4}
+1.59\times 10^{-41}
+9.18\times 10^{-38}\left(\f{f}{1~{\rm
Hz}}\right)^2\right]~{\rm Hz}^{-1}\,.
\label{Snsa}
\eeq
Rescaling frequencies in terms of the dimensionless frequency
${\cal F}_{lmn} = M\omega_{lmn}$, and inserting redshift factors
suitably, we obtain
\beq
\rho_{\rm FH} = \frac{5.1\times 10^3}{{\cal F}_{lmn}} 
	\left (\frac{\epsilon_{\rm rd}}{0.03} \right )^{1/2}
	\left (\frac{(1+z)M}{10^6 \,M_\odot} \right )^{3/2}
	 \left (\frac{1 \,{\rm Gpc}}{D_L(z)} \right )
	 \left (\frac{S_0}{S_h(f_{lmn})} \right )^{1/2}
	\frac{2Q_{lmn}}{\sqrt{1+4Q_{lmn}^2}}
\,,
\label{rhonumbers1}
\eeq
where
\beq
\frac{S_h(f_{lmn})}{S_0} = \frac{5.4 \times 10^{-5}}{{\cal F}_{lmn}^4}
	\left ( \frac{(1+z)M}{10^6 M_\odot} \right )^4
	+1+
	6.0 \,{\cal F}_{lmn}^2 
	\left ( \frac{10^6 M_\odot}{(1+z)M} \right )^2
\,.
\label{rhonumbers2}
\eeq
The dimensionless, mode-dependent quantities ${\cal F}_{lmn}$ and
$Q_{lmn}$ are of order unity and vary relatively weakly from mode to
mode; for low order modes they can be determined from the analytic
fits discussed in Appendix \ref{app:QNM}.

So far we have confined attention to the FH convention.
If we follow the alternative 
EF convention of keeping the waveform only for $t>0$,
and use the $\delta$-function limit,
we get the SNR
\beq\label{rhoFinn}
\rho_{\rm EF}^2=\langle \rho^2 \rangle =
\f{\Qlm}{40\pi^2\flm (1+4\Qlm^2)S}
&\times&
\left\{
\left(\f{M{\cal A}^{+}_{lmn}}{r}\right)^2 \left[1+\cos(2\ph^+)-2\Qlm\sin(2\ph^+)+4\Qlm^2\right]
\right. \nonumber \\
&+&\left.\left(\f{M{\cal A}^{\times}_{lmn}}{r}\right)^2 \left[1-\cos(2\ph^\times)+2\Qlm\sin(2\ph^\times)+4\Qlm^2\right]
\right\}\,,
\eeq
where the additional phase-dependent term comes from the lack of time
symmetry imposed on the waveform.  The rest of the formulae in this section
can be recast simply using this convention.

\subsection{Numerical results}
\label{singlemodenum}

We first compute SNRs for both inspiral and ringdown for events
at $D_L=3~$Gpc, corresponding to a redshift $z\simeq 0.54$, based on our
choice for cosmological parameters~\cite{cosmology}.
To compute the inspiral SNR we adopt the method discussed in
Ref.~\cite{BBW}. We perform an angle-average over pattern functions,
assuming that we observe the last year of inspiral and that we can
extrapolate the \lisa~noise curve down to a frequency $f\simeq
10^{-5}~$Hz. Following the common practice, we truncate the
signal-to-noise ratio integral, Eq.~(\ref{SNRdef}), using an upper
cutoff frequency determined by the conventional Schwarzschild ISCO for
a black hole of mass $M$.

We compute the ringdown SNR for the fundamental mode with $l=m=2$; 
calculations for different values of $l$ and $m$ yield similar
results, the SNR depending mainly on the ringdown efficiency in a
given mode.  
We use the FH SNR (\ref{rhoFHmain})  
and
adopt the $\delta$-function approximation (\ref{rhodeltafn}). 
Performing the ``full''
integral over the Breit-Wigner distribution (\ref{breitwigner})
we obtain essentially
indistinguishable results, except for a small ($\lesssim~10\%$)
disagreement in the mass/frequency region dominated by the white-dwarf
confusion noise (this statement is made more quantitative in
Fig.~\ref{DeltaError} of Appendix \ref{deltafun}).

\begin{figure*}[t]
\begin{center}
\begin{tabular}{cc}
\epsfig{file=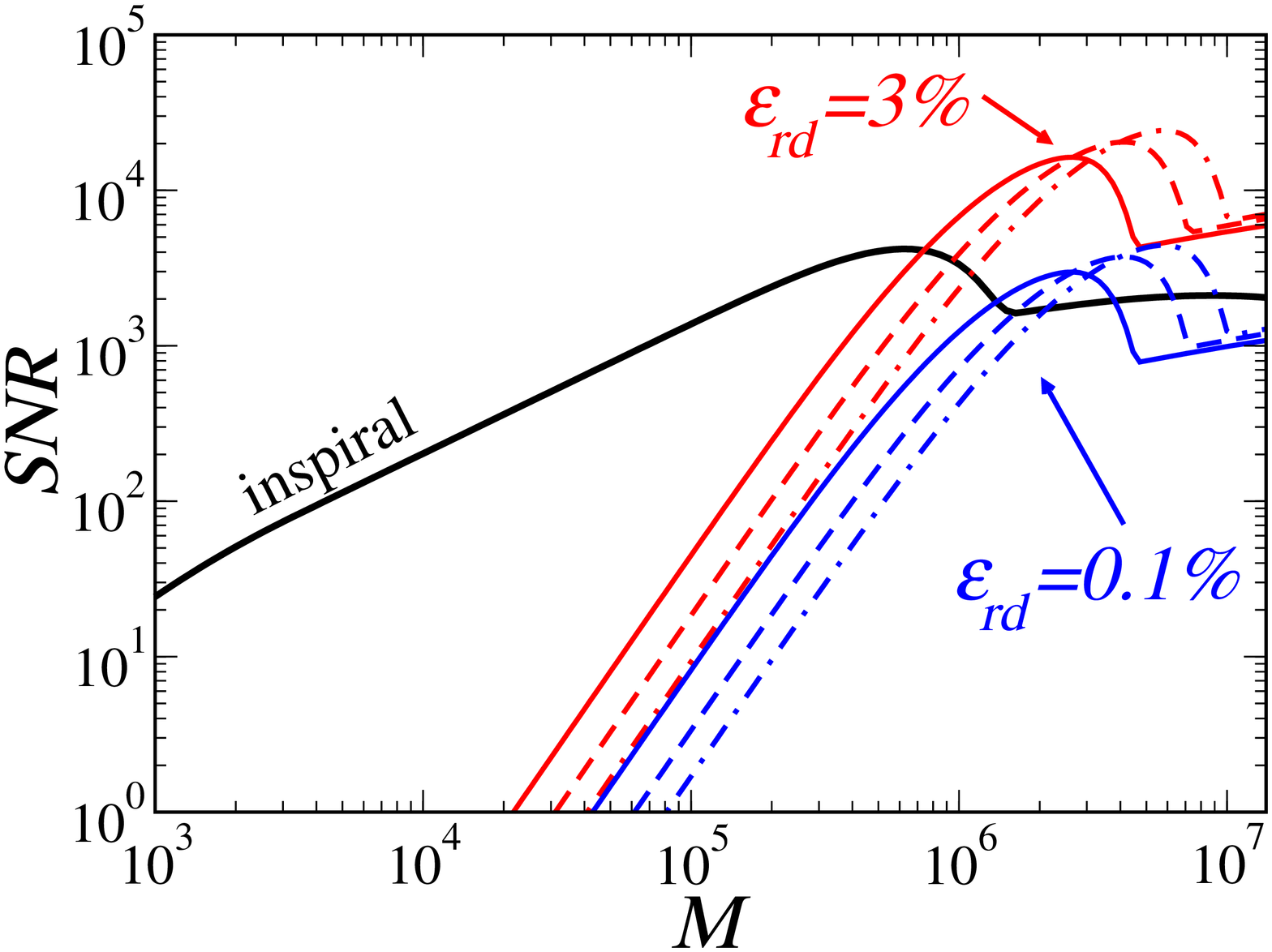,width=6cm,angle=-90} &
\epsfig{file=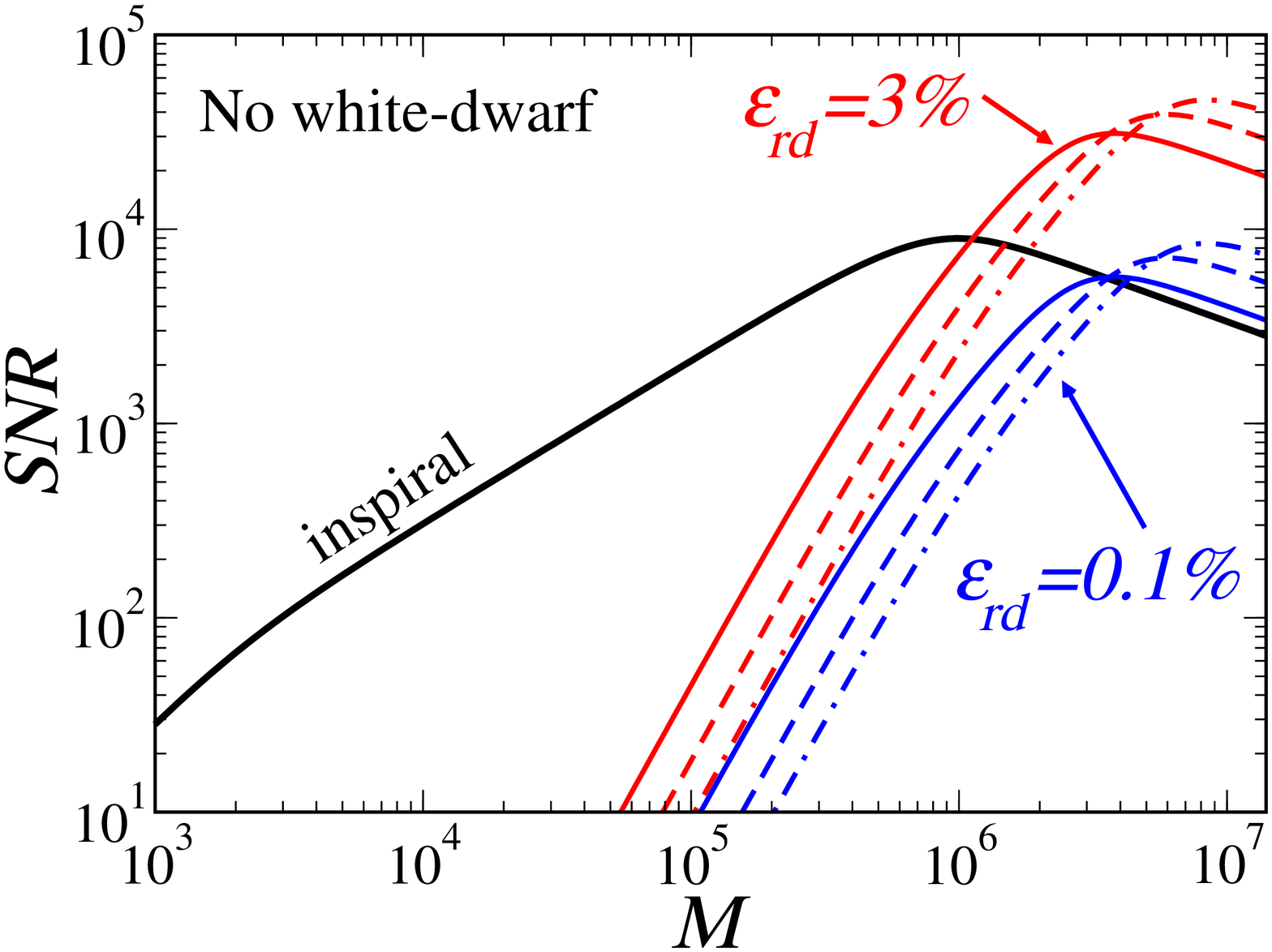,width=6cm,angle=-90} \\
\end{tabular}
\caption{Comparison of the SNR for inspiral and ringdown waveforms. In
the left panel we use the Barack-Cutler noise-curve; in the right
panel we use the same noise curve, but we do not include white-dwarf
confusion noise. The thick (black) line marked by ``inspiral'' is the
(angle-averaged) SNR for the inspiral of two equal-mass black holes at
$D_L=3~$Gpc. The other sets of lines (red and blue in color versions) 
show the SNR for the $l=m=2$ mode using the
$\delta$-function approximation,
assuming a ringdown efficiency
$\epsilon_{\rm rd}=3 \%$ and $0.1 \%$ respectively. 
Solid, dashed and dot-dashed lines correspond to 
$j=0$ (Schwarzschild), $j=0.8$ 
and $j=0.98$ respectively.
\label{insp-rd}}
\end{center}
\end{figure*}

The results are shown in Fig.~\ref{insp-rd}.
These plots can be viewed as an updated version of Fig.~6 in
FH.  Compared  to FH we use a better model of the \lisa~noise
curve (cf. Appendix \ref{app:noise}). In particular, a comparison of
the left and right panels illustrates the effect of white-dwarf
confusion noise on the expected SNR.
In both panels of Fig.~\ref{insp-rd}, the thick curve marked by
``inspiral'' represents the inspiral SNR for two equal-mass black
holes with total mass $M$ equal to the mass of the final black
hole.

The ringdown SNR in Fig.~\ref{insp-rd} is shown as sets of solid,
dashed and dot-dashed lines, corresponding to the limiting case of a
Schwarzschild black hole with $j=0$, an intermediate rotation rate
$j=0.8$, and a near extremal rate $j=0.98$, respectively.  The
intermediate value seems astrophysically quite plausible, based on the
best available astrophysical observations and on theoretical models of
merger and accretion (see, e.g. \cite{gammie}).  FH considered only
($j=0.98$).  For the ringdown efficiency we show the optimistic value
$\epsilon_{\rm rd}=3~\%$ considered by FH, as well as a pessimistic
value $\epsilon_{\rm rd}=0.1~\%$.  The latter value corresponds to the
present best estimates for the energy emitted in a maximally symmetric
merger, i.e. in the head-on collision of equal-mass black holes (see
\cite{sperhake} and references therein).  For unequal-mass mergers, FH
suggest (interpolating between numerical and perturbative results) an
energy scaling of the form $(4\mu/M)^2$, where $\mu$ is the reduced
mass.

The general features of the SNR curves are easy to understand. The SNR
is basically proportional to the inverse of the noise power spectral
density $S_h(f)$. It has a maximum in the mass range $M\sim 10^6
M_\odot$ corresponding to the frequency $f\sim 10^{-2}~$Hz at which
\lisa~is most sensitive. If we include white-dwarf confusion noise
(left panel) we observe the appearance of a dip in the SNR at masses
$M$ of the order of a few times $10^6 M_\odot$. The black hole at our
Galactic Center has an estimated mass $M\simeq 3.7\pm 0.2\times
10^6~M_\odot$ (see eg. \cite{narayan}), so an accurate modelling of
the white-dwarf confusion noise might be very important for detection
of black hole ringdown from galactic centers.  In this paper,
unless otherwise stated, we will include white-dwarf confusion
noise in all of our numerical calculations.

Fig.~\ref{insp-rd} illustrates that, even under pessimistic
assumptions, the ringdown SNR is generally comparable to the
inspiral SNR.  Reducing the rotation rate does not have a dramatic
effect, degrading the SNR of corotating modes by factors of order
unity. 
The crucial element for detectability is the fraction of
mass-energy $\epsilon_{\rm rd}$ going into each mode. Note that 
Eq.~(\ref{rhoanalytic}) implies that
$\rho\sim \sqrt{\epsilon_{\rm rd}}$.

FH and Ref.~\cite{rhook} pointed out that, depending on the ringdown
efficiency, there might be a ``critical mass'' at which black hole
ringdown becomes dominant over inspiral. Assuming an efficiency
$\epsilon_{\rm rd}=1 \%$ and a final black hole angular momentum
$j=0.997$, Ref.~\cite{rhook} found that the SNR is greater in the
ringdown signal for $M\gtrsim 10^6 M_\odot$ when $z=1$. Their result
is consistent with our Fig.~\ref{insp-rd}; in addition, we find that
this ``critical mass'' for the transition from inspiral to ringdown
dominance depends only weakly on $j$, being more sensitive to the
efficiency $\epsilon_{\rm rd}$.

Ringdown efficiency plays a very important role in the ringdown SNR.
Unfortunately, numerical relativity simulations do not provide us with
reliable estimates of $\epsilon_{\rm rd}$ \cite{BCW2}. Ref.~\cite{BCD}
provides alternative, semianalytical estimates of the energy radiated
in the plunge and ringdown phases for Earth-based detectors;
unfortunately, an extrapolation of those results to binaries of
relevance for \lisa~is not available. Given our ignorance of
$\epsilon_{\rm rd}$, it makes sense to ask the following question: how
much energy must be channelled into a given mode in order for it to be
detectable?

The answer to this question is provided by Fig.~\ref{epsrd} in Sec
\ref{intro}, where we
assume $\rho=10$ as a criterion for detectability and we show
the fundamental mode of a Kerr black hole with
$l=m=2$. The result is encouraging: even in the pessimistic
situation of a head-on collision $\epsilon_{\rm rd}=10^{-3}$, we
can reasonably expect \lisa~to detect all mergers yielding a black
hole of mass $M\gtrsim 10^5 M_\odot$.  Even accounting for the fact
that $\epsilon_{\rm rd}$ may be lower for unequal mass mergers, the
prospects for detection are still encouraging.

\begin{figure*}[t]
\begin{center}
\begin{tabular}{ccc}
\epsfig{file=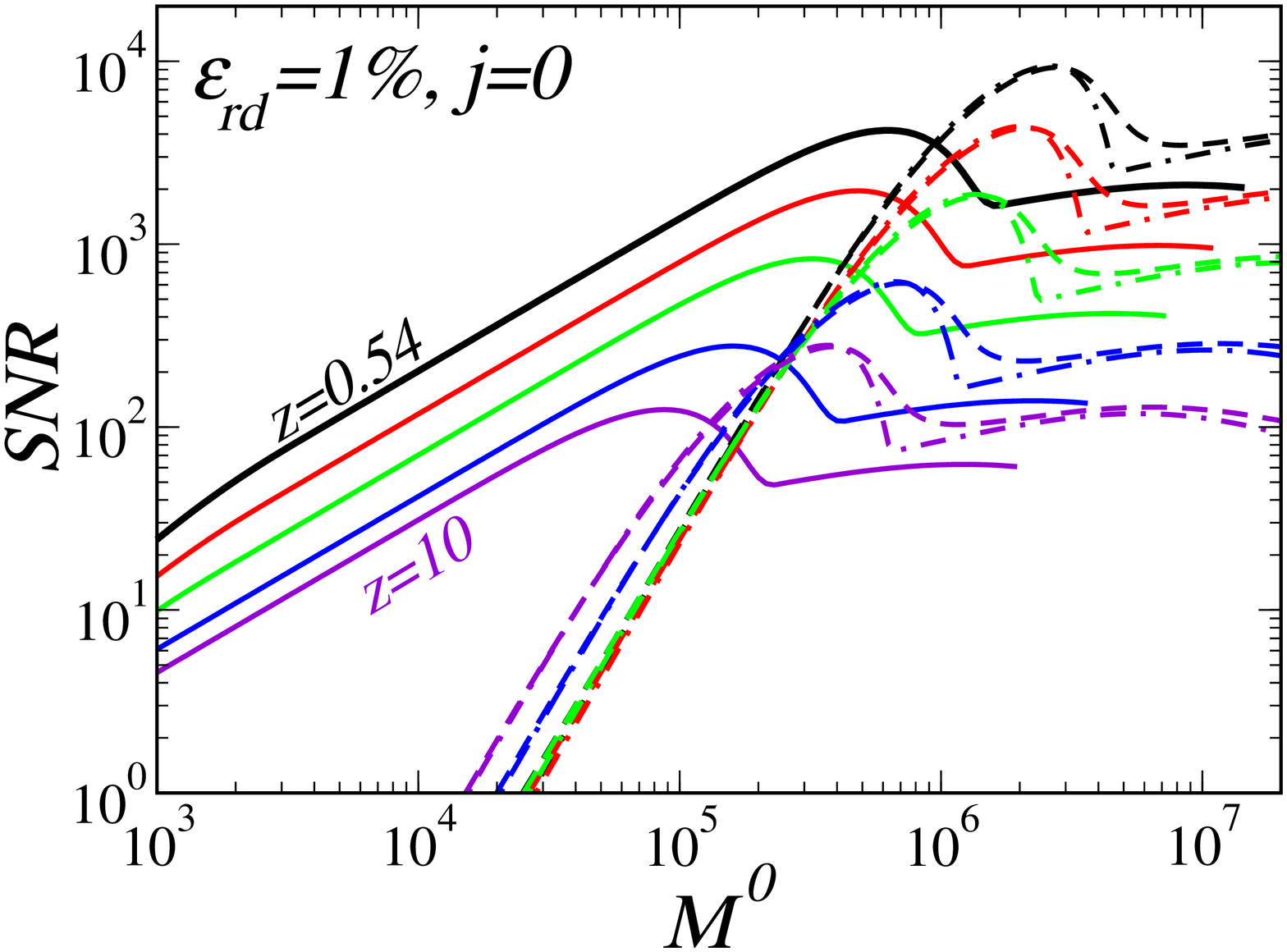,width=4.5cm,angle=-90} &
\epsfig{file=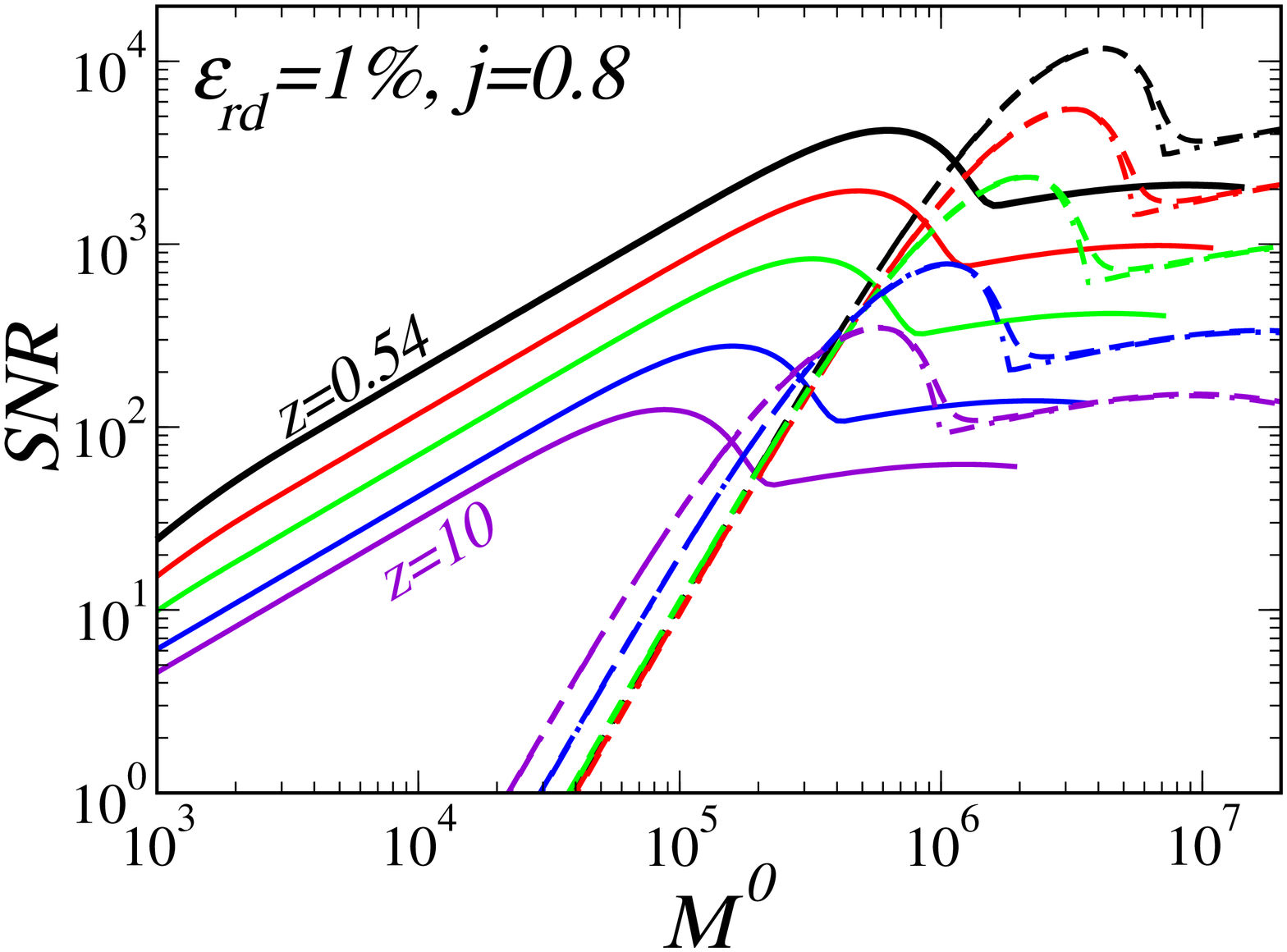,width=4.5cm,angle=-90} &
\epsfig{file=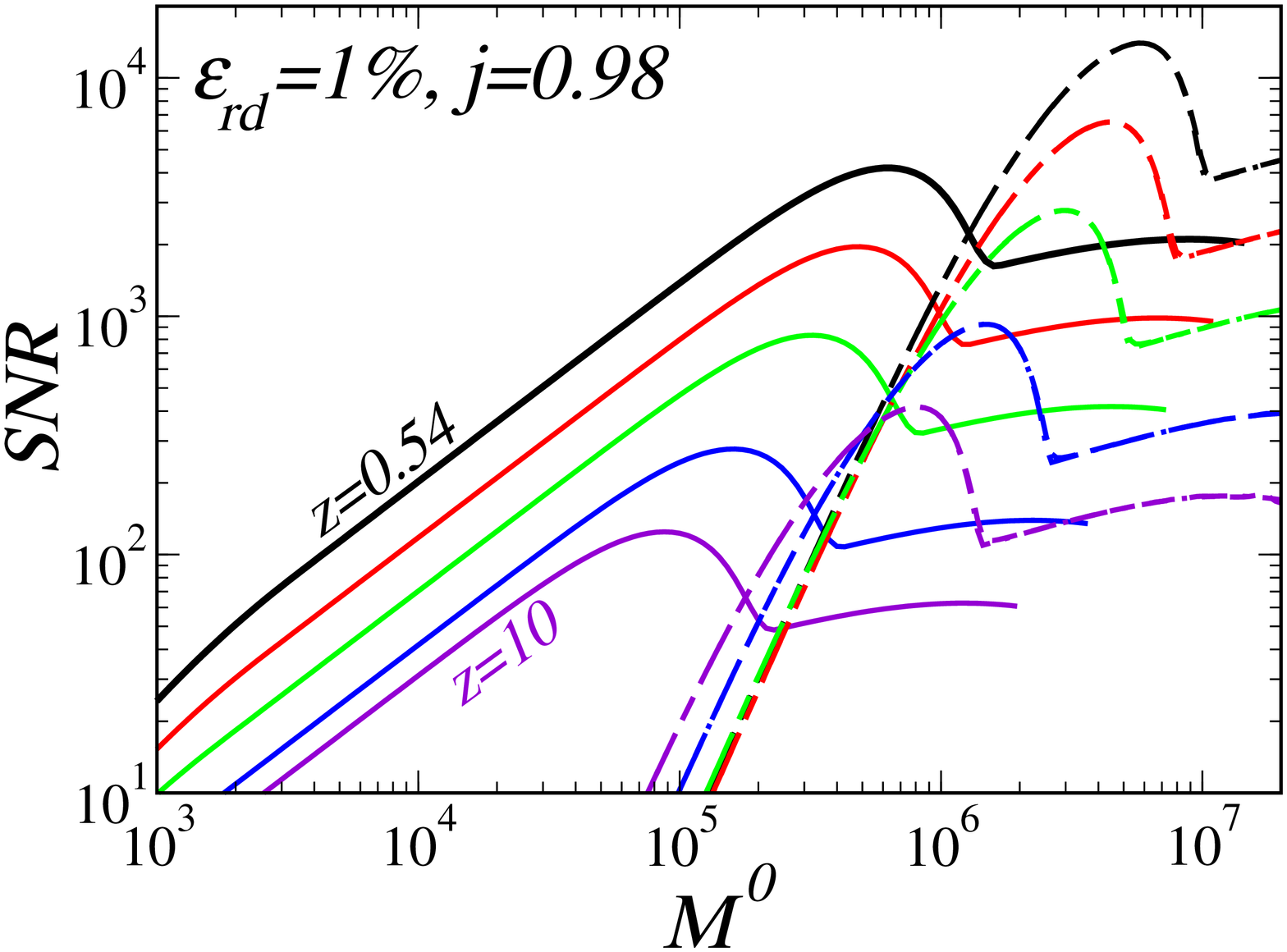,width=4.5cm,angle=-90} \\
\end{tabular}
\caption{Dependence of the SNR on redshift, for both inspiral
(continuous lines) and
ringdown (dashed and dot-dashed lines).   We choose a ringdown efficiency $\epsilon_{\rm rd} =
1 \%$ and consider the cases $j=0$, $j=0.8$ and $j=0.98$.  
From top to bottom the lines in each panel correspond to
$z=0.54$ (black in color versions; $D_L=3$~Gpc),
$z=1$ (red), $z=2$ (green), $z=5$ (blue)
and $z=10$ (purple).  The dashed lines are obtained from the full
integral, the dot-dashed lines
use the $\delta$-function approximation. 
\label{insp-rd-z}}
\end{center}
\end{figure*}

The cosmological redshift affects the detectability window and the SNR
both for inspiral and for ringdown waves, both through the decreasing
signal strength with distance and through shifting the relevant
frequencies to different parts of the \lisa~noise spectrum.
Fig.~\ref{insp-rd-z} shows the results of numerical calculations of
the SNR.  For the ringdown signal, we pick a "best guess" for the
efficiency $\epsilon_{\rm rd}$ of $1\%$, intermediate between the
$3\%$ of FH and the $0.1\%$ from head-on collisions.  Each plot gives
the SNR as a function of $M^0$ (mass in the source frame) for a
different value of $j$ (left to right, $j=0$, $j=0.8$ and $j=0.98$).
From top to bottom, the curves show sources at redshifts $z=0.54, \,
1, \, 2,\, 5$ and $10$.  The continuous lines are the inspiral SNR,
the dashed lines are obtained from the full integral, the dot-dashed
lines use the $\delta$-function approximation.  The inspiral SNR is
(somewhat arbitrarily) truncated at the (large) value of the mass for
which the starting frequency (which we pick to be one year before the
ISCO, as in \cite{BBW}) becomes lower than $10^{-5}$~Hz.

\section{Parameter estimation by detection of a single mode}
\label{onemode-pe}

\subsection{Analytic results}

In this Section we will go beyond the issue of detectability and try
to answer a different question: given the detection of a single QNM,
what can we learn about the black hole parameters?
To estimate the black hole parameters from ringdown waveforms, we use
the standard technique of parameter estimation in matched filtering.
By maximizing the correlation between a template gravitational
waveform that depends
on a set of parameters $\theta^a$ (for example, the black hole mass
and angular momentum) and a measured signal, matched filtering
provides a natural way to estimate the parameters of the signal and
their errors.  With a given noise spectral density for the detector,
$S_h(f)$, one defines the inner product between two signals $h_1(t)$
and $h_2(t)$ by 
\begin{equation}
(h_1|h_2) \equiv 2 \int_0^{\infty} \frac{ {\tilde{h}_1}^*\tilde{h}_2 +
{\tilde{h}_2}^*\tilde{h}_1 }{S_h(f)}df \,,
\label{innerproduct}
\end{equation}
where $\tilde{h}_1(f)$ and $\tilde{h}_2(f)$ are the Fourier transforms
of the respective gravitational waveforms $h(t)$.  The components of
the ``Fisher matrix'' $\Gamma_{ab}$ are then given by
\begin{equation}
\Gamma_{ab} \equiv \left( \frac{\partial h}{\partial\theta^a} \mid
\frac{\partial
h}{\partial\theta^b} \right) \,,
\label{fisher}
\end{equation}
In the limit of large SNR, if the noise is stationary and Gaussian,
the probability that the gravitational-wave 
signal $s(t)$ is characterized by a given
set of values of the source parameters $\theta^a$ is
\beq
p(\mbox{\boldmath$\theta$}|s)=p^{(0)}(\mbox{\boldmath$\theta$}) 
\exp\left[-\frac{1}{2}\Gamma_{ab}\delta \theta^a \delta \theta^b\right]\,.
\eeq
where $\delta \theta^a = \theta^a - {\hat \theta}^a$, and
$p^{(0)}(\mbox{\boldmath$\theta$})$ represents the distribution of prior
information.
An estimate of the rms error, $\Delta\theta^a = (\langle (\delta \theta^a)^2
\rangle )^{1/2}$, in measuring the
parameter $\theta^a$ can then be calculated, in the limit of large
SNR, by taking the square root of the diagonal elements of the inverse
of the Fisher matrix,
\begin{equation}
\Delta\theta^a = \sqrt{\Sigma^{aa}} \,, \qquad  \Sigma = \Gamma^{-1} \,.
\label{errors}
\end{equation}
The correlation coefficients between two parameters $\theta^a$ and
$\theta^b$ are given by
\begin{equation}
c_{ab} = \Sigma^{ab}/\sqrt{\Sigma^{aa}\Sigma^{bb}} \,.
\label{correlations}
\end{equation}


We consider a waveform given by Eq. (\ref{waveform0}), with the $S_{lmn}$
assumed to be real,
\begin{subequations}
\beq 
h_+&=& \frac{M}{r} {\cal A}_{lmn}^{+} e^{-\pi
f_{lmn} t/Q_{lmn}} \cos\left[2\pi f_{lmn} t+ \ph^+ \right]
S_{lmn}(\iota,\beta)\,,\\
h_{\times}&=&\frac{M}{r} {\cal A}_{lmn}^{\times} e^{-\pi f_{lmn}
t/Q_{lmn}} \sin\left[2\pi f_{lmn} t+ \ph^\times \right]
S_{lmn}(\iota,\beta)\,.
\eeq
\label{wf1} 
\end{subequations}
We also define
\beq \label{wf2}
\frac{M}{r} {\cal A}_{lmn_1}^{+}\equiv A^+ \,,\qquad
\frac{M}{r} {\cal A}_{lmn_1}^{\times}\equiv A^\times
\equiv A^+ {\rm N_\times} \,,
\eeq
where $N_\times$ is some numerical factor, and
\be \label{wf3}
\ph^\times=\ph^++\ph^0\,.
\ee
Assuming that we know $N_\times$ and $\ph^0$, this waveform is
dependent on four parameters ($A^+$, $\ph^+$, $M$, $j$); otherwise
it depends on six parameters ($A^+$, $A^\times$, $\ph^+$,
$\ph^\times$, $M$, $j$). 
A popular choice for $N_\times$ \cite{fhh,rhook} is to
assume that the distribution of the strain in the two polarizations
mimics that of the inspiral phase: $N_\times=-2(\hat{\mathbf{L}}\cdot
\hat{\mathbf{n}})/[1+(\hat{\mathbf{L}}\cdot \hat{\mathbf{n}})^2]$,
where $\hat{\mathbf{L}}$ is the orientation of the binary's angular
momentum and $\hat{\mathbf{n}}$ is a unit vector describing the
binary's position in the sky. Fortunately, we will see that the errors
have a very weak dependence on the number of parameters and on the
(uncertain) value of the parameters $N_\times$ and $\ph^0$.

Assuming constant noise over the bandwidth of
the signal, or taking the $\delta$-function approximation,
and using the FH doubling convention, we get the SNR
(\ref{rhoFH}).  In this
approximation, errors and correlation coefficients can be computed
analytically using Mathematica or Maple. The full expressions are lengthy and
unenlightening, and we have implemented them numerically in a Fortran
code. 

We first calculate the Fisher matrix in the parameter basis of $(A^{+},\,
\ph^+,\,\flm,\, \Qlm)$, where it takes on a simpler form:
\begin{subequations}
\beq
\Gamma_{A^+A^+} &=& \f{\gamma}{(A^+)^2} \left (1+4\Qlm^2-\beta \right) \,,
\\
\Gamma_{A^+\ph^+} &=& \f{\gamma}{A^+} \alpha \,,
\\
\Gamma_{A^+ \flm} &=& - \f{\gamma }{2 A^+ \flm} 
\left (1+4\Qlm^2-\beta \right) \,,
\\
\Gamma_{A^+ \Qlm} &=&  \f{\gamma}{2 A^+ \Qlm} 
\f{1}{1+4\Qlm^2} 
\left [(1+4\Qlm^2)^2 - (1-4\Qlm^2) \beta \right ]  \,,
\\
\Gamma_{\ph^+\ph^+} &=& \gamma \left (1+4\Qlm^2+\beta \right) \,,
\\
\Gamma_{\ph^+ \flm} &=& -\f{\gamma}{2\flm} \alpha \,,
\\
\Gamma_{\ph^+ \Qlm} &=& \f{\gamma}{2 \Qlm} 
\left (\f{1-4\Qlm^2}{1+4\Qlm^2} \right ) \alpha \,,
\\
\Gamma_{\flm \flm} &=& \f{\gamma}{2\flm^2} \left [ (1+4\Qlm^2)^2 - 
\beta \right ] \,,
\\
\Gamma_{\flm \Qlm} &=& -\f{\gamma}{2\flm \Qlm} 
\f{1}{1+4\Qlm^2} 
\left [ (1+4\Qlm^2)^2 - (1-4\Qlm^2)\beta \right ]  \,,
\\
\Gamma_{\Qlm \Qlm} &=& \f{\gamma}{2 \Qlm^2} 
\f{1}{(1+4\Qlm^2)^2}
\left [ (1+4\Qlm^2)^3 - (1-12\Qlm^2)\beta \right ]  \,,
\eeq
\label{fishermatrix}
\end{subequations}
\noindent
where
\begin{subequations}
\beq
\alpha &=& \sin^2 \psi \sin 2\ph^\times - \cos^2 \psi \sin 2\ph^+
\,,
\\
\beta &=& \sin^2 \psi \cos 2\ph^\times - \cos^2 \psi \cos 2\ph^+
\,,
\\
\gamma &=& \f{A^2 \Qlm }{40\pi^2 \flm (1+4\Qlm^2)} \,,
\eeq
\label{alphabetagamma}
\end{subequations}
with $\cos \psi \equiv 1/\sqrt{1+N_\times^2}$,
$\sin \psi \equiv N_\times/\sqrt{1+N_\times^2}$ and 
$A^2 = (A^+)^2 (1+N_\times^2) = (A^+)^2 + (A^\times )^2$.
Note that, in this notation, $\rho_{FH}^2 = \gamma (1+4\Qlm^2-\beta)$.

We note that the Fisher matrix written in terms of the
frequency and damping time is usually simpler \cite{finn,kaa} than
that in terms of mass and angular momentum; however
we prefer to deal directly with measurements of $j$ and $M$. In
Sec.~\ref{nohair}, to estimate QNM resolvability, we will work
in terms of frequencies and quality factors.

The transformation from the 
$(A^{+},\, \ph^+,\,\flm,\, \Qlm)$ basis to the $(A^{+},\, \ph^+,\, M, \,j)$
basis is straightforward, namely, for any index $k$,
\beq
\Gamma_{k\,M} &=& -(\flm/M) \Gamma_{k\,\flm} \,,
\nonumber \\
\Gamma_{k\,j}  &=& \flm'  \Gamma_{k\,\flm} + \Qlm' \Gamma_{k\,\Qlm} \,,
\label{Mjtransform}
\eeq
where $\flm' \equiv d\flm/dj$ and $\Qlm' \equiv d\Qlm/dj$.

Converting to this basis and inverting the Fisher matrix, we find, 
to leading order in $Q_{lmn}^{-1}$, the errors 
\begin{subequations}
\label{rii}
\beq
\sigma_j &=&
\frac{1}{\rho_{\rm FH}}
\left|2\frac{Q_{lmn}}{Q_{lmn}'}
\left(1+\f{1+4\beta}{16\Qlm^2}\right)\right|\,, \label{erra}\\
 \sigma _M &=&
\frac{1}{\rho_{\rm FH}}
\left|2\frac{MQ_{lmn}
f_{lmn}'}{f_{lmn}Q_{lmn}'}
\left(1+\f{1+4\beta}{16 \Qlm^2}\right)\right|\,, \label{errm}\\
 \sigma _{A^+} &=&
\frac{\sqrt{2} A^+}{\rho_{\rm FH}} 
\left |1+\f{3\beta}{8\Qlm^2} \right|\,,\\
 \sigma _{\ph^+} &=&
\frac{1}{\rho_{\rm FH}} \left|1-\f{\beta}{4\Qlm^2}\right|
 \,, \label{errph}
\eeq
\end{subequations}
and the correlation coefficients 
\begin{subequations}
\label{rij}
\beq
r_{jM}&=&
{\rm sgn}(\flm')\times
\left(1-\frac{f_{lmn}^2Q_{lmn}'^2}{16Q_{lmn}^4f_{lmn}'^2}\right)
+{\cal O}(1/Q^{6}) \,, \label{corrjm}\\
r_{jA^+}&=&-\frac{1}{\sqrt{2}}
\left(1-\f{1-6\beta}{16\Qlm^2}\right)
+{\cal O}(1/Q^{4})     \,,   \\
r_{MA^+}&=&-\frac{1}{\sqrt{2}}
\left(1-\f{1-6\beta}{16\Qlm^2}\right)
+{\cal O}(1/Q^{3})         \,,
  \\
r_{j\ph^+}&=&\frac{\alpha}{2Q_{lmn}^2} - \frac{7-8\beta}{32\Qlm^4}
+{\cal O}(1/Q^{6})
        \,,   \\
r_{M\ph^+}&=&\frac{\alpha}{2Q_{lmn}^2} - \frac{7-8\beta}{32\Qlm^4}
+{\cal O}(1/Q^{6})
 \,,   \\
r_{A^+ \ph^+}&=&
-\frac{3\alpha}{4\sqrt{2}Q_{lmn}^2} +
\frac{\alpha(10-11\beta)}{32\sqrt{2}Q_{lmn}^4}
+{\cal O}(1/Q^{6}) \,.
\eeq
\end{subequations}

In calculating derivatives of the waveforms (\ref{wf1}) with respect
to $M$ and $j$ (or with respect to $\flm$ and $\Qlm$), we have 
ignored derivatives of the spheroidal harmonics
themselves.  The $S_{lmn}$ are functions of $a\omega_{lmn} =
j{\cal F}_{lmn}$ which is a function of $j$ only.  However, the
$S_{lmn}$ may be expanded in powers of $j{\cal F}_{lmn}$, in the form
\be
S_{lmn} = Y_{lm} + (j{\cal F}_{lmn})\,
\sum_{l^\prime \ne l} c_{l^\prime lm}
Y_{l^\prime m} + O(j{\cal F}_{lmn})^2 \,,
\label{sexpand}
\ee
where $Y_{lm}$ denotes a {\it spin-weighted}
spherical harmonic, and $ c_{l^\prime lm}$ are related to
Clebsch-Gordan coefficients.  As a result, derivatives of
the $S_{lmn}$ with respect to $j$ will be linear in derivatives of 
$j{\cal F}_{lmn}$, and because of the orthogonality of the spin-weighted
spherical harmonics, inner products of $S_{lmn}$ with $S_{lmn}^\prime$ and
of $S_{lmn}^\prime$ with itself will be at least quadratic in $j{\cal
F}_{lmn}$ and its derivatives.
At least for
small $j{\cal F}_{lmn}$, 
we may expect these contributions to be small relative to the main
contribution obtained by ignoring these derivatives.  Nevertheless, the effect
of this approximation should be explored further.

The diagonal elements of the correlation matrix $r_{ii}=1$ for all
$i$.  The ${\rm sgn}(\flm')$ in $r_{jM}$ comes from a
$\sqrt{\flm'^2}/\flm'=|\flm'|/\flm'$. It implies that $j$ and $M$ have
a positive correlation for corotating and axisymmetric modes ($m\geq
0$), but they are anticorrelated for counterrotating modes ($m<0$):
this is basically determined by the different sign of $\flm'$ for the
two classes of modes (see eg. Fig.~\ref{Qmodes}).

The calculation using
the EF convention proceeds along similar lines; the results are very
similar, and they are reported in Appendix \ref{app:FisherOne}.  The
calculation for a six-dimensional Fisher matrix is straightforward,
and is also relegated to Appendix \ref{app:FisherOne}.

The large-$\Qlm$
expansions are typically accurate as long as $Q_{lmn}'/Q_{lmn}$
is not very large (see Fig.~\ref{errs-fh} below, where this statement
is made more quantitative).  An analytic parametrization for $\sigma$
can be obtained by combining the SNR formula (\ref{rhonumbers1}) 
and (\ref{rhonumbers2}) with the QNM
fits, whose coefficients are provided in Tables \ref{tab:fitQNMsl2},
\ref{tab:fitQNMsl3} and \ref{tab:fitQNMsl4}.

An important check is that the errors on $M$ and $j$ (and the
correlation between these parameters), as predicted by Finn
\cite{finn}, agree with ours. This is not a trivial check, since Finn
uses a different parametrization of the waveform. In
particular, it is easy to check that Finn's expressions for the
errors, his Eqs.~(4.20a) and (4.20b), agree with ours to leading order
in $\Qlm^{-1}$; so does Finn's expression for the correlation
coefficient, his Eq.~(4.20e). The fact that only high-order
corrections in $\Qlm^{-1}$ depend on the parametrization is a
reassuring feature of the Fisher matrix calculation (see Appendix
\ref{app:FisherOne}).

Some general comments on Eqs. (\ref{rii}) and (\ref{rij}) are in
order. 
First, by combining Eqs.~(\ref{erra}) and (\ref{errm}) with
Eq.~(\ref{rhonumbers1}), 
we see that the accuracy in measuring $M$ and $j$ can be very high
under the right circumstances, namely,
\begin{subequations}
\beq
\label{parametrizeM}
\f{\sigma_M}{M}&\simeq&
6.8\times 10^{-3}\times
h_{lmn}(j) {\cal F}_{lmn} \times
\left(\f{10^{-4}}{\epsilon_{\rm rd}}\right)^{1/2}
\left(\f{S_h(f_{lmn})}{S_0}\right)^{1/2}
\left(\f{D_L(z)}{1~{\rm Gpc}}\right)
\left(\f{10^6M_\odot}{(1+z)M}\right)^{3/2}\,,
\\
\label{parametrizej}
\sigma_j&\simeq&
6.8\times 10^{-3}\times
g_{lmn}(j) {\cal F}_{lmn} \times
\left(\f{10^{-4}}{\epsilon_{\rm rd}}\right)^{1/2}
\left(\f{S_h(f_{lmn})}{S_0}\right)^{1/2}
\left(\f{D_L(z)}{1~{\rm Gpc}}\right)
\left(\f{10^6M_\odot}{(1+z)M}\right)^{3/2}\,.
\eeq
\end{subequations}
Notice that the measurement error is small (less than a percent),
even under the very pessimistic assumption that a SMBH 
with $M \sim 10^6~M_\odot$ radiates only a modest fraction $E_{\rm
GW}\sim 10^{-4} M$ of its mass.  The functions $g_{lmn}(j)=\Qlm/\Qlm'$
and $h_{lmn}(j)=(\Qlm\flm')/(\Qlm'\flm)$ depend on the particular mode
we consider and on the black hole's angular momentum; they are
typically of order unity. For example, using the fitting relations in
Appendix \ref{app:QNM}, we find that, for the fundamental mode with
$l=m=2$ of a Schwarzschild black hole, these factors take the values
$g_{lmn}(0)=2.992$, $h_{lmn}(0)=1.214$. So, for a non-rotating black
hole the error in angular momentum is slightly larger than the error
on the mass. For a near-extremal black hole we have
$g_{lmn}(0.98)=0.043$, $h_{lmn}(0.98)=0.234$, and the error in angular
momentum is now smaller than the error in the mass. We will see that
this reversal of the magnitude of the errors for near-extremal black
holes is typical of corotating modes, but does not hold true for
counterrotating modes.

\begin{figure*}[t]
\begin{tabular}{cc}
\epsfig{file=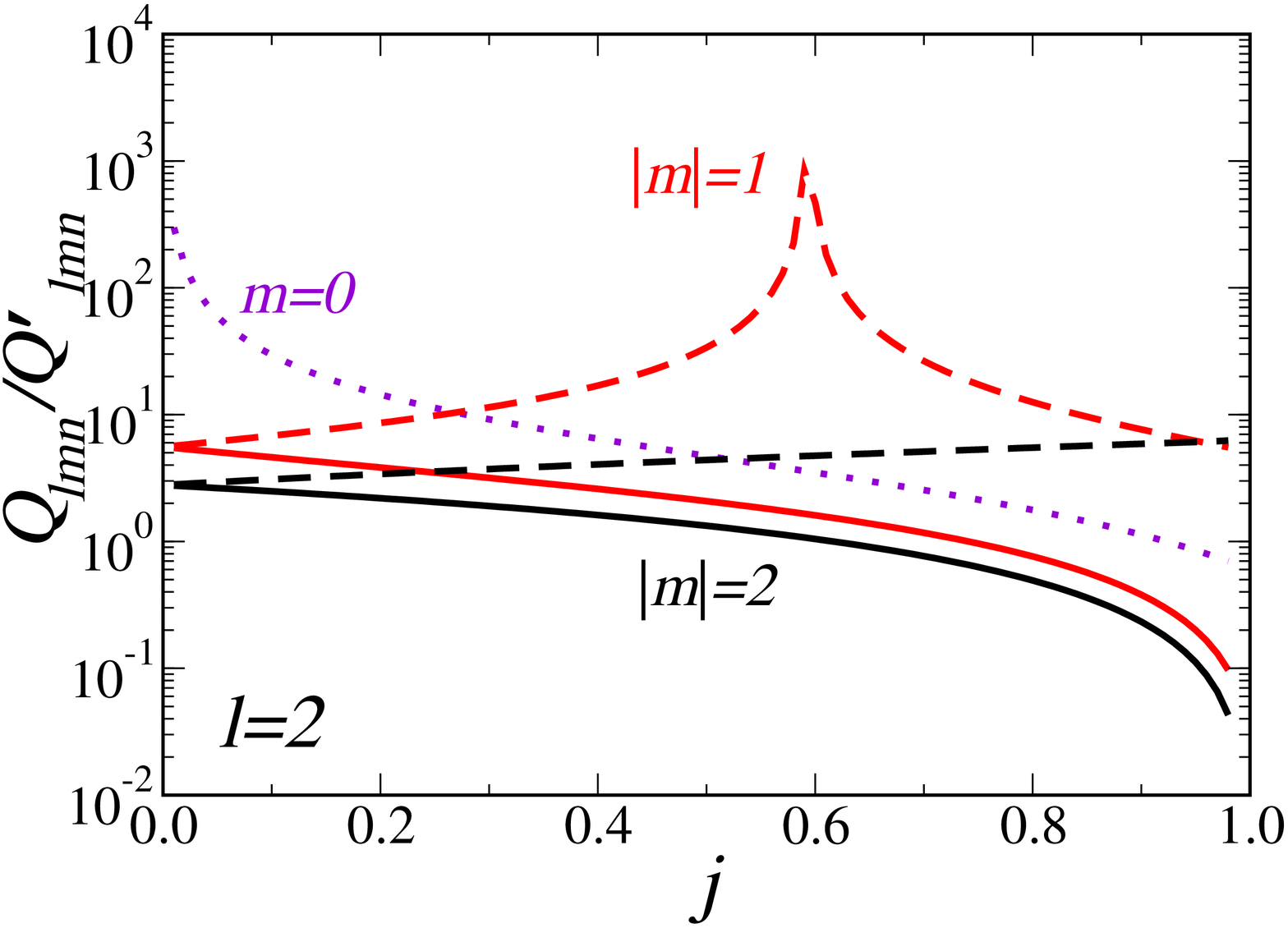,width=6cm,angle=-90} &
\epsfig{file=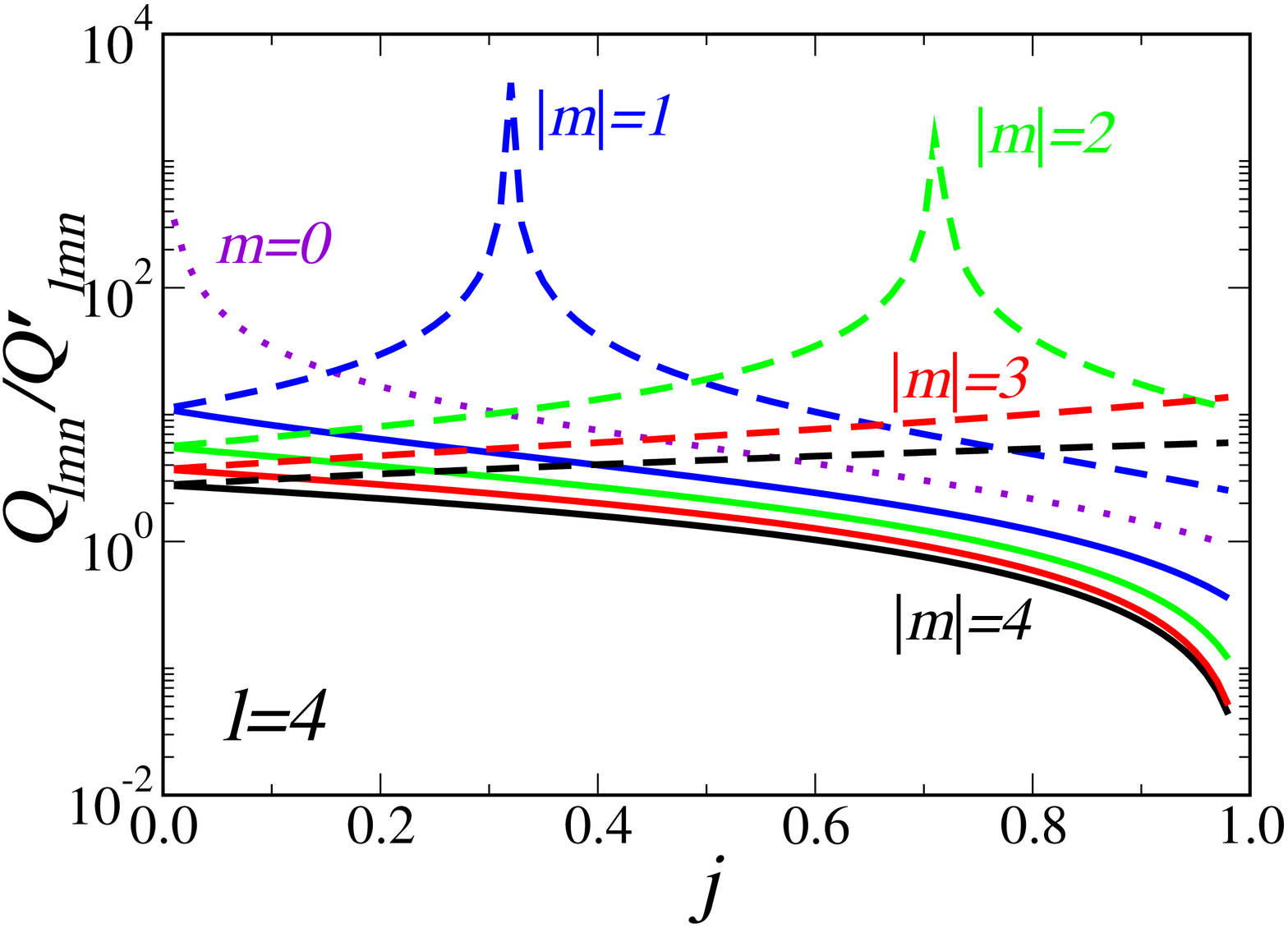,width=6cm,angle=-90} \\
\end{tabular}
\caption{$\Qlm/\Qlm'$ for the fundamental mode with $l=2$ (left) and
$l=4$. Solid lines refer to $m>0$, dotted lines to $m=0$, dashed lines
to $m<0$, and different shades (colors) denote different values of $m$, as
indicated.  Notice
that this factor increases with $j$ when $m>0$ and decreases with $j$
when $m=-2$. The factor blows up as $j\to 0$ for $m=0$, and as $j\to
j_{\rm crit}$ for certain values of $m<0$. This explains
most qualitative features of the error plots.  \label{QQa} }
\end{figure*}

To leading order in a large-$\Qlm$ expansion the errors on
angular momentum and mass, Eqs.~(\ref{erra}) and (\ref{errm}), are
proportional to $\Qlm/\Qlm'$. In Fig.~\ref{QQa} we plot this quantity
as a function of $j$ for different modes. From the plot we can
anticipate a few salient features. First of all, errors should
decrease with rotation for corotating modes ($m>0$). This was already
pointed out in Refs.~\cite{echeverria,finn}. 
However, errors should {\it increase} with
rotation for counterrotating modes ($m<0$); even worse, at those
``critical values'' of $j$ for which $\Qlm'=0$ the errors for
counterrotating modes blow up. Fig.~\ref{QQa} shows that, typically,
this phenomenon is present for counterrotating modes with $|m|\leq
l/2$. Finally, we can anticipate that $\Qlm/\Qlm'$ (hence the error)
will blow up as $j\to 0$ for modes with $m=0$.

\begin{figure*}[t]
\begin{center}
\begin{tabular}{cc}
\epsfig{file=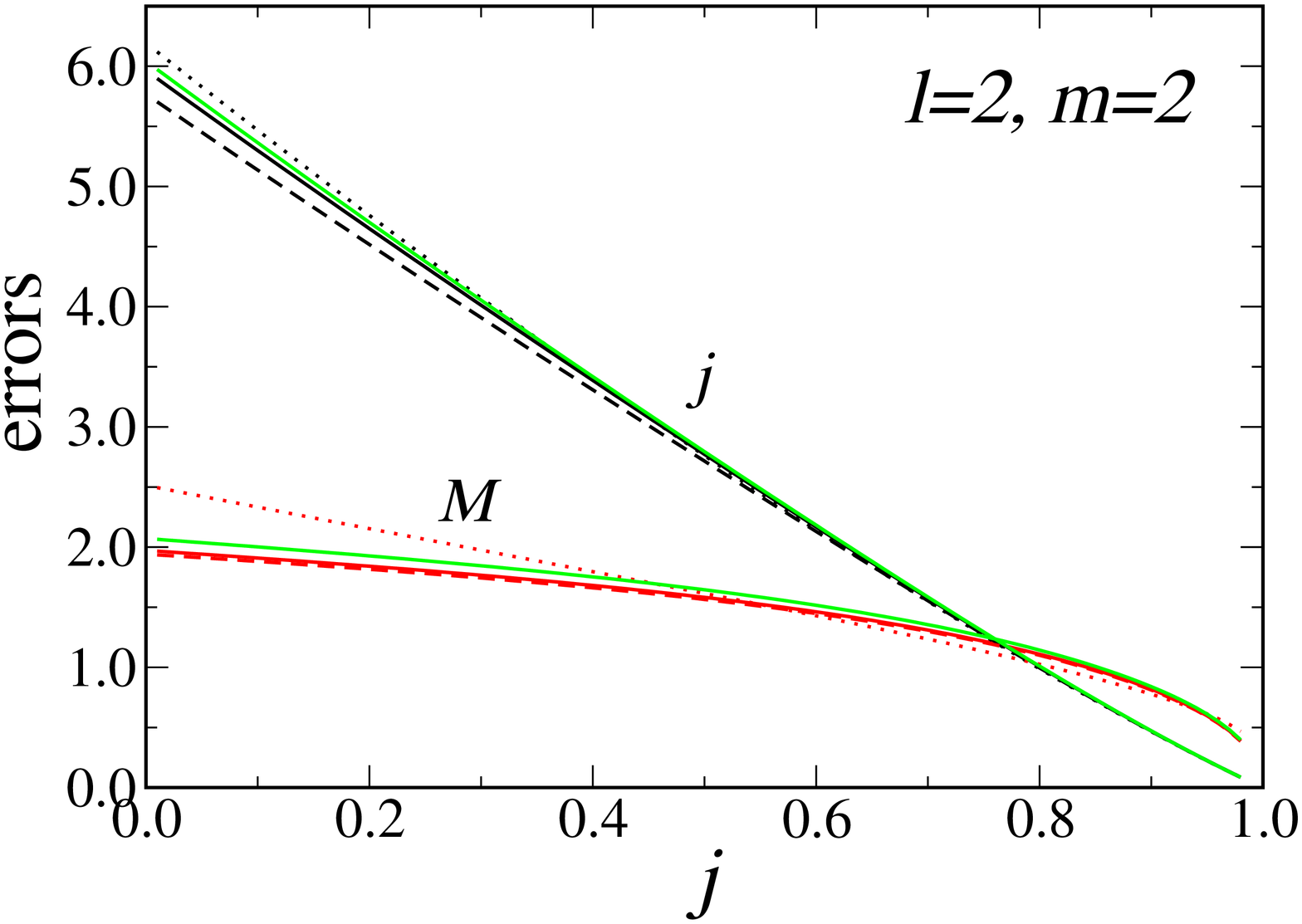,width=6cm,angle=-90} &
\epsfig{file=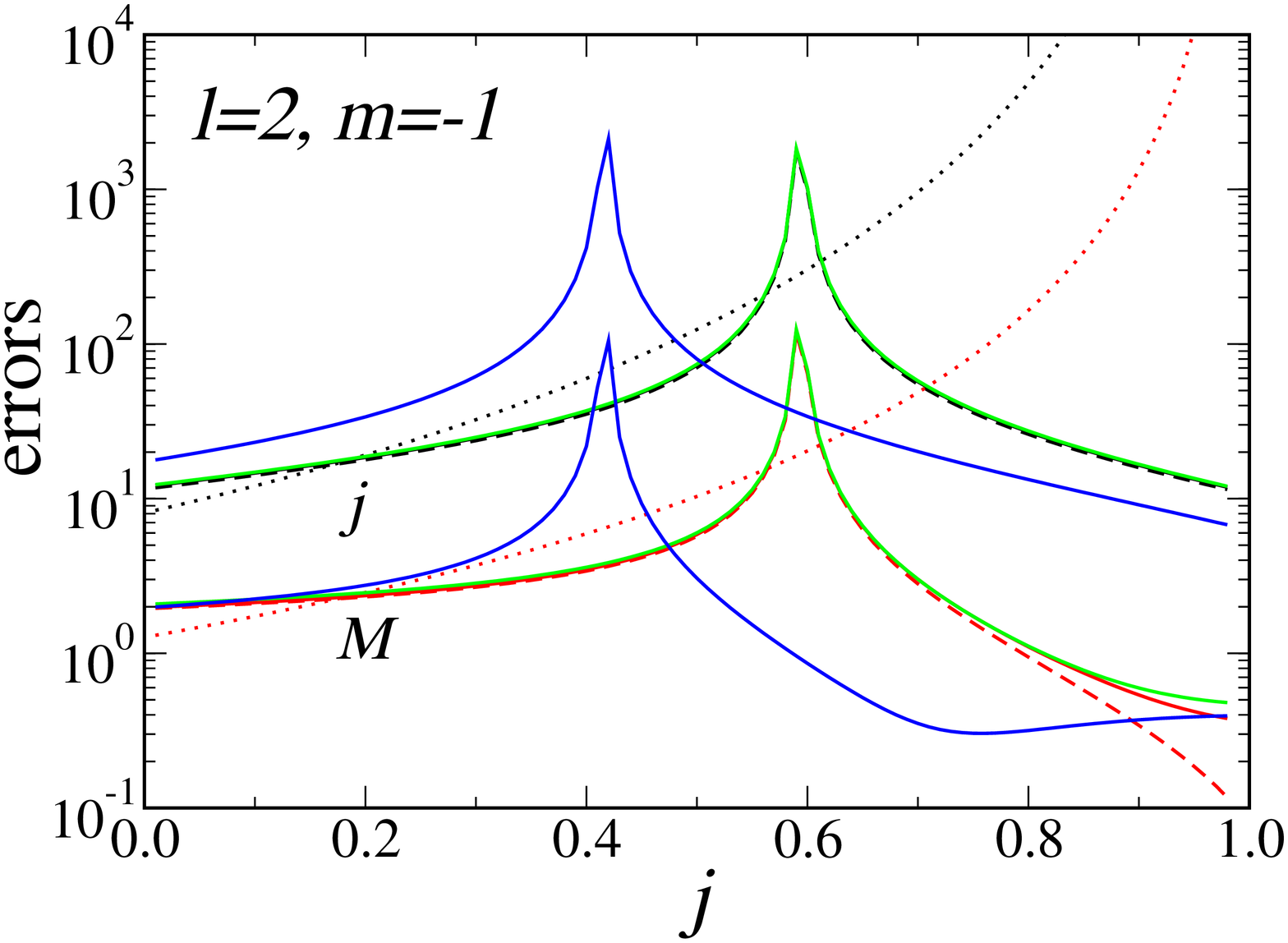,width=6cm,angle=-90} \\
\end{tabular}
\caption{Comparison between different approaches to compute the
error (multiplied by the SNR) for the fundamental mode for 
$(l,m)= (2,2)$ and $(2,-1)$.  
Solid lines use the numerical implementation of
the full expressions obtained by Mathematica, and a numerical
calculation of the derivatives of $\Qlm$ and $\flm$.  Dashed and dotted
lines
use expressions for the Fisher matrix
to leading order in $1/\Qlm$;
derivatives
of $\Qlm$ and $\flm$ are evaluated numerically for the former, and 
using
the fitting functions given in Appendix \ref{app:QNM} for the latter. 
Black (red) lines refer to calculations
of the error in $j$ ($M$) using the FH ``doubling
prescription''.  
The plot shows that all these different calculations are
in excellent agreement with each other. However, sometimes using the
fits can produce only order-of-magnitude estimates of the errors: this
happens for counterrotating modes, where an accurate calculation of
the derivatives of $\Qlm$ and $\flm$ is more important (dotted lines
in the right panel deviate significantly from the ``true'' answer
obtained by taking numerical derivatives of the QNM tables). Finally (and
only in the right panel) we plot, in blue, results obtained using
Eqs. (4.20a), (4.20b) in Finn's paper \cite{finn}, a numerical
implementation of the full expressions obtained by Mathematica, and a
numerical calculation of the derivatives of $\Qlm$ and $\flm$. Finn's
formula would lie on top of the other lines in the left plot (for
$l=m=2$), but it gives a slightly different prediction for the errors
on the counterrotating mode with $l=2$, $m=-1$.
\label{errs-fh}}
\end{center}
\end{figure*}

\subsection{Numerical results}

Our expectations are validated by an explicit numerical calculation of
the errors. We carry out this calculation in different ways, and
in Fig.~\ref{errs-fh} we show that results obtained from these
different methods are usually in very good agreement. The most
reliable calculation is ``fully numerical'' (solid lines in
Fig.~\ref{errs-fh}), in the sense that it involves no semianalytical
approximations. In this calculation we use the ``complete''
expressions for the errors obtained using Mathematica. For increased
accuracy, for any given value of $j$ we interpolate our numerical
tables of the QNMs by fifth-order polynomials and evaluate
``numerically'' the derivatives $\flm'$ and $\Qlm'$ by taking
derivatives of these interpolating polynomials at the given
$j$. Dashed lines investigate the accuracy of leading order Taylor
expansions for large $\Qlm$ of the errors, such as Eqs.~(\ref{rii}),
and use these ``local'' interpolating polynomials to compute $\flm'$
and $\Qlm'$. Finally, dotted lines use Taylor expansions of the errors
and evaluate the derivatives $\flm'$ and $\Qlm'$ using the (somehow
less accurate) ``global'' fits of the QNM tables we provide in
Appendix \ref{app:QNM}.

Overall, different choices for the doubling convention and/or for the
number of parameters yield consistent results for the errors.
In isolated, unfortunate cases the ``global'' fits of Appendix
\ref{app:QNM}, being valid in the whole range $j\in[0,1]$ but not
very accurate for certain values of $j$, provide only an
order-of-magnitude estimate of the errors. This happens, for example,
when we consider counterrotating modes with $l=2$ and $|m|<l/2$ (see
eg. the right panel of Fig.~\ref{errs-fh}). It also happens for modes
with $m=0$, in the limit $j\to 0$. The reason is that in these cases
$\Qlm$ has a minimum, and $\Qlm/\Qlm'$ blows up. This behavior is only
captured by an accurate {\it local} fit of the QNM data (such as the
polynomial fit we use to compute derivatives). The bottom line is that
the {\it global} fits of Appendix \ref{app:QNM} are generally
inaccurate to compute measurement errors whenever the numerical QNM
tables are such that $\Qlm'\simeq 0$ for some value of $j$.

\begin{figure*}[t]
\begin{center}
\begin{tabular}{cc}
\epsfig{file=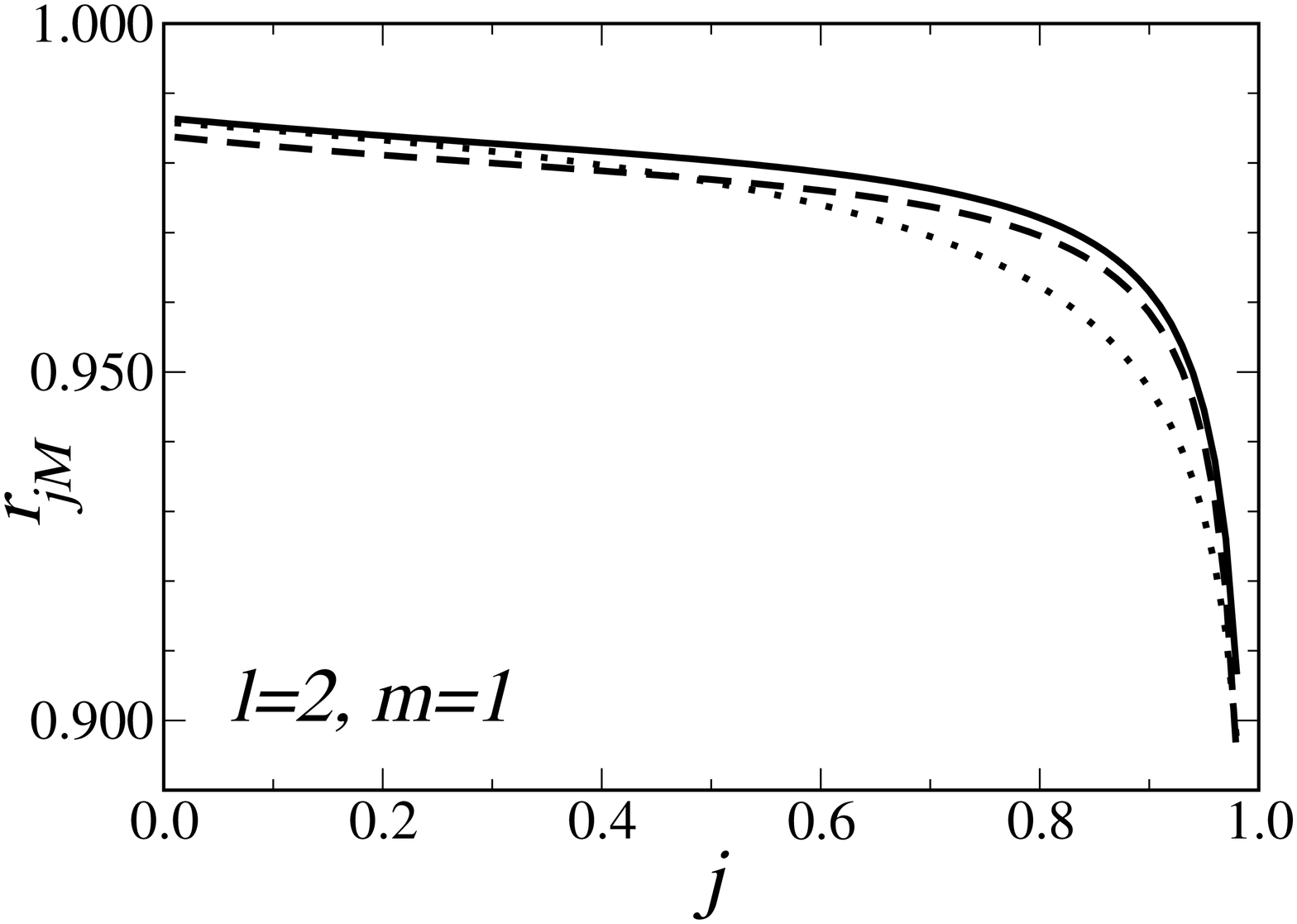,width=4.5cm,angle=-90} &
\epsfig{file=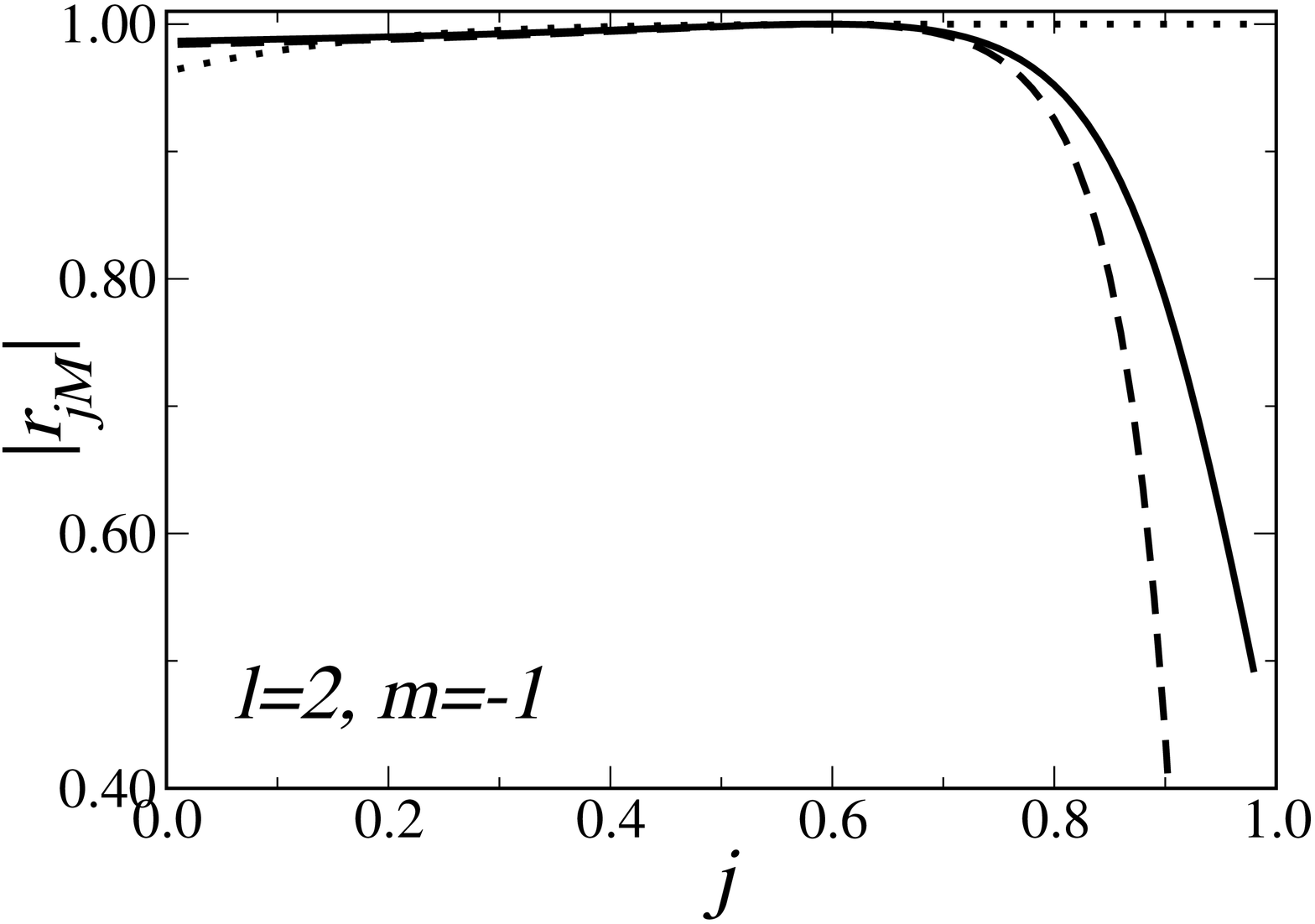,width=4.5cm,angle=-90} \\
\end{tabular}
\begin{tabular}{ccc}
\epsfig{file=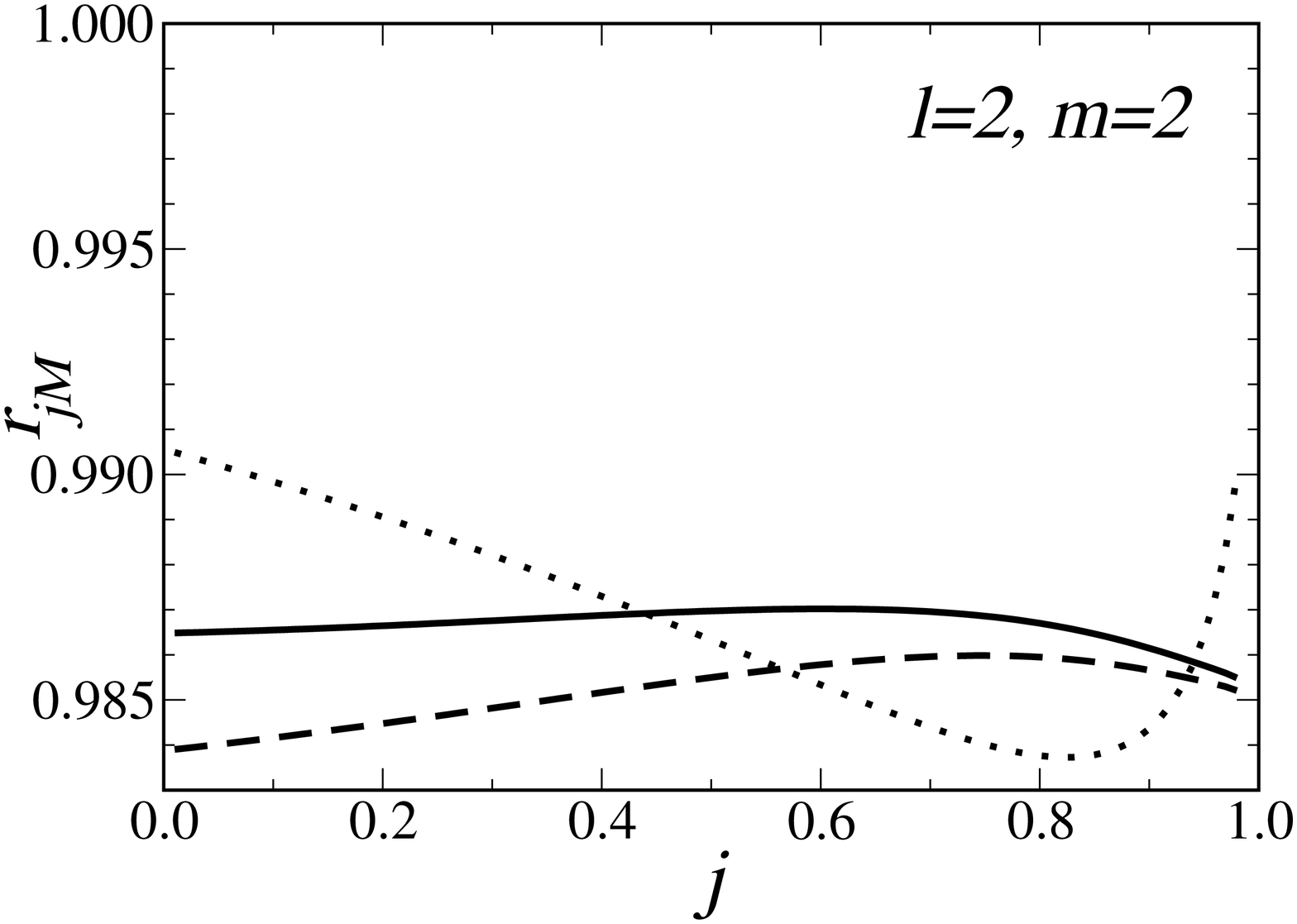,width=4.5cm,angle=-90} &
\epsfig{file=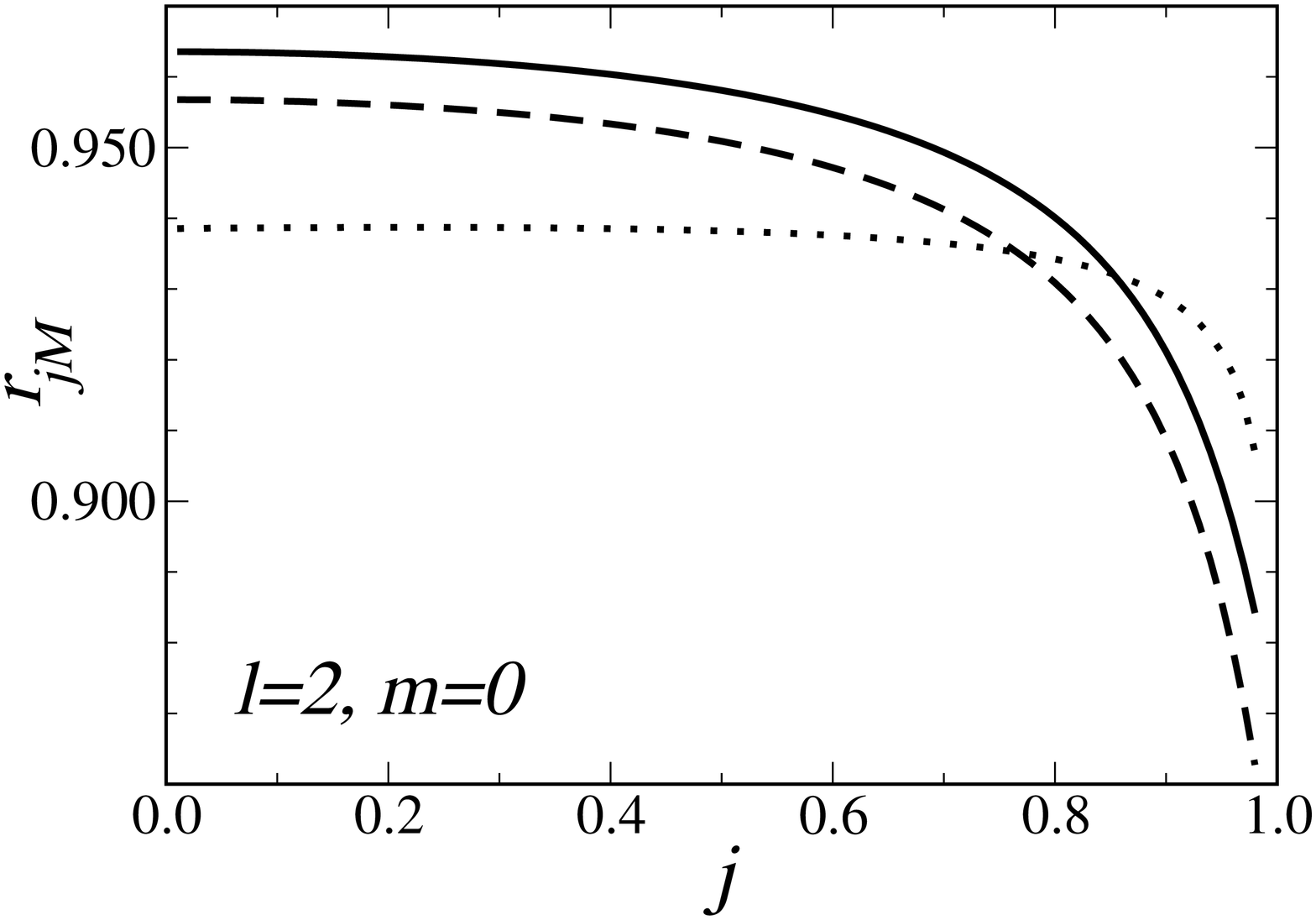,width=4.5cm,angle=-90} &
\epsfig{file=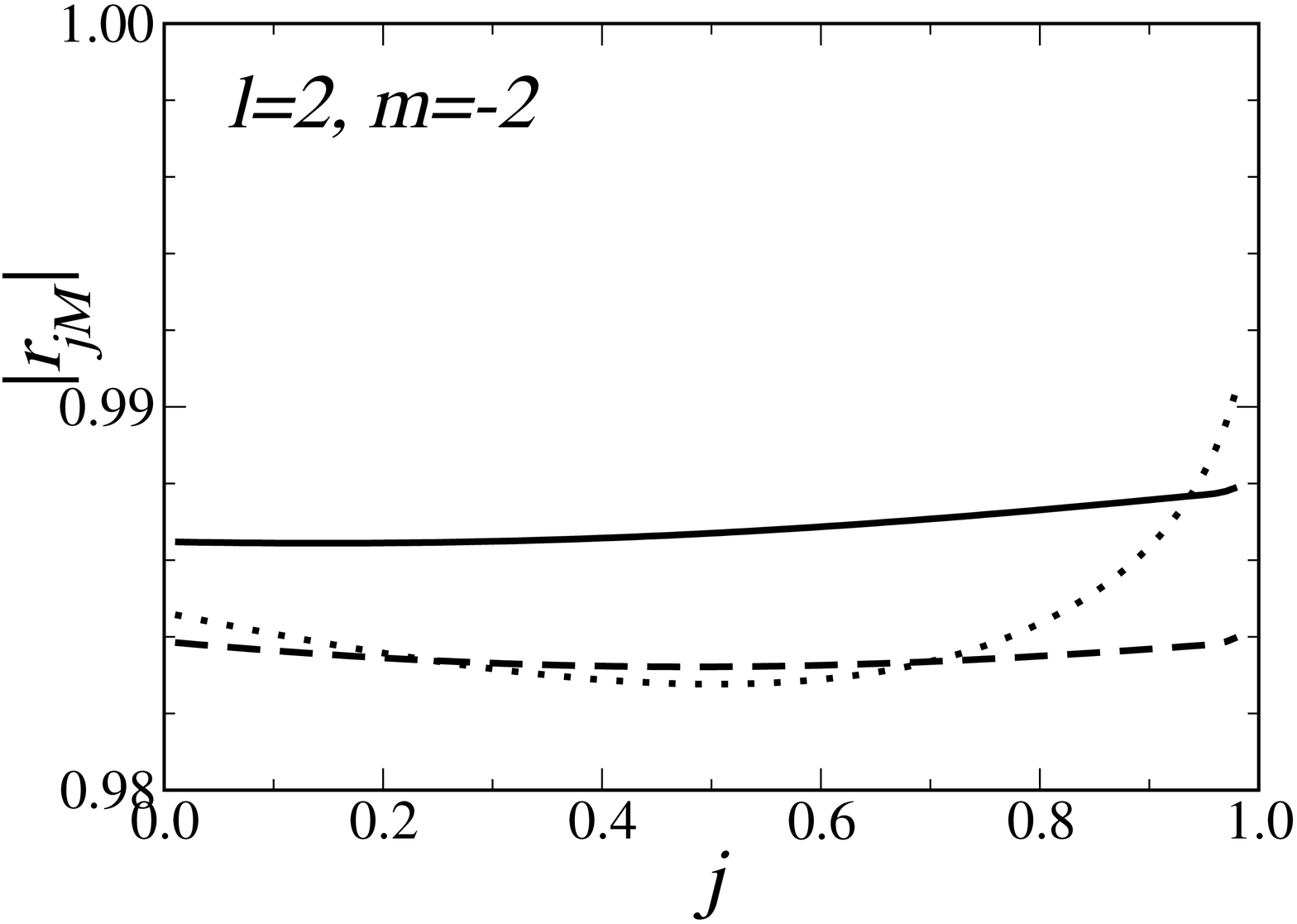,width=4.5cm,angle=-90} \\
\end{tabular}
\caption{Correlation between the mass and angular momentum parameters
for the fundamental mode and different values of $(l,m)$.  For
counterrotating modes $j$ and $M$ are actually {\it anticorrelated}
($r_{jM}<0$), and we plot the modulus of the correlation. The solid
line uses the numerical implementation of the full expressions
obtained by Mathematica, and a numerical calculation of the
derivatives of $\Qlm$ and $\flm$. The dashed line uses an expansion to
leading order in $1/\Qlm$ of the correlation matrix,
and a numerical calculation of the derivatives
of $\Qlm$ and $\flm$.  The dotted line uses again a Taylor expansion,
but this time we evaluate the derivatives of $\Qlm$ and $\flm$ using
the fitting functions given in Appendix \ref{app:QNM}.  The minimum
in the correlation between mass and angular momentum (for $l=m=2$)
disappears when we use the numerical derivatives. The behavior of
$r_{jM}$, as computed numerically, is qualitatively consistent with
the entry marked by $f_{Ma}$ in Echeverria's Table II
\cite{echeverria}, so the fake minimum really seems to be due to the
inaccuracy of the fit.
\label{corr-fh-all}}
\end{center}
\end{figure*}

Let us now turn to correlation coefficients. To leading order, the
correlation coefficient between mass and angular momentum $r_{jM}=1$
for all modes. This high correlation between mass and angular momentum
was first pointed out by Echeverria \cite{echeverria} for the
fundamental mode with $l=m=2$. Echeverria suggested that, if we have
some independent and more precise measurement of either the mass or
the angular momentum (but not both) we could exploit this strong
correlation to obtain an almost equally better estimate of the other
parameter. This means that mass measurements of SMBHs 
(as inferred from the Keplerian orbits of the surrounding stars, for
example)
could be used in conjuction with gravitational-wave observations to
provide accurate determinations of the hole's angular momentum. Mass
and angular momentum are also strongly correlated with the wave's
amplitude: to leading order, $|r_{jA}|=|r_{MA}|=1/\sqrt{2}\simeq
0.707$. On the contrary, the polarization phase $\ph$ is very weakly
correlated with the other parameters: the leading-order term is
proportional to $\Qlm^{-2}$ when we use the FH convention, and to
$\Qlm^{-1}$ when we use the EF convention,
Eq.~(\ref{finncorrs4}). Independently of our convention and of the
parametrization of the waveform, this small correlation implies that
we can expect the phase to be irrelevant in measuring the black hole
mass and angular momentum.


In Fig.~\ref{corr-fh-all} we present results from a numerical
calculation of the correlation coefficient $r_{jM}$ (which is, again,
independent of different choices on the doubling convention and/or on
the number of parameters).

\begin{figure*}[t]
\begin{center}
\begin{tabular}{cc}
\epsfig{file=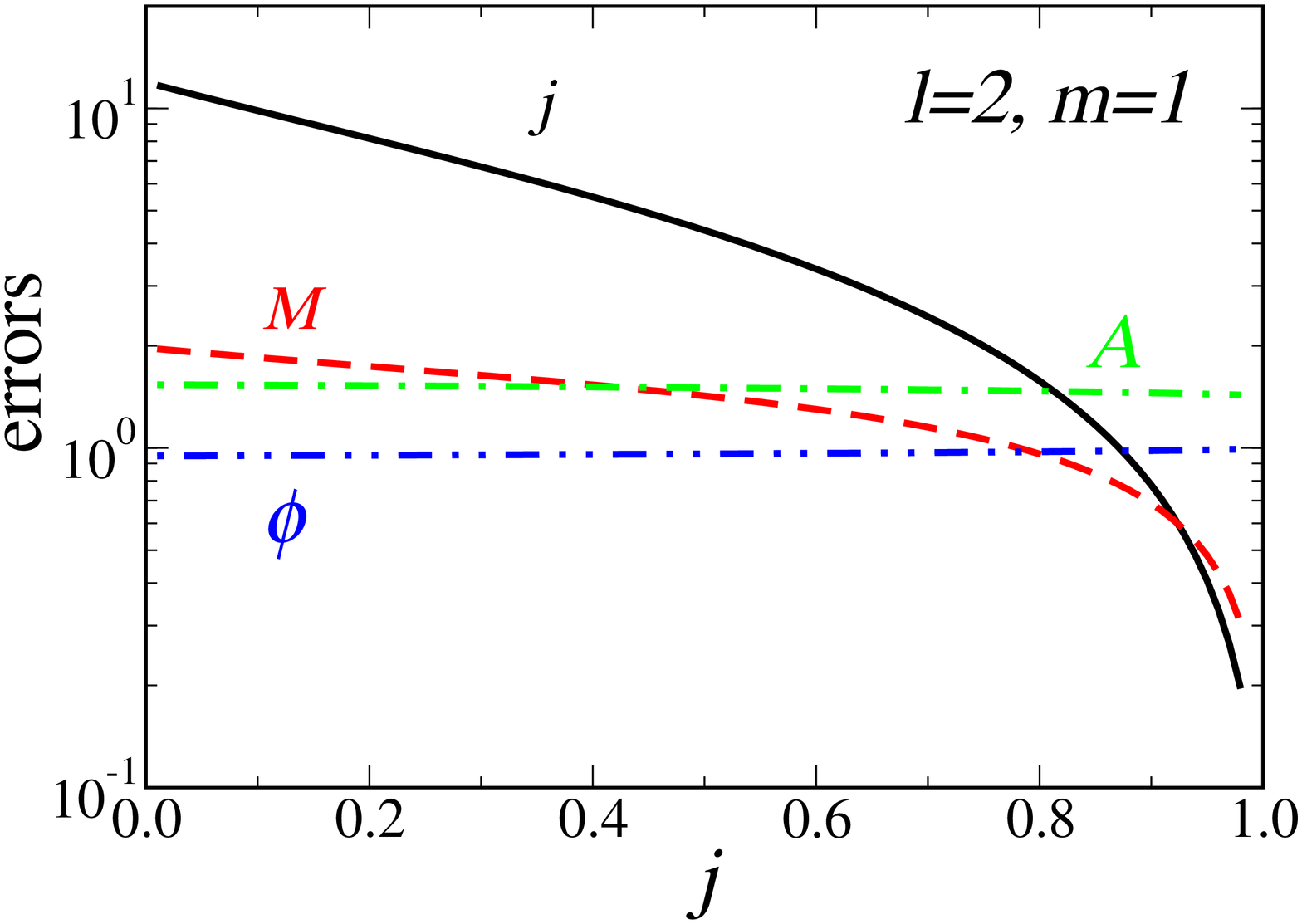,width=4.5cm,angle=-90} &
\epsfig{file=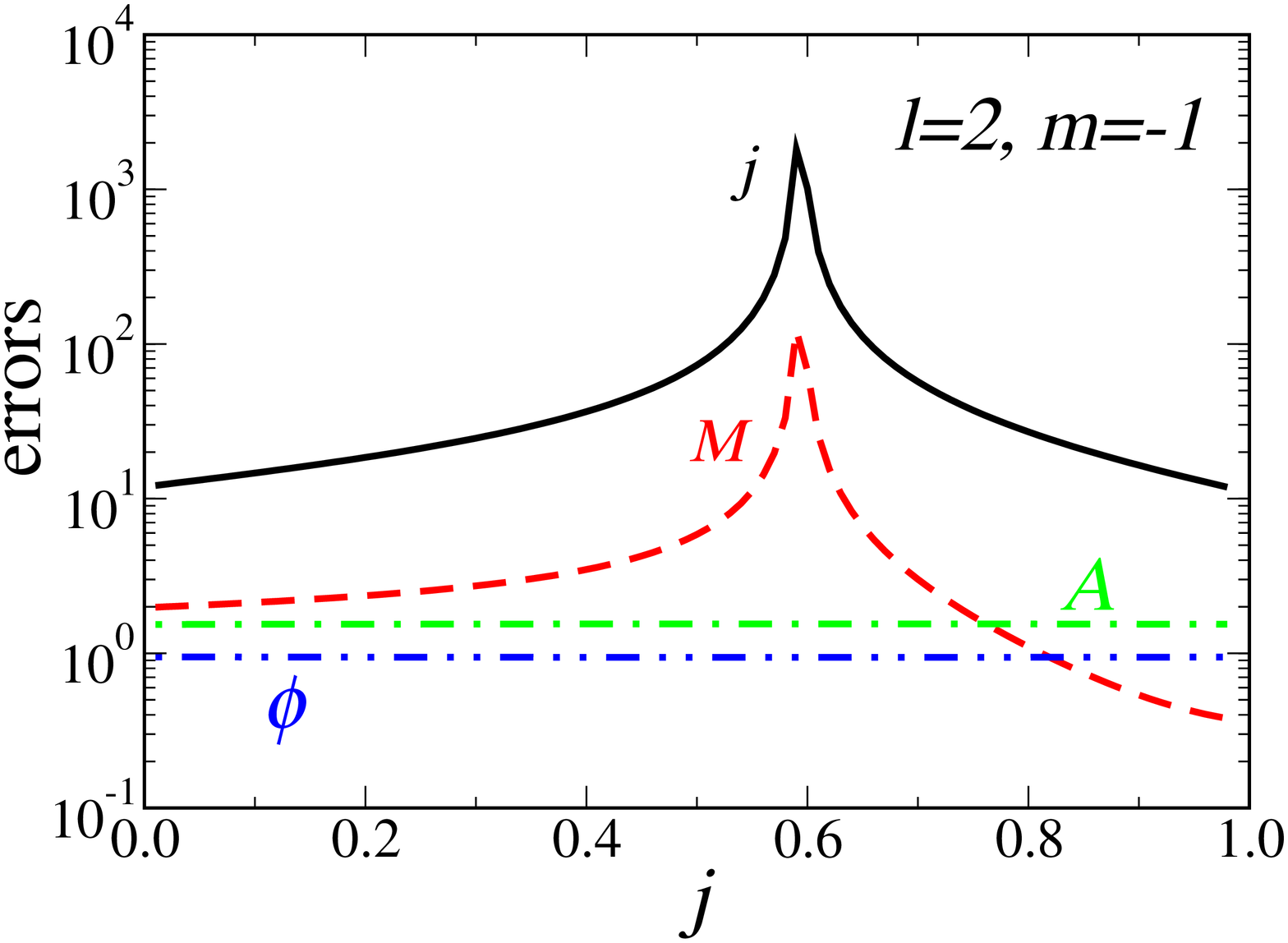,width=4.5cm,angle=-90} \\
\end{tabular}
\begin{tabular}{ccc}
\epsfig{file=l2m2all.ps,width=4.5cm,angle=-90} &
\epsfig{file=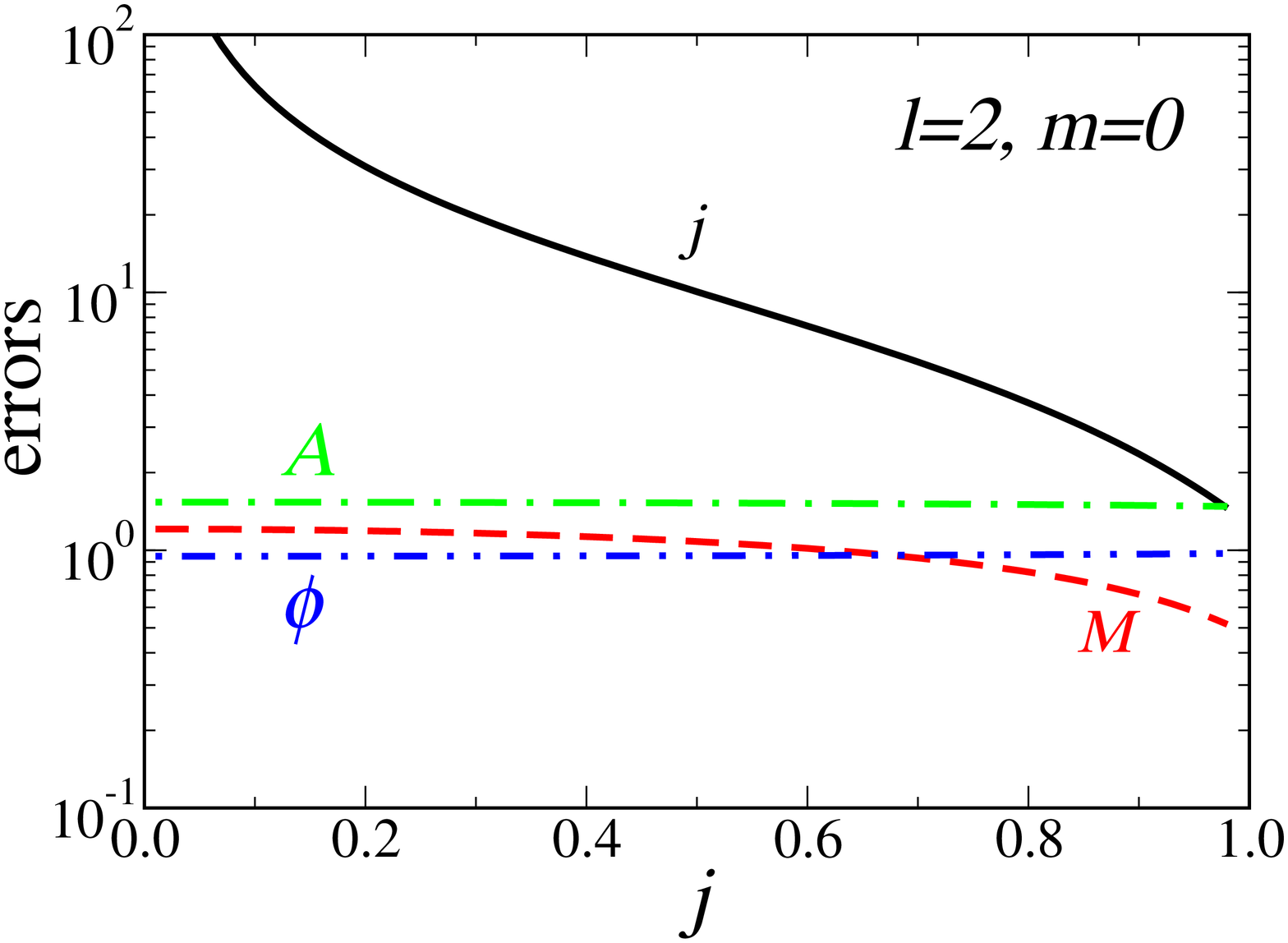,width=4.5cm,angle=-90} &
\epsfig{file=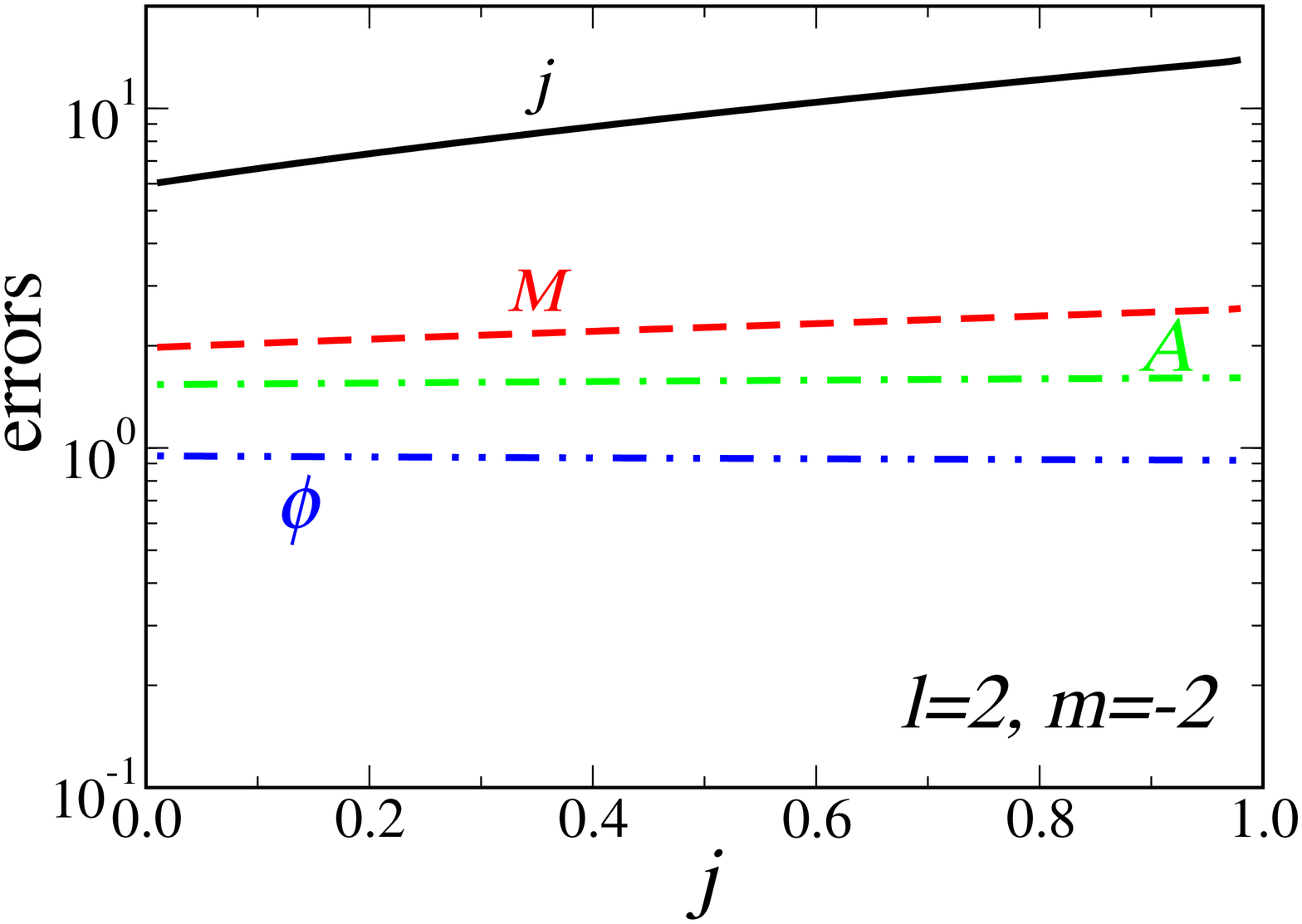,width=4.5cm,angle=-90} \\
\epsfig{file=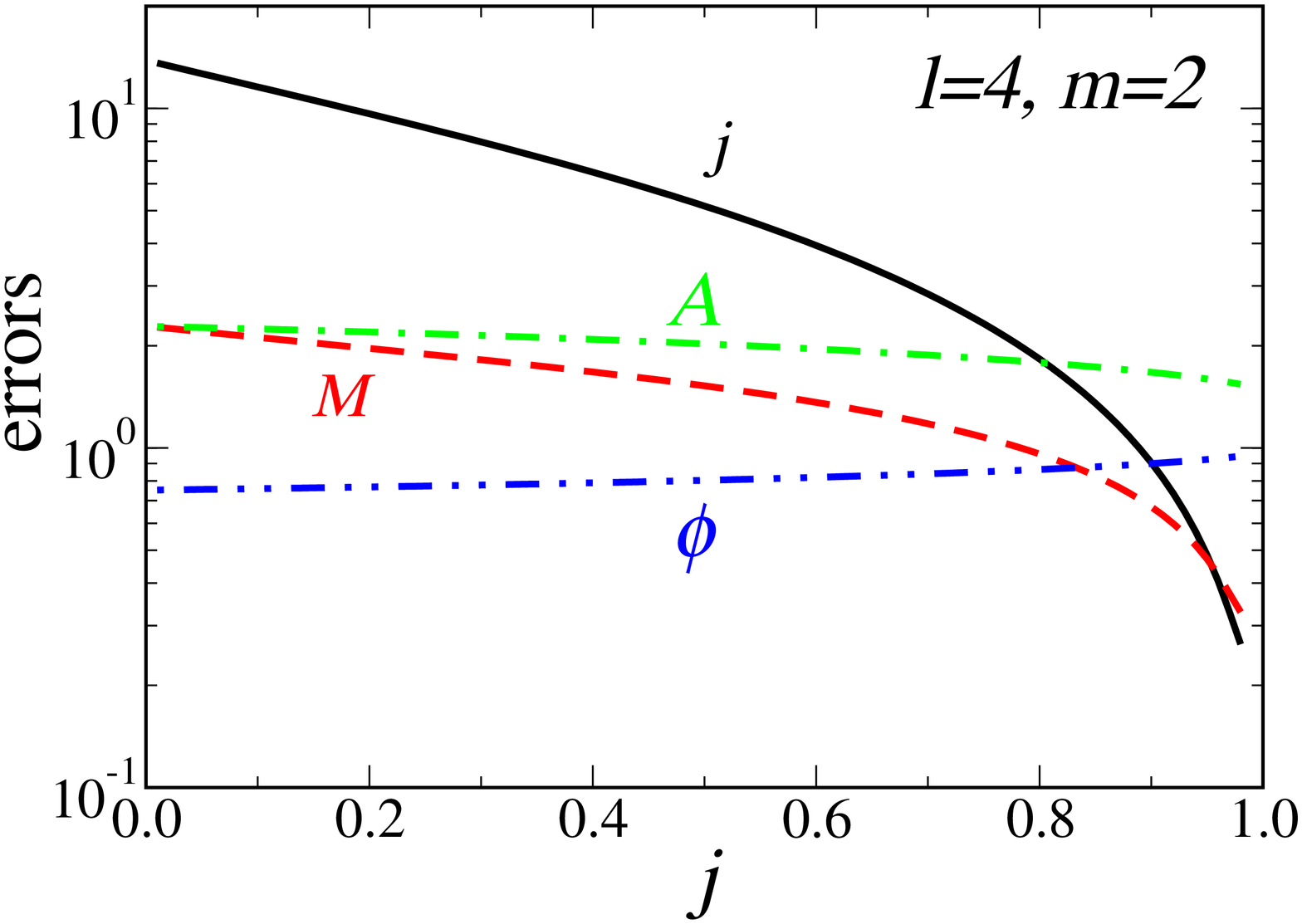,width=4.5cm,angle=-90} &
\epsfig{file=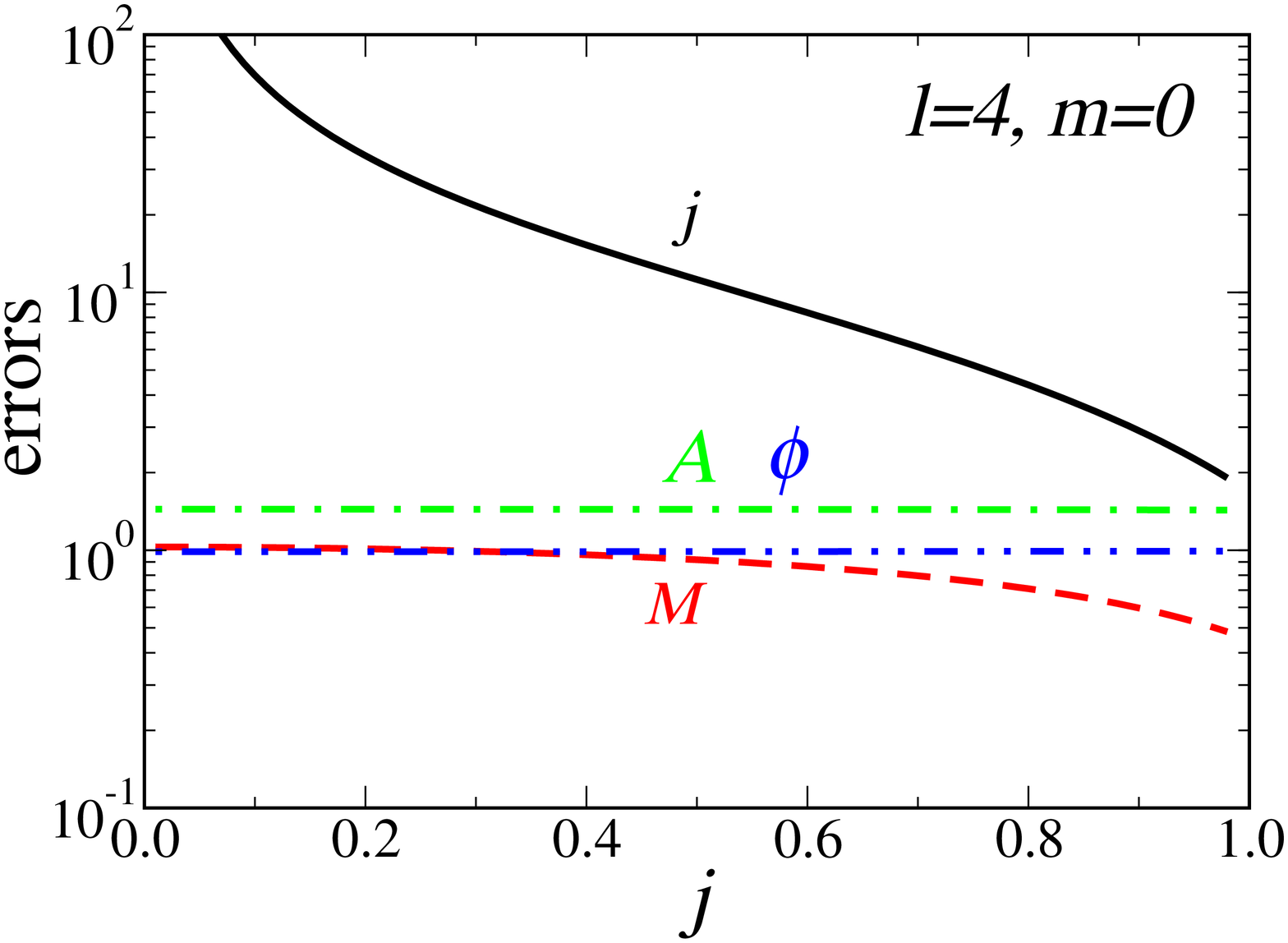,width=4.5cm,angle=-90} &
\epsfig{file=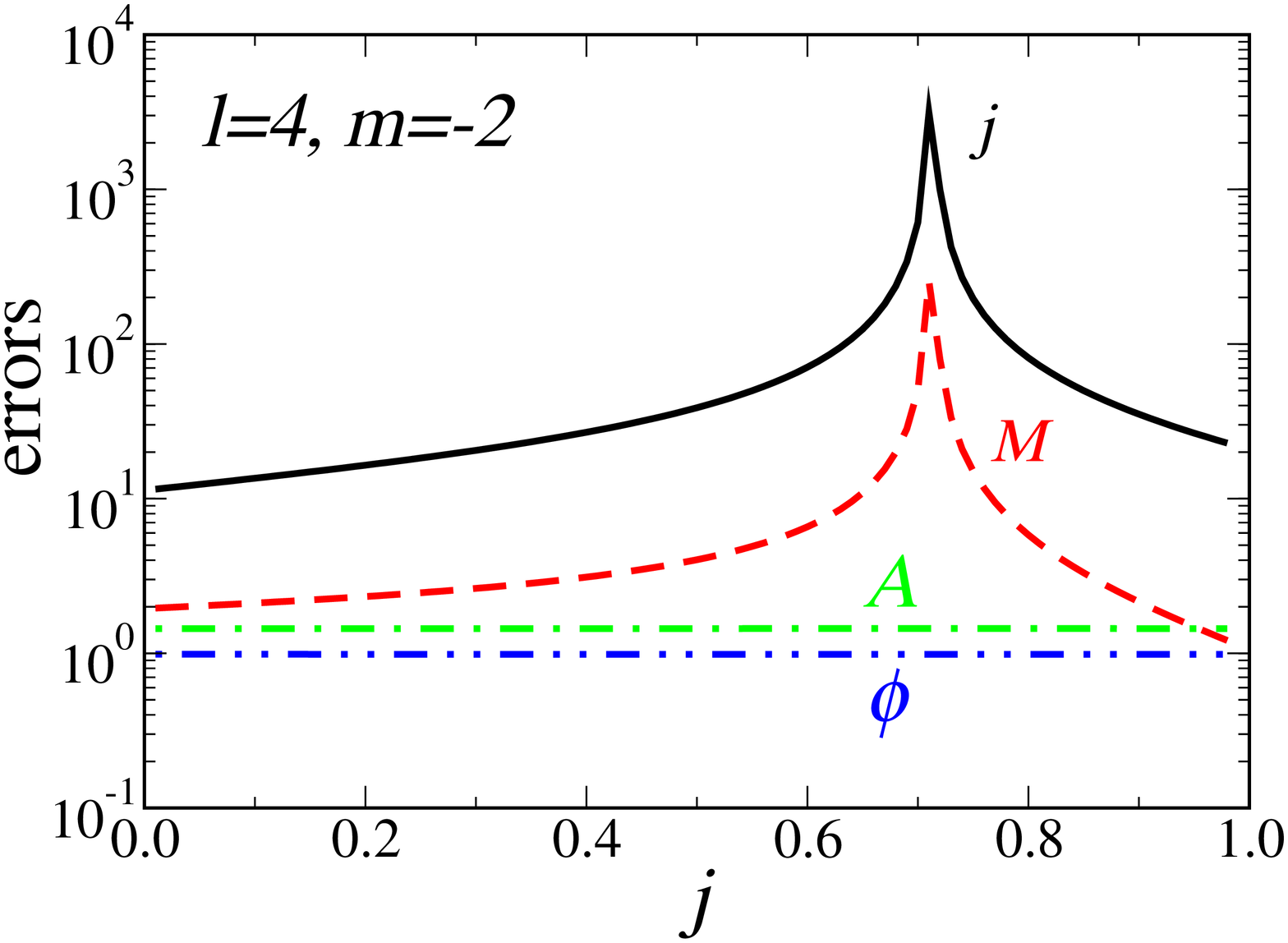,width=4.5cm,angle=-90} \\
\end{tabular}
\caption{Errors in measurements of different parameters for the
fundamental mode with different values of $(l,m)$, as functions of the
angular momentum parameter $j$. Solid (black) lines give $\rho
\sigma_j$, dashed (red) lines $\rho \sigma_M/M$, dot-dashed (green)
lines $\rho \sigma_A/A$, dot-dot-dashed (blue) lines $\rho
\sigma_{\ph}$.
\label{errs-fh-all}}
\end{center}
\end{figure*}

Fig.~\ref{errs-fh-all} shows the errors in different parameters
rescaled by the SNR for different QNMs. All errors scale with the
inverse of the SNR, $\sigma_x\sim \rho^{-1}$, and all information
about the detector is contained in the SNR. Therefore we plot the
quantities $(\rho \sigma_j,~\rho \sigma_M/M,~\rho \sigma_A/A,~\rho
\sigma _{\ph})$, which are, in some
sense,``universal'': they do not depend on the specifics of the
\lisa~noise curve, but only on intrinsic features of the gravitational
waveform emitted by the perturbed black hole.

In the plot, we use the FH convention and consider a
four-dimensional correlation matrix, but results would not have changed
much had we used the EF convention and/or a six dimensional
correlation matrix.  We use the numerical implementation of the full
expressions obtained by Mathematica, and a numerical calculation of
the derivatives of $\Qlm$ and $\flm$. Once again, results do not change
appreciably if we use a Taylor expansion to leading order in $1/\Qlm$
of the correlation matrix, and a
numerical calculation of the derivatives of $\Qlm$ and $\flm$.  Even
if we use the QNM fits of Appendix \ref{app:QNM} we get very similar
results: only for counterrotating modes do those fits fail to
reproduce the location
of the peak we can see for $(l=2,~m=-1)$ and $(l=4,~m=-2)$.

The general features emerging from Fig.~\ref{errs-fh-all} agree with
the expectations drawn from Fig.~\ref{QQa}.  Errors
on mass and angular momentum decrease with rotation for corotating
modes ($m>0$), but they {\it increase} with rotation for
counterrotating modes ($m<0$), blowing up at those ``critical values''
of $j$ for which $\Qlm'=0$ (which occurs for counterrotating modes
with $|m|\leq l/2$). The error on $j$ goes to infinity as $j\to 0$
when we
consider modes with $m=0$, but the error on $M$
stays finite in this same limit. Errors on the amplitude $A$ and
phase $\ph$ usually have a very weak dependence on $j$.

\begin{figure*}[t]
\begin{center}
\begin{tabular}{ccc}
\epsfig{file=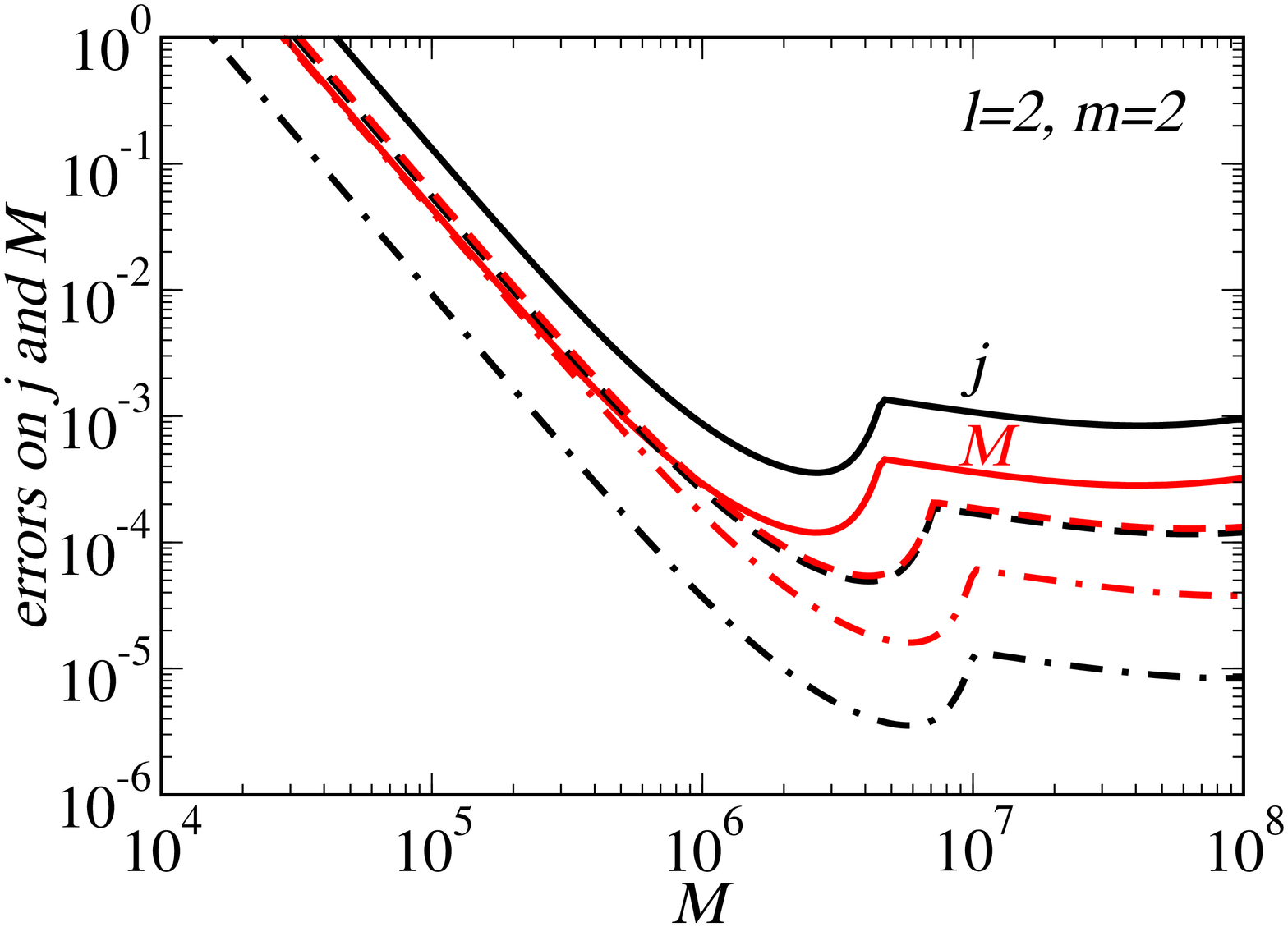,width=4.5cm,angle=-90} &
\epsfig{file=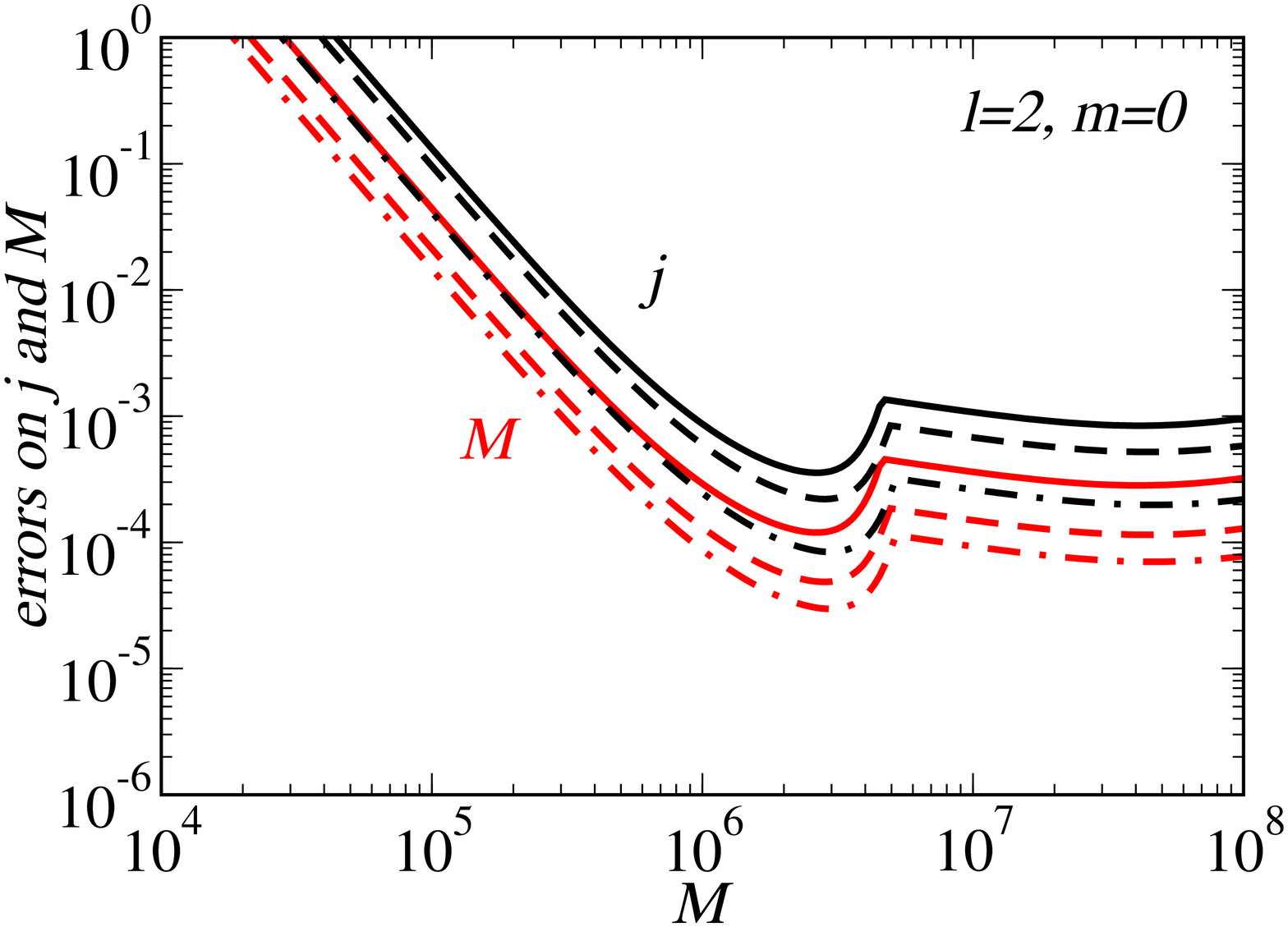,width=4.5cm,angle=-90} &
\epsfig{file=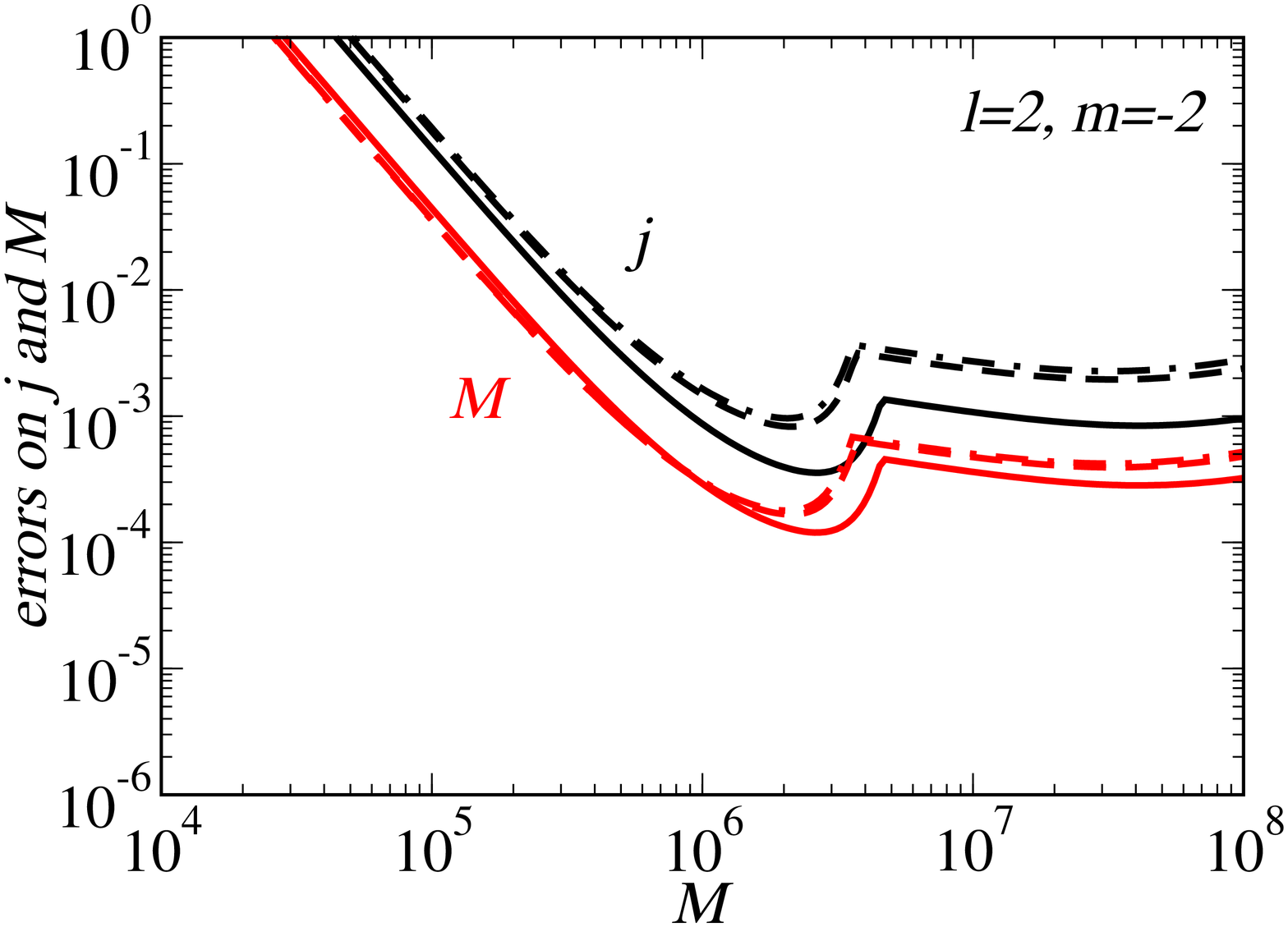,width=4.5cm,angle=-90}\\
\end{tabular}
\begin{tabular}{ccc}
\epsfig{file=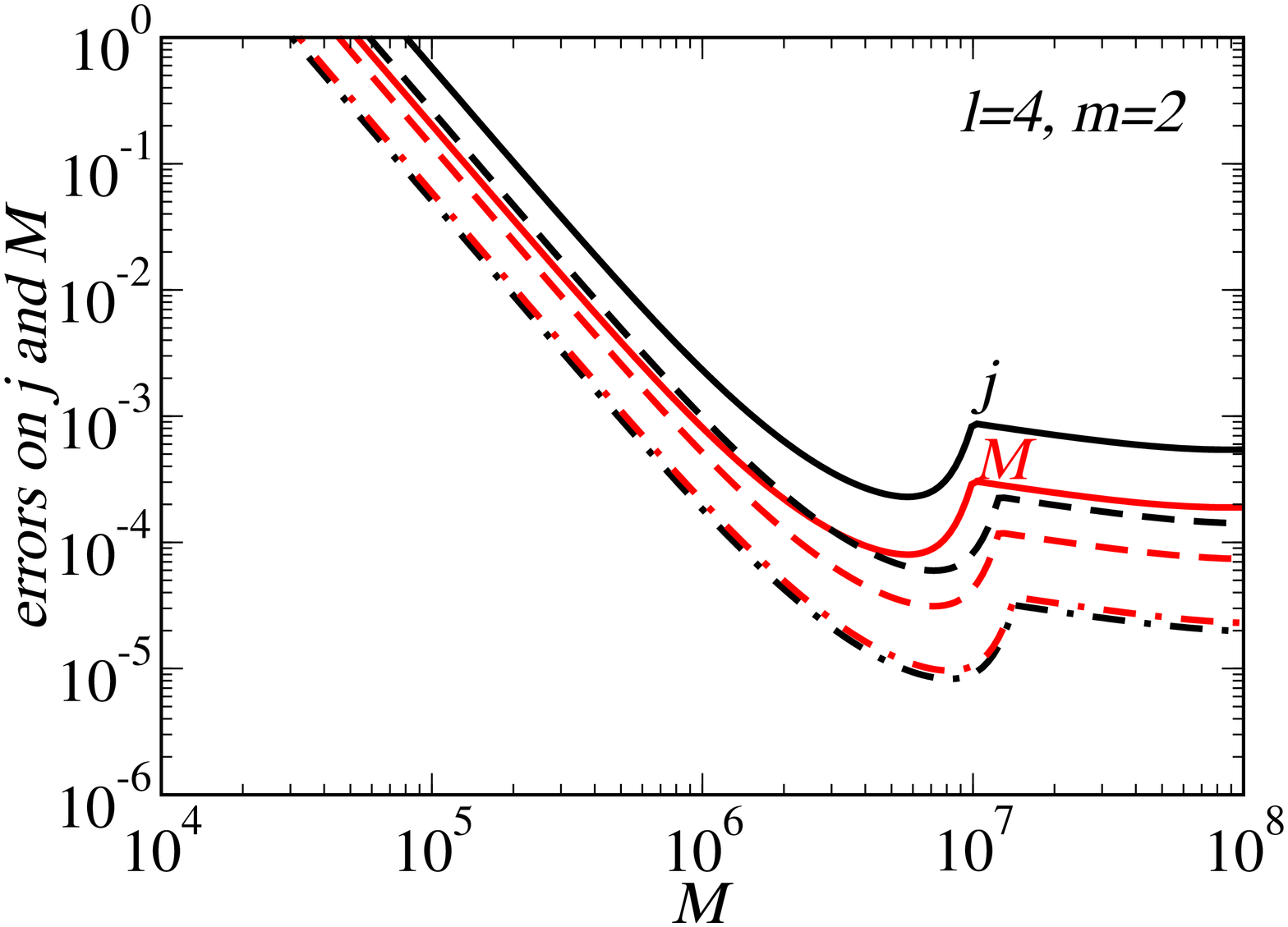,width=4.5cm,angle=-90} &
\epsfig{file=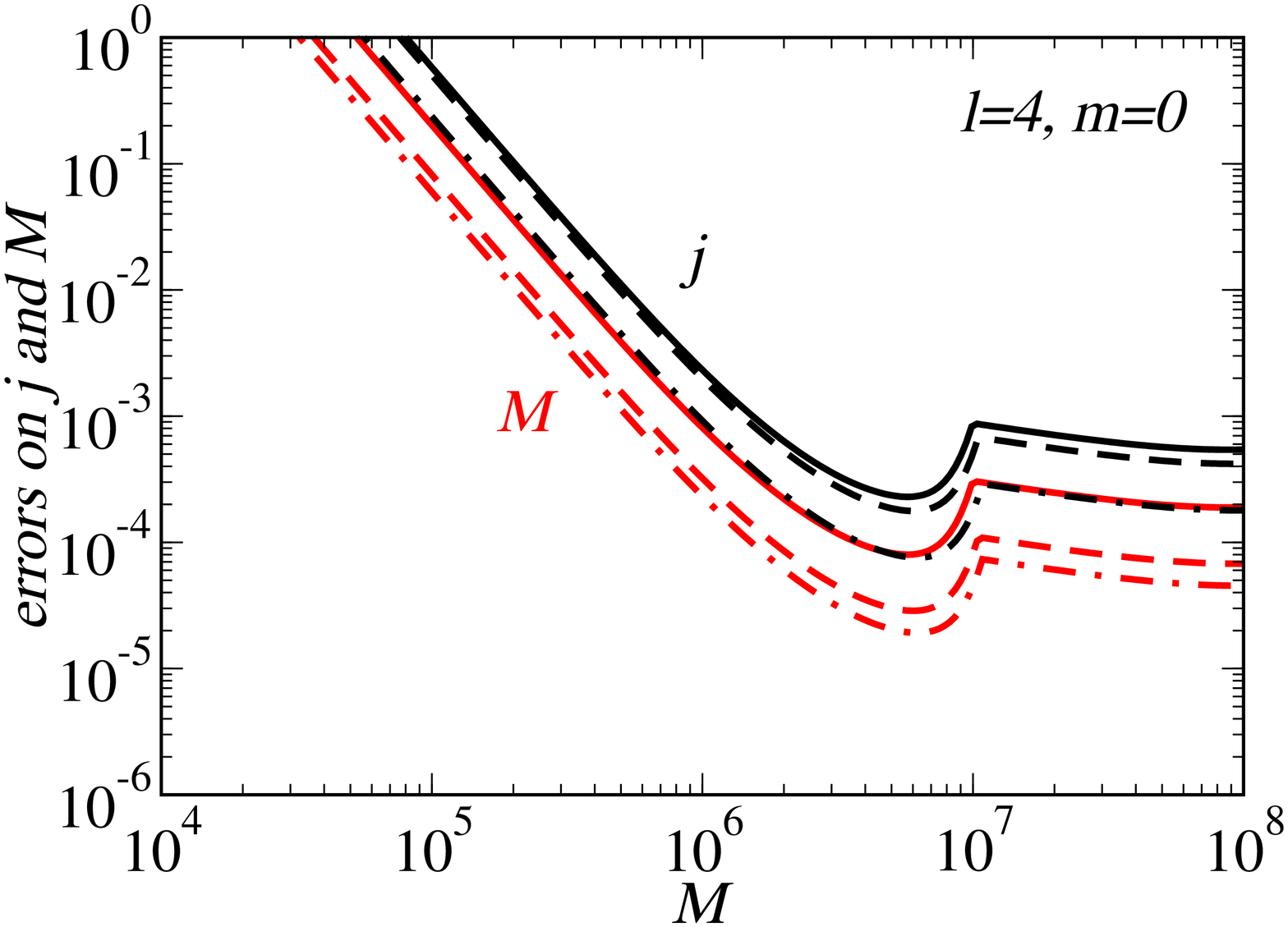,width=4.5cm,angle=-90} &
\epsfig{file=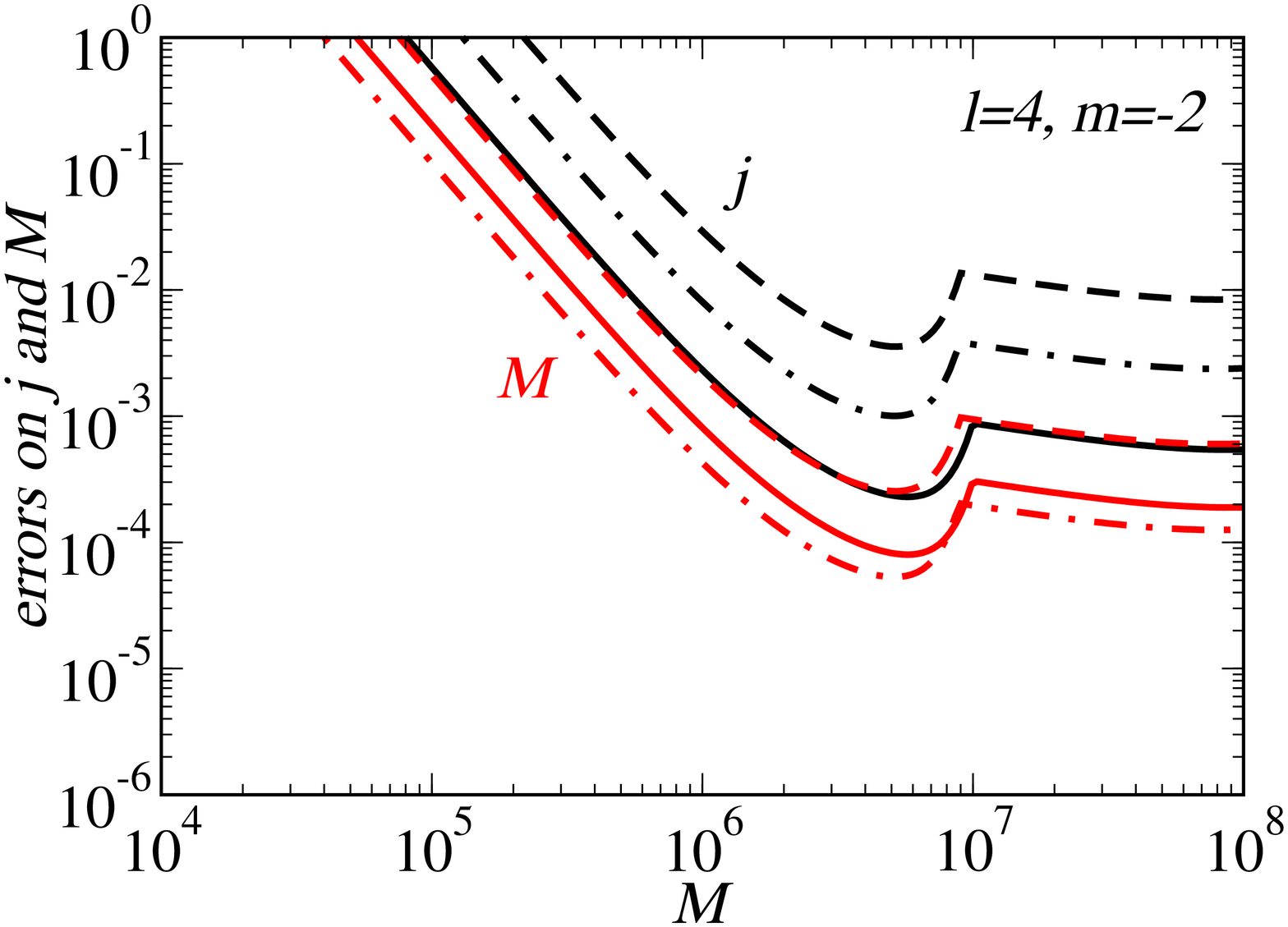,width=4.5cm,angle=-90}\\
\end{tabular}
\caption{Errors in angular momentum $\sigma_j$ (black) and mass
$\sigma_M/M$ (red) for: $l=m=2$ (top left), $l=2$, $m=0$ (top center),
$l=2$, $m=-2$ (top right); $l=4$, $m=2$ (bottom left), $l=4$, $m=0$
(bottom center), $l=4$, $m=-2$ (bottom right), for a source at $D_L=3 \,{\rm
Gpc}$ with $\epsilon_{\rm rd}=3\%$.  Solid lines refer to
$j=0$, dashed lines to $j=0.8$, dot-dashed lines to $j=0.98$. Errors
scale with $\rho^{-1}$, hence they are proportional to $\epsilon_{\rm
rd}^{-1/2}$.
\label{errs-aM}}
\end{center}
\end{figure*}

Of course Fig.~\ref{errs-fh-all} tells only part of the story, because
it does not involve any information about the actual specifics of the
\lisa~noise. Fig.~\ref{errs-aM} shows the actual errors computed using the
Barack-Cutler noise curve including white-dwarf confusion noise
described in Appendix \ref{app:noise}.  We
use the $\delta$-function approximation to compute
$\rho$, and consider different angular indices $(l,m)$. For
concreteness we assume that our source is located at $D_L=3$~Gpc and
has a ringdown efficiency $\epsilon_{\rm rd}=3 \%$ for each
mode. However, even in the worst-case equal-mass merger 
scenario ($\epsilon_{\rm
rd}=0.1 \%$) errors would only increase by a modest factor
$\sqrt{30}\simeq 5.5$.

The angular momentum dependence can most easily be understood looking
at the ``universal'' Fig.~\ref{errs-fh-all}. As an example, focus on
the mode with $l=m=2$. From Fig.~\ref{errs-fh-all} we see that the
error on $j$ is larger than the error on $M$ for $j\lesssim 0.8$,
comparable to it for $j\simeq 0.8$, and smaller than the error on $M$
for $j\gtrsim 0.8$: this is precisely what we see in the top-left
panel of Fig.~\ref{errs-aM}. Including information on the noise,
Fig.~\ref{errs-aM} gives a good quantitative idea of the kind of
accuracy we can achieve if we try to measure mass and angular momentum
of black holes with \lisa. Numerical results are in good agreement
with our expectations based on Eqs.~(\ref{parametrizeM}) and
(\ref{parametrizej}). Errors become unacceptably large only for black
hole masses $M\lesssim 10^5 M_\odot$, but in general we can expect
excellent accuracies: the measurement of a single ringing event can
provide the mass {\it and} angular momentum of black holes with
$M\gtrsim 5\times 10^5 M_\odot$ with errors smaller than one part in
$10^2$, and (in the most optimistic cases, eg. for black holes of mass
$\sim 5\times 10^6 M_\odot$) smaller than one part in $10^5$.

To generalize our results to other values of $\epsilon_{\rm rd}$ it is
enough to recall that the SNR scales with the square root of
$\epsilon_{\rm rd}$, $\rho\sim \epsilon_{\rm rd}^{1/2}$, so the errors
scale like $\epsilon_{\rm rd}^{-1/2}$. Numerical simulations suggest
that for the first overtone $\epsilon_{\rm rd}$ should be smaller by a
factor $10^2-10^4$ when compared to the fundamental mode. This means
that the error will be $\sim 10-10^2$ times larger than the value
plotted in Fig.~\ref{errs-aM}. Accurate tests of the no-hair theorem
involve the measurement of two QNM frequencies, so they {\it could} be
possible only in a very limited mass range. This conclusion could be
made even more pessimistic when we consider interference effects in
the multi-mode situation.

All of our numerical results on the errors have been obtained assuming
$N_\times=1$, $\ph^\times=0$ and $\ph^+=0$ (values have been
chosen to agree with FH), but the results we have presented so far can be
considered more general and robust.  In fact, the errors are almost
completely independent of $N_\times$ and $\ph^{+,\times}$.  This is
not only because these parameters only enter in subleading
corrections, but also because the variability of the correction
coefficients with $N_\times$ and $\ph^{+,\times}$ is extremely
weak.  To see this, notice that
the leading-order corrections to the errors
on mass and angular momentum are given in the FH convention
by $(1+4\beta)/16\Qlm^2$, 
and in the EF convention, 
by $(1+\alpha^2 + 2\beta^2)/8\Qlm^2$ [see Eqs. (\ref{erraEF}) and
(\ref{errmEF})] and use the fact that $|\alpha| \le 1$
and $|\beta| \le 1$.

In conclusion, not only are our results on the errors independent of
the way we fold the waveform (EF/FH convention) and independent of
the way we parametrize the waveform (see Fig.~\ref{errs-fh}): they are
also largely independent of $N_\times$ and $\ph^{+,\times}$.

\subsection{Bounding the black hole's mass and angular momentum through
detection of a single mode}

What kind of information can we extract from the detection of the
frequency and damping time of a single QNM?
Although we have parametrized our Fisher matrix formalism in terms of $M$
and $j$, what we really measure are the frequency $f$ and damping time
$\tau$ or quality factor $Q$ of the ringdown wave.  Unfortunately this is
not sufficient to tell us
the values of $(l,m,n)$
corresponding to the mode detected, so we cannot determine the mass
and spin of the black hole uniquely.  
The problem is that
there are several values of the parameters $(M,j,l,m,n)$ that yield the
same frequencies and damping times.  However, if we make the
plausible assumption that the only modes likely to be detected are the
first two overtones with $l=2,~3,~4$, we can narrow the possibilities.

\begin{table}
\begin{center}
\begin{tabular}{|c|c|c|c|c|} \hline
{$Q_{lmn}$} & \multicolumn{1}{c|}{
$(j,l,m,n)$}  \\
\hline
20 & $ (0.988,4,3,0)\,(0.987,3,3,0)\,(0.976,4,4,0) $\\
\hline
19&$ (0.986,3,3,0)\,(0.986,4,3,0)\,(0.973,4,4,0) $\\
\hline
18 &$ (0.984,3,3,0)\,(0.984,4,3,0)\,(0.970,4,4,0)$ \\
\hline
17 &$ (0.982,3,3,0)\,(0.982,4,3,0)\,,(0.965,4,4,0)$ \\
\hline
16&$ (0.979,3,3,0)\,(0.979,4,3,0)\,(0.960,4,4,0) $ \\
\hline
15 &$(0.976,3,3,0)\,(0.975,4,3,0)\,(0.954,4,4,0)$ \\
\hline
13&$ (0.989,2,2,0)\,(0.972,4,4,0)\,(0.970,4,3,0)\,(0.946,4,4,0) $ \\
\hline
12&$ (0.986,3,2,0)\,(0.984,2,2,0)\,(0.961,3,3,0)\,(0.954,4,3,0)\,(0.924,4,4,0)
$ \\
\hline
11&$(0.985,4,2,0)\,
(0.981,2,2,0)\,(0.981,3,2,0)\,(0.952,3,3,0)\,(0.943,4,3,0)\,(0.907,4,4,0) $ \\
\hline
10&$(0.990,4,4,1)\,
(0.977,2,2,0)\,(0.975,3,2,0)\,(0.940,3,3,0)\,(0.926,4,3,0)\,(0.884,4,4,0) $ \\
\hline
9&$(0.988,4,4,1)\,(0.971,2,2,0)\,(0.966,3,2,0)\,(0.961,4,2,0)\,(0.924,3,3,0)\,(0.900,4,3,0)\,(0.851,4,4,0)
$ \\
\hline
8&$(0.984,4,4,1)\,
(0.962,2,2,0)\,(0.951,3,2,0)\,(0.932,4,2,0)\,(0.900,3,3,0)\,(0.861,4,3,0)\,(0.802,4,4,0)
$ \\
\hline
7&$(0.993,2,1,0)\,(0.989,4,3,1)\,(0.988,3,3,1)\,(0.978,4,4,1)\,
(0.976,4,1,0)\,(0.949,2,2,0)\,(0.926,3,2,0)\,(0.875,4,2,0)$\,
\\
&$(0.863,3,3,0)\,(0.794,4,3,0)\,(0.724,4,4,0) $ \\
\hline 6&$
(0.987,2,1,0)\,(0.984,4,3,1)\,(0.984,3,3,1)\,(0.969,4,4,1)\,(0.929,2,2,0)\,
(0.897,4,1,0)\,(0.881,3,2,0)$
\\
&$(0.802,3,3,0)\,(0.757,4,2,0)\,(0.671,4,3,0)\,(0.592,4,4,0) $ \\
\hline 5&$(0.990,2,2,1)
\,(0.976,3,3,1)\,(0.974,4,3,1)\,(0.972,2,1,0)\,(0.954,4,4,1)\,
(0.892,2,2,0)\,(0.859,4,0,0)$
\\
&
(0.785,3,2,0)\,(0.688,3,3,0)\,(0.620,4,1,0)\,(0.492,4,2,0)\,(0.410,4,3,0)\,(0.338,4,4,0)\\
\hline 4&$ (0.985,3,2,1)\,(0.984,2,2,1)\,
(0.961,3,3,1)\,(0.954,4,3,1)\,(0.936,3,0,0)
$ \\
&
(0.929,2,1,0)\,(0.924,4,4,1)\,(0.816,2,2,0)\,(0.677,3,1,0)\,
(0.544,3,2,0)\,(0.441,3,3,0)
  \\ \hline
3&
(0.971,2,2,1)\,(0.965,3,2,1)\,(0.961,4,2,1)\,(0.924,3,3,1)\,(0.901,4,3,1)\,
(0.851,4,4,1)\\
&
(0.772,2,1,0)\,(0.620,2,2,0)\,(0.409,3,-2,0)\,(0.247,3,-3,0)
 \\ \hline
2&(0.990,2,-1,0)\,(0.987,2,1,1)\,(0.981,3,1,1)\,(0.929,2,2,1)\,(0.906,4,1,1)\\
&
(0.883,3,2,1)\,(0.805,3,3,1)\,(0.765,4,2,1)\,(0.681,4,3,1)\,(0.601,4,4,1)
\,(0.150,2,-2,0)\\
\hline 1& (0.803,2,1,1)\, (0.656,2,2,1)\,
(0.148,3,-2,1)\,(0.088,3,-3,1) \\
\hline 0.5& (0.628,2,-2,1) \\ \hline
\end{tabular}
\end{center}
\caption{\label{tab:Quadruples} Different quadruples $(j,l,m,n)$
yielding the same $\Qlm$.}
\end{table}

Suppose that we measure a specific value of $Q_{lmn}$.  In Table
\ref{tab:Quadruples} we list the different values of $(j,l,m,n)$
yielding the same $\Qlm$.  The number of modes in our constrained set that
correspond to that value ranges from one to about a dozen.
Each of these modes then corresponds to a unique value of 
${\cal F}_{lmn} = 2\pi
M f_{lmn}$.  From the measured $f_{lmn}$, we can obtain a discrete list of
{\it provisional}, accurately measured masses $M$.  
This list cannot be narrowed further without additional information, such as
an estimate or bound obtained from the inspiral waveform, or the detection
of an additional QNM.  

Nevertheless, some potentially useful bounds may be 
obtained from detection of a single mode.
Suppose we observe a $Q_{lmn}$ larger than
(say) 10.  According to Table~\ref{tab:Quadruples}
we cannot determine $(l,m,n)$, but we
can impose a lower bound on $j$ of about $0.88$.

Prospects improve if we assume that we can measure two modes. For
definiteness, suppose one mode has $\Qlm\sim 6$ and the other has
$\Qlm\sim 3$. Since they must belong to a quadruple with the same $j$
(within a measurement error of, say, one percent),
the only possible pairs are, according to Table \ref{tab:Quadruples}:
$(0.969,4,4,1)$, $(0.929,2,2,0)$, $(0.897,4,1,0)$
with $\Qlm=6$; and $(0.971,2,2,1)$,
$(0.965,3,2,1)$, $(0.961,4,2,1)$, $(0.924,3,3,1)$, $(0.901,4,3,1)$
with $\Qlm=3$.
Consider the hypothesis that we have detected the pair of modes
$(0.929,2,2,0)$ with $\Qlm=6$ and $(0.924,3,3,1)$ with $\Qlm=3$. We
can test this by computing the mass of the black hole from
the two different measured ringdown frequencies.  Given the measured
frequency $f_{220}(j=0.929)$, we can invert the relation
$f_{220}(j=0.929)=0.703/(2\pi M)$ to compute $M$, and then repeat
the procedure for the $(0.924,3,3,1)$ mode.  If they yield the same
mass, our hypothesis is correct, and the measurement is compatible
with general relativity.  If the masses don't match, then we may have
detected the mode $(0.969,4,4,1)$ with $\Qlm=6$ and any one of the
modes $(0.971,2,2,1)$, $(0.965,3,2,1)$, $(0.961,4,2,1)$ with
$\Qlm=3$.  To proceed we take the two different measured values of
$f_{lmn}$ and we test the compatibility of the resulting masses.  In
this particular example, a determination of the modes involved could
be possible.  The frequencies corresponding to the three different
modes, $f_{221}(0.971)\sim 0.79/(2\pi M)\,,\,f_{321}(0.965)\sim
0.94/(2\pi M)\,,\,f_{421}(0.961)\sim 1.13/(2\pi M)$, are different
within the required accuracy. 

\section{Multi-mode ringdown waveforms: preliminaries}
\label{excitation}

An accurate measurement of QNM frequencies can provide conclusive
proof of the astrophysical reality of the black hole solutions of
general relativity. 
Most existing studies of ringdown detection
\cite{fh,echeverria,finn,kaa,creighton,nakano,tsunesada}
assume that the waveform can be described using only the fundamental
($n=0$) $l=m=2$ mode of a Kerr black hole.  In many cases, numerical
and perturbative models of astrophysical gravitational wave sources
show that, given radiation with a certain angular dependence - that
is, given $(l,m)$ - a good fit of the waveform requires the two lowest
modes ($n=0$ and $n=1$) \cite{BCW2}.

Even more importantly, as stressed by Dreyer {\it et al.}
\cite{dreyer}, a test of the general relativistic no-hair theorem 
requires the identification of at least {\it two} QNM frequencies in the
ringdown waveform.

The extension of the formalism to multi-mode situations 
raises a number of important questions. Which modes should we
expect to be most relevant in the ringdown waveform?  How much energy
should we attribute to each QNM when a black hole is formed, following
either a galaxy merger or the collapse of a supermassive star? Are the
energy and quality factor of the more rapidly damped modes large enough
for them to be detected? Does it make sense to talk about the SNR of
each QNM, given that (in general) they are not orthonormal in any
well-defined sense? Can we really discriminate between different QNMs,
given that consecutive overtones usually have very similar oscillation
frequencies?
The rest of this paper is devoted to providing 
preliminary answers to these questions.

\subsection{Mathematical issues in the definition of mode excitation}

QNMs are relevant to all systems with radiative boundary
conditions. For some of these systems QNMs are actually a natural
extension of an underlying {\it normal mode} system. Consider for
example the nonradial oscillations of a star. The short periods of
these oscillations are driven by fluid pressures, and their long
damping times are due to the (weak) emission of gravitational
waves. If we omit gravitational radiation damping we end up with a
system that can be analyzed in normal modes. In this case we can
identify the radiated energy coming from each separate oscillation
frequency, and decompose the total radiative power into the fraction
assigned to each frequency.

Black hole QNMs (and, for that matter, also the ``pure spacetime''
$w$-modes of a star) are different. In this case there is a single
timescale (given, in geometrical units, by the black hole mass)
determining both the frequency {\it and} the damping time of the
oscillations. There is no meaningful way to switch off the radiation
damping, and no underlying normal mode system. Mathematically, this is
reflected in QNMs being eigenfunctions of a non-self-adjoint problem
\cite{Beyer,Beyer2,NollertPrice,Nollert}.  

This poses difficulties in the definition of a useful and
rigorous notion of QNM excitation. In fact, Nollert and Price
\cite{NollertPrice,Nollert} conjectured that there is no quantitative
measure of QNM oscillations satisfying {\it all} of the following
three criteria: (1) the measure is independent of a simple (time) shift
of the waveform, (2) the measure can be quantified individually for any
number of modes, so that the single measures add up to the total norm
of the waveform, and (3) the measure is useful to quantify the
excitation (in particular it lies between, say, $0~\%$ and $100~\%$).

Andersson \cite{Nils97} advocated a more practical viewpoint on this
issue. He introduced the following, useful ``asymptotic
approximation'': we require spacetime to be essentially flat in the
region of both the observer {\it and the initial data}, so that
initial data should have (compact) support only far from the black
hole. Under this assumption we can define a mode-decomposition of the
time-domain Green's function which provides an accurate representation
of the mode excitation and is {\it convergent} at late times (see
Fig.~2 of \cite{Nils97} and the related discussion).  Notice however
that we expect QNM excitations to arise from data located close to the
peak of the perturbative (Zerilli or Regge-Wheeler) potential, where
spacetime is certainly not flat.  For this reason it is not clear how
relevant the asymptotic approximation is to realistic scenarios.
The asymptotic approximation was extended to rotating (Kerr) black
holes by Glampedakis and Andersson~\cite{GA}. 
Unfortunately they only computed excitation
coefficients for {\it scalar} perturbations.

\subsection{Physical predictions of the energy distribution between
different modes}
\label{energy-main}

A rough attempt to deal with a two-mode ringdown waveform in the
context of gravitational collapse leading to black hole formation can
be found in \cite{fhh}. Those authors considered modes with $l=m=2$
and $l=2$, $m=0$, distributing energy between the two through a
phenomenological parameter. Is this assumption correct, or should
other modes be considered as well? How does the energy distribution
between modes depend on the physical process deforming the black hole?
Does the present knowledge of black hole ringdown waveforms provide
any information on this energy distribution?  These are the questions
we will try to answer in this Section.

Ideally, to estimate the relative QNM excitation we would like to have
full general relativistic simulations of black hole merger and
ringdown under different assumptions (different black hole mass
ratios, different initial angular momenta, realistic initial
conditions at the orbital innermost stable orbit).
Unfortunately, present state-of-the-art
simulations in numerical relativity do not provide us with long-term
evolutions of black hole mergers, nor with reliable estimates of their
gravitational wave emission (with the exception of a few,
unrealistically symmetric situations). 
Even if we had clean, general relativistic simulations
of black hole mergers, there would be additional complications of
astrophysical nature. Supermassive black hole mergers take place in a
very ``dirty'' galactic environment, and a detailed theoretical model
of the dynamical interaction between black holes and their
surroundings is out of reach, given the present understanding of
galaxy mergers \cite{merritt}.

Despite these difficulties, we can obtain some insight by considering
existing studies of QNM excitation in different {\it idealized}
processes related with black hole formation.  
The following is a brief summary of 
interesting results from the point of view of black hole
excitation. More details can be found in \cite{BCW2}.

\subsubsection{Evolution of distorted black holes in full general
relativity} 

In these simulations, the distortion is sometimes introduced
by considering Misner initial data (corresponding to two black holes at
some given separation), and sometimes evolving initial data
corresponding to gravitational waves (``Brill waves''). The amplitude
of these waves can be large, allowing for the introduction of
non-linear effects not amenable to perturbation theory. Distorted
black hole simulations invariably show that, after a transient
depending on the details of the initial distortion, quasi-normal
ringing dominates the emitted radiation. Most importantly for our
analysis, they provide some insight into how initial data affect the
energy distribution between different QNMs. The simulations usually
monitor the distortion (more precisely, the ratio of polar and
equatorial circumferences) of the black hole horizon, fitting the
numerical data with the fundamental mode and the first overtone. They
typically find that the $l=4$ horizon distortion is $\sim 10^{-4}$
smaller than the $l=2$ component (see eg. Figs. 2 and 3 in
\cite{bernstein2}). 

Quite independently of the initial data and of the black hole spin,
the $l=2$ component carries away $\gtrsim 95\%$ of the gravitational
wave energy, but the character of the initial distortion strongly
affects the energy distribution between the subdominant modes (Table
IV of \cite{brandt}).  To our knowledge, results from only one {\it
nonaxisymmetric} ($m\neq 0$) simulation have been produced so far
\cite{allen}.  Full three-dimensional simulations of distorted black
holes are now computationally feasible, and more work in this
direction is definitely required.

\subsubsection{Simulations of head-on black hole collisions} 

Because of its high degree of symmetry, this process has been studied in
great detail in full general relativity. Existing simulations deal
with equal- as well as unequal-mass black holes, either starting from
rest or having non-zero initial momentum (``boosted'' black
holes). The resulting gravitational waveforms and energy emission are
in surprisingly good agreement with predictions from linear
perturbation theory.  These simulations provide at least
well-motivated and reliable {\it lower bounds} on the energy emitted
in a realistic merger. 

The present best estimates for equal mass black holes starting from
rest predict that the $l=2$ component radiates $\simeq 0.13\%$ of the
black hole mass in gravitational waves \cite{sperhake}. Unfortunately
wave extraction for the $l=4$ component has not been carried out yet,
and in the past the extraction of this component presented a
significant challenge (see Fig.~12 of Ref.~\cite{anninosl4}). Even if
the black holes are not initially at rest, the emitted energy should
be less than about $\sim 0.16 \%$ of the mass \cite{boosted}.

\subsubsection{Black hole formation in gravitational collapse} 

With a few exceptions \cite{shibata}, perturbative and numerical
simulations of gravitational collapse usually concentrate on {\it
stellar mass} black holes, so they are not directly relevant to the
SMBHs observable by~\lisa. Nonetheless, some predictions of the QNM
energy distribution from stellar collapse could carry over (at least
qualitatively) to SMBH formation induced by the
collapse of a supermassive star, and perhaps even to black hole
formation following SMBH merger. 

Perturbative calculations show that a typical core collapse radiates
very little energy (up to $\simeq 10^{-7} M$) in gravitational waves,
and that the energy radiated in $l=3$ is typically two to three orders of
magnitude smaller than the $l=2$ radiation (see eg. Fig. 9 in
\cite{cpm1}). 

A classical, nonperturbative axisymmetric simulation by Stark and
Piran found that the waveform is very similar to the waveform produced
by a particle falling into a black hole, and that the radiated energy
scales with angular momentum $j$ as $E/M\simeq 1.4\times 10^{-3}j^4$,
saturating at $E/M\sim 10^{-4}$ for some critical value of $j$ close to the
extremal value of $j=1$. This
simulation has recently been extended to the three-dimensional case
\cite{baiotti}, still keeping a high degree of axisymmetry (only $m=0$
modes are excited). The new simulations are closer to perturbation
theory than the original calculation by Stark and Piran, predicting a
radiated energy $E\simeq 1.45\times 10^{-6} (M/M_\odot)$. 
They also show that the cross component of the strain
is suppressed by roughly one order of magnitude with respect to the
plus component: in our notation, they predict that the parameter
$N_\times\equiv {\cal A}_\times/{\cal A}_+\simeq 0.06$. We stress
again that extrapolation of these results to SMBHs
is not justified, but the simulations could contain interesting
indications on the possible outcome of a full, general relativistic
black hole merger simulation.

\section{Multi-mode signal-to-noise ratio and parameter estimation}
\label{multimode}

Let us now turn to a multi-mode analysis. For simplicity we assume (as
we did in the single-mode case) that spin-weighted spheroidal
harmonics are real, and we consider a waveform given by the
superposition of only two QNMs, say
\beq
h_{+,\times}&=&
\f{M}{r}
{\cal A}^{+,\times}_{lmn}
\sin(\om t+\ph^{+,\times}) e^{-t/\ta} 
\Slm(\iota,\beta) \nonumber \\
&&+
\f{M}{r}
{\cal A'}^{+,\times}_{l'm'n'}
\sin(\omega_{l'm'n'} t+\phi_{l'm'n'}^{+,\times}) e^{-t/\tau_{l'm'n'}} 
S_{l'm'n'}(\iota,\beta) 
\,,
\eeq
where we have chosen the phases arbitrarily
so that the $+$ and $\times$ waveforms for
both modes are sine functions.
We denote the mode with indices $(l,m,n)$ by a
``1'', and the mode with indices $(l',m',n')$ by a ``2''. We have seen
in the single-mode analysis that phases do not play a significant role
in parameter estimation, so in principle we could simplify things
by assuming all phases to be zero. To be able to check and confirm these
expectations we only assume that $\ph^{+,\times}=0$, but we allow for
a nonzero relative phase of the second mode
$\phi=\phi_{l'm'n'}^{+}=\phi_{l'm'n'}^{\times}\neq 0$.  Then the waveforms
take the form
\beq\label{h12}
h_{+,\times}&=&
\f{M}{r}
\left[
{\cal A}^{+,\times}_{1}
\sin(\omega_1 t) e^{-t/\tau_1} 
S_1(\iota,\beta)+
{\cal A}^{+,\times}_{2}
\sin(\omega_2 t+\phi) e^{-t/\tau_2} 
S_2(\iota,\beta) \right]
\,.
\eeq
Again, the measured waveform is given by $h=h_{+}F_{+} + h_{\times}F_{\times}$.

In \cite{berticardoso} we show by explicit calculations that, to a
good approximation, the angular scalar products between spheroidal harmonics
for the
modes we are interested in are orthonomal on $l$ and $m$, namely
\be\label{ortho}
\int~{S_{lmn}}^*(\iota,\beta) ~
S_{l'm'n'}(\iota,\beta) d\Omega\simeq 
\delta_{l,l'}
\delta_{m,m'}\,.
\ee
This suggests that we make a separate treatment of the
following two cases:
\begin{itemize}
\item[(A)] 
Either $l\neq l'$ or $m\neq m'$, i.e. we look at modes with {\it
different angular dependence}. Then the angular scalar product between the
different modes is zero
to a good approximation, and the Fisher matrix can be
expressed approximately as the sum of the Fisher matrix for mode ``1'' and the
Fisher matrix for mode ``2''. This case is dealt with in
Section~\ref{twom}.
\item[(B)] 
The angular indices $l=l'$ and $m=m'$, but $n\neq n'$, so we are
looking at {\it different overtones with the same angular
dependence}. In this case, as long as we limit attention to the first
few modes, the angular scalar product is very close to unity for any
($n,\,n'$) pairs. This
case is dealt with in Section~\ref{overtone}.

\end{itemize}

\subsection{Quasi-orthonormal waveforms ($l\neq l'$ or $m\neq m'$)}
\label{twom}

We use the FH convention and consider, for simplicity, equal
polarization amplitudes ${\cal A}^+_i={\cal A}^\times_i={\cal A}_i$
for each mode $(i=1,2)$.  We further assume the noise to be constant
over each mode's bandwidth. In principle we should assume $S_1\neq
S_2$, but if $f_1$ and $f_2$ are close enough (which is true in all
cases we consider) we can set $S_1\simeq S_2\simeq S$. Then, a simple
calculation based on the waveform (\ref{h12}) shows that the total
signal-to-noise ratio can be expressed as a sum in quadrature,
\be\label{rhoNoMod}
\rho^2=\rho_1^2+\rho_2^2\,,
\ee
of the single-mode SNRs
\be\label{rho12}
\rho_1^2=\left(\f{M {\cal A}_1}{r}\right)^2
\f{Q_1^3}
{5\pi^2 f_1 (1+4 Q_1^2)S}\,,\qquad
\rho_2^2=\left(\f{M {\cal A}_2}{r}\right)^2
\f{Q_2(\sin^2\phi+2Q_2^2)}
{10\pi^2 f_2 (1+4 Q_2^2)S}\,.
\ee
Notice that $\rho_2$ becomes symmetric with $\rho_1$, as it should, in
the limit $\phi\to 0$. These single-mode SNRs can be obtained as
appropriate limits of the general expression in the $\delta$-function or
constant noise approximation in the
FH convention, Eq.~(\ref{rhoFH}), by choosing the phases there to be 
$\phi_{1}^{+}=-\pi/2$, 
$\phi_{1}^{\times}=0$, $\phi_{2}^{+}=-\pi/2 + \phi$, and
$\phi_{2}^{\times}=\phi$.  The generalization to waveforms involving more
quasi-orthonormal modes is straightforward.

Under our simplifying assumptions, the Fisher matrix for a two-mode
waveform depends on five parameters: $\{j,~M,~{\cal A}_1,~{\cal
A}_2,~\phi\}$.  For concreteness, we computed the errors on mass and
angular momentum obtained by setting $\phi=0$. 
The analytical expressions are too lengthy to present
here.  As physical intuition suggests, the crucial element in
determining the total error is the relative amplitude of the two
modes. Without loss of generality, we pick ${\cal A}_1=1$ and vary
the ratio ${\cal A}_1/{\cal A}_2$.

As a second step, we must choose which particular pair of modes we
want to compare. We first make the reasonable assumption that the
dominant modes have $n=n'=0$, so we are left with a four-dimensional
parameter space, the parameters being $(l,m,l',m')$. Given the
standard lore that $l=m=2$ modes should in some sense be dominant, we
examine in detail three cases.

First, we fix $l=m=2$, then pick $l'=m'=3$ or $l'=m'=4$. In this case
considering additional modes does not significantly affect the error
with respect to the single-mode case, whatever the amplitude of the
second mode. The reason is that the functional dependence on $j$ of
the mass and angular momentum errors for modes with $l=m$ is basically
the same, whatever the value of $l$ (or $m$).  As we change the
relative amplitude of the modes there is a smooth transition from,
say, the errors on $j$ and $M$ corresponding to $l=m=2$ to the errors
corresponding to $l'=m'=3$, but this transition is almost
imperceptible to the eye, the functional behaviors of
$(\rho\sigma_j)(j)$ and $(\rho\sigma_M/M)(j)$ being so similar in the
two extreme cases. For this reason we decided not to show any plot,
since they would be almost indistinguishable from (say) the left panel
of Fig.~\ref{errs-fh}. Furthermore, for these corotating modes the
fitting functions of Appendix \ref{app:QNM} do an excellent job at
approximating the ``true'' errors obtained by a numerical calculation
of the derivatives.

\begin{figure*}[t]
\begin{center}
\begin{tabular}{ccc}
\epsfig{file=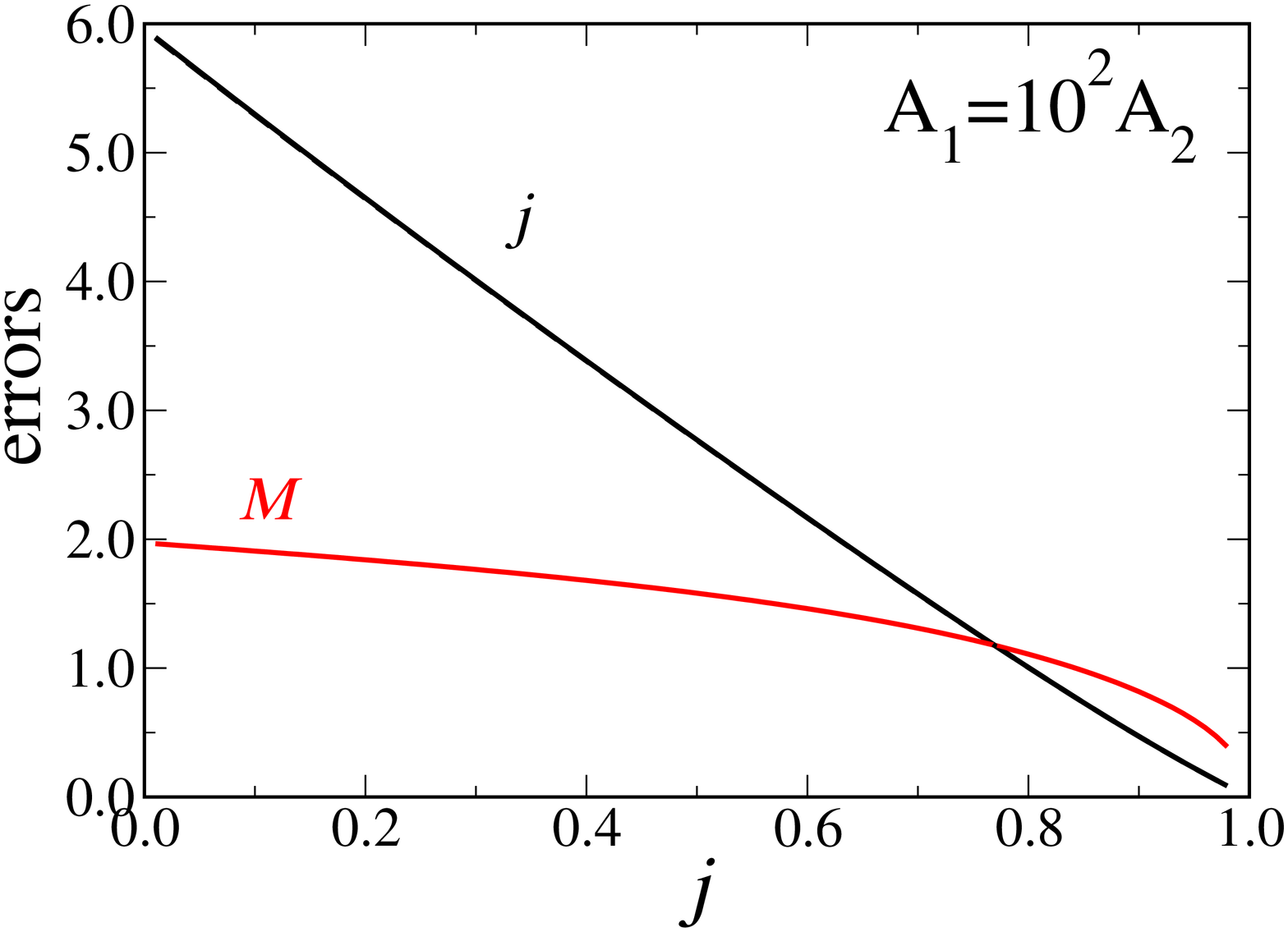,width=4.5cm,angle=-90} &
\epsfig{file=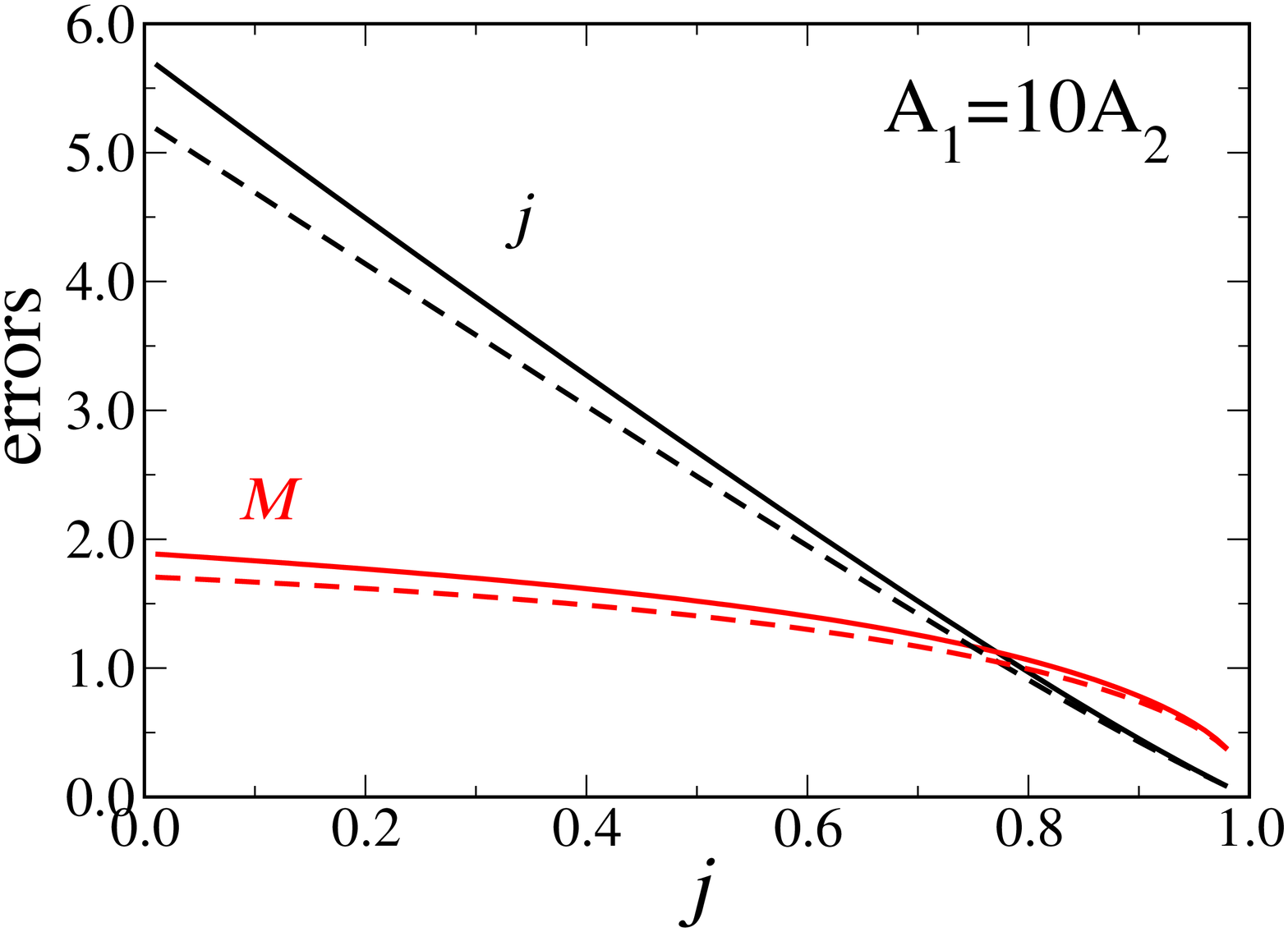,width=4.5cm,angle=-90} &
\epsfig{file=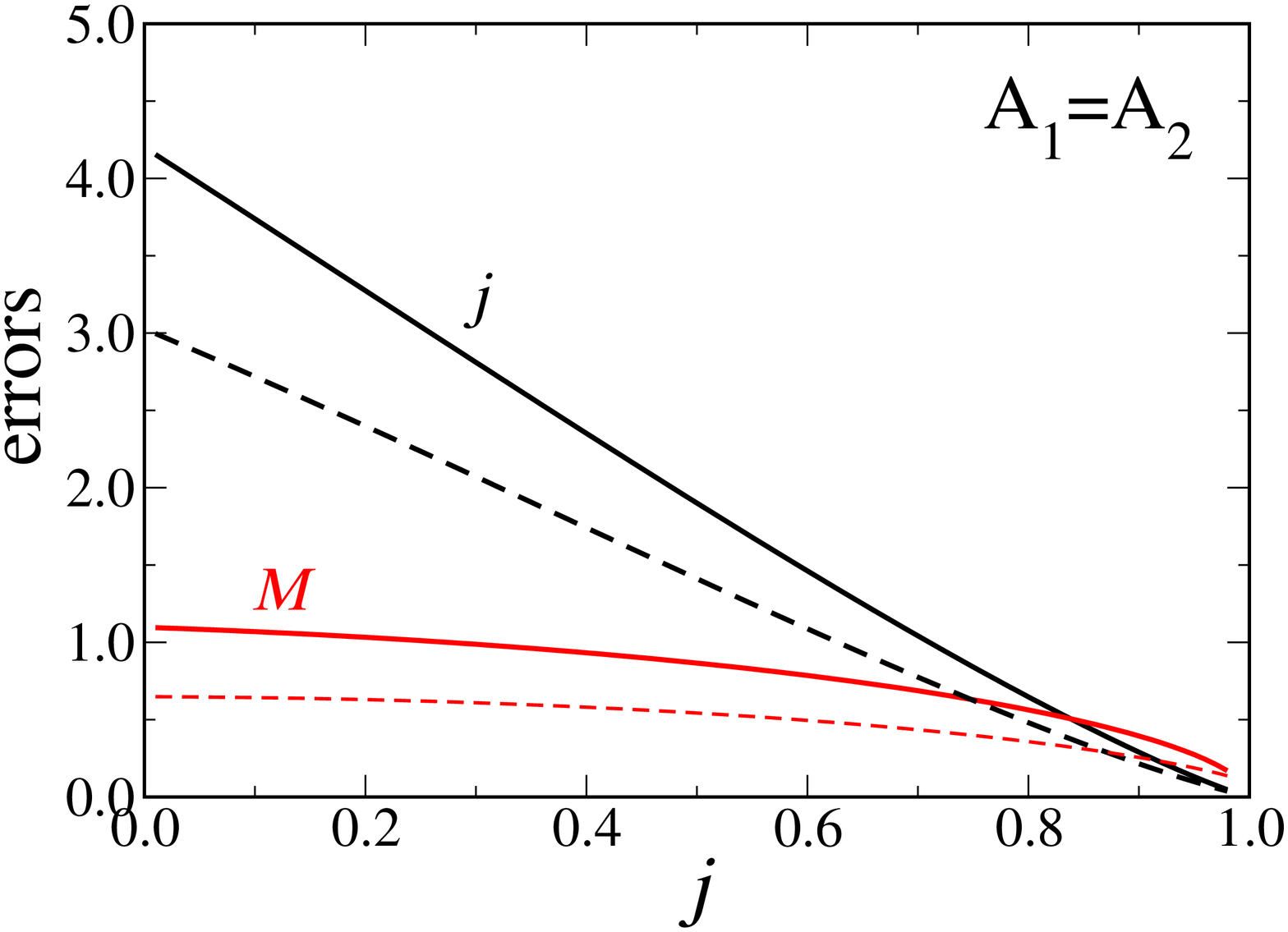,width=4.5cm,angle=-90} \\
\end{tabular}
\begin{tabular}{cc}
\epsfig{file=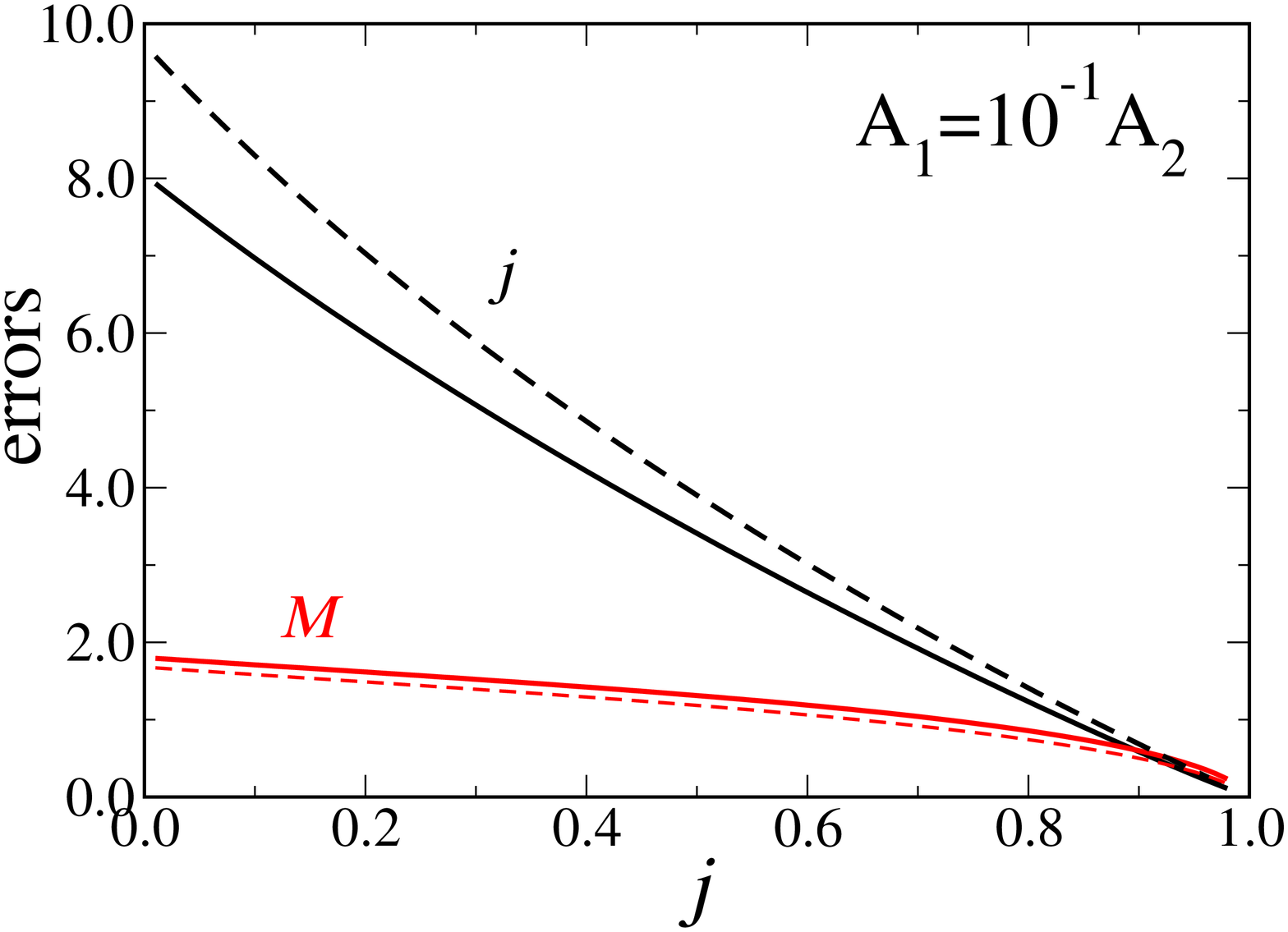,width=4.5cm,angle=-90} &
\epsfig{file=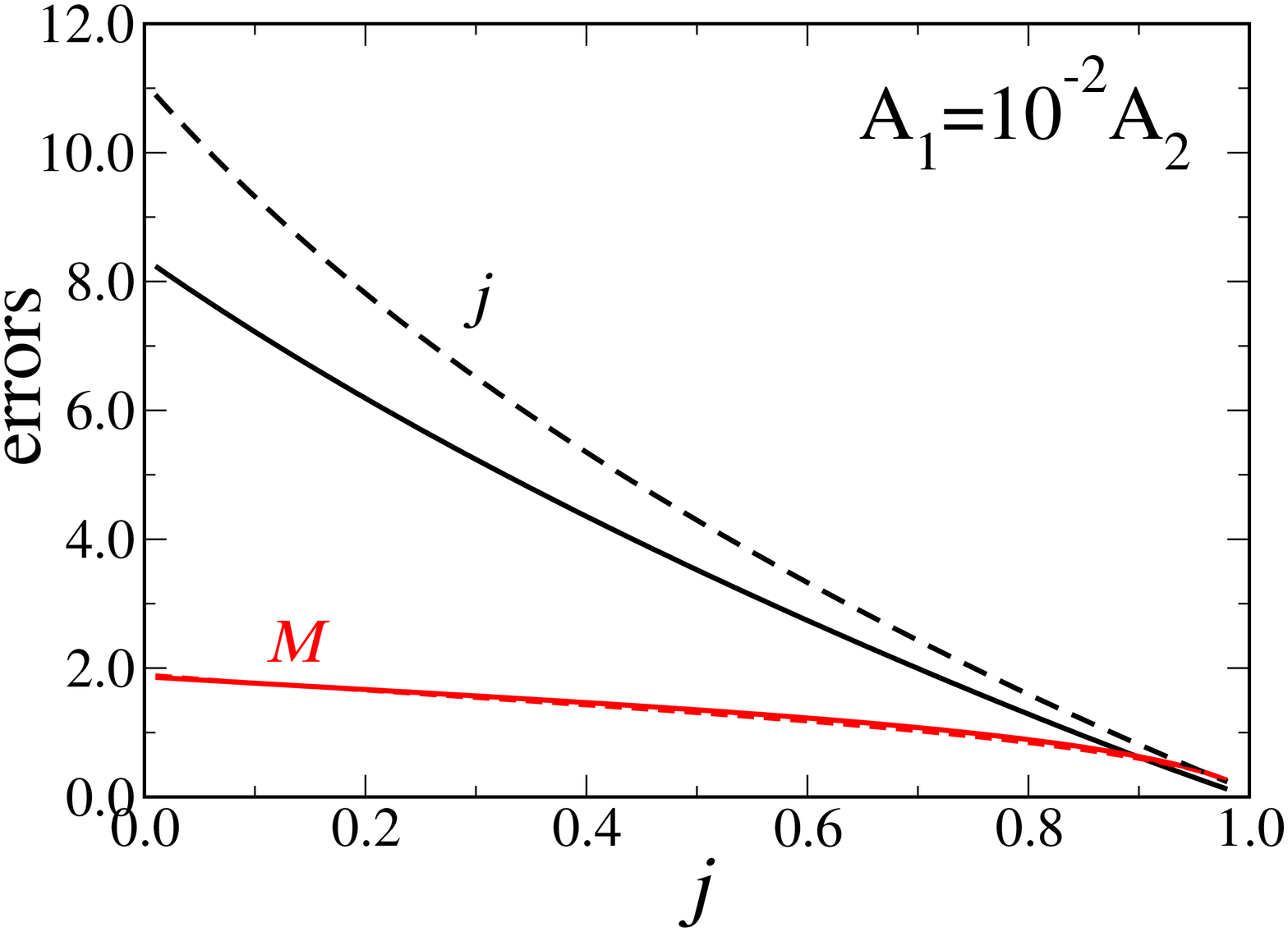,width=4.5cm,angle=-90} \\
\end{tabular}
\caption{Scaled errors ($\rho \sigma_M$, $\rho \sigma_j$) 
for two-mode measurements of mass and
angular momentum, with $l=m=m'=2$. Solid lines refer to $l'=3$,
dashed lines to $l'=4$.
\label{mm-m2mp2}}
\end{center}
\end{figure*}

\begin{figure*}[t]
\begin{center}
\begin{tabular}{ccc}
\epsfig{file=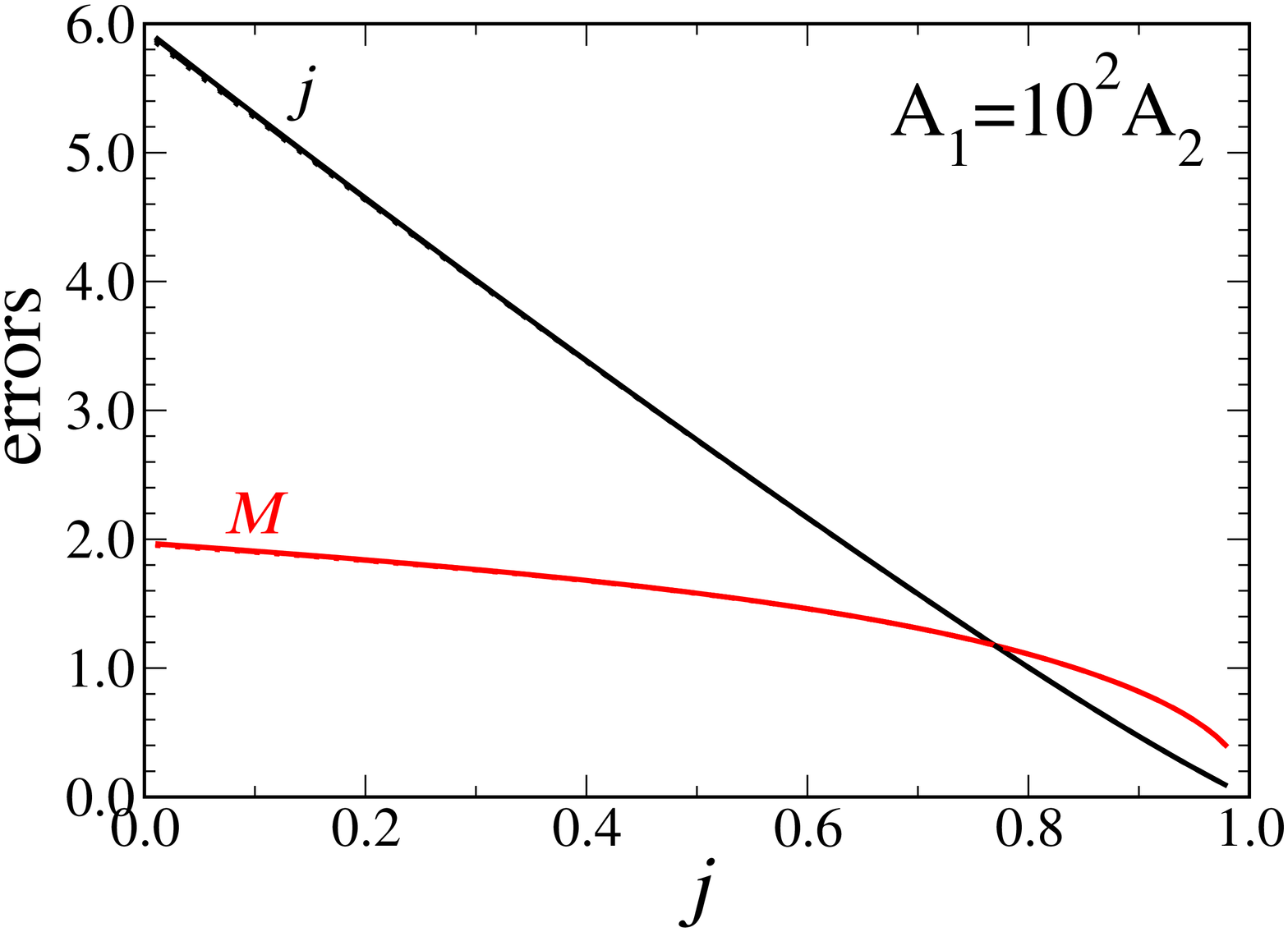,width=4.5cm,angle=-90} &
\epsfig{file=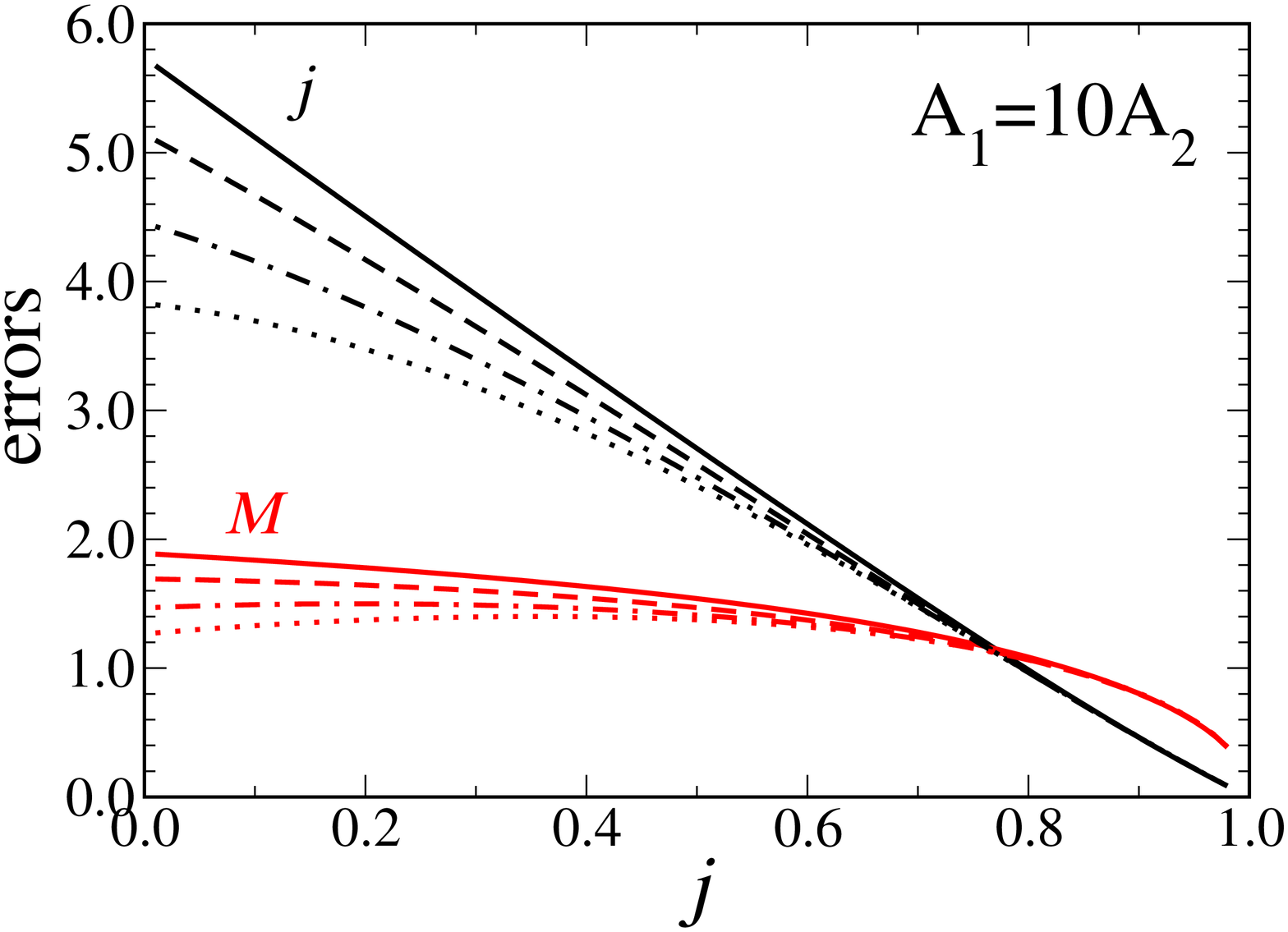,width=4.5cm,angle=-90} &
\epsfig{file=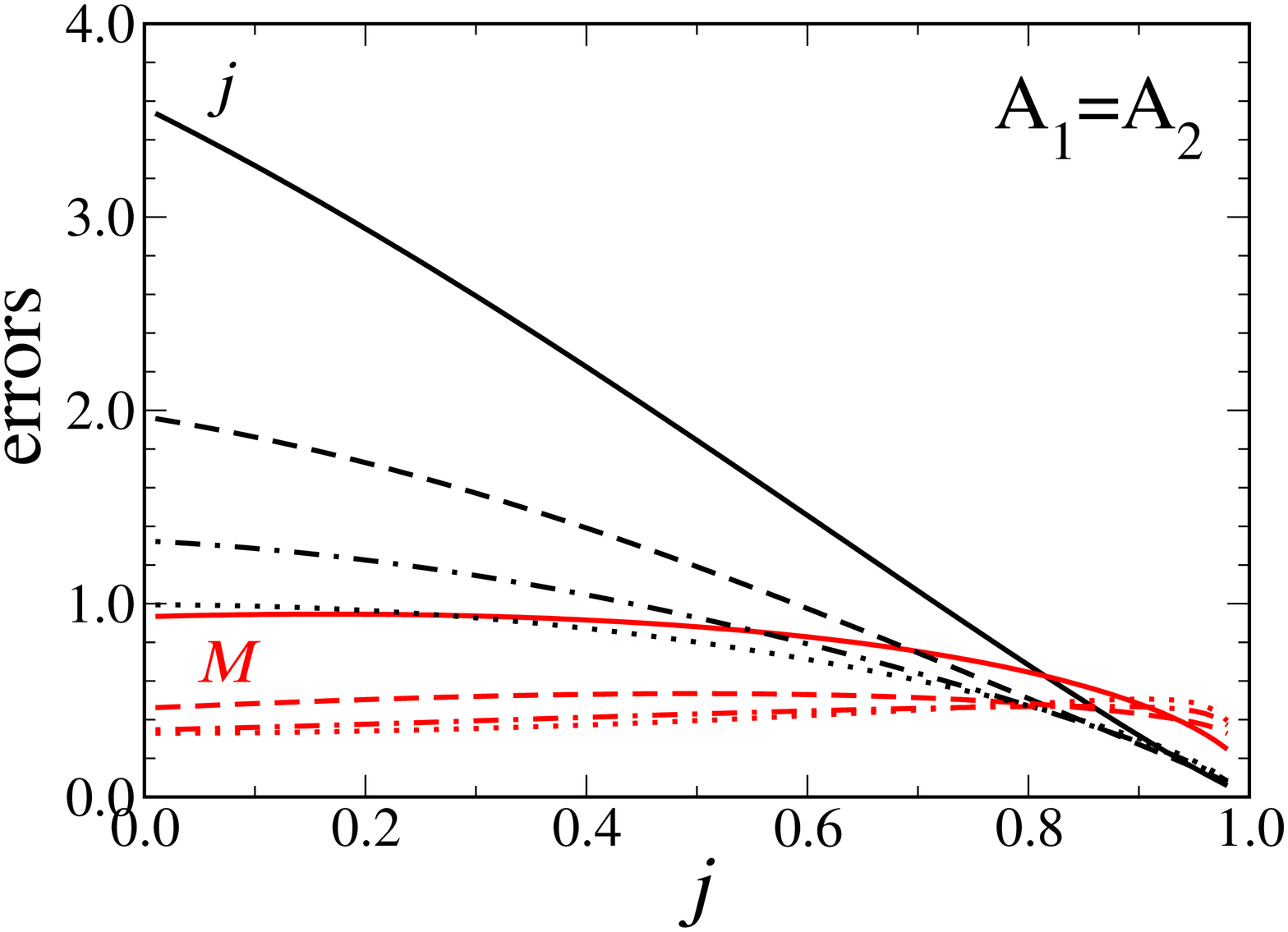,width=4.5cm,angle=-90} \\
\end{tabular}
\begin{tabular}{ccc}
\epsfig{file=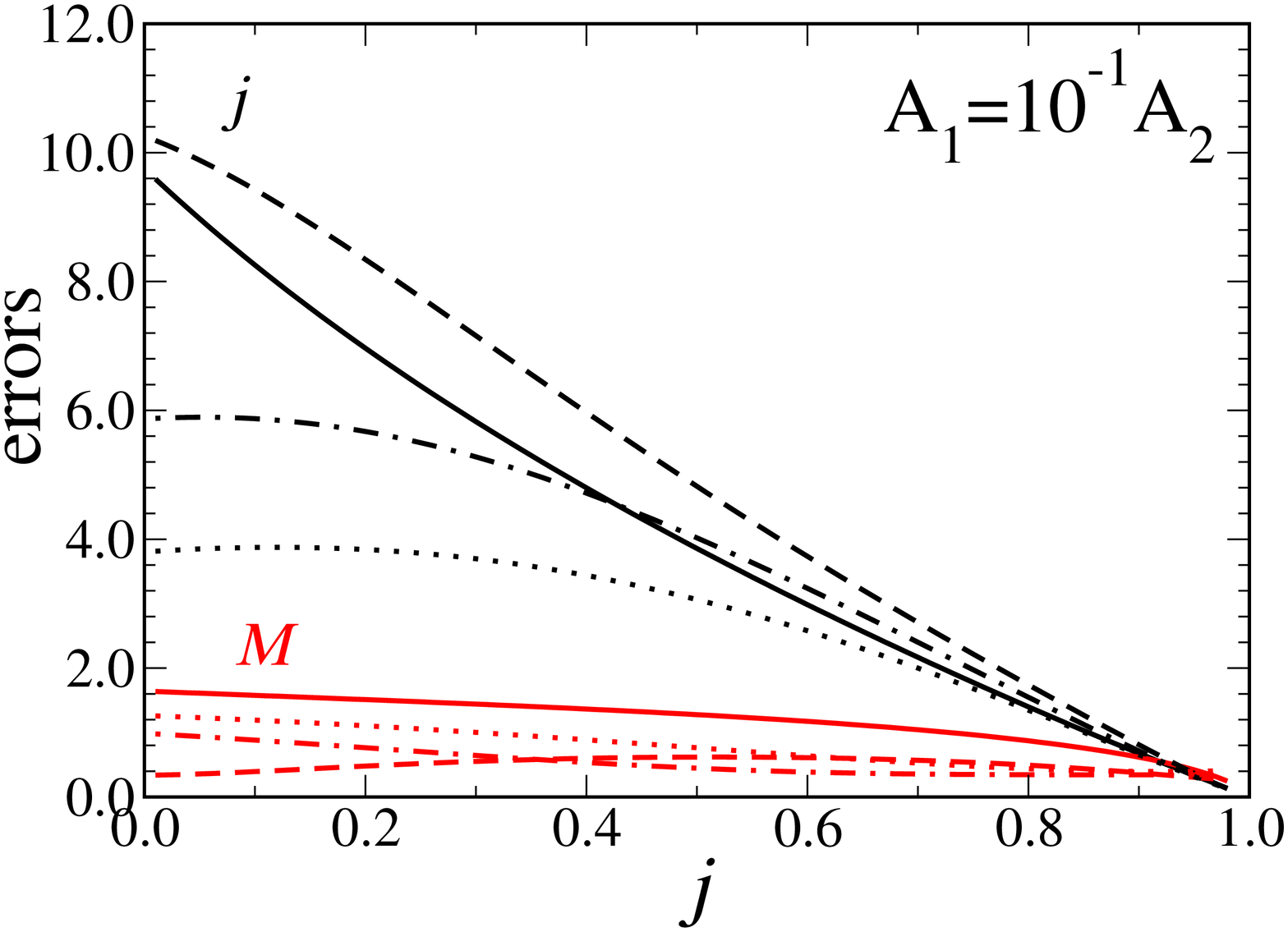,width=4.5cm,angle=-90} &
\epsfig{file=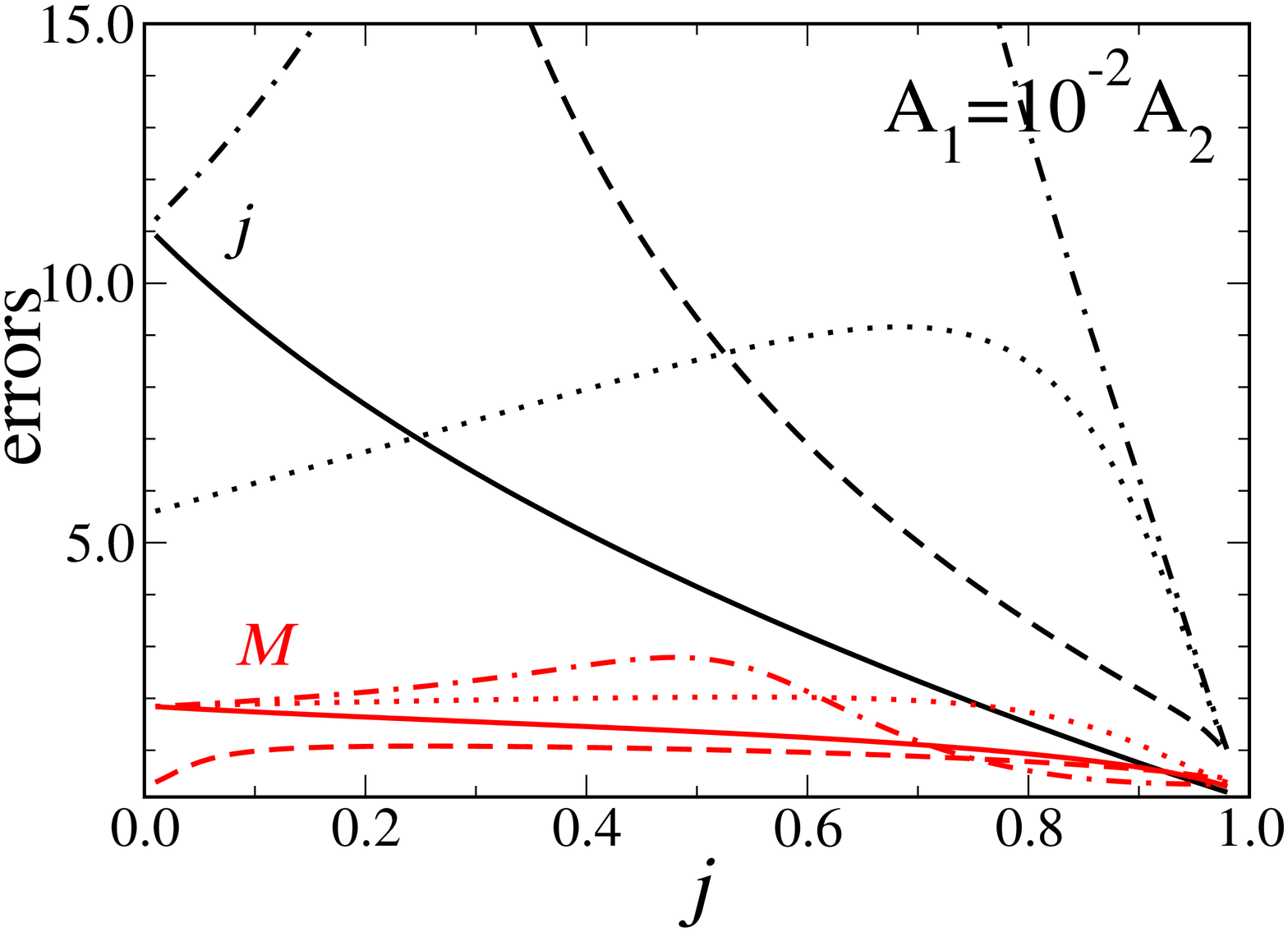,width=4.5cm,angle=-90} &
\epsfig{file=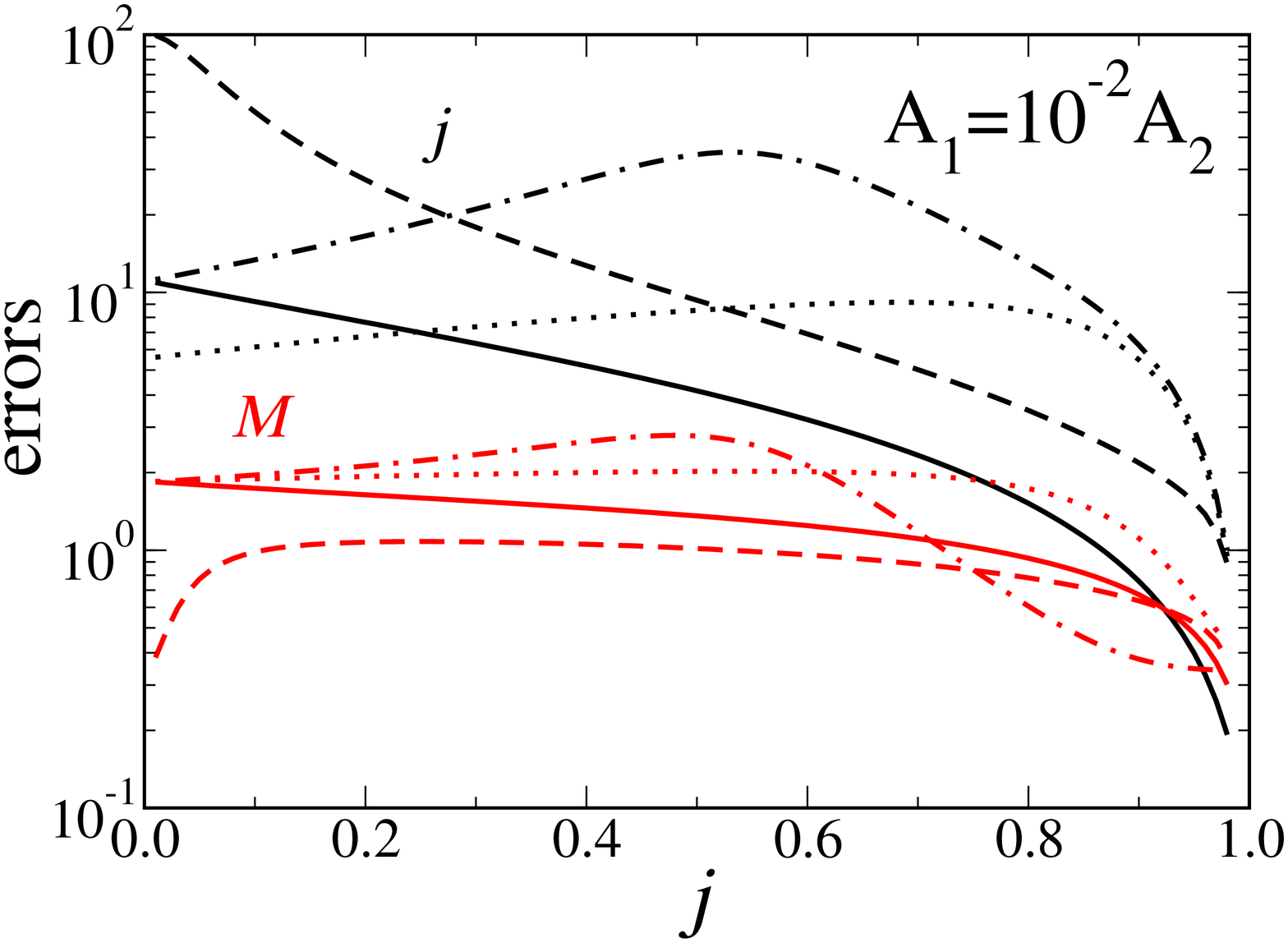,width=4.5cm,angle=-90} \\
\end{tabular}
\caption{Scaled errors ($\rho \sigma_M$, $\rho \sigma_j$)
for two-mode measurements of mass and
angular momentum, with $l=m=l'=2$.  Solid lines refer to $m'=1$,
dashed lines to $m'=0$, dot-dashed lines to $m'=-1$, dotted lines to
$m'=-2$.  In the case ${\cal A}_1=10^{-2}{\cal A}_2$ the error for
$m'=0$ and $m'=-1$ becomes quite large, and for clarity we show results
using both a linear and a log scale.
\label{mm-l2lp2}}
\end{center}
\end{figure*}

Secondly, we consider a more
physically motivated case in which we fix $l=m=2$,
$m'=2$, so that both modes correspond to a ``bar-shaped'' deformation.
We then 
look at the effect of setting $l'=3,\,4$.  Results for these combinations
of angular indices are shown in Fig.~\ref{mm-m2mp2}. From left to
right and from top to bottom we assume that the ratio of the two mode
amplitudes ${\cal A}_1/{\cal
A}_2=\left\{10^2,~10,~1,~10^{-1},~10^{-2}\right\}$.  Now the second
mode plays some role, and we can observe a smooth deformation from the
errors corresponding to $l=m=2$ (top left panel) to the errors
corresponding to $l'=3,~m'=2$ or $l'=4,~m'=2$ (bottom right
panel). Notice that the error on mass is roughly independent of $l$,
but, when a single $l$ dominates, the error on angular momentum 
scales (roughly) as $l$. For
$l=m=2$ the mass error becomes larger than the angular momentum error
at $j\simeq 0.8$. As a consequence of the (rough) scaling with $l$
of $\rho \sigma_j$, this transition moves to larger and larger values
of $j$ as the relative amplitude of the second mode grows. When $l'=3$
dominates, the transition still occurs at $j\simeq 0.9$, but for
$l'=4$ the angular momentum error becomes subdominant (if at all) only
for $j\gtrsim 0.95$.

In Fig.~\ref{mm-m2mp2} we compute the derivatives $\flm'$
and $\Qlm'$ ``numerically'' (that is, by local interpolating
polynomials).  However, we verified that the fits typically do a very
good job, even though they are not as accurate as in the case considered
above. This is expected, since we are considering corotating
modes. For all of these modes $\Qlm'$ does not cross zero in the range
$j\in [0,1]$ (see eg. Fig.~\ref{QQa}), and the fits of Appendix
\ref{app:QNM} are reasonably accurate.

The third, and presumably the most physically realistic case,  results
from fixing $l=m=2$, $l'=2$ and looking at the effect of changing
$m'$.  In Fig.~\ref{mm-l2lp2} we show the smooth deformation of the
errors induced by changing ${\cal A}_1/{\cal A}_2$ in this
case. Different linestyles correspond to $m'=1,0,-1,-2$, as explained
in the caption. The plot is better understood as a series of
snapshots. The top left panel is an almost-pure $l=m=2$ waveform
(closely approximating the top left panel of Fig.~\ref{errs-fh-all},
or equivalently the left panel of Fig.~\ref{errs-fh}). Each line
becomes more and more dominated by the second mode, until (bottom
right panels) it approximates quite closely the errors with $l=2$ and
$m=1,0,-1,-2$ we displayed in Fig.~\ref{errs-fh-all}. In the bottom
right panels we clearly see signatures left by the dominance of the
second mode. For example, the error becomes very large as $j\to 0$ for
$m'=0$, and as $j\to 0.6$ for $m'=-1$.  

In Fig.~\ref{mm-l2lp2} the derivatives $\flm'$ and $\Qlm'$ are
computed numerically. The fits of Appendix \ref{app:QNM} provide a
good approximation of the numerical results, except for those cases
(i.e. dominance of $m'=0$ and $m'=-1$) where the single-mode fits fail
to reproduce the location of $\Qlm'=0$.

\subsection{Overtones with the same $l$ and $m$}
\label{overtone}

Consider now modes for which $l=l'$ and $m=m'$. To reduce the number
of parameters to be estimated, 
we assume that the plus- and cross-components
have equal amplitudes (the generalization to unequal amplitudes is
trivial).  Approximating the scalar product between the different
spin-weighted harmonics in this case
by one, assuming that the frequencies are close enough
that $S_1\simeq S_2\simeq S$, and using the expression (\ref{h12}) for
the two-mode waveform, we get the following SNR:
\be\label{SNRovertone}
\rho^2=\rho_1^2+\rho_2^2+
\left(\f{M}{r}\right)^2\f{{\cal A}_1 {\cal A}_2}{5\pi^2 S}
\left\{
\f{16f_1 f_2 Q_1^3 Q_2^3 (f_1 Q_2+f_2 Q_1)\cos \phi}
{\Lambda_+\Lambda_-}
\right\}\,,
\ee
where 
\be
\Lambda_\pm=f_2^2 Q_1^2+2 f_1 f_2 Q_1 Q_2+Q_2^2
\left[f_1^2+4(f_1\pm f_2)^2 Q_1^2\right]\,,
\ee
and $\rho_1$ and $\rho_2$ are defined by Eq.~(\ref{rho12}).

\begin{figure*}[t]
\begin{center}
\epsfig{file=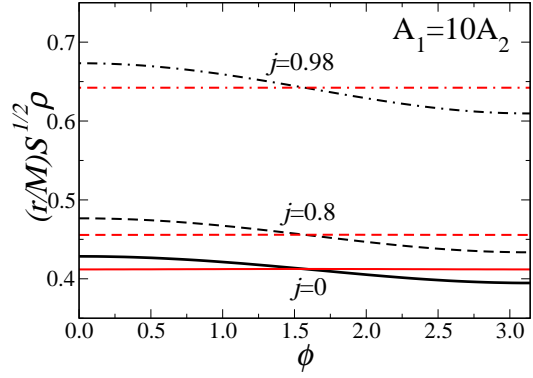,width=6cm,angle=-90}
\caption{Modulation of the total SNR, Eq.~(\ref{SNRovertone}), induced
by the second overtone. We consider an $l=m=2$ perturbation and three
different values of the angular momentum $j$ (solid lines: $j=0$,
dashed lines: $j=0.8$, dot-dashed lines: $j=0.98$). For concreteness,
we assume that the first overtone has an amplitude ${\cal
A}_2=10^{-1}{\cal A}_1$. The nearly horizontal (red) lines correspond
to the SNR in the absence of modulations, Eq.~(\ref{rhoNoMod}). 
\label{SNRphi}}
\end{center}
\end{figure*}

Fig.~\ref{SNRphi} shows a rescaled SNR given by $(r/M)S^{1/2}\rho$ as
a function of $\phi$ for a reasonable amplitude ratio ${\cal
A}_2/{\cal A}_1=10^{-1}$. The modulation induced by the mixed term is
modest, and it scales (roughly) like $\left({\cal A}_2/{\cal
A}_1\right)^{1/2}$.

\begin{figure*}[t]
\begin{center}
\begin{tabular}{cc}
\epsfig{file=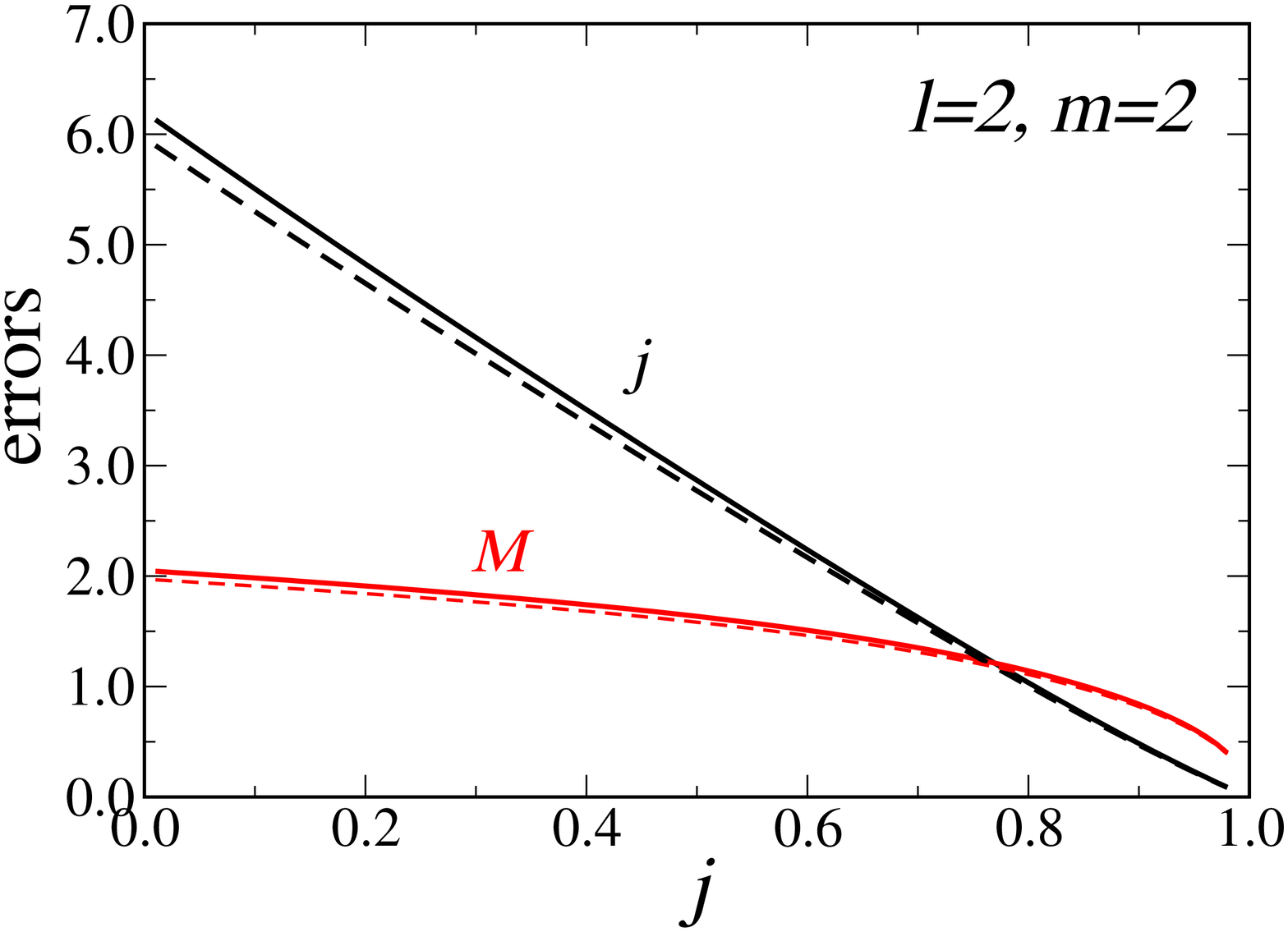,width=6cm,angle=-90} &
\epsfig{file=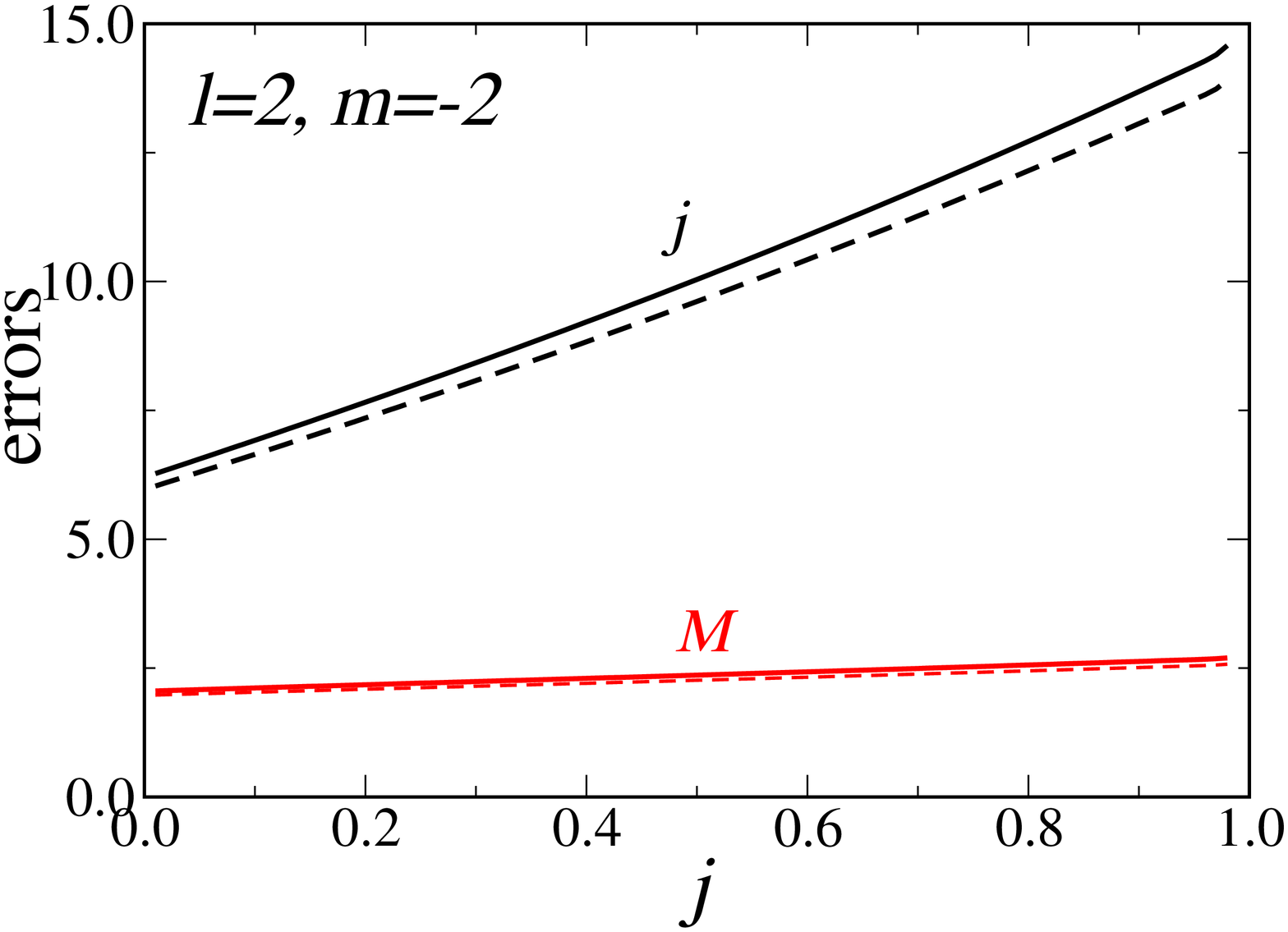,width=6cm,angle=-90} \\
\end{tabular}
\caption{Scaled errors ($\rho \sigma_M$, $\rho \sigma_j$)
for the two-mode waveform (solid lines) versus a
one-mode waveform (dashed line). Here we assume ${\cal A}_1=10 {\cal
A}_2$ and consider perturbations with $l=m=2$ (left) and $l=2$, $m=-2$
(right).
\label{overtones}}
\end{center}
\end{figure*}

In Fig.~\ref{overtones} we compute the errors on mass and angular
momentum for a two-mode waveform with ${\cal A}_1=10 {\cal A}_2$, and
compare the results with a single-mode waveform. We show two
representative cases ($l=2,~m=2$ and $l=2,~m=-2$) but we looked at all
modes with $l=2,~3,~4$ and all possible values of $m$. We always found
that the correction induced by the addition of the second overtone is
very small. Indeed, it is so small that it is comparable to the
variations induced by different prescriptions to compute the error
(compare the left panel with the left panel of Fig.~\ref{errs-fh}).

A more realistic treatment would not assume that the second mode has
amplitude 1/10 of the first, but would attempt to account for the mode
excitation using an explicit calculation of the excitations coefficients as
functions of $j$.  We are currently working on such an approach.

\section{Resolving quasi-normal modes: critical signal-to-noise ratio to test the no-hair theorem}
\label{nohair}

So far we have been discussing multi-mode detection without
considering whether a discrimination between two different frequencies
(in a signal buried in noise) is possible or not.  A common rule of
thumb to resolve the frequencies of two sinusoidal signals with equal
amplitudes and quality factors is the so-called Rayleigh criterion
(see eg. \cite{rayleigh}): two frequencies $f_1\,,\,f_2$ are
resolvable if 
\be
\label{rayleigh}
|f_1-f_2|\Delta t>1\,, 
\ee
where $\Delta t$ is the duration of the signal. This (purely
classical) Rayleigh limit can be beaten, given a signal with
sufficiently large SNR, as discussed in
\cite{smilanfar,milanfars}. Here we introduce a slightly different
resolvability criterion that allows for different amplitudes (${\cal
A}_1\neq {\cal A}_2$) and quality factors ($Q_1\neq Q_2$). In
Appendix~\ref{recover-rayleigh} we will show that
\cite{smilanfar,milanfars} actually deal with a very special case of
our own resolvability criterion.

To determine whether two quasi-normal mode frequencies are resolvable
we first need to determine the measurement errors.  
Here we work in terms of frequency and damping time, instead of $M$ and $j$.
Because $\Qlm = \pi \flm \tau_{lmn}$, the Fisher matrix in terms of 
$\tau_{lmn}$ may be written, for each index $k$,
$\Gamma_{k\,\tau_{lmn}} = \pi \flm \Gamma_{k\,\Qlm}$ and
$\Gamma_{k\,\flm} = \Gamma_{k\,\flm} +  \pi \tau_{lmn} \Gamma_{k\,\Qlm}$.
Inverting this Fisher matrix, 
one can show that the errors in
$f_{lmn}$ and $\tau_{lmn}$ are given,
to leading order for large Q, by
\be\label{approx}
\sigma_f\simeq \f{1}{\sqrt{2}\pi \tau \rho}\,,\qquad
\sigma_\tau\simeq \f{2\tau}{\rho}\,.
\ee
These relations reproduce to leading-order the large-$Q$ behavior
predicted by Refs.~\cite{finn,kaa} (which use a different
parametrization of the waveform), except for a factor $\sqrt{2}$ in
the leading term of $\sigma_f$.

Then, for detection of two modes, a natural criterion ({\it \'a la} Rayleigh) 
to resolve frequencies and damping times is
\be\label{criterion}
|f_1-f_2|>{\rm max}(\sigma_{f_1},\sigma_{f_2})\,,\qquad
|\tau_1-\tau_2|>{\rm max}(\sigma_{\tau_1},\sigma_{\tau_2})\,.
\ee
This means, for example, that the frequencies are (barely) resolvable
if ``the maximum of the diffraction pattern of object 1 is located at
the minimum of the diffraction pattern of object 2''.  In other words,
given two Gaussians with variance $\sigma$ we can only distinguish
between the peaks if they are separated by a distance larger than
$\sigma$. We can introduce two ``critical'' SNRs required to resolve
frequencies and damping times,
\begin{subequations}
\beq
\label{rhof}
\rho_{\rm crit}^f
&=&
\f{{\rm max}(\rho \sigma_{f_1},\rho \sigma_{f_2})}{|f_1-f_2|}\,,\\
\label{rhotau}
\rho_{\rm crit}^\tau
&=&
\f{{\rm max}(\rho \sigma_{\tau_1},\rho \sigma_{\tau_2})}{|\tau_1-\tau_2|}\,,
\eeq
\end{subequations}
and recast our resolvability conditions as
\begin{subequations}
\beq
\label{minimal}
\rho&>&\rho_{\rm crit}=
{\rm min}(\rho_{\rm crit}^f,\rho_{\rm crit}^\tau)\,,\\
\label{both}
\rho&>&\rho_{\rm both}=
{\rm max}(\rho_{\rm crit}^f,\rho_{\rm crit}^\tau)\,.
\eeq
\end{subequations}
The first condition implies resolvability of either the frequency or
the damping time, the second implies resolvability of both.

We now need to compute the 
errors on frequencies ($\sigma_{f_1},~\sigma_{f_2}$) and damping times
($\sigma_{\tau_1},~\sigma_{\tau_2}$) in a two-mode situation.
We again use the Fisher matrix
formalism.  

We first consider the waveform (\ref{h12}) in the
quasi-orthonormal case of Sec.~\ref{twom}, and for simplicity we also
pick $\phi=0$. Then the analytic expressions for the errors turn out
to be very simple:
\begin{subequations}\label{sigft-qo-both}
\beq\label{sigft-qo}
\rho \sigma_{f_1}&=&
\f{1}{2\sqrt{2}}
\left\{
\f{f_1^3\left(3+16Q_1^4\right)}{{\cal A}_1^2 Q_1^7}
\left[
\f{{\cal A}_1^2 Q_1^3}{f_1\left(1+4Q_1^2\right)}+
\f{{\cal A}_2^2 Q_2^3}{f_2\left(1+4Q_2^2\right)}
\right] \right\}^{1/2}\,,\\
\rho \sigma_{\tau_1}&=&
\f{2}{\pi}
\left\{
\f{\left(3+4Q_1^2\right)}{{\cal A}_1^2 f_1 Q_1}
\left[
\f{{\cal A}_1^2 Q_1^3}{f_1\left(1+4Q_1^2\right)}+
\f{{\cal A}_2^2 Q_2^3}{f_2\left(1+4Q_2^2\right)}
\right] \right\}^{1/2}\,.
\eeq
\end{subequations}
Since we consider the ``symmetric'' case $\phi=0$, the errors on $f_2$
and $\tau_2$ are simply obtained by exchanging indices ($1\leftrightarrow
2$). Notice also that the term in square parentheses is nothing but
$5\pi^2S(r/M)^2\rho^2$, with $\rho^2$ given by Eq.~(\ref{rhoNoMod}).
In the single-mode  (${\cal A}_2 \to 0$), large-$Q$ 
limit, these errors reproduce Eq. (\ref{approx}).

\begin{figure*}[t]
\begin{center}
\begin{tabular}{ccc}
\epsfig{file=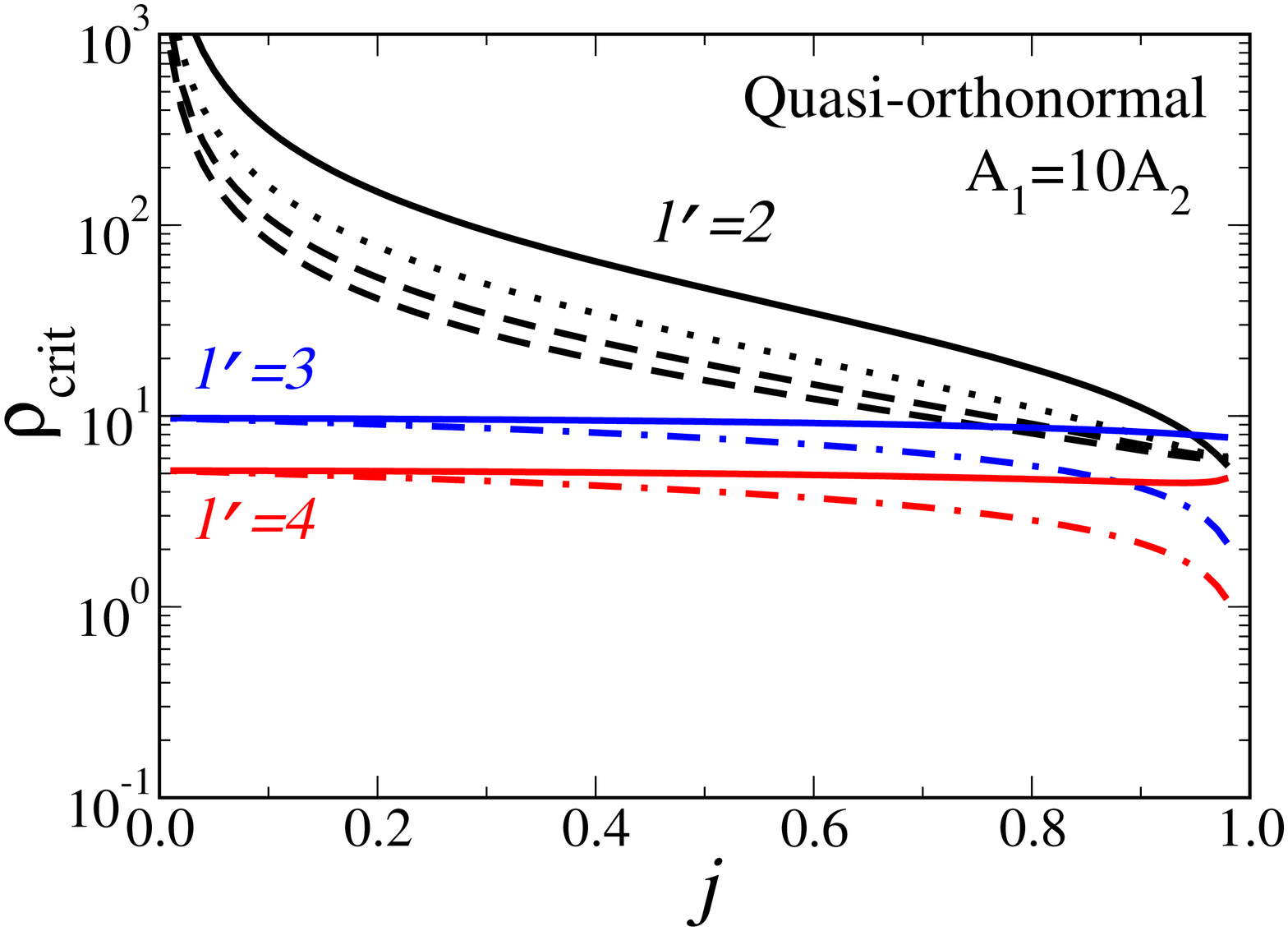,width=4.5cm,angle=-90} &
\epsfig{file=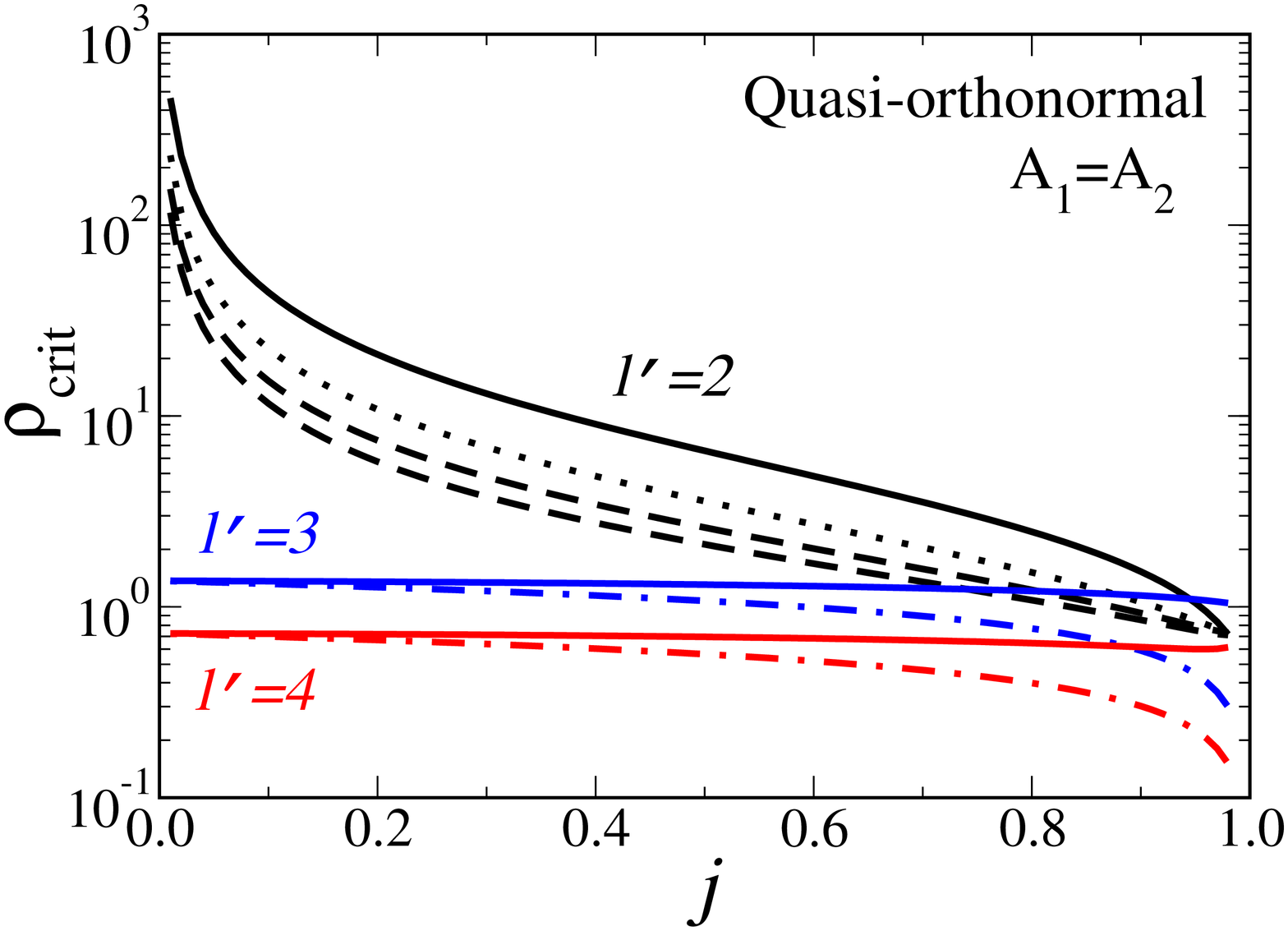,width=4.5cm,angle=-90} &
\epsfig{file=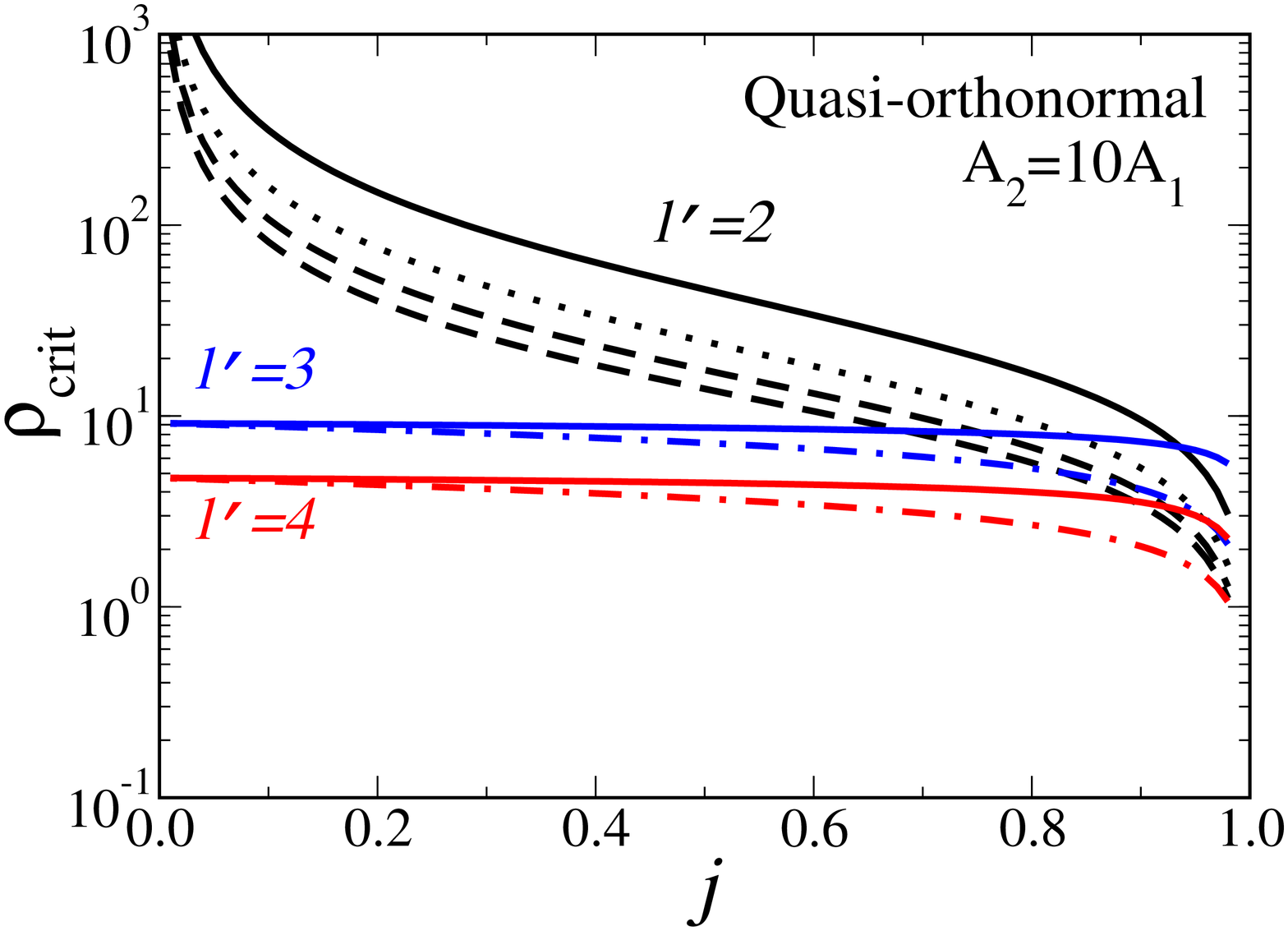,width=4.5cm,angle=-90} \\
\epsfig{file=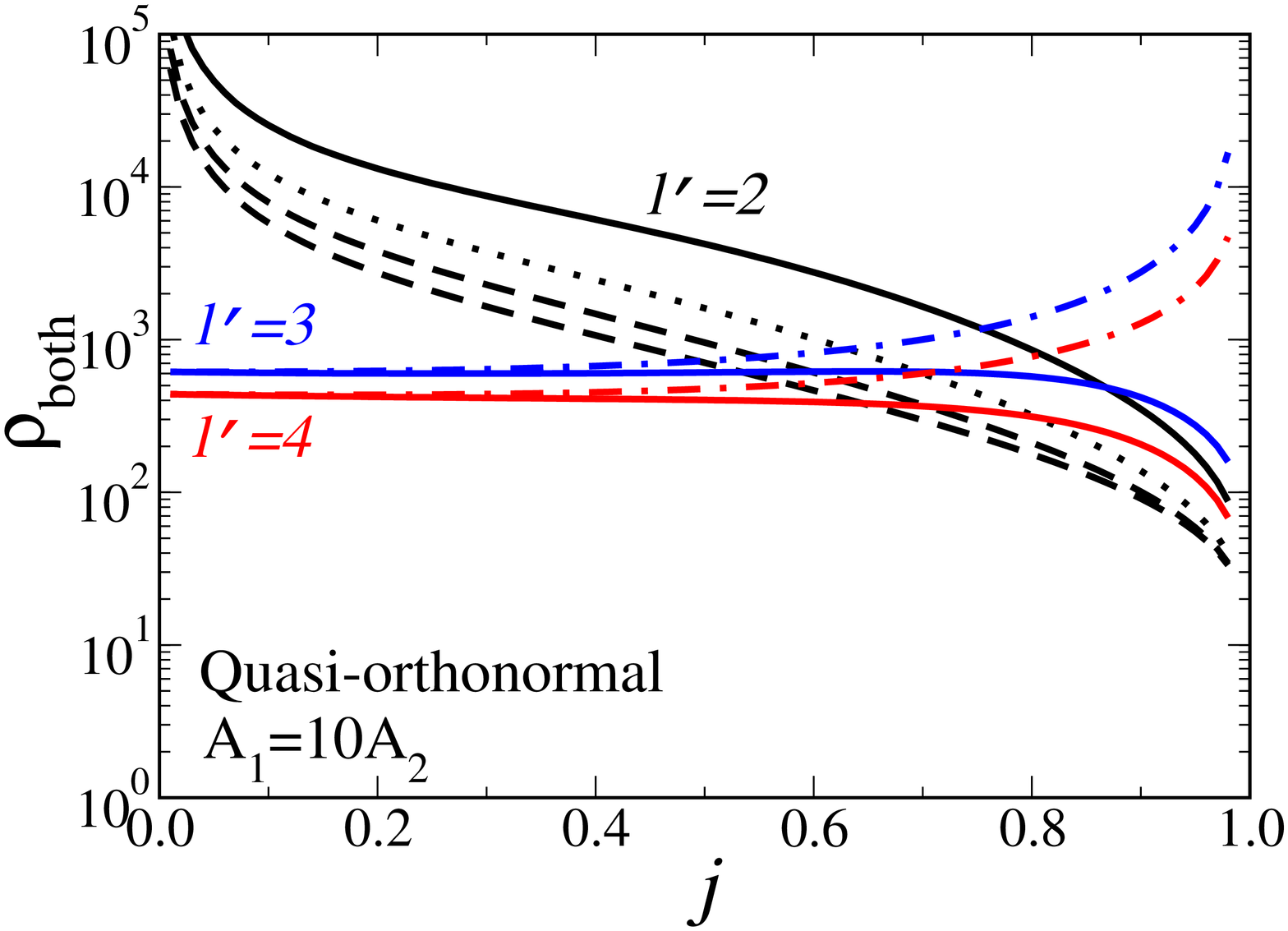,width=4.5cm,angle=-90} &
\epsfig{file=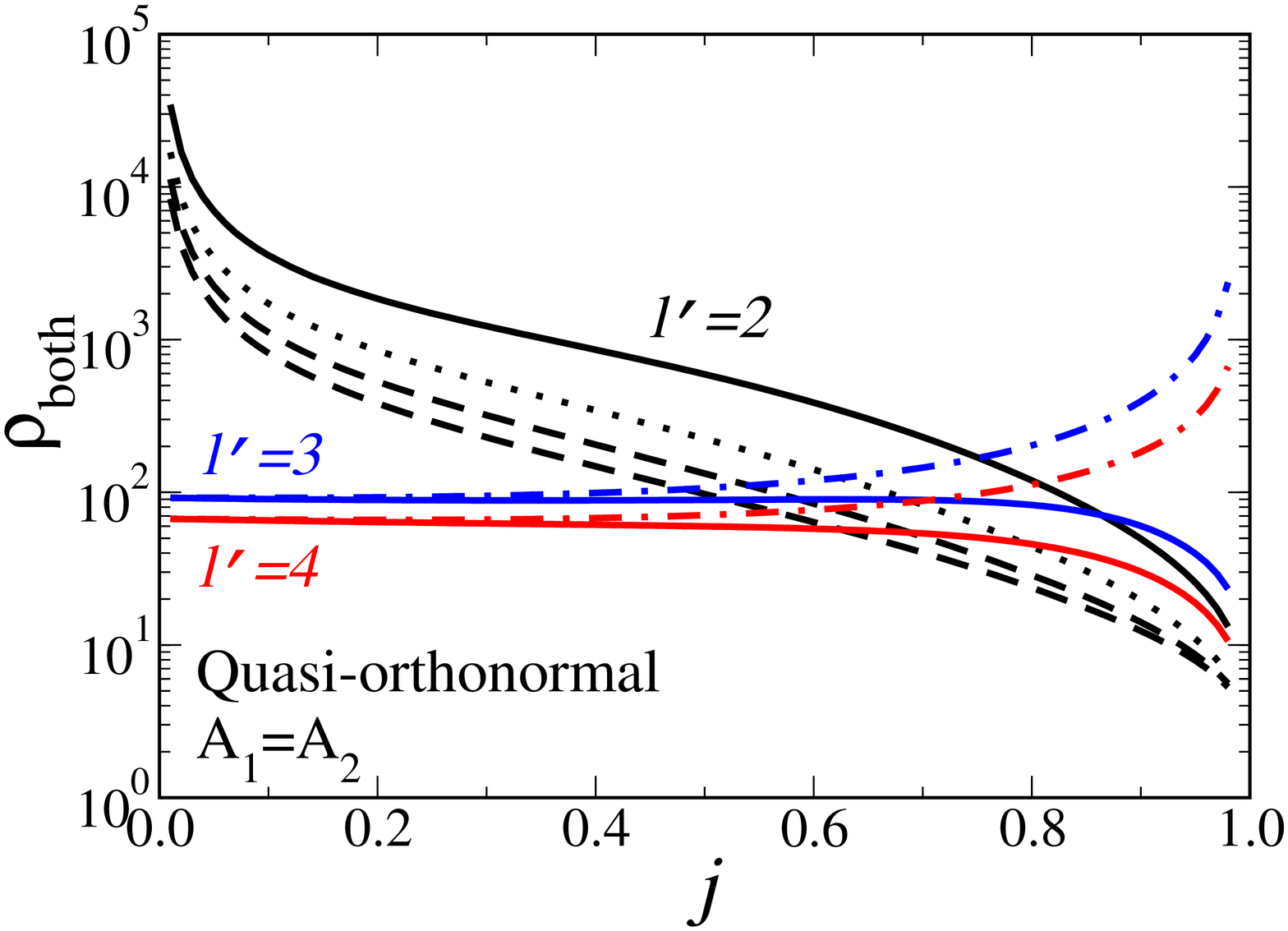,width=4.5cm,angle=-90} &
\epsfig{file=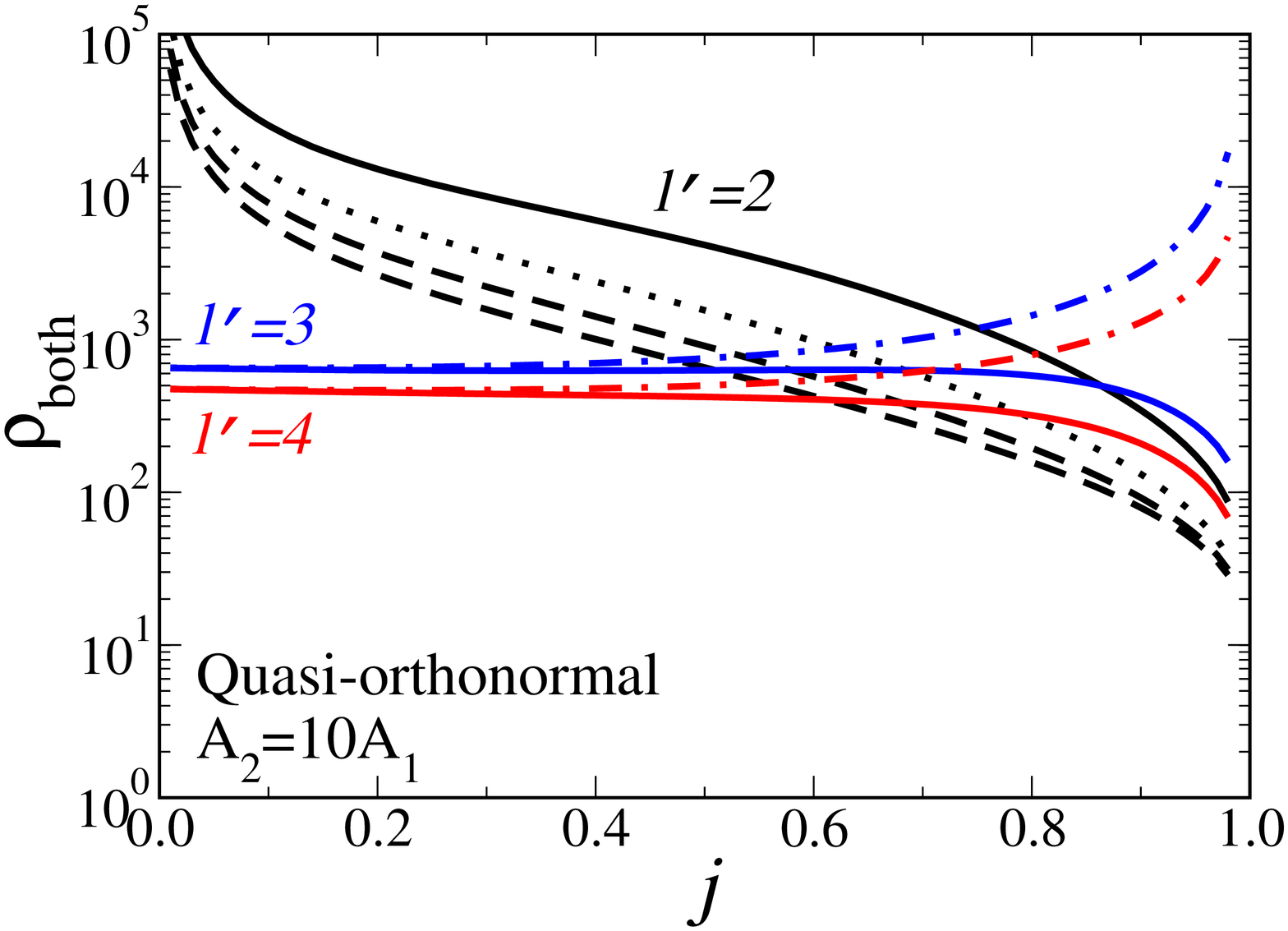,width=4.5cm,angle=-90} \\
\end{tabular}
\caption{``Critical'' SNR $\rho_{\rm crit}$ (top) and $\rho_{\rm
both}$ (bottom) required to resolve different combinations of
modes. Here we consider the ``quasi-orthonormal'' perturbations of
Sec.~\ref{twom} (either $l\neq l'$ or $m\neq m'$). Following the
treatment in that Section, we always assume $l=m=2$ for the first
mode. Different colors/linestyles refer to different values of $l'$
and $m'$ for the second mode. Black: $l'=2$ and $m'=1$ (solid), $m'=0$
(dotted), $m'=-1,~-2$ (dashed, top to bottom). Blue: $l'=3$ and $m'=2$
(solid), $m'=3$ (dot-dashed). Red: $l'=4$ and $m'=2$ (solid), $m'=4$
(dot-dashed). Different panels refer to different ratios ${\cal
A}_1/{\cal A}_2$, as indicated.
\label{milanfar-fig}}
\end{center}
\end{figure*}

A calculation of the critical SNRs $\rho_{\rm crit}$ and $\rho_{\rm
both}$ for different QNM pairs is shown in Fig.~\ref{milanfar-fig}.
Most qualitative features of this plot can be explained by looking at
Fig.~\ref{Qmodes}.  First, it is natural that modes with
$l=l'=2$ are harder to resolve than modes with $l'\neq l$: in the
latter case both frequencies and damping times are different.  For the
same reason, modes with $l=l'$ and $m\simeq m'$ are harder to
resolve. Another predictable feature is that, for modes with $l=l'$,
the Schwarzschild limit is very bad in terms of resolvability: as
$j\to 0$ all frequencies and damping times are degenerate with respect
to $m$, so $\rho_{\rm crit}$ blows up.  In particular, if
modes with $l=2,~m=2$ and $l=2,~m=0$ are dominant (as suggested in
\cite{fhh}), resolving them requires large SNRs for black holes with
spin $j\lesssim 0.5$, especially if the amplitude ratio ${\cal
A}_1/{\cal A}_2$ is not close to unity.  The almost-flat $j$-dependence
of $\rho_{\rm crit}$ and $\rho_{\rm both}$ for modes with $m'=2$,
$l'=3,~4$ and $l'=m'=3,~4$ is in line with the corresponding
discussion in Sec.~\ref{twom}: the errors on these modes have a very
similar functional dependence on $j$, whatever the value of $l$ (or
$m$), so it is natural for the critical SNR to be quite insensitive to
$j$. As a rule of thumb, a large rotation parameter $j$ usually helps
to resolve modes, mainly because of the larger quality factor
(remember that we chose the first mode in the pair to have $l=m=2$).
The only exception to this rule is the growth of $\rho_{\rm both}$ as
$j\to 1$ for modes with $l=m=2$ and $l'=m'$($=3,4$). This growth is
easy to understand: for all modes with $l=m$ the damping tends to zero
in the extremal limit, so the denominator of Eq.~(\ref{rhotau}) goes
to zero and the corresponding critical SNR blows up.

We now consider the resolvability of
two  overtones of modes with the same $l$ and $m$. 
Here an explicit calculation of the errors on frequency and
damping time is not simple, because of the presence of ``cross-terms'' in
the inner products.  The final expressions are much more lengthy and
involved than Eq.~(\ref{sigft-qo-both}), so we do not include them
here.  Instead we carried out the calculation of
Eqs.~(\ref{rhof})-(\ref{both}) numerically. We assumed that the
waveform is a superposition of the fundamental mode and the first
overtone with some given values of $(l,m)$, and made the plausible
educated guess that the amplitude of the overtone is a factor ten
smaller than the amplitude of the fundamental mode (${\cal A}_1=10
{\cal A}_2$), meaning that the energy carried by the first overtone is
roughly $1 \%$ of the total energy radiated in the ringdown.

Our results for $\rho_{\rm crit}$ and $\rho_{\rm both}$ are shown in
Figs.~\ref{milanfar-fig2} and \ref{milanfar-fig3}, respectively.  Once
again, most features of these plots can be explained by looking at
Fig.~\ref{Qmodes}. As long as we are content with resolving either the
frequency {\it or} the damping time, higher $l$'s and positive $m$'s
are easier to resolve because the frequency separations are similar,
but the quality factor is larger. In fact, as $j$ grows the quality
factor for corotating modes grows, but it decreases for
counterrotating modes, and this affects resolvability in the
corresponding way. If we want to distinguish both frequency and
damping time, the behavior of the critical SNR changes. The corotating
fundamental mode with $l=m$ is impossible to resolve from the first
overtone in the extremal limit $j\to 1$ because they become
degenerate, tending to the Detweiler frequency. 

Fig.~\ref{milanfar-fig2} shows that, quite
independently of $(l,~m)$ and of the angular momentum $j$, {\it
resolving either the frequency or the damping time requires a SNR
$\rho>\rho_{\rm crit}\sim 10^2$}. Since the addition of a
small-amplitude overtone does not significantly alter the SNR, we can
use the SNR predictions of Fig.~\ref{insp-rd} to deduce that {\it tests
of the no-hair theorem should be feasible even under the most
pessimistic assumptions on the ringdown efficiency $\epsilon_{\rm rd}$
(at least for equal-mass mergers) as long as the first overtone
radiates a fraction $\sim 10^{-2}$ of the total ringdown
energy}. However, resolving {\it both} frequencies and damping times
typically requires a SNR $\rho>\rho_{\rm both}\sim 10^3$. This is only
possible under rather optimistic assumptions on the radiative
efficiency, and it can be impossible if the dominant mode has $l=m=2$
and the black hole is rapidly spinning (solid black line in the left
panel of Fig.~\ref{milanfar-fig3}).  We hope to refine this analysis using
computed estimates of excitation coefficients.

\section{Conclusions }
\label{conclusions}

We have presented a general framework for analysing the detectability of
quasi-normal ringdown gravitational waves from massive black holes, and for
using them to estimate parameters of the hole and to test the general
relativistic no-hair theorem.  In this initial work, we made a number of
simplifying assumptions, including the reality of the spin-weighted
spheroidal harmonics (thus simplifying the angle averaging),  
the restriction to one-mode or two-mode situations, and
the assumption of large quality factors in deriving analytic
expressions.  In future work, we
hope to explore the consequences of relaxing some of these assumptions.
Although we focused on detection of modes using \lisa, our framework is
equally applicable to ground-based interferometers.

For example, the non-angle-averaged 
case can be studied using Monte Carlo simulations in place of angle
averaging, thereby permitting inclusion of the fully
complex spheroidal harmonics.   
Multi-mode calculations involving more than two modes should be carried out.
Our discussion of testing the no-hair theorem has focused only on
resolvability of modes; we hope to use
our tools to perform in detail the no-hair test suggested in
\cite{dreyer}. 
A crucial point is to determine the excitation of modes.  When they have
different angular dependence, we must use numerical relativistic simulations
of mergers, or minimally simulations of distorted black holes with initial
data that mimics the merger.  For overtones with the same angular
dependence, we can use perturbative and analytical techniques to estimate
the relative amplitudes; this work is in progress.

It will be important to extend the analysis from matched-filtering to 
more sophisticated data-analysis
techniques (eg. the tiling method used by the 
TAMA group~\cite{nakano,tsunesada}) and combine
them with time-delay interferometry (TDI), especially in the case of
short damping times. 

Another important question is the effect of 
combining parameter estimation from ringdown with parameter estimation
from inspiral.  This may make it possible to
improve the measurement of mass or angular momentum by taking
advantage of ``prior'' information, and may permit one to
deduce how much energy and angular momentum is radiated in the merger
phase.

\section*{Acknowledgements}
We are grateful to Nils Andersson, Alessandra Buonanno, Sam Finn, Kostas
Glampedakis, Kostas Kokkotas, Pablo Laguna, Shane Larson, Hans-Peter
Nollert, Deirdre Shoemaker and Uli Sperhake for discussions and
suggestions. We also thank Hisashi Onozawa for kindly sharing with us
his results on Kerr quasinormal frequencies.
This work was supported in part by the National Science
Foundation under grant PHY 03-53180.

\appendix

\section{Accuracy of the $\delta$-function approximation}
\label{deltafun}

\begin{figure*}[t]
\begin{center}
\begin{tabular}{cc}
\epsfig{file=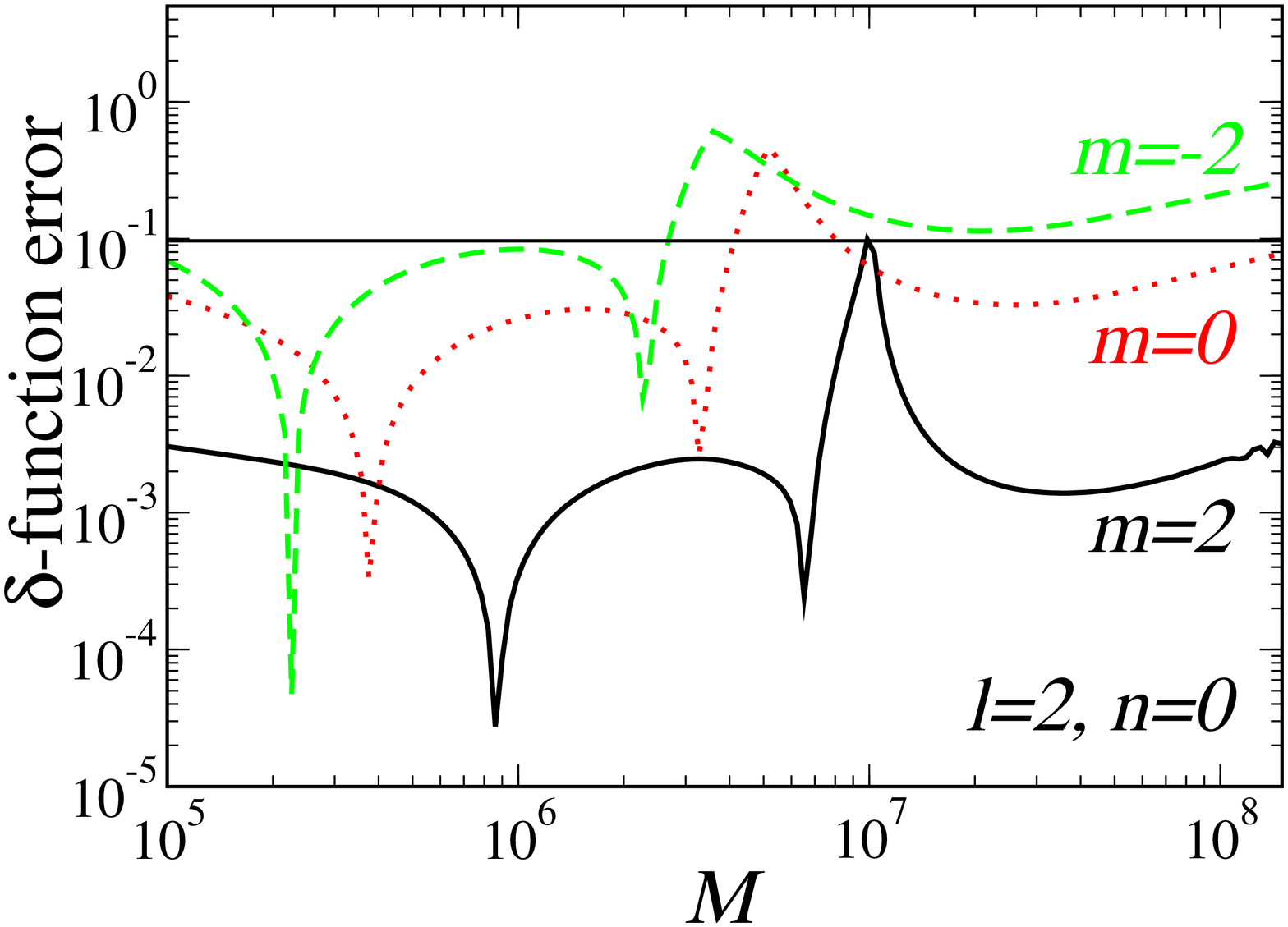,width=6cm,angle=-90} &
\epsfig{file=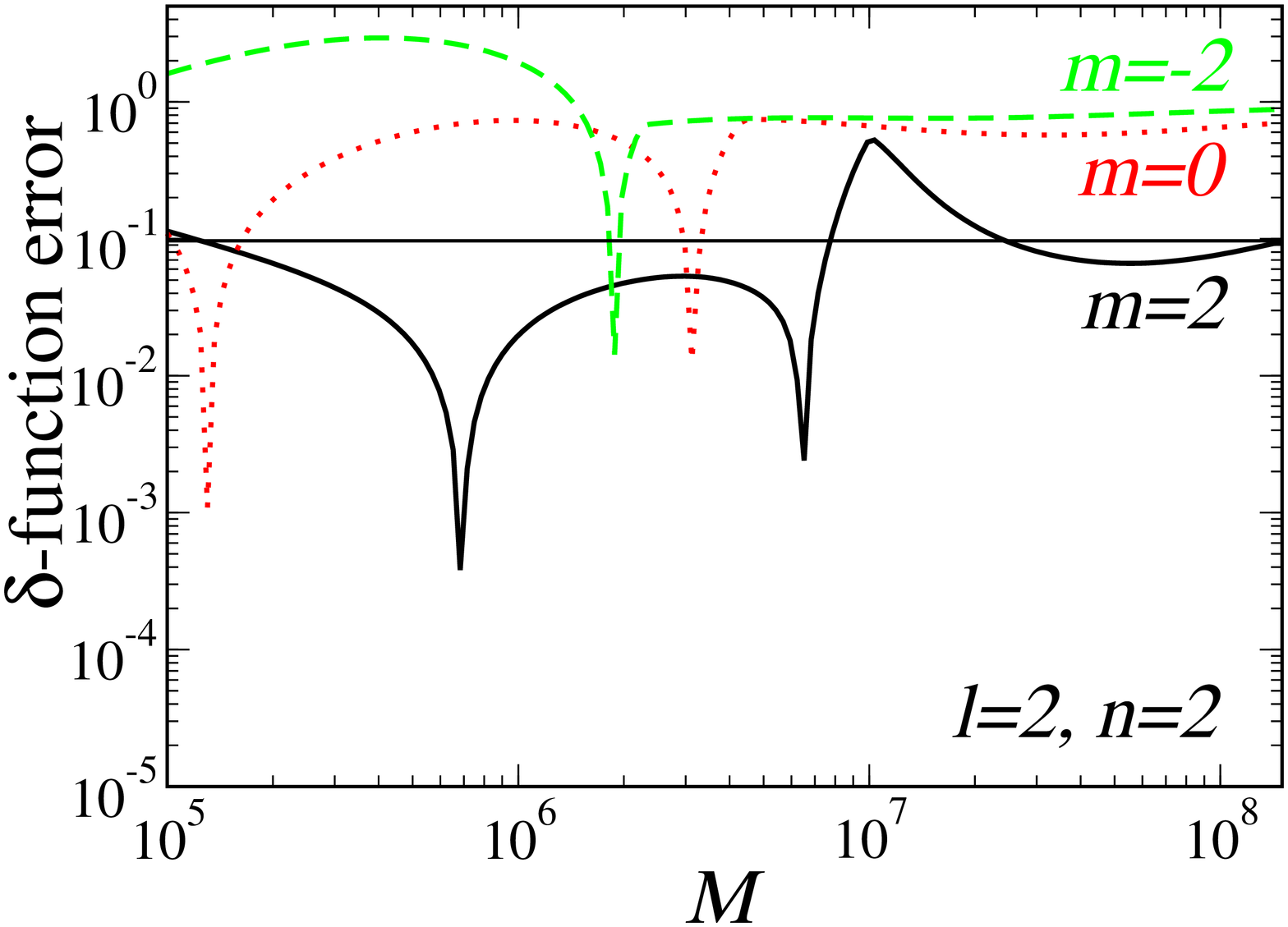,width=6cm,angle=-90} \\
\epsfig{file=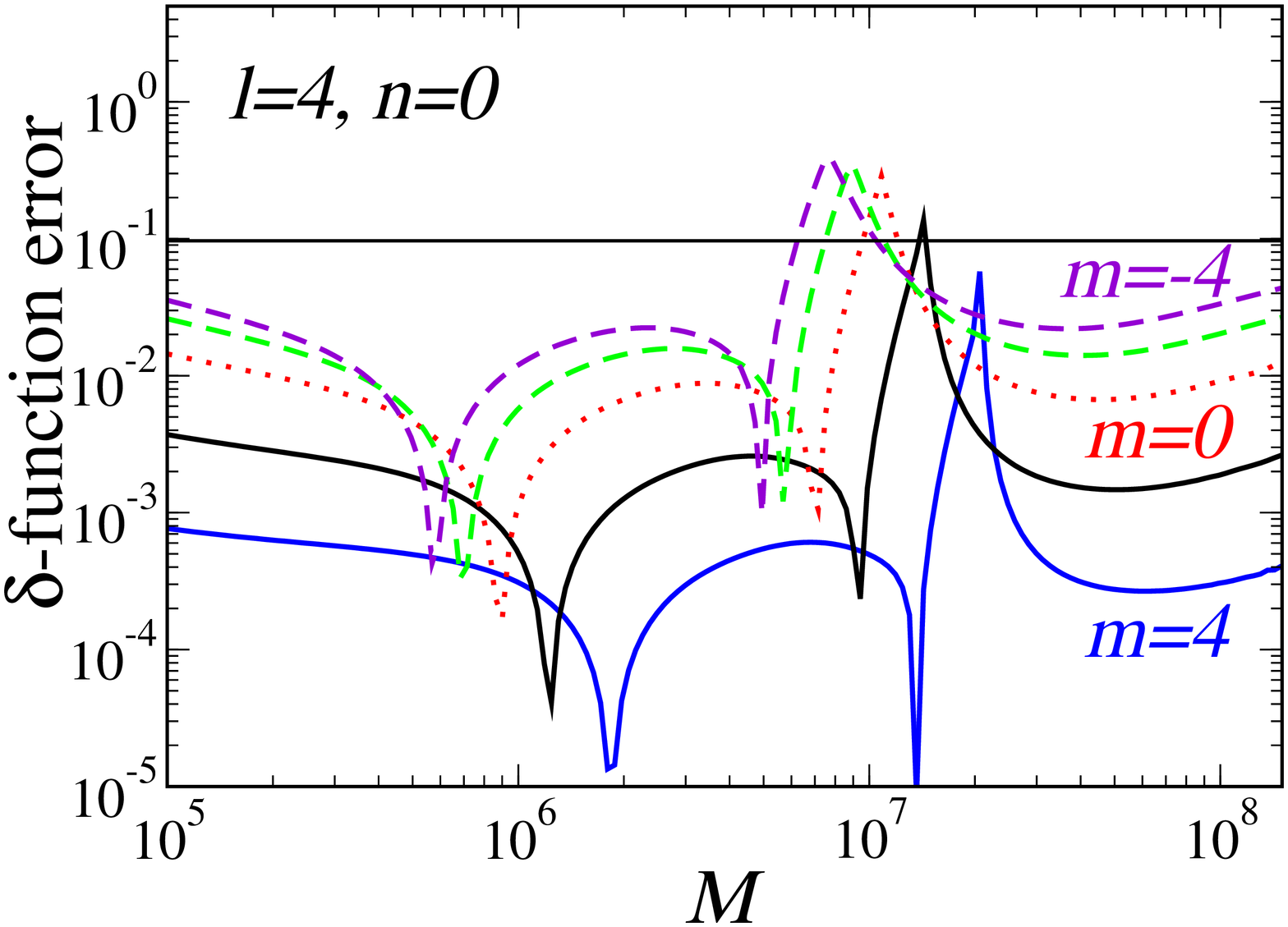,width=6cm,angle=-90} &
\epsfig{file=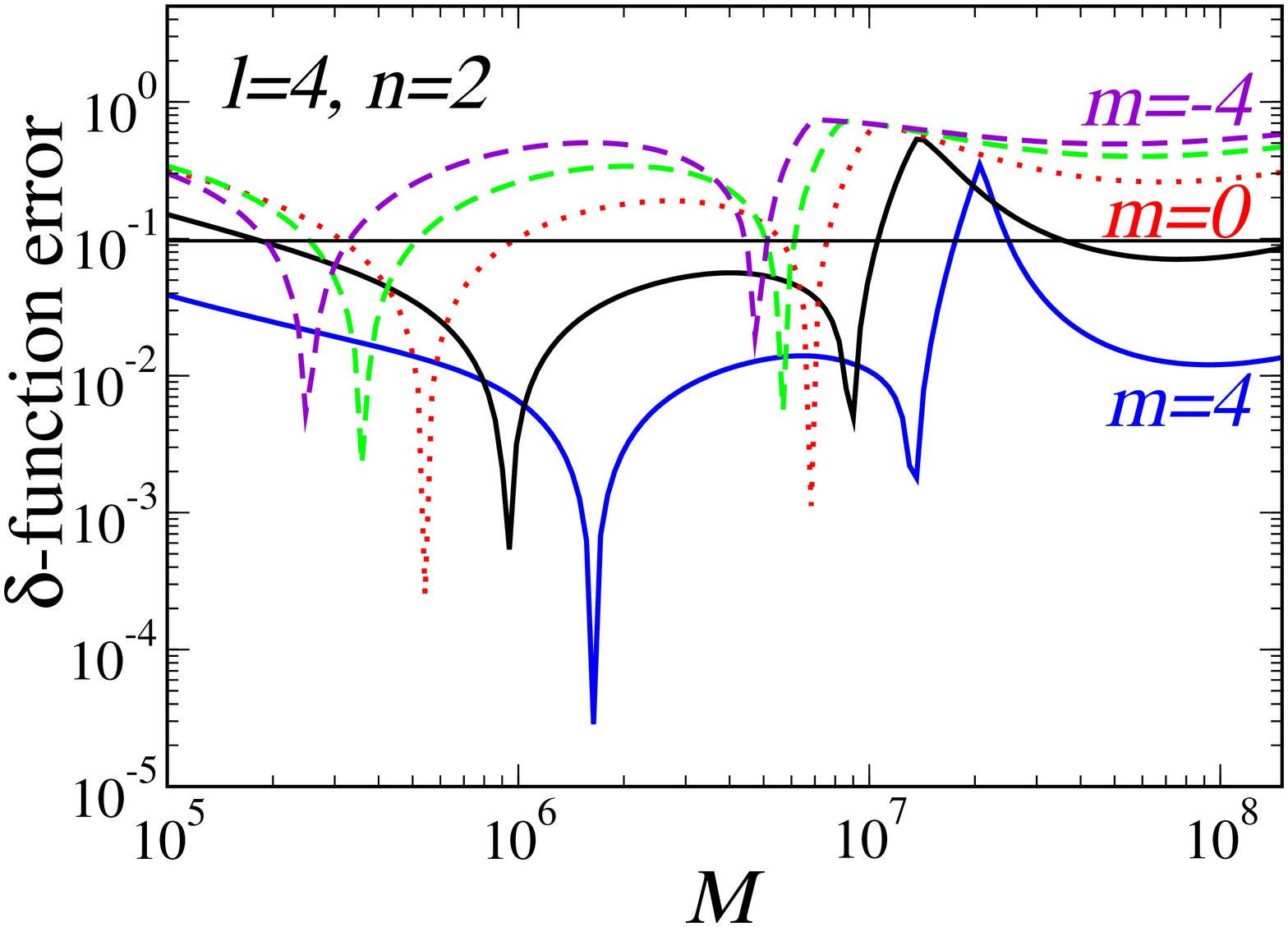,width=6cm,angle=-90} \\
\end{tabular}
\caption{Modulus of the error in the SNR due to the $\delta$-function
approximation, $|(\rho_{\delta}-\rho_{\rm{exact}})/\rho_{\rm{exact}}|$,
for modes with $l=2$ and $l=4$, for $j=0.98$ and $D_L = 3$ Gpc. 
For each $l$ we consider the
fundamental mode ($n=0$) and the second overtone ($n=2$). The
horizontal line marks a $10 \%$ deviation from the ``exact'' result:
for those modes lying below the horizontal line the $\delta$-function
approximation can be considered very accurate. The ``inverted peaks'' are just
zero-crossings of $|(\rho_{\delta}-\rho_{\rm{exact}})/\rho_{\rm{exact}}|$.
\label{DeltaError}}
\end{center}
\end{figure*}

How accurate is the $\delta$-function approximation we employed in
Sec.~\ref{snranalytic}? 
The answer to this question depends on $(l,m,n)$ and
on the angular momentum $j$ of the black hole.  In general, we expect
the approximation to hold better in the high-$\Qlm$ limit: that is,
for low overtones and (when the black hole is spinning) for modes with
$m>0$. This expectation is made more quantitative in
Fig.~\ref{DeltaError}, where we plot the relative error on the SNR due
to the $\delta$-function approximation for different modes as a
function of the black hole mass, for $D_L=3$ Gpc. 
To compute the ``exact'' SNR, 
we use the FH 
SNR in the Breit-Wigner form [Eq. (\ref{rhosqanalytic}) or 
(\ref{breitwigner})]. 
The plot refers to a near-extremal black hole
($j=0.98$), but the qualitative features we observe would be the same
for different (non-zero) values of the rotation rate. In this
near-extremal case the numerical integration of the ``full'' SNR
becomes problematic for corotating modes ($m>0$),
because the Breit-Wigner function effectively approaches a
$\delta$-function. This problem usually shows up for very large masses
(this is the origin of the small wiggles in the high-mass tail of the
curve marked by $m=2$ in the top-left panel). To achieve some
required numerical accuracy it is sufficient to double the number of
integration points until the high-mass wiggles disappear and the
integration routine converges.

The plot confirms our expectations. As a general trend, the
$\delta$-function approximation is more accurate for corotating modes
($m>0$), for the simple reason that the quality factor $\Qlm$ of
corotating modes tends to infinity (or becomes very large) as $j\to
1$. The approximation is more accurate for slowly damped modes (small
values of the overtone index $n$), which have a larger quality
factor.  As a rule of thumb, we can say that the $\delta$-function
approximation is reasonably accurate whenever it induces deviations
smaller than $\sim 10 \%$ with respect to the exact result: all modes
lying below the horizontal lines in Fig.~\ref{DeltaError} satisfy this
criterion. Fig.~\ref{DeltaError} shows that the approximation is very
accurate for the fundamental ($n=0$) mode, whatever the values of
$(l,m)$. However, when we consider the $n=2$
overtone, it is only marginally accurate for corotating modes,
and inaccurate for modes with $m\leq 0$. 

The large deviation at mass $M\sim 10^7 M_\odot$ corresponds to the
frequency band where white-dwarf confusion noise dominates over
instrumental noise. In this frequency band the noise curve is not a
very smooth function, and the SNR (especially for corotating modes) is
more sensitive to the ``full'' shape of the noise curve. We verified
that the peak disappears when we omit the white-dwarf confusion noise
from the Barack-Cutler noise curve. The larger disagreement in this
frequency regime corresponds, unfortunately, to the most promising
mass range for detection of black hole ringdown from galactic
centers. This means that we probably need a good control of the
white-dwarf confusion noise (and a rather accurate model of the
ringdown waveform) to detect these events.

\section{Fisher matrices for a single-mode waveform}
\label{app:FisherOne}

Taking the Fourier transform of ringdown waveforms requires some
care. For ringdown waves from SMBH mergers, the waveform emitted
before ringdown starts (say, for $t<0$) is the poorly known merger
waveform. To obtain the SNR from the ringdown signal alone, a possible
guess is to assume that the waveform $h(t)=0$ for $t<0$. Under this
assumption we can calculate the Fourier transform of the waveform by
integrating only over the range $t>0$. This method was used, among
others, by Echeverria \cite{echeverria} and Finn \cite{finn}, and for
this reason we call it the Echeverria-Finn (EF) convention. More
rigorously, we would really like to know the probability that a
ringdown waveform is present in the data stream starting (say) at
$t=0$. This probability should be computed integrating over all
possible realizations of the noise for $t<0$. Appendix A of FH shows
that this is equivalent to minimizing the SNR over all choices of the
function $h(t)$ on the negative $t$ axis, and that the SNR obtained by
minimizing over these choices is always within a few tens of a percent
of the SNR obtained using the ``FH prescription'', as described in
Section \ref{snranalytic}. In the body of the paper we usually adopt
the FH prescription. In this Appendix we present, for completeness,
the Fisher matrix calculation in the EF convention. We also list the
elements of the Fisher matrix adopting the FH convention, but assuming
that we do not know {\it a priori} the relative amplitudes or phases
of the two polarizations.

\subsection{Echeverria-Finn convention, four parameters}

Here we compute the single-mode Fisher matrix following the Echeverria-Finn
convention for calculating the Fourier transforms.  We start with the
waveform as defined in Eqs. (\ref{wf1}) -- (\ref{wf3}).  But in this case,
instead of 
taking $|t|$ in the damped exponential, integrating over $t$ from $-\infty$ to
$+\infty$
and dividing the resulting spectrum by $\sqrt{2}$ to compensate for
the doubling, we assume that the waveform vanishes for $t<0$, and integrate
only over positive $t$.

Assuming that the noise can be considered constant over
the bandwidth of the signal, we get the
angle-averaged SNR $\rho_{\rm EF} = \gamma (1+4\Qlm^2 -\beta + 2\Qlm
\alpha)$, where $\alpha$, $\beta$ and $\gamma$ are 
given by Eqs.~(\ref{alphabetagamma}).
The result is equivalent to 
Eq.~(\ref{rhoFinn}).  

The Fisher matrix in the $(A^{+},\, \ph^+,\,\flm,\, \Qlm)$ basis is given
by
\begin{subequations}
\beq
\Gamma_{A^+A^+} &=& \f{\gamma}{(A^+)^2} \left (1+4\Qlm^2-\beta + 2\Qlm
\alpha \right) \,,
\\
\Gamma_{A^+\ph^+} &=& \f{\gamma}{A^+} \left (\alpha +2\Qlm \beta \right ) \,,
\\
\Gamma_{A^+ \flm} &=& - \f{\gamma }{2 A^+ \flm}
\left (1+4\Qlm^2-\beta +2\Qlm \alpha \right) \,,
\\
\Gamma_{A^+ \Qlm} &=&  \f{\gamma}{2 A^+ \Qlm}
\f{1}{1+4\Qlm^2}
\left [(1+4\Qlm^2)^2 - (1-4\Qlm^2) \beta + 4\Qlm \alpha \right ]  \,,
\\
\Gamma_{\ph^+\ph^+} &=& \gamma \left (1+4\Qlm^2+\beta - 2\Qlm \alpha \right) \,,
\\
\Gamma_{\ph^+ \flm} &=& -\f{\gamma}{2\flm} \left [\alpha 
 -2\Qlm (1+4\Qlm^2 - \beta) \right ]\,,
\\
\Gamma_{\ph^+ \Qlm} &=& \f{\gamma}{2 \Qlm}
\left (\f{1}{1+4\Qlm^2} \right ) \left [ (1-4\Qlm^2)\alpha +4\Qlm
\beta \right ]
\,,
\\
\Gamma_{\flm \flm} &=& \f{\gamma}{2\flm^2} \left [ (1+4\Qlm^2)^2 -
\beta + 2\Qlm \alpha \right ] \,,
\\
\Gamma_{\flm \Qlm} &=& -\f{\gamma}{2\flm \Qlm}
\f{1}{1+4\Qlm^2}
\left [ (1+4\Qlm^2)^2 - (1-4\Qlm^2)\beta +4\Qlm \alpha \right ]  \,,
\\
\Gamma_{\Qlm \Qlm} &=& \f{\gamma}{2 \Qlm^2}
\f{1}{(1+4\Qlm^2)^2}
\left [ (1+4\Qlm^2)^3 - (1-12\Qlm^2)\beta + (6-8\Qlm^2)\Qlm\alpha \right ]  \,.
\eeq
\label{fishermatrixEF}
\end{subequations}
\noindent
In the $(A^{+},\, \ph^+,\, M, \,j)$ basis, 
an expansion of the errors in powers of $Q_{lmn}^{-1}$ gives
\begin{subequations}
\label{riiEF}
\beq
\sigma_j &=&
\frac{1}{\rho_{\rm EF}}
\left|2\frac{Q_{lmn}}{Q_{lmn}'}
\left(1+\f{1+\alpha^2+2\beta^2}{8\Qlm^2}\right)\right|\,, \label{erraEF}\\
\sigma _M &=&
\frac{1}{\rho_{\rm EF}}
\left|2\frac{MQ_{lmn}
f_{lmn}'}{f_{lmn}Q_{lmn}'}
\left(1+\f{1+\alpha^2+2\beta^2}{8 \Qlm^2}\right)\right|\,, \label{errmEF}\\
\sigma _{A^+} &=&
\frac{\sqrt{2} A^+}{\rho_{\rm EF}}
\left |1-\f{\alpha}{4\Qlm} \right|\,,\\
\sigma _{\ph^+} &=&
\frac{\sqrt{2}}{\rho_{\rm EF}} \left|1+\f{3\alpha}{4\Qlm}\right|
\,. \label{errphEF}
\eeq
\end{subequations}
~For the correlation coefficients we get
\allowdisplaybreaks{
\begin{subequations}
\label{finncorrs4}
\beq r_{jM}&=&{\rm sgn}(\flm')\times
\left(
1-\frac{f_{lmn}^2Q_{lmn}'^2}{8f_{lmn}'^2Q_{lmn}^4}
\right)
+{\cal O}(1/Q^{5}) \,, \\
r_{jA}&=&-\frac{1}{\sqrt{2}}
\left(1-\f{\alpha}{4Q_{lmn}}\right)
+{\cal O}(1/Q^{2})\,,  \\
r_{MA}&=&-\frac{1}{\sqrt{2}}
\left(1-\f{\alpha}{4Q_{lmn}}\right)
+{\cal O}(1/Q^{2})\,,  \\
r_{j\ph^+}&=&-\frac{1}{2\sqrt{2}}\frac{1-2\beta}{Q_{lmn}}
+{\cal O}(1/Q^{2})\,,  \\
r_{M\ph^+}&=&-\frac{1}{2\sqrt{2}}\frac{1-2\beta}{Q_{lmn}}
+{\cal O}(1/Q^{2})\,,  \\
r_{A\ph^+}&=&
-\frac{\beta}{Q_{lmn}}-
\frac{\alpha}{Q_{lmn}^2}
+{\cal O}(1/Q^{3})\,.
\eeq
\end{subequations}
}

\subsection{Flanagan-Hughes convention, six parameters}

Here we list the elements of the Fisher matrix for the case where we do not
know {\it a priori} the relative amplitudes or phases of the two
polarizations.  In the six-parameter basis 
$(A^{+},\, A^{\times},\, \ph^{+},\, \ph^{\times},\, \flm,\, \Qlm)$, they
are
\allowdisplaybreaks{
\begin{subequations}
\beq
\Gamma_{A^+A^+} &=& \f{\gamma}{A^2} \left (1+4\Qlm^2+ \cos 2\ph^{+} \right) \,,
\\
\Gamma_{A^\times A^\times} &=& \f{\gamma}{A^2} \left (1+4\Qlm^2 \
-\cos 2\ph^{\times} \right) \,,
\\
\Gamma_{A^+ A^\times}  &=& \Gamma_{A^+\ph^{\times}}
=\Gamma_{A^\times\ph^{+}} = 0 \,,
\\
\Gamma_{A^+ \ph^+} &=& - \f{\gamma}{A} \sin 2\ph^+ \cos \psi \,,
\\
\Gamma_{A^\times\ph^\times} &=& \f{\gamma}{A} \sin 2\ph^\times \sin \psi \,,
\\
\Gamma_{A^+ \flm} &=& - \f{\gamma }{2 A \flm}
\left (1+4\Qlm^2 + \cos 2 \ph^+ \right) \cos \psi \,,
\\
\Gamma_{A^\times \flm} &=& - \f{\gamma }{2 A \flm}
\left (1+4\Qlm^2 - \cos 2 \ph^\times \right) \sin \psi \,,
\\
\Gamma_{A^+ \Qlm} &=&  \f{\gamma}{2 A \Qlm}
\f{1}{1+4\Qlm^2}
\left [(1+4\Qlm^2)^2 + (1-4\Qlm^2) \cos 2 \ph^+ \right ] \cos \psi \,,
\\
\Gamma_{A^\times \Qlm} &=&  \f{\gamma}{2 A \Qlm}
\f{1}{1+4\Qlm^2}
\left [(1+4\Qlm^2)^2 - (1-4\Qlm^2) \cos 2 \ph^\times \right ] \sin \psi \,,
\\
\Gamma_{\ph^+ \ph^+ } &=& \gamma \left (1+4\Qlm^2 - \cos 2 \phi^+ \right)
\cos^2 \psi\,,
\\
\Gamma_{\ph^\times \ph^\times } &=& \gamma \left (1+4\Qlm^2 + 
\cos 2 \phi^\times \right) \sin^2 \psi\,,
\\
\Gamma_{\ph^+ \ph^\times } &=& 0 \,,
\\
\Gamma_{\ph^+ \flm} &=& \f{\gamma}{2\flm} \sin 2 \ph^+ \cos^2 \psi \,,
\\
\Gamma_{\ph^\times \flm} &=& -\f{\gamma}{2\flm} \sin 2 \ph^\times \sin^2 \psi \,,
\\
\Gamma_{\ph^+ \Qlm} &=& -\f{\gamma}{2 \Qlm}
\left (\f{1-4\Qlm^2}{1+4\Qlm^2} \right ) \sin 2 \phi^+ \cos^2 \psi \,,
\\
\Gamma_{\ph^\times \Qlm} &=& \f{\gamma}{2 \Qlm}
\left (\f{1-4\Qlm^2}{1+4\Qlm^2} \right ) \sin 2 \phi^\times \sin^2 \psi \,,
\\
\Gamma_{\flm \flm} &=& \f{\gamma}{2\flm^2} \left [ (1+4\Qlm^2)^2 -
\beta \right ] \,,
\\
\Gamma_{\flm \Qlm} &=& -\f{\gamma}{2\flm \Qlm}
\f{1}{1+4\Qlm^2}
\left [ (1+4\Qlm^2)^2 - (1-4\Qlm^2)\beta \right ]  \,,
\\
\Gamma_{\Qlm \Qlm} &=& \f{\gamma}{2 \Qlm^2}
\f{1}{(1+4\Qlm^2)^2}
\left [ (1+4\Qlm^2)^3 - (1-12\Qlm^2)\beta \right ]  \,,
\eeq
\label{fishermatrix6}
\end{subequations}
}
\noindent
where $A^2 = (A^+)^2 + (A^\times)^2$, $\cos \psi \equiv A^+/A$ and 
$\sin \psi \equiv A^\times /A$, and where
$\alpha$, $\beta$ and $\gamma$ are given by
Eqs. (\ref{alphabetagamma}).  Also, as in the
four-parameter case, $\rho_{FH}^2 = \gamma (1+4\Qlm^2-\beta)$.

As before, one can convert to the $(M,j)$ basis using Eqs.
(\ref{Mjtransform}), and then invert the Fisher matrix to obtain the errors
and correlation coefficients.  The resulting formulae are lengthy and
unenlightening, and we do not display them here.

\section{\lisa~noise curve }\label{app:noise}

\begin{figure*}[t]
\epsfig{file=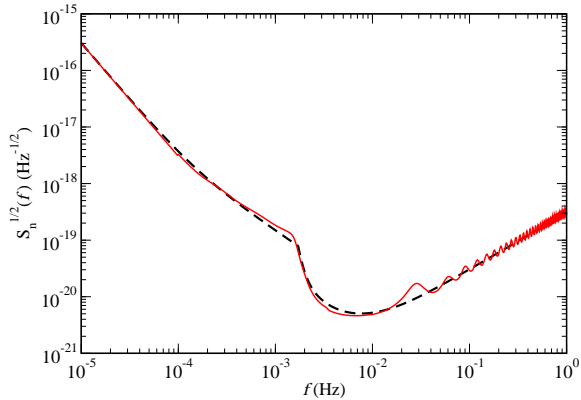,width=0.75\sizeonefig,angle=-90}
\caption{Analytic approximation to the \lisa~ root noise spectral
density curve used in this paper and in Ref.~\cite{BC} (dashed line)
and the curve produced using the {\em LISA} Sensitivity Curve
Generator~\cite{SCG} (solid line).  The SCG curve has been multiplied
by a factor of $\sqrt{3/20}$ to obtain an {\em effective} non-sky
averaged noise spectral density.  The SCG noise curve does not include
the extragalactic white dwarf confusion noise while the analytical
approximation curve does.
\label{noise}}
\end{figure*}

Generally, the \lisa~ community has been using the so-called sky-averaged
spectral noise density $S_h^{\rm SA}$ [see e.g., Ref.~\cite{FT} and the
\lisa~Pre-Phase A Report],
computed by a combination of three factors, including: (i) the raw
spectral noise density $S_n$, (ii) the gravitational-wave transfer
(response) function $R$ and (iii) the noise transfer (response)
function $R_n$.  They combine together in~\cite{LHH}
\beq
S_h^{\rm SA} = \frac{S_n\,R_n}{R}\,.
\label{SA}
\eeq
In the low frequency
limit, the GW transfer function used in the {\em LISA} Sensitivity
Curve Generator~\cite{SCG} is $R=4(\sqrt{3}/2)^2 1/5=3/5$, where the
factor $(\sqrt{3}/2)^2$ comes from the \lisa~arms being at $60^{o}$,
the factor 1/5 is due to the sky-average of the pattern functions
($\langle F_{+,\times}^2 \rangle = 1/5$) and the factor 4 depends on
the particular read-out variable used.  Since our definition of the GW
signal already includes the factor $\sqrt{3}/2$, and since we choose to
sky-average the signal, we use an
effective non-sky-averaged spectral density, obtained by multplying
$S_h^{\rm
SA}$ by $(\sqrt{3}/2)^2/5=3/20$. The final result is Eq. (\ref{Snsa}),
and has been obtained also in Ref.~\cite{BC}.  We estimate white-dwarf
confusion noise following~\cite{BC}, which uses results
from~\cite{conf1,conf2}: the galactic contribution is approximated as
\beq
S_h^{\rm gal}(f)=
2.1\times 10^{-45}\left(\f{f}{1~{\rm Hz}}\right)^{-7/3}~{\rm Hz}^{-1}\,,
\eeq
and the contribution from extra-galactic white dwarfs as
\beq
S_h^{\rm ex-gal}(f)=
4.2\times 10^{-47}\left(\f{f}{1~{\rm Hz}}\right)^{-7/3}~{\rm Hz}^{-1}\,.
\eeq
We compute the total (instrumental plus confusion) noise as
\beq
S_h(f)={\rm min}\left\{
S_h^{\rm NSA}(f)/{\rm exp}
\left(-\kappa T^{-1}_{\rm mission} dN/df\right),~
S_h^{\rm NSA}(f)+S_h^{\rm gal}(f)
\right\}+S_h^{\rm ex-gal}(f)\,.
\label{Shtot}
\eeq
Here $dN/df$ is the number density of galactic white-dwarf binaries
per unit gravitational-wave frequency, for which we adopt the estimate
\beq
\f{dN}{df}=2\times 10^{-3}~{\rm Hz}^{-1}
\left(\f{1~{\rm Hz}}{f}\right)^{11/3}\,;
\eeq
$\Delta f=T^{-1}_{\rm mission}$ is the bin size of the discretely
Fourier transformed data for a {\em LISA} mission lasting a time
$T_{\rm mission}$; and $\kappa\simeq 4.5$ is the average number of
frequency bins that are lost when each galactic binary is fitted
out. The factor ${\rm exp}\left(-\kappa T^{-1}_{\rm mission}
dN/df\right)$ thus represents the fraction of ``uncorrupted'' bins
where instrumental noise still dominates. The analytic root noise spectral
density curve (\ref{Shtot}) used in this paper is shown in
Fig.~\ref{noise} together with the corresponding root noise spectral
density curve from the {\em LISA} Sensitivity Curve
Generator~\cite{SCG}.  The SCG curve shown is obtained using the
nominal values SNR=1, arm length $=5 \times 10^9$ m, telescope
diameter $=0.3$ m, laser wavelength $=1064$ nanometers, laser power
$=1.0$ Watts, optical train efficiency $=0.3$, acceleration noise $=3
\times 10^{-15} \,{\rm m}\,{\rm s}^{-2}\,{\rm Hz}^{-1/2}$, and
position noise budget $=2 \times 10^{-11} \,{\rm m}\,{\rm Hz}^{-1/2}$,
with position noise setting the floor at high frequency.  The data
returned by the SCG is then multiplied by $\sqrt{3/20}$ to obtain the
effective non-sky averaged curve shown in Fig. ~\ref{noise}.

\section{The Shahram-Milanfar criterion for resolvability} 
\label{recover-rayleigh}

For damped sinusoids it makes sense to assume that the signal duration
$\Delta t$ is given by the e-folding time $\tau$, so the ``standard''
Rayleigh criterion (\ref{rayleigh}) becomes $(f_1-f_2)\tau>1$.  In
\cite{smilanfar} (see also \cite{milanfars}) the authors discuss how
to go beyond this (purely classical) Rayleigh limit. They show how to
bypass this limit for two sinusoids with nearby frequencies
$f_1\,,f_2$ such that $2\delta\equiv f_1-f_2\ll 1/B$, where $B$ is the
duration of the signal.  Assuming white noise with zero mean and unit
variance $\sigma ^2$, they get a condition for the minimum SNR
required for the frequencies of two equal-amplitude sinusoids to be
resolved \cite{notemilanfar}:
\be
\rho^2 \geq \frac{640}{\pi^4}\frac{\left (T^{-1}(P_f)-T^{-1}(P_d)\right
)^2}{(2\delta B)^4}\,.\label{mil}
\ee
Here $T(x)$ is the right-tail probability function for a Gaussian
random variable $X$ with zero mean and unit variance
\cite{noterighttail}. $P_d$ and $P_f$ denote the detection and
false-alarm rates, typical values for these being, say, $P_d
=0.99\,,\,P_f=0.01$.
Their analysis can be carried over to our case, assuming large $Q$
factors, and therefore a signal with duration $\tau \sim Q/f$.  We
would get
\be
\rho^2 \geq \frac{640}{\pi^4}\frac{\left (T^{-1}(P_f)-T^{-1}(P_d)\right )^2}
{\tau^4(f_1-f_2)^4}\sim \frac{1}{\tau^4(f_1-f_2)^4}\,.
\label{milour}
\ee

For overtones (``nearly-parallel signals'', in the sense of the scalar
product of SWSHs being close to unity) with equal amplitudes and
quality factors the frequency error has a large-$Q$ behavior
\be 
\sigma_{f}\sim \frac{1}{\rho \tau^2 |f_1-f_2|}\,,\label{sigmaour}
\ee
so our frequency resolvability criterion, $|f_1-f_2|>{\rm max}\left
(\sigma_{f_1},\sigma_{f_2}\right )$ agrees with the Shahram-Milanfar
result (\ref{milour}). Notice however that the situation considered in
\cite{smilanfar} is very special: overtones with the same $(l,m)$,
equal amplitudes ${\cal A}_1={\cal A}_2$ and equal quality factors
$Q_1=Q_2$ are the {\it only} class of modes for which $\sigma_f \sim
1/|f_1-f_2|$.

\section{Quasinormal frequencies for rotating black holes}
\label{app:QNM}

There is a vast literature on Kerr QNMs
\cite{leaver,sai,onozawa,cardosoberti}, but the QNM frequencies which
are relevant for detection have never been systematically tabulated.
Here we list for reference the values of the complex frequencies and
separation constants for selected QNMs.  These were calculated
using two different numerical implementations of the continued
fraction method of Leaver~\cite{leaver}, one written by one of us
(EB), and the other provided by Hisashi Onozawa.  Both methods give
results in excellent agreement with each other.  Shown in Tables
\ref{tab:l2n0}, \ref{tab:l3n0} and \ref{tab:l4n0} are the real and
imaginary parts of the complex QNM frequencies, separated by commas,
for $l=2$, $3$ and $4$, respectively, and for the three lowest
overtones.  Tables \ref{tab:l2Alm}, \ref{tab:l3Alm} and
\ref{tab:l4Alm} display the angular separation constants $A_{lmn}$ in
the same format.  
Following Leaver, the numerical values of the $A_{lmn}$'s were
obtained assuming that the perturbations have a time dependence of the
form $e^{-\ii \omega t}$, while in this paper we use the opposite
convention on the Fourier transform.

\begin{table}[t] 
\centering \caption{\label{tab:l2n0} First three overtones for $l=2$.
A comma separates the real part from the imaginary part of
$M\omega$. To save space, in this and the following Tables we omit
leading zeros.}
\begin{tabular}{cccccc}  \hline
\multicolumn{6}{c}{$l=2$, $n=0$} \\
\hline
$j$ &$m=2$ &$m=1$ &$m=0$ &$m=-1$ &$m=-2$\\
\hline
0.00& .3737,.0890& .3737,.0890& .3737,.0890& .3737,.0890& .3737,.0890\\
0.10& .3870,.0887& .3804,.0888& .3740,.0889& .3678,.0890& .3618,.0891\\
0.20& .4021,.0883& .3882,.0885& .3751,.0887& .3627,.0889& .3511,.0892\\
0.30& .4195,.0877& .3973,.0880& .3770,.0884& .3584,.0888& .3413,.0892\\
0.40& .4398,.0869& .4080,.0873& .3797,.0878& .3546,.0885& .3325,.0891\\
0.50& .4641,.0856& .4206,.0862& .3833,.0871& .3515,.0881& .3243,.0890\\
0.60& .4940,.0838& .4360,.0846& .3881,.0860& .3489,.0876& .3168,.0890\\
0.70& .5326,.0808& .4551,.0821& .3941,.0845& .3469,.0869& .3098,.0887\\
0.80& .5860,.0756& .4802,.0780& .4019,.0822& .3454,.0860& .3033,.0885\\
0.90& .6716,.0649& .5163,.0698& .4120,.0785& .3444,.0849& .2972,.0883\\
0.98&.8254,.0386& .5642,.0516& .4223,.0735& .3439,.0837& .2927-.0881\\
\hline
\hline
\multicolumn{6}{c}{$l=2$, $n=1$} \\
\hline
$j$ &$m=2$ &$m=1$ &$m=0$ &$m=-1$ &$m=-2$\\
\hline
0.00& .3467,.2739& .3467,.2739& .3467,.2739& .3467,.2739& .3467,.2739\\
0.10& .3619,.2725& .3545,.2731& .3472,.2737& .3400,.2744& .3330,.2750\\
0.20& .3790,.2705& .3635,.2717& .3486,.2730& .3344,.2744& .3206,.2759\\
0.30& .3984,.2680& .3740,.2698& .3511,.2718& .3296,.2741& .3093,.2765\\
0.40& .4208,.2647& .3863,.2670& .3547,.2700& .3256,.2734& .2989,.2769\\
0.50& .4474,.2602& .4009,.2631& .3594,.2674& .3225,.2723& .2893,.2772\\
0.60& .4798,.2538& .4183,.2575& .3655,.2638& .3201,.2708& .2803,.2773\\
0.70& .5212,.2442& .4399,.2492& .3732,.2585& .3184,.2686& .2720,.2773\\
0.80& .5779,.2281& .4676,.2358& .3826,.2507& .3173,.2658& .2643,.2772\\
0.90& .6677,.1953& .5059,.2097& .3935,.2385& .3167,.2620& .2570,.2770\\
0.98& .8249,.1159& .5477,.1509& .4014,.2231& .3164,.2581& .2515,.2768\\
\hline
\hline
\multicolumn{6}{c}{$l=2$, $n=2$} \\
\hline
$j$ &$m=2$ &$m=1$ &$m=0$ &$m=-1$ &$m=-2$\\
\hline
0.00& .3011,.4783& .3011,.4783& .3011,.4783& .3011,.4783& .3011,.4783\\
0.10& .3192,.4735& .3104,.4756& .3017,.4778& .2932,.4801& .2846,.4825\\
0.20& .3393,.4679& .3214,.4719& .3038,.4764& .2866,.4811& .2697,.4862\\
0.30& .3619,.4613& .3342,.4671& .3074,.4739& .2813,.4814& .2559,.4895\\
0.40& .3878,.4533& .3492,.4607& .3124,.4701& .2772,.4808& .2433,.4925\\
0.50& .4179,.4433& .3669,.4522& .3190,.4647& .2741,.4794& .2316,.4952\\
0.60& .4542,.4303& .3878,.4407& .3273,.4571& .2721,.4768& .2207,.4977\\
0.70& .4999,.4123& .4133,.4241& .3374,.4464& .2709,.4729& .2107,.4999\\
0.80& .5622,.3839& .4451,.3984& .3488,.4307& .2703,.4674& .2013,.5019\\
0.90& .6598,.3275& .4867,.3502& .3591,.4067& .2697,.4600& .1925,.5038\\
0.98& .8238,.1933& .5201,.2331& .3599,.3808& .2686,.4527& .1858,.5051\\
\hline
\end{tabular}
\end{table}

\begin{table}[t]
\centering \caption{\label{tab:l3n0} First three overtones for $l=3$.}
\begin{tabular}{cccccccc}  \hline
\multicolumn{8}{c}{$l=3$, $n=0$} \\
\hline
$j$ &$m=3$ &$m=2$ &$m=1$ &$m=0$ &$m=-1$ &$m=-2$ &$m=-3$\\
\hline
0.00& .5994,.0927& .5994,.0927& .5994,.0927& .5994,.0927& .5994,.0927& .5994,.0927& .5994,.0927\\
0.10& .6208,.0924& .6137,.0925& .6067,.0926& .5999,.0926& .5932,.0927& .5867,.0928& .5802,.0928\\
0.20& .6448,.0920& .6297,.0921& .6153,.0923& .6014,.0924& .5880,.0926& .5752,.0927& .5628,.0929\\
0.30& .6721,.0913& .6480,.0915& .6252,.0918& .6038,.0921& .5837,.0923& .5647,.0926& .5469,.0928\\
0.40& .7037,.0902& .6689,.0906& .6369,.0911& .6074,.0915& .5802,.0920& .5553,.0924& .5323,.0927\\
0.50& .7409,.0887& .6934,.0893& .6506,.0900& .6121,.0908& .5776,.0915& .5467,.0920& .5188,.0925\\
0.60& .7862,.0864& .7228,.0873& .6670,.0885& .6183,.0897& .5758,.0908& .5388,.0916& .5063,.0922\\
0.70& .8437,.0829& .7592,.0842& .6870,.0861& .6261,.0882& .5749,.0899& .5316,.0912& .4946,.0919\\
0.80& .9219,.0770& .8068,.0791& .7121,.0824& .6360,.0860& .5749,.0888& .5250,.0906& .4837,.0916\\
0.90& 1.0446,.0655& .8762,.0689& .7455,.0758& .6487,.0826& .5758,.0872& .5191,.0900& .4735,.0913\\
0.98& 1.2602,.0387& .9769,.0453& .7833,.0643& .6615,.0782& .5773,.0856& .5146,.0894& .4657,.0910\\
\hline
\hline
\multicolumn{8}{c}{$l=3$, $n=1$} \\
\hline
$j$ &$m=3$ &$m=2$ &$m=1$ &$m=0$ &$m=-1$ &$m=-2$ &$m=-3$\\
\hline
0.00& .5826,.2813& .5826,.2813& .5826,.2813& .5826,.2813& .5826,.2813& .5826,.2813& .5826,.2813\\
0.10& .6053,.2802& .5978,.2805& .5904,.2808& .5832,.2811& .5761,.2814& .5691,.2817& .5623,.2820\\
0.20& .6306,.2785& .6148,.2791& .5995,.2798& .5848,.2804& .5706,.2811& .5569,.2817& .5437,.2823\\
0.30& .6593,.2761& .6341,.2771& .6102,.2782& .5876,.2793& .5662,.2804& .5459,.2814& .5268,.2823\\
0.40& .6924,.2726& .6563,.2741& .6227,.2758& .5916,.2776& .5627,.2793& .5360,.2808& .5112,.2822\\
0.50& .7312,.2676& .6821,.2698& .6375,.2724& .5969,.2751& .5602,.2777& .5270,.2800& .4967,.2818\\
0.60& .7782,.2604& .7130,.2634& .6550,.2674& .6038,.2717& .5587,.2756& .5188,.2789& .4833,.2813\\
0.70& .8375,.2494& .7510,.2538& .6762,.2600& .6124,.2668& .5580,.2728& .5113,.2775& .4708,.2807\\
0.80& .9176,.2315& .8004,.2379& .7025,.2484& .6230,.2598& .5582,.2692& .5045,.2758& .4590,.2800\\
0.90& 1.0425,.1966& .8714,.2068& .7362,.2277& .6357,.2491& .5593,.2644& .4983,.2739& .4480,.2792\\
0.98& 1.2599,.1161& .9706,.1341& .7679,.1930& .6468,.2359& .5606,.2593& .4938,.2720& .4396,.2784\\
\hline
\hline
\multicolumn{8}{c}{$l=3$, $n=2$} \\
\hline
$j$ &$m=3$ &$m=2$ &$m=1$ &$m=0$ &$m=-1$ &$m=-2$ &$m=-3$\\
\hline
0.00& .5517,.4791& .5517,.4791& .5517,.4791& .5517,.4791& .5517,.4791& .5517,.4791& .5517,.4791\\
0.10& .5766,.4763& .5684,.4771& .5603,.4779& .5523,.4787& .5445,.4795& .5368,.4803& .5292,.4812\\
0.20& .6043,.4725& .5872,.4741& .5705,.4758& .5543,.4775& .5386,.4792& .5234,.4809& .5086,.4826\\
0.30& .6356,.4674& .6084,.4699& .5825,.4725& .5577,.4753& .5340,.4781& .5114,.4809& .4897,.4836\\
0.40& .6714,.4605& .6328,.4640& .5966,.4679& .5625,.4721& .5305,.4763& .5005,.4804& .4723,.4842\\
0.50& .7131,.4511& .6612,.4558& .6131,.4614& .5689,.4675& .5282,.4736& .4907,.4793& .4561,.4845\\
0.60& .7632,.4379& .6948,.4441& .6327,.4522& .5769,.4611& .5270,.4698& .4819,.4778& .4410,.4846\\
0.70& .8258,.4185& .7358,.4268& .6561,.4387& .5868,.4521& .5267,.4648& .4739,.4757& .4269,.4843\\
0.80& .9096,.3874& .7884,.3988& .6844,.4177& .5984,.4391& .5272,.4581& .4667,.4730& .4136,.4839\\
0.90& 1.0383,.3282& .8622,.3449& .7177,.3806& .6106,.4197& .5282,.4492& .4600,.4696& .4011,.4834\\
0.98& 1.2592,.1935& .9605,.2181& .7349,.3224& .6176,.3976& .5286,.4403& .4550,.4665& .3916,.4828\\
\hline
\end{tabular}
\end{table}

\begin{table}[t]
\centering \caption{\label{tab:l4n0} First three overtones for $l=4$.}
\begin{tabular}{cccccccccc}  \hline
\multicolumn{10}{c}{$l=4$, $n=0$} \\
\hline
$j$ &$m=4$ &$m=3$ &$m=2$ &$m=1$ &$m=0$ &$m=-1$ &$m=-2$ &$m=-3$ &$m=-4$\\
\hline
0.00& .8092,.0942& .8092,.0942& .8092,.0941& .8092,.0942& .8092,.0942& .8092,.0942& .8092,.0942& .8092,.0942& .8092,.0942\\
0.10& .8387,.0940& .8313,.0940& .8240,.0940& .8168,.0941& .8098,.0941& .8028,.0941& .7960,.0942& .7893,.0942& .7826,.0942\\
0.20& .8717,.0935& .8560,.0936& .8407,.0937& .8259,.0938& .8116,.0939& .7977,.0940& .7842,.0941& .7712,.0941& .7585,.0942\\
0.30& .9092,.0929& .8839,.0930& .8597,.0932& .8366,.0934& .8146,.0935& .7937,.0937& .7737,.0938& .7546,.0940& .7365,.0941\\
0.40& .9525,.0918& .9159,.0921& .8815,.0924& .8493,.0927& .8191,.0930& .7908,.0933& .7643,.0935& .7395,.0937& .7162,.0939\\
0.50& 1.0034,.0903& .9532,.0907& .9069,.0912& .8642,.0917& .8250,.0922& .7890,.0927& .7559,.0931& .7255,.0934& .6974,.0936\\
0.60& 1.0650,.0879& .9978,.0885& .9369,.0893& .8820,.0902& .8326,.0911& .7883,.0919& .7485,.0925& .7125,.0930& .6800,.0933\\
0.70& 1.1427,.0842& 1.0530,.0852& .9734,.0865& .9035,.0881& .8423,.0896& .7889,.0909& .7420,.0919& .7005,.0926& .6637,.0930\\
0.80& 1.2475,.0781& 1.1254,.0796& 1.0198,.0820& .9301,.0848& .8546,.0874& .7907,.0895& .7363,.0911& .6894,.0921& .6485,.0926\\
0.90& 1.4104,.0662& 1.2322,.0685& 1.0836,.0733& .9647,.0791& .8702,.0841& .7941,.0877& .7314,.0901& .6790,.0915& .6343,.0922\\
0.98& 1.6919,.0388& 1.3944,.0425& 1.1631,.0558& 1.0019,.0703& .8860,.0800& .7979,.0858& .7282,.0892& .6712,.0910& .6235,.0919\\
\hline
\hline
\multicolumn{10}{c}{$l=4$, $n=1$} \\
\hline
$j$ &$m=4$ &$m=3$ &$m=2$ &$m=1$ &$m=0$ &$m=-1$ &$m=-2$ &$m=-3$ &$m=-4$\\
0.00& .7966,.2843& .7966,.2843& .7966,.2843& .7966,.2843& .7966,.2843& .7966,.2843& .7966,.2843& .7966,.2843& .7966,.2843\\
0.10& .8271,.2835& .8194,.2837& .8119,.2838& .8046,.2840& .7973,.2841& .7901,.2843& .7830,.2844& .7761,.2846& .7692,.2847\\
0.20& .8611,.2821& .8449,.2824& .8293,.2828& .8140,.2831& .7992,.2835& .7848,.2838& .7709,.2841& .7574,.2844& .7442,.2847\\
0.30& .8997,.2799& .8738,.2805& .8489,.2811& .8252,.2817& .8025,.2823& .7808,.2829& .7601,.2835& .7403,.2840& .7214,.2845\\
0.40& .9441,.2766& .9067,.2775& .8715,.2785& .8384,.2796& .8072,.2806& .7780,.2816& .7505,.2826& .7247,.2834& .7003,.2840\\
0.50& .9962,.2717& .9451,.2731& .8977,.2747& .8539,.2765& .8136,.2782& .7763,.2799& .7420,.2813& .7103,.2825& .6809,.2834\\
0.60& 1.0591,.2645& .9909,.2665& .9287,.2691& .8725,.2720& .8217,.2748& .7759,.2775& .7345,.2797& .6969,.2814& .6628,.2826\\
0.70& 1.1382,.2532& 1.0474,.2562& .9663,.2604& .8947,.2653& .8319,.2701& .7767,.2743& .7279,.2777& .6846,.2801& .6459,.2817\\
0.80& 1.2445,.2347& 1.1212,.2391& 1.0137,.2464& .9221,.2551& .8446,.2634& .7788,.2702& .7222,.2752& .6732,.2787& .6300,.2807\\
0.90& 1.4090,.1986& 1.2293,.2055& 1.0779,.2199& .9565,.2376& .8602,.2532& .7822,.2646& .7174,.2723& .6625,.2770& .6152,.2796\\
0.98& 1.6917,.1165& 1.3917,.1268& 1.1519,.1663& .9899,.2113& .8746,.2407& .7857,.2586& .7140,.2694& .6545,.2756& .6039,.2787\\
\hline
\hline
\multicolumn{10}{c}{$l=4$, $n=2$} \\
\hline
$j$ &$m=4$ &$m=3$ &$m=2$ &$m=1$ &$m=0$ &$m=-1$ &$m=-2$ &$m=-3$ &$m=-4$\\
0.00& .7727,.4799& .7727,.4799& .7727,.4799& .7727,.4799& .7727,.4799& .7727,.4799& .7727,.4799& .7727,.4799& .7727,.4799\\
0.10& .8049,.4780& .7969,.4784& .7890,.4788& .7812,.4791& .7734,.4795& .7658,.4799& .7583,.4803& .7509,.4807& .7436,.4811\\
0.20& .8409,.4751& .8239,.4759& .8074,.4767& .7913,.4775& .7756,.4784& .7604,.4792& .7455,.4800& .7311,.4808& .7170,.4816\\
0.30& .8815,.4708& .8544,.4721& .8283,.4735& .8033,.4749& .7794,.4763& .7563,.4778& .7342,.4792& .7130,.4805& .6927,.4817\\
0.40& .9280,.4647& .8892,.4666& .8524,.4687& .8176,.4710& .7847,.4733& .7536,.4756& .7243,.4777& .6965,.4797& .6702,.4814\\
0.50& .9824,.4559& .9296,.4586& .8802,.4618& .8344,.4654& .7918,.4690& .7522,.4725& .7155,.4757& .6813,.4785& .6493,.4808\\
0.60& 1.0478,.4431& .9775,.4469& .9131,.4518& .8543,.4573& .8008,.4630& .7522,.4683& .7078,.4730& .6672,.4769& .6299,.4800\\
0.70& 1.1295,.4237& 1.0365,.4291& .9526,.4367& .8780,.4456& .8119,.4546& .7534,.4627& .7012,.4696& .6542,.4750& .6117,.4789\\
0.80& 1.2387,.3921& 1.1129,.3998& 1.0020,.4123& .9065,.4276& .8253,.4426& .7558,.4554& .6954,.4654& .6422,.4727& .5946,.4777\\
0.90& 1.4061,.3313& 1.2236,.3428& 1.0667,.3665& .9402,.3972& .8404,.4247& .7591,.4455& .6904,.4602& .6310,.4701& .5786,.4764\\
0.98& 1.6913,.1942& 1.3871,.2096& 1.1283,.2730& .9655,.3535& .8520,.4037& .7619,.4352& .6868,.4552& .6225,.4678& .5664,.4752\\
\hline
\end{tabular}
\end{table}

\begin{table}[t]
\centering \caption{\label{tab:l2Alm} Angular separation constants for
the first three overtones with $l=2$. To save space, in this and the
following Tables we omit leading zeros.}
\begin{tabular}{cccccc}  \hline
\multicolumn{6}{c}{$l=2$, $n=0$} \\
\hline
$j$ &$m=2$ &$m=1$ &$m=0$ &$m=-1$ &$m=-2$\\
\hline
0.00 &4.0000,.0000 &4.0000,.0000 &4.0000,.0000 &4.0000,.0000  &4.0000,.0000\\
0.10 &3.8957,.0242 &3.9485,.0122 &3.9993,.0003 &4.0483,-.0115 &4.0956,-.0233\\
0.20 &3.7810,.0492 &3.8932,.0252 &3.9972,.0014 &4.0939,-.0222 &4.1839,-.0457\\
0.30 &3.6531,.0751 &3.8332,.0389 &3.9937,.0031 &4.1371,-.0322 &4.2659,-.0674\\
0.40 &3.5087,.1019 &3.7676,.0532 &3.9886,.0056 &4.1784,-.0415 &4.3427,-.0882\\
0.50 &3.3423,.1292 &3.6947,.0682 &3.9817,.0088 &4.2178,-.0500 &4.4147,-.1083\\
0.60 &3.1454,.1567 &3.6125,.0835 &3.9730,.0126 &4.2558,-.0577 &4.4827,-.1277\\
0.70 &2.9032,.1832 &3.5172,.0986 &3.9619,.0172 &4.2924,-.0645 &4.5470,-.1463\\
0.80 &2.5853,.2053 &3.4023,.1122 &3.9480,.0223 &4.3279,-.0705 &4.6081,-.1643\\
0.90 &2.1098,.2111 &3.2534,.1195 &3.9304,.0276 &4.3623,-.0755 &4.6662,-.1816\\
0.98 &1.3336,.1500 &3.0797,.1022 &3.9127,.0315 &4.3892,-.0787 &4.7108,-.1950\\
\hline
\hline
\multicolumn{6}{c}{$l=2$, $n=1$} \\
\hline
$j$ &$m=2$ &$m=1$ &$m=0$ &$m=-1$ &$m=-2$\\
\hline
0.00 &4.0000,.0000 &4.0000,.0000 &4.0000,.0000 &4.0000,.0000  &4.0000,.0000\\
0.10 &3.9031,.0741 &3.9524,.0375 &3.9998,.0010 &4.0451,-.0355 &4.0885,-.0720\\
0.20 &3.7958,.1504 &3.9017,.0771 &3.9990,.0040 &4.0883,-.0690 &4.1702,-.1420\\
0.30 &3.6756,.2287 &3.8469,.1185 &3.9977,.0090 &4.1300,-.1004 &4.2461,-.2100\\
0.40 &3.5386,.3091 &3.7869,.1617 &3.9956,.0161 &4.1707,-.1296 &4.3172,-.2761\\
0.50 &3.3791,.3908 &3.7197,.2065 &3.9925,.0252 &4.2105,-.1566 &4.3841,-.3405\\
0.60 &3.1882,.4726 &3.6429,.2520 &3.9880,.0364 &4.2497,-.1813 &4.4476,-.4031\\
0.70 &2.9498,.5513 &3.5523,.2968 &3.9815,.0496 &4.2884,-.2034 &4.5080,-.4641\\
0.80 &2.6313,.6171 &3.4405,.3365 &3.9722,.0645 &4.3266,-.2227 &4.5658,-.5236\\
0.90 &2.1455,.6342 &3.2918,.3564 &3.9587,.0800 &4.3642,-.2389 &4.6213,-.5815\\
0.98 &1.3457,.4499 &3.1248,.2952 &3.9442,.0906 &4.3937,-.2494 &4.6643,-.6268\\
\hline
\hline
\multicolumn{6}{c}{$l=2$, $n=2$} \\
\hline
$j$ &$m=2$ &$m=1$ &$m=0$ &$m=-1$ &$m=-2$\\
\hline
0.00 &4.0000,.0000 &4.0000,.0000 &4.0000,.0000 &4.0000,.0000  &4.0000,.0000\\
0.10 &3.9158,.1285 &3.9594,.0651 &4.0007,.0015 &4.0399,-.0624 &4.0770,-.1266\\
0.20 &3.8222,.2589 &3.9171,.1329 &4.0028,.0061 &4.0798,-.1219 &4.1486,-.2516\\
0.30 &3.7162,.3912 &3.8721,.2031 &4.0061,.0137 &4.1202,-.1785 &4.2160,-.3750\\
0.40 &3.5935,.5253 &3.8226,.2756 &4.0104,.0246 &4.1616,-.2319 &4.2805,-.4970\\
0.50 &3.4480,.6602 &3.7666,.3498 &4.0150,.0388 &4.2041,-.2817 &4.3428,-.6178\\
0.60 &3.2697,.7944 &3.7006,.4246 &4.0193,.0563 &4.2479,-.3276 &4.4039,-.7375\\
0.70 &3.0405,.9229 &3.6195,.4972 &4.0222,.0771 &4.2928,-.3691 &4.4643,-.8560\\
0.80 &2.7230,1.0312 &3.5141,.5602 &4.0219,.1005 &4.3384,-.4056 &4.5245,-.9736\\
0.90 &2.2175,1.0596 &3.3653,.5869 &4.0163,.1237 &4.3835,-.4366 &4.5851,-1.0903\\
0.98 &1.3696,.7498 &3.1982,.4474 &4.0087,.1378 &4.4184,-.4576 &4.6340,-1.1830\\
\hline
\end{tabular}
\end{table}

\begin{table}[t]
\centering \caption{\label{tab:l3Alm} Angular separation constants for
the first three overtones with $l=3$.}
\begin{tabular}{cccccccc}  \hline
\multicolumn{8}{c}{$l=3$, $n=0$} \\
\hline
$j$ &$m=3$ &$m=2$ &$m=1$ &$m=0$ &$m=-1$ &$m=-2$ &$m=-3$\\
\hline
0.00 &1.0000,.0000 &1.0000,.0000 &1.0000,.0000 &1.0000,.0000 &1.0000,.0000 &1.0000,.0000 &1.0000,.0000\\
0.10 &9.8739,.0191 &9.9167,.0128 &9.9583,.0066 &9.9988,.0004 &1.0383,-.0058 &1.0769,-.0119 &1.1144,-.0180\\
0.20 &9.7338,.0392 &9.8257,.0265 &9.9128,.0139 &9.9953,.0015 &1.0736,-.0108 &1.1481,-.0230 &1.2190,-.0351\\
0.30 &9.5764,.0605 &9.7257,.0410 &9.8629,.0220 &9.9893,.0033 &1.1062,-.0150 &1.2145,-.0332 &1.3153,-.0512\\
0.40 &9.3970,.0828 &9.6146,.0563 &9.8080,.0307 &9.9808,.0059 &1.1361,-.0185 &1.2766,-.0426 &1.4043,-.0665\\
0.50 &9.1890,.1060 &9.4895,.0722 &9.7470,.0401 &9.9695,.0092 &1.1637,-.0211 &1.3348,-.0511 &1.4870,-.0809\\
0.60 &8.9414,.1299 &9.3461,.0884 &9.6785,.0500 &9.9552,.0132 &1.1890,-.0229 &1.3896,-.0587 &1.5642,-.0946\\
0.70 &8.6358,.1534 &9.1773,.1043 &9.6006,.0601 &9.9375,.0179 &1.2121,-.0238 &1.4412,-.0656 &1.6364,-.1075\\
0.80 &8.2345,.1739 &8.9702,.1181 &9.5096,.0697 &9.9157,.0231 &1.2330,-.0239 &1.4900,-.0716 &1.7043,-.1198\\
0.90 &7.6364,.1815 &8.6940,.1235 &9.3992,.0767 &9.8889,.0285 &1.2518,-.0231 &1.5362,-.0768 &1.7682,-.1314\\
0.98 &6.6687,.1319 &8.3443,.0951 &9.2872,.0749 &9.8630,.0326 &1.2653,-.0218 &1.5713,-.0803 &1.8167,-.1403\\
\hline
\hline
\multicolumn{8}{c}{$l=3$, $n=1$} \\
\hline
$j$ &$m=3$ &$m=2$ &$m=1$ &$m=0$ &$m=-1$ &$m=-2$ &$m=-3$\\
\hline
0.00 &1.0000,.0000 &1.0000,.0000 &1.0000,.0000 &1.0000,.0000 &1.0000,.0000 &1.0000,.0000 &1.0000,.0000\\
0.10 &9.8775,.0578 &9.9192,.0388 &9.9597,.0199 &9.9991,.0011 &1.0375,-.0176 &1.0749,-.0363 &1.1113,-.0548\\
0.20 &9.7413,.1186 &9.8312,.0800 &9.9161,.0420 &9.9965,.0044 &1.0726,-.0329 &1.1448,-.0700 &1.2132,-.1069\\
0.30 &9.5881,.1825 &9.7344,.1237 &9.8686,.0663 &9.9920,.0098 &1.1056,-.0460 &1.2104,-.1013 &1.3072,-.1563\\
0.40 &9.4131,.2495 &9.6268,.1697 &9.8165,.0927 &9.9854,.0175 &1.1366,-.0566 &1.2723,-.1301 &1.3944,-.2033\\
0.50 &9.2094,.3192 &9.5053,.2174 &9.7585,.1209 &9.9766,.0273 &1.1658,-.0649 &1.3310,-.1565 &1.4758,-.2480\\
0.60 &8.9659,.3905 &9.3652,.2661 &9.6931,.1505 &9.9651,.0392 &1.1934,-.0707 &1.3870,-.1805 &1.5522,-.2905\\
0.70 &8.6633,.4607 &9.1992,.3136 &9.6180,.1807 &9.9504,.0531 &1.2193,-.0740 &1.4405,-.2020 &1.6240,-.3310\\
0.80 &8.2629,.5217 &8.9935,.3544 &9.5294,.2090 &9.9316,.0685 &1.2435,-.0747 &1.4918,-.2212 &1.6919,-.3696\\
0.90 &7.6604,.5443 &8.7162,.3698 &9.4211,.2293 &9.9077,.0846 &1.2658,-.0726 &1.5410,-.2378 &1.7563,-.4064\\
0.98 &6.6779,.3956 &8.3645,.2808 &9.3156,.2228 &9.8841,.0963 &1.2820,-.0690 &1.5790,-.2494 &1.8055,-.4346\\
\hline
\hline
\multicolumn{8}{c}{$l=3$, $n=2$} \\
\hline
$j$ &$m=3$ &$m=2$ &$m=1$ &$m=0$ &$m=-1$ &$m=-2$ &$m=-3$\\
\hline
0.00 &1.0000,.0000 &1.0000,.0000 &1.0000,.0000 &1.0000,.0000 &1.0000,.0000 &1.0000,.0000 &1.0000,.0000\\
0.10 &9.8842,.0980 &9.9238,.0658 &9.9623,.0337 &9.9997,.0018 &1.0361,-.0301 &1.0713,-.0619 &1.1056,-.0937\\
0.20 &9.7555,.2006 &9.8415,.1355 &9.9225,.0710 &9.9989,.0071 &1.0710,-.0566 &1.1389,-.1200 &1.2028,-.1833\\
0.30 &9.6105,.3078 &9.7512,.2089 &9.8798,.1118 &9.9974,.0159 &1.1051,-.0794 &1.2035,-.1743 &1.2933,-.2692\\
0.40 &9.4442,.4196 &9.6505,.2857 &9.8331,.1559 &9.9950,.0283 &1.1385,-.0984 &1.2657,-.2248 &1.3780,-.3515\\
0.50 &9.2492,.5353 &9.5360,.3653 &9.7811,.2030 &9.9911,.0443 &1.1715,-.1134 &1.3262,-.2716 &1.4580,-.4305\\
0.60 &9.0138,.6534 &9.4027,.4460 &9.7218,.2522 &9.9853,.0637 &1.2040,-.1245 &1.3853,-.3146 &1.5341,-.5064\\
0.70 &8.7175,.7691 &9.2422,.5244 &9.6524,.3020 &9.9767,.0864 &1.2359,-.1314 &1.4432,-.3538 &1.6068,-.5794\\
0.80 &8.3192,.8696 &9.0394,.5910 &9.5687,.3485 &9.9641,.1116 &1.2667,-.1339 &1.5002,-.3892 &1.6766,-.6498\\
0.90 &7.7083,.9064 &8.7596,.6140 &9.4645,.3796 &9.9459,.1375 &1.2960,-.1319 &1.5562,-.4207 &1.7439,-.7176\\
0.98 &6.6964,.6591 &8.3986,.4546 &9.3745,.3656 &9.9274,.1558 &1.3177,-.1274 &1.6004,-.4431 &1.7963,-.7701\\
\hline
\end{tabular}
\end{table}

\begin{table}[t]
\centering \caption{\label{tab:l4Alm} Angular separation constants for
the first three overtones with $l=4$.}
\begin{tabular}{cccccccccc}  \hline
\multicolumn{10}{c}{$l=4$, $n=0$} \\
\hline
$j$ &$m=4$ &$m=3$ &$m=2$ &$m=1$ &$m=0$ &$m=-1$ &$m=-2$ &$m=-3$ &$m=-4$\\
\hline
0.00 &18.0000,.0000 &18.0000,.0000 &18.0000,.0000 &18.0000,.0000 &18.0000,.0000 &18.0000,.0000 &18.0000,.0000 &18.0000,.0000 &18.0000,.0000\\
0.10 &17.8633,.0156 &17.8978,.0118 &17.9317,.0081 &17.9650,.0043 &17.9977,.0006 &18.0298,-.0032 &18.0614,-.0070 &18.0925,-.0108 &18.1231,-.0146\\
0.20 &17.7101,.0323 &17.7842,.0248 &17.8555,.0172 &17.9243,.0097 &17.9906,.0022 &18.0547,-.0053 &18.1167,-.0129 &18.1766,-.0205 &18.2347,-.0281\\
0.30 &17.5365,.0502 &17.6569,.0388 &17.7703,.0275 &17.8774,.0162 &17.9787,.0049 &18.0749,-.0064 &18.1663,-.0177 &18.2535,-.0292 &18.3366,-.0408\\
0.40 &17.3369,.0694 &17.5128,.0539 &17.6743,.0387 &17.8235,.0237 &17.9618,.0088 &18.0905,-.0063 &18.2109,-.0215 &18.3239,-.0370 &18.4302,-.0527\\
0.50 &17.1033,.0897 &17.3475,.0698 &17.5655,.0509 &17.7616,.0323 &17.9394,.0137 &18.1017,-.0051 &18.2508,-.0242 &18.3885,-.0438 &18.5164,-.0637\\
0.60 &16.8229,.1110 &17.1544,.0865 &17.4405,.0637 &17.6903,.0417 &17.9111,.0197 &18.1083,-.0028 &18.2862,-.0259 &18.4480,-.0497 &18.5963,-.0741\\
0.70 &16.4735,.1324 &16.9225,.1029 &17.2943,.0768 &17.6075,.0518 &17.8762,.0266 &18.1104,.0006 &18.3175,-.0265 &18.5029,-.0547 &18.6706,-.0838\\
0.80 &16.0106,.1519 &16.6313,.1174 &17.1183,.0888 &17.5099,.0619 &17.8335,.0343 &18.1076,.0050 &18.3448,-.0261 &18.5536,-.0588 &18.7399,-.0929\\
0.90 &15.3146,.1609 &16.2310,.1230 &16.8945,.0959 &17.3918,.0707 &17.7814,.0425 &18.0997,.0105 &18.3682,-.0247 &18.6003,-.0622 &18.8046,-.1013\\
0.98 &14.1819,.1190 &15.6920,.9097 &16.6463,.0850 &17.2747,.0734 &17.7313,.0487 &18.0894,.0155 &18.3843,-.0228 &18.6352,-.0642 &18.8535,-.1078\\
\hline
\hline
\multicolumn{10}{c}{$l=4$, $n=1$} \\
\hline
$j$ &$m=4$ &$m=3$ &$m=2$ &$m=1$ &$m=0$ &$m=-1$ &$m=-2$ &$m=-3$ &$m=-4$\\
0.00 &18.0000,.0000 &18.0000,.0000 &18.0000,.0000 &18.0000,.0000 &18.0000,.0000 &18.0000,.0000 &18.0000,.0000 &18.0000,.0000 &18.0000,.0000\\
0.10 &17.8655,.0471 &17.8995,.0357 &17.9330,.0244 &17.9658,.0130 &17.9980,.0016 &18.0296,-.0097 &18.0607,-.0211 &18.0913,-.0326 &18.1212,-.0440\\
0.20 &17.7148,.0974 &17.7881,.0746 &17.8586,.0519 &17.9265,.0293 &17.9919,.0065 &18.0550,-.0162 &18.1159,-.0391 &18.1747,-.0621 &18.2314,-.0852\\
0.30 &17.5441,.1512 &17.6633,.1168 &17.7756,.0827 &17.8816,.0487 &17.9817,.0147 &18.0764,-.0194 &18.1662,-.0539 &18.2514,-.0886 &18.3323,-.1237\\
0.40 &17.3476,.2086 &17.5220,.1620 &17.6823,.1164 &17.8301,.0713 &17.9669,.0261 &18.0939,-.0194 &18.2120,-.0655 &18.3222,-.1122 &18.4251,-.1598\\
0.50 &17.1173,.2695 &17.3596,.2100 &17.5762,.1529 &17.7710,.0969 &17.9473,.0408 &18.1075,-.0160 &18.2538,-.0740 &18.3879,-.1332 &18.5110,-.1936\\
0.60 &16.8400,.3331 &17.1693,.2597 &17.4539,.1914 &17.7027,.1251 &17.9221,.0586 &18.1172,-.0094 &18.2919,-.0794 &18.4490,-.1514 &18.5908,-.2254\\
0.70 &16.4932,.3974 &16.9396,.3091 &17.3102,.2305 &17.6228,.1552 &17.8906,.0792 &18.1229,.0005 &18.3264,-.0817 &18.5061,-.1671 &18.6654,-.2552\\
0.80 &16.0315,.4557 &16.6493,.3522 &17.1358,.2663 &17.5279,.1856 &17.8514,.1023 &18.1241,.0135 &18.3575,-.0809 &18.5595,-.1802 &18.7352,-.2833\\
0.90 &15.3329,.4824 &16.2473,.3686 &16.9127,.2872 &17.4121,.2115 &17.8028,.1265 &18.1204,.0296 &18.3853,-.0771 &18.6095,-.1910 &18.8009,-.3097\\
0.98 &14.1893,.3570 &15.7029,.2713 &16.6706,.2524 &17.2991,.2191 &17.7555,.1447 &18.1135,.0443 &18.4051,-.0719 &18.6474,-.1979 &18.8507,-.3296\\
\hline
\hline
\multicolumn{10}{c}{$l=4$, $n=2$} \\
\hline
$j$ &$m=4$ &$m=3$ &$m=2$ &$m=1$ &$m=0$ &$m=-1$ &$m=-2$ &$m=-3$ &$m=-4$\\
0.00 &18.0000,.0000 &18.0000,.0000 &18.0000,.0000 &18.0000,.0000 &18.0000,.0000 &18.0000,.0000 &18.0000,.0000 &18.0000,.0000 &18.0000,.0000\\
0.10 &17.8697,.0793 &17.9029,.0601 &17.9355,.0410 &17.9674,.0219 &17.9987,.0027 &18.0293,-.0165 &18.0594,-.0358 &18.0889,-.0551 &18.1178,-.0744\\
0.20 &17.7240,.1637 &17.7958,.1255 &17.8647,.0873 &17.9309,.0490 &17.9946,.0107 &18.0558,-.0277 &18.1146,-.0664 &18.1711,-.1053 &18.2254,-.1444\\
0.30 &17.5589,.2537 &17.6760,.1960 &17.7862,.1387 &17.8899,.0815 &17.9876,.0241 &18.0796,-.0336 &18.1662,-.0919 &18.2477,-.1507 &18.3244,-.2102\\
0.40 &17.3686,.3494 &17.5403,.2715 &17.6981,.1950 &17.8434,.1191 &17.9773,.0429 &18.1009,-.0341 &18.2148,-.1123 &18.3197,-.1916 &18.4161,-.2722\\
0.50 &17.1448,.4505 &17.3836,.3513 &17.5976,.2559 &17.7899,.1616 &17.9632,.0669 &18.1196,-.0293 &18.2607,-.1276 &18.3877,-.2281 &18.5017,-.3308\\
0.60 &16.8739,.5561 &17.1987,.4341 &17.4808,.3200 &17.7275,.2086 &17.9444,.0962 &18.1356,-.0191 &18.3042,-.1379 &18.4524,-.2604 &18.5820,-.3861\\
0.70 &16.5324,.6625 &16.9734,.5159 &17.3418,.3849 &17.6535,.2586 &17.9198,.1302 &18.1485,-.0034 &18.3454,-.1432 &18.5142,-.2886 &18.6577,-.4383\\
0.80 &16.0730,.7590 &16.6851,.5871 &17.1708,.4440 &17.5639,.3087 &17.8876,.1681 &18.1578,.0174 &18.3844,-.1435 &18.5735,-.3129 &18.7293,-.4879\\
0.90 &15.3693,.8032 &16.2796,.6133 &16.9490,.4770 &17.4528,.3507 &17.8458,.2076 &18.1626,.0431 &18.4210,-.1389 &18.6306,-.3333 &18.7975,-.5348\\
0.98 &14.2040,.5948 &15.7228,.4476 &16.7200,.4106 &17.3486,.3620 &17.8045,.2369 &18.1625,.0666 &18.4484,-.1317 &18.6748,-.3469 &18.8498,-.5707\\
\hline
\end{tabular}
\end{table}

The numerical values of the quasi-normal frequencies we listed in
Tables \ref{tab:l2n0},  \ref{tab:l3n0} and \ref{tab:l4n0}
can be fitted to reasonable accuracy by
simple functions. which can easily be inverted to yield $j\,,\,M$
once the ringing frequencies are known.  The fitting functions all
have the form 
\beq
{\cal F}_{lmn}=M\omega_{lmn}&=&f_1+f_2(1-j)^{f_3}\,,\\
Q_{lmn}&=&q_1+q_2(1-j)^{q_3}\,.
\eeq
In Tables \ref{tab:fitQNMsl2}, \ref{tab:fitQNMsl3} and
\ref{tab:fitQNMsl4} we list the fitting coefficients $f_i$ (for the
frequency) and $q_i$ (for the quality factor).
\begin{table}[t]
\centering \caption{\label{tab:fitQNMsl2} Fitting coefficients for the
dimensionless frequency, ${\cal F}_{lmn}=f_1+f_2(1-j)^{f_3}$, and the quality
factor $Q_{lmn}=q_1+q_2(1-j)^{q_3}$ of a Kerr black hole. We give
coefficients for $l=2$, all values of $m$, and the three lowest overtones
($n=0,~1,~2$). For each fit we also give the
maximum percentage error in the range $j\in[0,0.99]$.}
\begin{tabular}{|cc|ccc|c|ccc|c|}
\hline
\hline
$m$    & $n$ & $f_1$   &$f_2$   &$f_3$   &$\%$   &$q_1$   & $q_2$   &  $q_3$
&$\%$\\
\hline
$2$    & $0$ & $1.5251$ &$ -1.1568$ &$ 0.1292$ &$1.85$ &$ 0.7000 $  &$ 1.4187$
 &$ -0.4990  $ &$0.88$ \\
       & $1$ & $1.3673$ &$ -1.0260$ &$ 0.1628$ &$1.56$ &$ 0.1000 $  &$ 0.5436$
 &$ -0.4731  $ &$1.69$ \\
       & $2$ & $1.3223$ &$ -1.0257$ &$ 0.1860$ &$1.91$ &$ -0.1000 $  &$ 0.4206$
 &$ -0.4256  $ &$2.52$ \\
\hline
$1$    & $0$ & $0.6000$ &$ -0.2339$ &$ 0.4175$ &$2.03$ &$ -0.3000$  &$ 2.3561$
 &$ -0.2277 $ &$3.65$ \\
       & $1$ & $0.5800$ &$ -0.2416$ &$ 0.4708$ &$2.40$ &$ -0.3300$  &$ 0.9501$
 &$ -0.2072 $ &$3.18$ \\
       & $2$ & $0.5660$ &$ -0.2740$ &$ 0.4960$ &$4.04$ &$ -0.1000$  &$ 0.4173$
 &$ -0.2774 $ &$2.46$ \\
\hline
$0$    & $0$ & $0.4437$ &$ -0.0739$ &$ 0.3350$ &$1.04$ &$  4.0000$  &$
-1.9550$ &$ 0.1420 $  &$2.63$ \\
       & $1$ & $0.4185$ &$ -0.0768$ &$ 0.4355$ &$1.50$ &$  1.2500$  &$
-0.6359$ &$ 0.1614 $  &$4.01$ \\
       & $2$ & $0.3734$ &$ -0.0794$ &$ 0.6306$ &$2.72$ &$  0.5600$  &$
-0.2589$ &$ 0.3034 $  &$4.33$ \\
\hline
$-1$   & $0$ & $0.3441$ &$  0.0293$ &$ 2.0010$ &$0.07$ &$  2.0000$  &$ 0.1078$
 &$ 5.0069 $  &$2.82$ \\
       & $1$ & $0.3165$ &$  0.0301$ &$ 2.3415$ &$0.05$ &$  0.6100$  &$ 0.0276$
 &$ 13.1683$  &$0.67$ \\
       & $2$ & $0.2696$ &$  0.0315$ &$ 2.7755$ &$0.43$ &$  0.2900$  &$ 0.0276$
 &$ 6.4715 $  &$2.40$ \\
\hline
$-2$   & $0$ & $0.2938$ &$  0.0782$ &$ 1.3546$ &$0.63$ &$  1.6700$  &$ 0.4192$
 &$ 1.4700 $  &$0.71$ \\
       & $1$ & $0.2528$ &$  0.0921$ &$ 1.3344$ &$0.87$ &$  0.4550$  &$ 0.1729$
 &$ 1.3617 $  &$0.79$ \\
       & $2$ & $0.1873$ &$  0.1117$ &$ 1.3322$ &$1.34$ &$  0.1850$  &$ 0.1266$
 &$ 1.3661 $  &$1.16$ \\
\hline
\hline
\end{tabular}
\end{table}

\begin{table}[t]
\centering \caption{\label{tab:fitQNMsl3}
Same as Table \ref{tab:fitQNMsl2}, but for $l=3$.
}
\begin{tabular}{|cc|ccc|c|ccc|c|}
\hline
\hline
$m$    & $n$ & $f_1$   &$f_2$   &$f_3$   &$\%$   &$q_1$   & $q_2$   &  $q_3$
&$\%$\\
\hline
$3$     & $0$ & $1.8956$ &$ -1.3043 $ &$ 0.1818$ &$1.36$ &$ 0.9000 $  &$
2.3430 $ &$ -0.4810$ &$0.42$ \\
        & $1$ & $1.8566$ &$ -1.2818 $ &$ 0.1934$ &$1.35$ &$ 0.2274 $  &$
0.8173 $ &$ -0.4731$ &$0.88$ \\
        & $2$ & $1.8004$ &$ -1.2558 $ &$ 0.2133$ &$1.28$ &$ 0.0400 $  &$
0.5445 $ &$ -0.4539$ &$1.52$ \\
\hline
$2$     & $0$ & $1.1481$ &$ -0.5552 $ &$ 0.3002$ &$1.09$ &$ 0.8313 $  &$
2.3773 $ &$ -0.3655$ &$1.28$ \\
        & $1$ & $1.1226$ &$ -0.5471 $ &$ 0.3264$ &$1.23$ &$ 0.2300 $  &$
0.8025 $ &$ -0.3684$ &$0.51$ \\
        & $2$ & $1.0989$ &$ -0.5550 $ &$ 0.3569$ &$1.41$ &$ 0.1000 $  &$
0.4804 $ &$ -0.3784$ &$0.81$ \\
\hline
$1$     & $0$ & $0.8345$ &$ -0.2405 $ &$ 0.4095$ &$1.12$ &$ 23.8450$ 
&$-20.7240$ &$ 0.03837$ &$3.47$ \\
        & $1$ & $0.8105$ &$ -0.2342 $ &$ 0.4660$ &$1.55$ &$ 8.8530 $  &$
-7.8506$ &$ 0.03418$ &$3.64$ \\
        & $2$ & $0.7684$ &$ -0.2252 $ &$ 0.5805$ &$2.67$ &$ 2.1800 $  &$
-1.6273$ &$ 0.1136 $ &$4.04$ \\
\hline
$0$     & $0$ & $0.6873$ &$ -0.09282$ &$ 0.3479$ &$0.83$ &$ 6.7841 $  &$
-3.6112$ &$ 0.09480$ &$3.99$ \\
        & $1$ & $0.6687$ &$ -0.09155$ &$ 0.4021$ &$0.95$ &$ 2.0075 $  &$
-0.9930$ &$ 0.12297$ &$4.18$ \\
        & $2$ & $0.6343$ &$ -0.08915$ &$ 0.5117$ &$1.28$ &$ 0.9000 $  &$
-0.3409$ &$ 0.2679 $ &$2.89$ \\
\hline
$-1$    & $0$ & $0.5751$ &$ 0.02508 $ &$ 3.1360$ &$0.42$ &$ 3.0464 $  &$
0.1162 $ &$ -0.2812$ &$2.65$ \\
        & $1$ & $0.5584$ &$ 0.02514 $ &$ 3.4154$ &$0.42$ &$ 1.2000 $  &$
-0.1928$ &$ 0.1037 $ &$2.75$ \\
        & $2$ & $0.5271$ &$ 0.02561 $ &$ 3.8011$ &$0.29$ &$ 1.0000 $  &$
-0.4424$ &$ 0.02467$ &$3.15$ \\
\hline
$-2$    & $0$ & $0.5158$ &$ 0.08195 $ &$ 1.4084$ &$0.35$ &$ 2.9000 $  &$
0.3356 $ &$ 2.3050 $ &$0.72$ \\
        & $1$ & $0.4951$ &$ 0.08577 $ &$ 1.4269$ &$0.41$ &$ 0.9000 $  &$
0.1295 $ &$ 1.6142 $ &$0.80$ \\
        & $2$ & $0.4567$ &$ 0.09300 $ &$ 1.4469$ &$0.53$ &$ 0.4900 $  &$
0.0848 $ &$ 1.9737 $ &$0.52$ \\
\hline
$-3$    & $0$ & $0.4673$ &$ 0.1296  $ &$ 1.3255$ &$0.61$ &$ 2.5500 $  &$
0.6576 $ &$ 1.3378 $ &$0.79$ \\
        & $1$ & $0.4413$ &$ 0.1387  $ &$ 1.3178$ &$0.68$ &$ 0.7900 $  &$
0.2381 $ &$ 1.3706 $ &$0.73$ \\
        & $2$ & $0.3933$ &$ 0.1555  $ &$ 1.3037$ &$0.82$ &$ 0.4070 $  &$
0.1637 $ &$ 1.3819 $ &$0.88$ \\
\hline
\hline
\end{tabular}
\end{table}

\begin{table}[t]
\centering \caption{\label{tab:fitQNMsl4}
Same as Table \ref{tab:fitQNMsl2}, but for $l=4$.
}
\begin{tabular}{|cc|ccc|c|ccc|c|}
\hline
\hline
$m$    & $n$ & $f_1$   &$f_2$   &$f_3$   &$\%$   &$q_1$   & $q_2$   &  $q_3$
&$\%$\\
\hline
$4$     & $0$ & $2.3000$ &$ -1.5056$ &$  0.2244$ &$1.83$ &$ 1.1929 $  &$
3.1191  $ &$-0.4825$ &$0.37$ \\
        & $1$ & $2.3000$ &$ -1.5173$ &$  0.2271$ &$1.75$ &$ 0.3000 $  &$
1.1034  $ &$-0.4703$ &$0.77$ \\
        & $2$ & $2.3000$ &$ -1.5397$ &$  0.2321$ &$1.61$ &$ 0.1100 $  &$
0.6997  $ &$-0.4607$ &$0.10$ \\
\hline
$3$     & $0$ & $1.6869$ &$ -0.8862$ &$  0.2822$ &$1.05$ &$ 1.4812 $  &$
2.8096  $ &$-0.4271$ &$0.14$ \\
        & $1$ & $1.6722$ &$ -0.8843$ &$  0.2923$ &$1.10$ &$ 0.4451 $  &$
0.9569  $ &$-0.4250$ &$0.37$ \\
        & $2$ & $1.6526$ &$ -0.8888$ &$  0.3081$ &$1.15$ &$ 0.2200 $  &$
0.5904  $ &$-0.4236$ &$0.66$ \\
\hline
$2$     & $0$ & $1.2702$ &$ -0.4685$ &$  0.3835$ &$1.11$ &$ -3.6000$  &$
7.7749  $ &$-0.1491$ &$0.97$ \\
        & $1$ & $1.2462$ &$ -0.4580$ &$  0.4139$ &$1.39$ &$ -1.5000$  &$
2.8601  $ &$-0.1392$ &$0.12$ \\
        & $2$ & $1.2025$ &$ -0.4401$ &$  0.4769$ &$2.26$ &$ -1.5000$  &$
2.2784  $ &$-0.1124$ &$0.31$ \\
\hline
$1$     & $0$ & $1.0507$ &$ -0.2478$ &$  0.4348$ &$0.97$ &$ 14.0000$  &$
-9.8240 $ &$0.09047$ &$0.81$ \\
        & $1$ & $1.0337$ &$ -0.2439$ &$  0.4695$ &$1.15$ &$ 4.2000 $  &$
-2.8399 $ &$0.1081 $ &$0.91$ \\
        & $2$ & $1.0019$ &$ -0.2374$ &$  0.5397$ &$1.53$ &$ 2.2000 $  &$
-1.4195 $ &$0.1372 $ &$0.53$ \\
\hline
$0$     & $0$ & $0.9175$ &$ -0.1144$ &$  0.3511$ &$0.75$ &$ 7.0000 $  &$
-2.7934 $ &$0.1708 $ &$0.26$ \\
        & $1$ & $0.9028$ &$ -0.1127$ &$  0.3843$ &$0.82$ &$ 2.2000 $  &$
-0.8308 $ &$0.2023 $ &$0.26$ \\
        & $2$ & $0.8751$ &$ -0.1096$ &$  0.4516$ &$0.96$ &$ 1.2000 $  &$
-0.4159 $ &$0.2687 $ &$0.60$ \\
\hline
$-1$    & $0$ & $0.7908$ &$ 0.02024$ &$  5.4628$ &$0.96$ &$ 4.6000 $  &$
-0.4038 $ &$0.4629 $ &$2.52$ \\
        & $1$ & $0.7785$ &$ 0.02005$ &$  5.8547$ &$0.98$ &$ 1.6000 $  &$
-0.2323 $ &$0.2306 $ &$2.37$ \\
        & $2$ & $0.7549$ &$ 0.01985$ &$  6.5272$ &$0.96$ &$ 1.6000 $  &$
-0.8136 $ &$0.03163$ &$2.32$ \\
\hline
$-2$    & $0$ & $0.7294$ &$ 0.07842$ &$  1.5646$ &$0.23$ &$ 4.0000 $  &$ 
0.2777 $ &$ 2.0647$ &$2.11$ \\
        & $1$ & $0.7154$ &$ 0.07979$ &$  1.5852$ &$0.25$ &$ 1.3200 $  &$ 
0.08694$ &$ 4.3255$ &$0.75$ \\
        & $2$ & $0.6885$ &$ 0.08259$ &$  1.6136$ &$0.32$ &$ 0.7500 $  &$ 
0.05803$ &$ 3.7971$ &$0.66$ \\
\hline
$-3$    & $0$ & $0.6728$ &$ 0.1338 $ &$  1.3413$ &$0.43$ &$ 3.700  $  &$ 
0.5829 $ &$ 1.6681$ &$0.45$ \\
        & $1$ & $0.6562$ &$ 0.1377 $ &$  1.3456$ &$0.46$ &$ 1.1800 $  &$ 
0.2111 $ &$ 1.4129$ &$0.70$ \\
        & $2$ & $0.6244$ &$ 0.1454 $ &$  1.3513$ &$0.52$ &$ 0.6600 $  &$ 
0.1385 $ &$ 1.3742$ &$0.82$ \\
\hline
$-4$    & $0$ & $0.6256$ &$ 0.1800 $ &$  1.3218$ &$0.62$ &$ 3.4000 $  &$ 
0.8696 $ &$ 1.4074$ &$0.63$ \\
        & $1$ & $0.6061$ &$ 0.1869 $ &$  1.3168$ &$0.67$ &$ 1.0800 $  &$ 
0.3095 $ &$ 1.3279$ &$0.81$ \\
        & $2$ & $0.5686$ &$ 0.2003 $ &$  1.3068$ &$0.74$ &$ 0.5980 $  &$ 
0.2015 $ &$ 1.3765$ &$0.69$ \\
\hline
\hline
\end{tabular}
\end{table}

Our fits (and the quoted values for the errors) refer to the range
$j\in [0,0.99]$.  The only such fits carried out by other authors
to date treated only the
$l=2\,,\,m=2$ and $l=2\,,\,m=0$ modes.  For $l=2\,,\,m=2$ Echeverria
\cite{echeverria} proposed the fit
\beq
{\cal F}_{220}&\simeq& \left[1-0.63(1-j)^{0.3}\right]\,,\\
Q_{220}&\simeq& 2(1-j)^{-0.45}\,.
\eeq
For $l=2\,,\,m=0$ Fryer, Holz and Hughes \cite{fhh} found
\beq
{\cal F}_{200}&\simeq& \frac{7}{16}\left[1-0.13(1-j)^{0.6}\right]\,,\\
Q_{200}&\simeq& 3-(1-j)^{0.4}\,.
\eeq

\begin{figure*}[t]
\begin{center}
\begin{tabular}{cc}
\epsfig{file=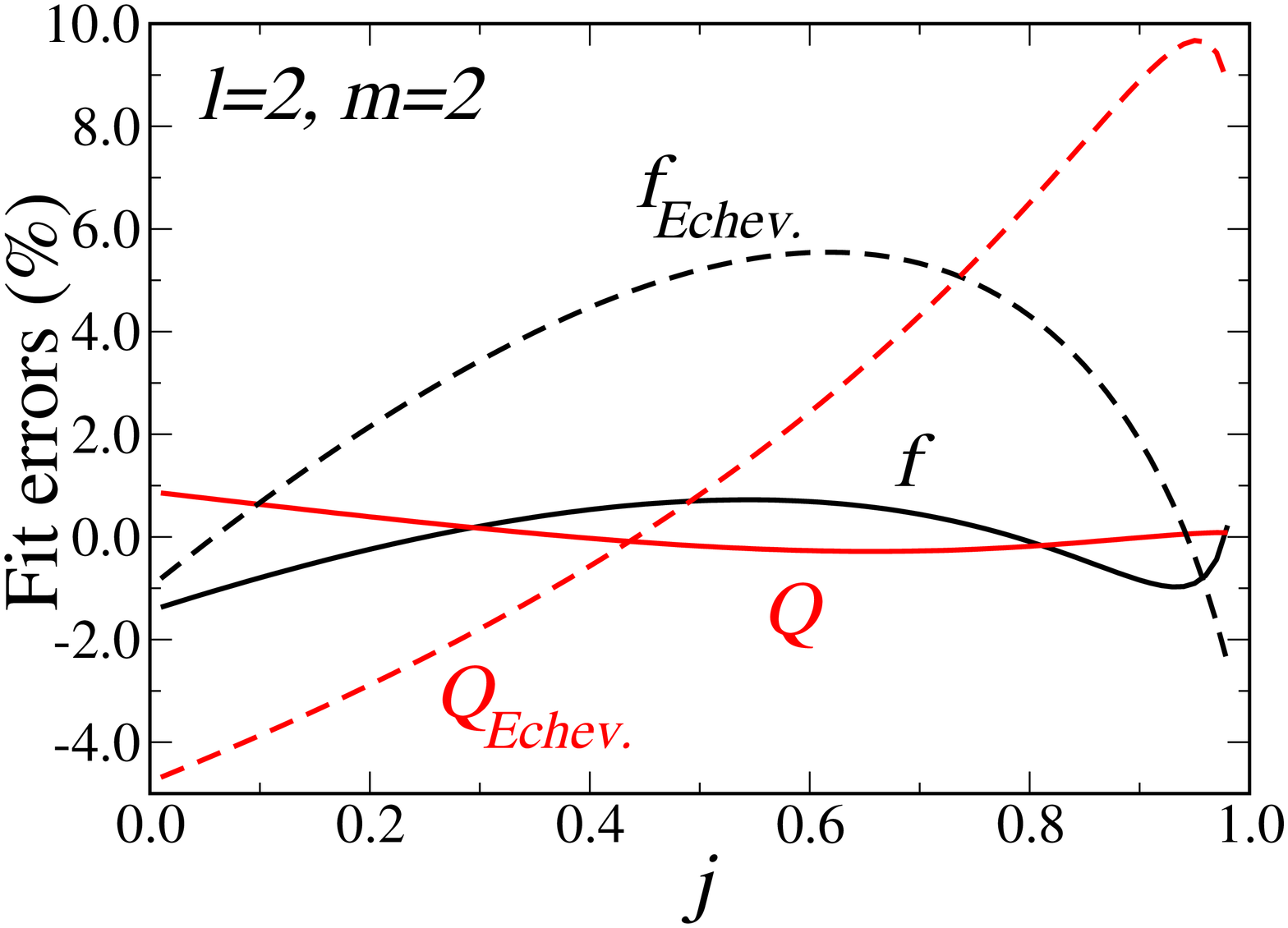,width=6cm,angle=270} &
\epsfig{file=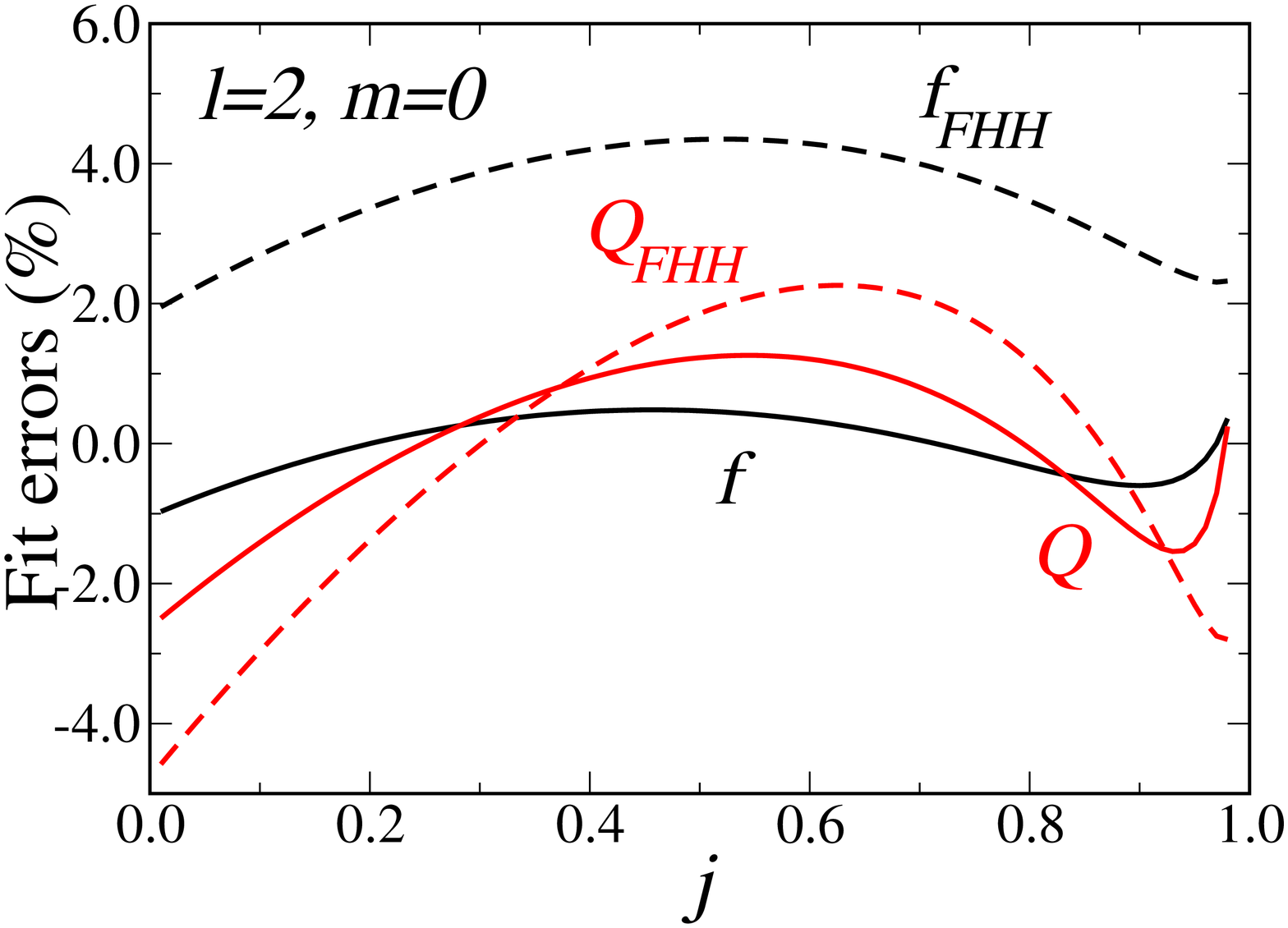,width=6cm,angle=270} \\
\end{tabular}
\caption{Accuracy of the fits of frequency (black) and quality factor
(red) for the fundamental mode with $l=m=2$ (left) and $l=2$, $m=0$
(right). Solid lines refer to our fits, dashed lines to the fits
proposed in Refs.~\cite{echeverria} and \cite{fhh}.
\label{fitaccuracy}}
\end{center}
\end{figure*}

The first few quasi-normal frequencies for extremal black holes
asymptote to $m/2$ for $m>0$, but imposing this behavior (for instance
by assuming fitting functions of the form ${\cal F} \sim
m/2+f_2(1-j)^{f_3}$) usually decreases the accuracy of the fits.
Indeed our fit for $l=2\,,\,m=2$ performs better than Echeverria's,
and the one for $l=2\,,\,m=0$ performs better than the fit provided in
\cite{fhh}.  This is shown in Fig.~\ref{fitaccuracy}, where we plot
the percentage error for each of the fits, in the range $0<j<0.99$.

For the $l=|m|$ modes we can also make contact with an approach
initiated by Press \cite{press} and Goebel \cite{goebel}, and
investigated in depth by Mashhoon \cite{mashhoon}, in which QNMs are
regarded as waves orbiting around the unstable photon orbit, and
slowly leaking out.  Mashhoon's analytical formulas were recently
compared to numerical results \cite{bkkn} and found to be in good
agreement for $l=|m|$.  This approach basically predicts that
${\cal F}_{l\pm l0} =|m|M\omega_0$, where $\omega_0$ is the
co-rotating (counter-rotating) orbital frequency at the photon orbit
for positive (negative) $m$ (see \cite{bkkn} for further details). The
counter-rotating frequency is well approximated by $M\omega _o \sim
-1/7+5(j-1)/147$. This implies that the fundamental $l=2,m=-2$ QNM
frequency should be ${\cal F}_{2-20}=2/7+10(1-j)/147\,\simeq
0.286+0.068(1-j)$. To compare with this formula, we can do a slightly
different fit to the data for $m=-2$. If we fit the data to ${\cal
F}_{2-20}=b+c(1-j)$ we get $b=0.287\,,\, c=0.0805$ with a maximum
error of $1.6\%$. This is in reasonable agreement with Mashhoon's
prediction.  For $l=3\,,\,m=-3$ the same fit yields ${\cal
F}_{3-30}=0.456+0.134(1-j)$ with $1.6\%$ maximum error, whereas
Mashhoon's prediction is ${\cal F}_{3-30}=0.428+0.102(1-j)$.  Thus for
negative $m\,,\,l=|m|$ we can approximate the frequencies by ${\cal
F}_{l-l0}\sim |m|\left [1/7+5(1-j)/147\right ]$.  For positive $m$ we
don't get such good agreement with a linear fit, because $M \omega$ is
a rapidly varying function of $j$. Nevertheless, for moderate to large
$j$ and $l=m$ we get, using Mashhoon's formula, ${\cal F}_{ll0} \sim
|l|\left (1/2 -\sqrt{3(1-j)/8}\right )$. This is in very good
agreement with the numerical data.


\clearpage
\begin{thebibliography}{99}

\bibitem{BBW} E. Berti, A. Buonanno and C. M. Will,
Phys. Rev. D {\bf 71}, 084025 (2005).

\bibitem{BC} L. Barack and C. Cutler, Phys. Rev. D {\bf 69}, 082005
(2004). See also L. Barack and C. Cutler, gr-qc/0310125 v3, where an
erroneous factor of 3/4 in the instrumental noise is corrected.

\bibitem{dreyer} O. Dreyer, B. Kelly, B. Krishnan, L. S. Finn,
D. Garrison and R. Lopez-Aleman,
Class. Quantum Grav. {\bf 21}, 787 (2004).

\bibitem{fh} \'E. \'E. Flanagan and S. A. Hughes,
Phys. Rev. D {\bf 57}, 4535 (1998).

\bibitem{echeverria} F. Echeverria,
Phys. Rev. D {\bf 40}, 3194 (1989).

\bibitem{finn} L. S. Finn,
Phys. Rev. D {\bf 46}, 5236 (1992).

\bibitem{kaa} K. D. Kokkotas, T. A. Apostolatos and N. Andersson,
Mon. Not. R. Astron. Soc. {\bf 320}, 307 (2001).

\bibitem{creighton}
J. D. E. Creighton, 
Phys. Rev. D {\bf 60}, 022001 (1999).

\bibitem{nakano}
H. Nakano, H. Takahashi, H. Tagoshi and M. Sasaki,
Phys. Rev. D {\bf 68}, 102003 (2003);
H. Nakano, H. Takahashi, H. Tagoshi and M. Sasaki,
Prog. Theor. Phys. {\bf 111}, 781 (2004).

\bibitem{tsunesada}
Y. Tsunesada, N. Kanda, H. Nakano, D. Tatsumi, M. Ando, M. Sasaki,
H. Tagoshi and H. Takahashi,
Phys. Rev. D {\bf 71}, 103005 (2005).

\bibitem{Beyer}
H. R. Beyer,
{\it Commun. Math. Phys.} {\bf 204}, 397 (1999).

\bibitem{Beyer2}
H. R. Beyer,
{\it Commun. Math. Phys.} {\bf 221}, 659 (2001).

\bibitem{NollertPrice}
H.-P. Nollert, R. H. Price,
J. Math. Phys. {\bf 40}, 980 (1999).

\bibitem{Nollert} 
H.-P. Nollert, 
{\it Characteristic Oscillations of Black Holes and Neutron Stars:
From Mathematical Background to Astrophysical Applications},
unpublished Habilitationsschrift (2000).

\bibitem{Szpak}
N. Szpak, unpublished (gr-qc/0411050).

\bibitem{leaver2}
E. W. Leaver,
Phys. Rev. D {\bf 34}, 384 (1986); Erratum, Phys. Rev. D {\bf 38}, 725
(1988).

\bibitem{A94}
N. Andersson,
Phys. Rev. D {\bf 51}, 353 (1995).

\bibitem{Nils97}
N. Andersson,
Phys. Rev. D {\bf 55}, 468 (1997).

\bibitem{GA}
K. Glampedakis, N. Andersson,
Phys. Rev. D {\bf 64}, 104021 (2001).

\bibitem{kokkotas} K. D. Kokkotas and B. G. Schmidt,
Living Rev. Rel. {\bf2}, 2 (1999); H.-P. Nollert, Class. Quantum
Grav. {\bf 16}, R159 (1999).

\bibitem{teukolsky}
S. A. Teukolsky,
Astrophys. J. {\bf 185}, 635 (1973).

\bibitem{berticardoso}
E. Berti, V. Cardoso and M. Casals,
Phys. Rev. D {\bf 73}, 024013 (2006).

\bibitem{NRezzolla} In the Schwarzschild limit, analogous expansions
of the gravitational wave amplitude in terms of the Zerilli and
Regge-Wheeler functions can be found in: A. Nagar and L. Rezzolla,
Class. Quant. Grav. {\bf 22}, R167 (2005).

\bibitem{krivan}
W. Krivan, P. Laguna, P. Papadopoulos and N. Andersson,
Phys. Rev. D {\bf 56}, 3395 (1997).

\bibitem{dorband}
N. Dorband, E. Berti, P. Diener, E. Schnetter and M. Tiglio
in preparation.

\bibitem{nakamura}
K. Oohara and T. Nakamura, 
Prog. Theor. Phys. {\bf 70,} 757--771 (1983); 
M. Sasaki and T. Nakamura, 
Prog. Theor. Phys. {\bf 67,} 1788--1809 (1982); 
Y. Kojima and T. Nakamura, 
Prog. Theor. Phys. {\bf 71,} 79--90 (1984).

\bibitem{lazarus}
J. Baker, M. Campanelli, C. O. Lousto and R. Takahashi,
Phys. Rev. D {\bf 65}, 124012 (2002).

\bibitem{AG} 
R. A. Araya-G\'ochez,
Mon. Not. R. Astron. Soc. {\bf 355}, 336 (2004).

\bibitem{markovic} 
D. Markovic, 
Phys. Rev. D {\bf 48}, 4738 (1993).

\bibitem{cosmology}
D. N. Spergel, et al.,
Astrophys. J. Suppl. {\bf 148}, 175 (2003).

\bibitem{gammie}
C. F. Gammie, S. L. Shapiro and J. C. McKinney,
Astrophys. J. {\bf 602}, 312 (2004).

\bibitem{sperhake}
U. Sperhake, B. Kelly, P. Laguna, K. L. Smith, E. Schnetter,
Phys. Rev. D {\bf 71}, 124042 (2005).

\bibitem{narayan} R. Narayan,
New J. Phys. {\bf 7}, 199 (2005).

\bibitem{rhook} K. J. Rhook and J. S. B. Wyithe,
Mon. Not. Roy. Astron. Soc. {\bf 361}, 1145 (2005).

\bibitem{BCW2}
E. Berti, V. Cardoso and C. M. Will,
gr-qc/0601077.

\bibitem{BCD} A. Buonanno, Y. Chen and T. Damour,
gr-qc/0508067.

\bibitem{fhh} C. L. Fryer, D. E. Holz and S. A. Hughes,
Astrophys. J. {\bf 565}, 430 (2002).

\bibitem{merritt}
D. Merritt and M. Milosavljevi\'c,
Living Rev. Relativ. {\bf 8}, 8 (2005).

\bibitem{bernstein2}
P. Anninos, D. Bernstein, S. R. Brandt, D. Hobill, E. Seidel and L. Smarr,
Phys. Rev. D {\bf 50}, 3801 (1994).

\bibitem{brandt} P. Anninos and S. Brandt,
Phys. Rev. Lett. {\bf 81}, 508 (1998).

\bibitem{allen}
G. Allen, K. Camarda and E. Seidel,
gr-qc/9806036.

\bibitem{anninosl4}
P. Anninos, R. H. Price, J. Pullin, E. Seidel and W. M. Suen,
Phys. Rev. D {\bf 52}, 4462 (1995).

\bibitem{boosted}
A. M. Abrahams and G. B. Cook,
Phys. Rev. D {\bf 50}, R2364 (1994).

\bibitem{shibata}
M. Saijo, T. W. Baumgarte, S. L. Shapiro and M. Shibata,
Astrophys. J. {\bf 569}, 349 (2002);
M. Shibata and S. L. Shapiro,
Astrophys. J. {\bf 572}, L39 (2002);
S. L. Shapiro and M. Shibata,
Astrophys. J. {\bf 577}, 904 (2002).

\bibitem{cpm1}
C. T. Cunningham, R. Price and V. Moncrief,
Astrophys. J. {\bf 224}, 643 (1978).

\bibitem{baiotti}
L. Baiotti, I. Hawke, L. Rezzolla and E. Schnetter,
Phys. Rev. Lett. {\bf 94}, 131101 (2005).

\bibitem{rayleigh} 
L. L. Scharf and P. H. Moose, 
IEEE Trans. Inf. Theory {\bf 22}, No. 1, 11 (1976).

\bibitem{smilanfar} 
M. Shahram and P. Milanfar,
IEEE Transactions on Signal Processing {\bf 53}, 2579 (2005).

\bibitem{milanfars} 
P. Milanfar and A. Shakouri, Proceedings of the International
Conference on Image Processing, 864 (2002).

\bibitem{FT}
S. Finn and K. S. Thorne, 
Phys. Rev. D {\bf 62}, 124021 (2000).

\bibitem{LHH} 
S. L. Larson, W. A. Hiscock and R. W. Hellings,
Phys. Rev. D {\bf 62}, 062001 (2000).

\bibitem{SCG} The Sensitivity Curve Generator was originally written
by Shane Larson and may be found online at

\noindent
\url{http://www.srl.caltech.edu/~shane/sensitivity/MakeCurve.html}

\bibitem{conf1} G. Nelemans, L. R. Yungelson and S. F. Portegies Zwart,
Astron. and Astrophys. {\bf 375}, 890 (2001).

\bibitem{conf2} 
A. J. Farmer and E. S. Phinney,
Mon. Not. R. Astron. Soc. {\bf 346}, 1197 (2003).

\bibitem{notemilanfar} Beside a factor 2 difference, the authors in
\cite{smilanfar} use a slightly different notation for the
signal-to-noise: What they call SNR is really $\rho^2$ and is defined
as ${h^2/\sigma^2}$.  They assume white noise, which is
certainly a good approximation for their sine signal, and 
is also a good approximation for our purposes. 
The noise probability distribution
function is a Gaussian, and therefore the power spectral density
(which
we call $S$) is equal to $\sigma^2$, for a sampled process.  Thus
their definition is the same, apart from these nuances.

\bibitem{noterighttail} The right-tail probability $T(X)$ (which is
usually denoted by $Q$, but we don't want the reader to confuse this
with the $Q$ factor) is defined as the probability of a random
variable $X$ exceeding a given value $x$,
%
\be
T(x)=P(X>x)=\int_x^{\infty}p(x)dx\,,
\ee
%
where $p(x)$ is the distribution function.
For a Gaussian with mean $\mu$ and variance $\sigma$ we have
%
\be
T(x)=\int_x^{\infty}\frac{1}{\sqrt{2\pi\sigma^2}}e^{-\frac{(t-\mu)^2}{2\sigma^2}}dt
\sim
\frac{\sigma}{(x-\mu)\sqrt{2\pi}}e^{-\frac{(x-\mu)^2}{2\sigma^2}}\,.
\ee

\bibitem{leaver}
E. W. Leaver,
Proc. R. Soc. London {\bf A402}, 285 (1985).

\bibitem{sai}
H. E. Seidel and S. Iyer,
Phys. Rev. D {\bf 41}, 374 (1990).

\bibitem{onozawa}
H. Onozawa, 
Phys. Rev. D {\bf 55}, 3593 (1997).

\bibitem{cardosoberti} 
E. Berti and K. D. Kokkotas, 
Phys. Rev. D {\bf 68}, 044027 (2003);
E. Berti, V. Cardoso, K. D. Kokkotas and H. Onozawa, 
Phys. Rev. D {\bf 68}, 124018 (2003); 
E. Berti, V. Cardoso and S. Yoshida,
Phys. Rev. D {\bf 69}, 124018 (2004). For a review see 
E. Berti,
gr-qc/0411025.

\bibitem{press}
W. Press, 
Astrophys. J. {\bf 170}, L105 (1971).

\bibitem{goebel}
C. G. Goebel, 
Astrophys. J. Lett. {\bf 172}, L95 (1972).

\bibitem{mashhoon} B. Mashhoon,
Phys. Rev. D {\bf 31}, 290 (1985).

\bibitem{bkkn}
E. Berti and K. D. Kokkotas,
Phys. Rev. D {\bf 71}, 124008 (2005).

\end{thebibliography}
\end{document}